%% file: kitaev_main_v2.tex
\documentclass[aps,prx,reprint,superscriptaddress,longbibliography,footinbib]{revtex4-2}

\usepackage{graphicx}
\usepackage{dcolumn}
\usepackage{bm}

\usepackage[figurename=Figure]{caption}
\usepackage{ragged2e}
\DeclareCaptionJustification{plain}{\justifying}
\usepackage[percent]{overpic}
\usepackage{float} 
\usepackage{verbatim} 
\usepackage{amsmath} 
\usepackage{bbold}
\usepackage{upgreek} 
\usepackage{booktabs}
\usepackage{amssymb} 
\usepackage{svg}
\usepackage{enumitem} 
\usepackage{latexsym,epsfig}
\usepackage{subcaption}
\usepackage{stackrel}
\usepackage{mathtools}
\usepackage[normalem]{ulem}
\usepackage{multirow}
\usepackage{makecell}

\usepackage{placeins} 

\usepackage[colorlinks=true,breaklinks=true,allcolors=blue]{hyperref}
\usepackage[capitalize]{cleveref}
\usepackage{lipsum}
\usepackage{dsfont}
\usepackage{coffeestains}

\usepackage[scr=boondox]{mathalfa}

\renewcommand{\emph}[1]{\textbf{\textit{#1}}}

\newcommand{\be}{\begin{equation}}
	\newcommand{\ee}{\end{equation}}
\newcommand{\bea}{\begin{equation}\begin{aligned}}
		\newcommand{\eea}{\end{aligned}\end{equation}}
\newcommand{\ben}{\begin{enumerate}}
	\newcommand{\een}{\end{enumerate}}

\DeclareDocumentCommand{\nint}{ O{} O{} m }{\ensuremath{ \int_{\mbox{\scriptsize $#1$}}^{\mbox{\scriptsize$#2$}}\!\!\! \mbox{\small $\,\mathrm{d}#3$\! }}}

\usepackage{tikz,bm} 
\usepackage{circuitikz}
\usetikzlibrary{decorations.markings}
\usetikzlibrary{decorations.pathreplacing,calligraphy}

\definecolor{mycolor}{rgb}{1,0.2,0.3}
\definecolor{brightgreen}{rgb}{0.4, 1.0, 0.0}
\definecolor{britishracinggreen}{rgb}{0.0, 0.26, 0.15}
\definecolor{cadmiumgreen}{rgb}{0.0, 0.42, 0.24}
\definecolor{ceruleanblue}{rgb}{0.16, 0.32, 0.75}
\definecolor{darkelectricblue}{rgb}{0.33, 0.41, 0.47}
\definecolor{darkpowderblue}{rgb}{0.0, 0.2, 0.6}
\definecolor{darktangerine}{rgb}{1.0, 0.66, 0.07}
\definecolor{emerald}{rgb}{0.31, 0.78, 0.47}
\definecolor{palatinatepurple}{rgb}{0.41, 0.16, 0.38}
\definecolor{pastelviolet}{rgb}{0.8, 0.6, 0.79}

\begin{document}

\title{Exact Temporal Criticalities of Entanglement and Negativity in Gaussian and Non-Gaussian Dissipative Kitaev Chains}
	
	\author{Shrey Chowdhary}
	\email{shreycho@usc.edu}
	\affiliation{Department of Physics and Astronomy, University of Southern California, Los Angeles, California 90089-0484, USA}
	
	\author{Sanket Chirame}
	\email{schirame@bard.edu}
	\affiliation{Physics Program, Bard College, 30 Campus Road, Annandale-on-Hudson, New York 12504, USA}
	
	\author{Stephan Haas}
	\email{shaas@usc.edu}
	\affiliation{Department of Physics and Astronomy, University of Southern California, Los Angeles, California 90089-0484, USA}
	
	\author{Rishabh Jha}
	\email{rishabh.jha@usc.edu}
	\affiliation{Department of Physics and Astronomy, University of Southern California, Los Angeles, California 90089-0484, USA}

\begin{abstract}
	When the two largest eigenvalues of an entanglement or negativity spectrum meet, the gap of the corresponding entanglement or negativity Hamiltonian closes while entropies stay smooth, an event we call a temporal criticality. We ask when such criticalities occur in open fermionic systems, in which order, and whether they precede the geometric decoherence time, at which the negativity starts to fall. We compare loss and gain, which respect fermion parity only on average (weak symmetry) and keep the state Gaussian, with density dephasing, which respects it in every jump (strong symmetry) and makes the state non-Gaussian. We find that Gaussianity, rather than the symmetry, fixes the criterion for a criticality. In a Gaussian state the entanglement gap of a subsystem closes exactly when its two parity sectors carry equal weight, and the negativity gap closes exactly when this happens for the subsystem or its complement, so the negativity criticality never comes after the entanglement one. Uniform balanced loss and gain only shrink the two parities and leave both criticalities in place at every rate, for any quadratic Hamiltonian. In an open Kitaev chain, weak imbalanced loss and gain shift the criticalities by amounts set by the occupations at the cut, and at strong rates they allow a closing near the relaxation time only when an odd number of neighboring sites pass half filling. For dephasing we solve the non-Gaussian dynamics exactly at the point where the bond terms of the chain commute. Weak dephasing delays both criticalities and, away from the chain ends, moves them ahead of the parity balance, where the parity rules of Gaussian states would place them. As the dephasing rate approaches critical damping, both criticalities recede to infinite time, and at strong dephasing the negativity starts to fall long before either gap begins to shrink.
\end{abstract}

\maketitle

	
\tableofcontents
	
\section{Introduction}
\label{sec:intro}

Entanglement tells us how quantum information is shared between the parts of a many-body system, and it often reorganizes while local observables change smoothly. After a quench in an isolated one-dimensional system, pairs of quasiparticles carry correlations across the system and entanglement spreads ballistically \cite{Calabrese2005Apr,Calabrese2009Dec}. In an open system, part of this picture survives, but the environment adds incoherent correlations and classical mixing \cite{Alba2021Jan,Sieberer2025Jun}. An entropy measures how much correlation is present but not how the correlations are organized.

The entanglement spectrum keeps this information. It is the set of eigenvalues $\lambda_k$ of the reduced density matrix $\rho_A$ of a subsystem $A$, written as Boltzmann weights $\lambda_k=e^{-\epsilon_k}$ of the entanglement Hamiltonian $K_E=-\ln\rho_A$ \cite{Li2008Jul,Peschel2009Dec,Dalmonte2022Nov}. In equilibrium its degeneracies reveal edge states and topological order \cite{Li2008Jul,Fidkowski2010Apr,Pollmann2010Feb}, and boundary dynamics carries related signatures of topology \cite{Jha2026Jun,menzler2026krylovedgespectroscopysymmetryprotected}. We study a dynamical event instead. At some instant the two largest eigenvalues of $\rho_A$ become equal, the lowest gap of $K_E$ closes, and the most probable state of $A$ changes. The lowest level of $K_E$ then has a kink and perturbation theory for the dominant eigenvector fails \cite{Kato1995,StewartSun1990}, while entropies, which are sums over the whole spectrum, stay smooth~\footnote{An entropy has the form $\sum_kf(\lambda_k)=\operatorname{Tr}f(\rho_A)$, which does not change when the levels are relabeled, and the trace norm behind the logarithmic negativity is the same kind of sum over singular values. Picking out the two largest levels requires sorting, and the sorting is what fails to be smooth when they meet. For a finite chain the levels continue as analytic functions of time through a simple crossing, so $\operatorname{Tr}f(\rho_A)$ stays analytic wherever $f$ is analytic at the levels involved, whereas the gap between the two leading levels has a kink.}. Following the temporal entanglement transitions found in quenched and driven systems \cite{Gong2018Dec,Poyhonen2021Nov,Sayyad2021Jul,Gadge2026Mar,wang2026criticaltouchingtemporalentanglement}, we call such an instant a temporal criticality and, more briefly, a closing. It is a sharp spectral event~\cite{Kehrein2024Jun,Jha2025Jun,Li2025Jul}, and complements dynamical quantum phase transitions, where the nonanalyticity appears in a return amplitude \cite{Heyl2013Mar,Heyl2018Apr}. In a dissipative Kitaev chain such closings can reveal Majorana physics that density relaxation hides \cite{Gandhi2026Aug}, but a closing is not topological by itself, and its interpretation depends on the dynamics that produce it.

In an open system $\rho_A$ mixes entanglement with classical noise from the bath, so a closing of its gap need not be a quantum event. Entanglement negativity discards the classical part, since it vanishes for every separable state \cite{Vidal2002Feb,Plenio2005Aug,Calabrese2013Feb}. For fermions it is built from the partial time reversal of the full density matrix \cite{Shapourian2017Apr,Shapourian2019Feb}, whose singular values form a negativity spectrum with its own Hermitian negativity Hamiltonian \cite{Murciano2022Apr,Rottoli2023Jan}. We follow three times, which we call clocks: the first closing of the entanglement gap at $t_E$, the first closing of the negativity gap at $t_N$, and the geometric decoherence time $\tau_d^g$, the first time at which the logarithmic negativity $E_N$ falls while the R\'enyi-$1/2$ entropy $S_{1/2}$ of $A$ still rises \cite{Jha2026Jul}. The first two mark a change of the dominant state, inside $\rho_A$ for $t_E$ and inside an operator that also keeps the complement $B$ for $t_N$. The third is a signature of decoherence, because a pure state has $E_N=S_{1/2}$ and cannot separate the two. A pure state also has a negativity spectrum built from pairs of entanglement eigenvalues, so $t_E=t_N$ exactly. Starting from a pure state, only a bath can separate the two clocks, and we ask how different baths do so.

Fermion parity is the natural symmetry for this question, because pairing conserves the particle number only modulo two. A Markovian bath, described by a Lindblad equation \cite{Gorini1976May,Lindblad1976Jun,Breuer2002,Daley2014Jul}, can respect parity in two ways \cite{Buca2012Jul,Albert2014Feb}. Under a weak symmetry only the averaged evolution respects it, and a single loss event removes one fermion and flips the parity of that quantum trajectory. Under a strong symmetry every jump respects it, as for density dephasing, which never changes an occupation. This distinction is central to current work on symmetry breaking, error correction, and topology in mixed states \cite{Lieu2020Dec,deGroot2022Nov,Lessa2025}, and dissipation can also prepare topological states \cite{Diehl2011Oct,Bardyn2013Aug}. The second property is Gaussianity. A fermionic state is Gaussian when all its correlations follow from two-point functions through Wick's theorem. Every Fock state is Gaussian. The equal mixture of $|10\rangle$ and $|01\rangle$, one fermion on either of two sites, is not. Its two-point functions coincide with those of the maximally mixed state, so for the site occupations $n_j=c_j^\dagger c_j$ Wick's theorem would predict $\langle n_1n_2\rangle=1/4$, whereas the state has $\langle n_1n_2\rangle=0$. Quadratic Hamiltonians with jumps linear in the fermions keep Gaussian states Gaussian, and such problems are solved exactly with covariance matrices \cite{Prosen2008Apr,Prosen2010Jul,Bravyi2012,Barthel2022Nov}. The two properties are linked in one direction only. A strong parity symmetry requires every jump to contain an even number of fermion operators, so every nontrivial jump is at least quadratic, and such jumps generally destroy Gaussianity. A weak symmetry allows linear jumps without requiring them, and the correlated loss $c_jn_{j+1}$ is weakly symmetric and non-Gaussian.

The two standard parity-respecting baths, particle loss and gain on one side and density dephasing on the other, therefore differ in symmetry and in Gaussianity at the same time. To determine which of the two properties shapes the clocks, we study an open Kitaev chain \cite{kitaev2001}, keeping the Hamiltonian fixed and changing only the bath. Neither label alone decides where a closing occurs. Adding a little loss to dephasing makes the symmetry weak, and adding a little dephasing to loss and gain makes the state non-Gaussian, yet in a finite chain either change moves a simple closing only slightly, because closings move continuously with the rates. Gaussianity, however, fixes the criterion for a closing. In a Gaussian state the entanglement gap closes exactly when the even and odd parity sectors of subsystem $A$ carry equal weight, a parity balance at which the expectation value $q_A$ of the fermion parity of $A$ vanishes. We show that the negativity gap obeys the same rule on both sides of the cut and closes exactly when $q_A$ or the parity $q_B$ of the complement vanishes. The negativity clock is therefore the earlier of the entanglement clocks of $A$ and $B$, and it never comes after $t_E$. Both rules hold for any Gaussian state, pure or mixed, and they fix both clocks under balanced loss and gain. For any quadratic Hamiltonian this bath only multiplies $q_A$ and $q_B$ by positive factors, so both clocks keep their closed-system values at every rate. The symmetry label enters the clocks only through the global parity. A strong symmetry conserves it, so the parities of subsystems $A$ and $B$ balance together, whereas loss and gain let them balance at different times. Beyond this, the symmetry decides where the evolution ends, because a strong symmetry keeps the initial parity forever and leaves classical correlations across the cut, which none of the three clocks detects.

We work mostly at the compact point of the Kitaev chain, where the chemical potential vanishes and the pairing equals the hopping, $\mu=0$ and $\Delta=J$. There every Majorana mode except the two at the chain ends pairs with one on the neighboring site, and the Hamiltonian becomes a sum of commuting terms, one per bond. When the rates are weak, imbalanced loss and gain shift the closings by amounts set by the initial occupations of the two sites at the cut. When the rates are strong, a gap closes near the relaxation time only as an odd block of neighboring sites passes half filling, so the negativity gap can close through $B$ while the entanglement gap of $A$ stays open. Dephasing makes the state non-Gaussian, and the parity rules no longer apply, but the commuting bonds make the dynamics exactly solvable. Each bond acts on its own, as a two-level system undergoing Rabi oscillations between two occupation patterns. Dephasing at rate $\gamma$ can be written as random kicks, each of which multiplies the state by the parity of one site and reverses the rotation of the bonds touching that site. Every history of kicks produces a pure Gaussian state, while the physical state, the average over all histories, is not Gaussian. From this picture we obtain the leading levels of both spectra to first order at weak dephasing and both complete spectra at leading order at strong dephasing. Weak dephasing delays both closings and, for a cut away from the chain ends, moves them ahead of the parity balance, where the parity rules would place them. Strong dephasing nearly freezes the occupations, lets each bond flip its two sites at random and independently of the others, and removes both closings at leading order in $J/\gamma$.

The same trajectory idea gives exact correlation functions of Ising models with spin flips \cite{FossFeig2013Apr}, and other exact results for dephased chains give Liouvillian spectra, steady states, or two-point functions \cite{Medvedyeva2016Sep,Shibata2019}. To our knowledge, no exact solution of non-Gaussian Lindblad dynamics in a fermionic chain with pairing has been used before to obtain its entanglement and fermionic negativity spectra. On the Gaussian side, loss and gain in Kitaev chains \cite{Starchl2022,Gandhi2026Aug}, the partial transpose of Gaussian states \cite{Eisler2015May,Shapourian2017Apr}, and the negativity of free fermions with gain and loss \cite{Alba2023} are well studied. Our contribution there is the parity rule for the negativity gap and, with it, exact criteria for when the two gaps close and in which order.

The entanglement clock can be measured with existing tools, since measurements of entanglement spectra have been proposed for cold atoms and performed on superconducting and trapped-ion platforms \cite{Pichler2016Nov,Choo2018Aug,Kokail2021Aug}. The negativity clock needs one further step. Randomized measurements give moments of the ordinary partial transpose of a mixed state \cite{Elben2020Nov}, and adapting them to fermionic partial time reversal would give access to the negativity spectrum. For Gaussian dynamics the parity rules lower the cost further, because measuring the parities of $A$ and $B$ locates both clocks without reconstructing either spectrum. \Cref{sec:summary} collects the results, \cref{sec:model} defines the model and the three clocks, \cref{sec:temporal_criticalities_weak,sec:temporal_criticalities_strong} derive the results for the two baths, and \cref{sec:conclusion} concludes.

\section{Summary of Results}
\label{sec:summary}

\begin{figure*}[htbp]
	\centering
	\begin{subfigure}{\textwidth}
		\includegraphics[width=\textwidth]{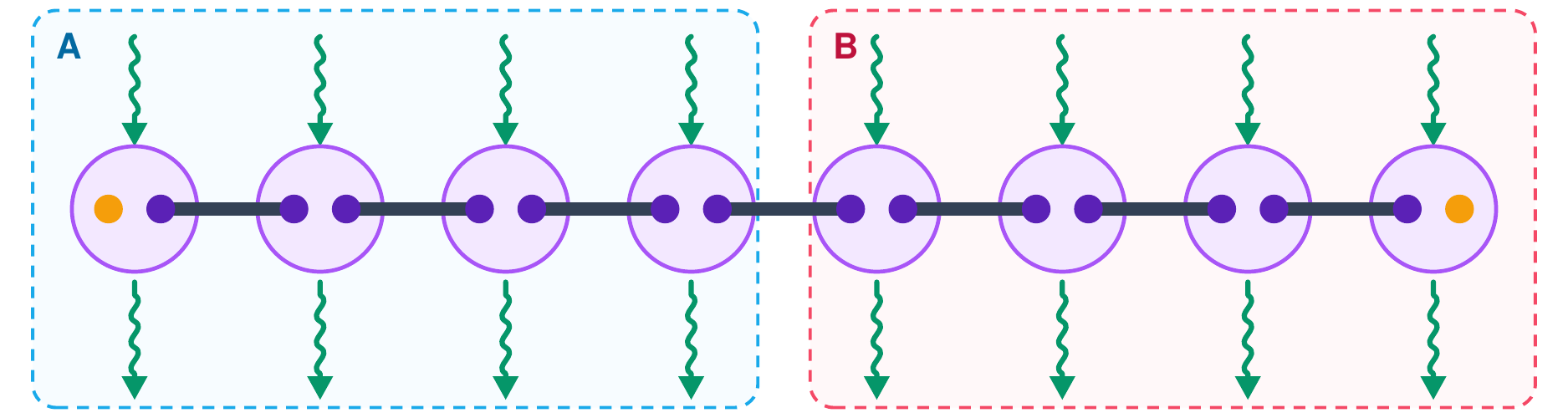}
	\end{subfigure}
	\begin{subfigure}{0.5\textwidth}
		\includegraphics[width=\textwidth]{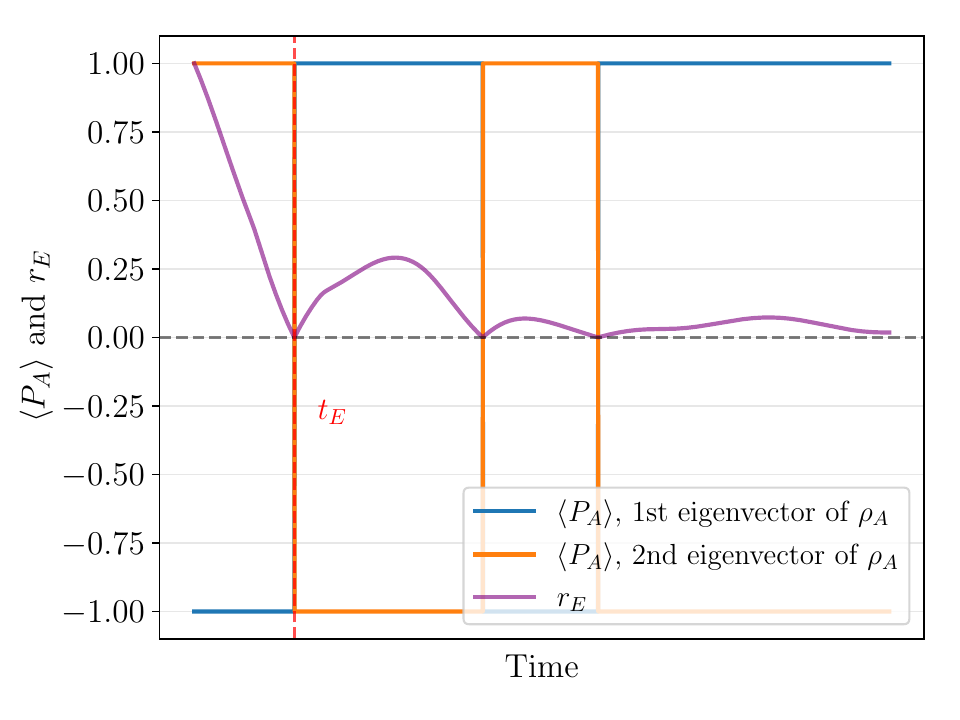}
	\end{subfigure}\hfill
	\begin{subfigure}{0.5\textwidth}
		\includegraphics[width=\textwidth]{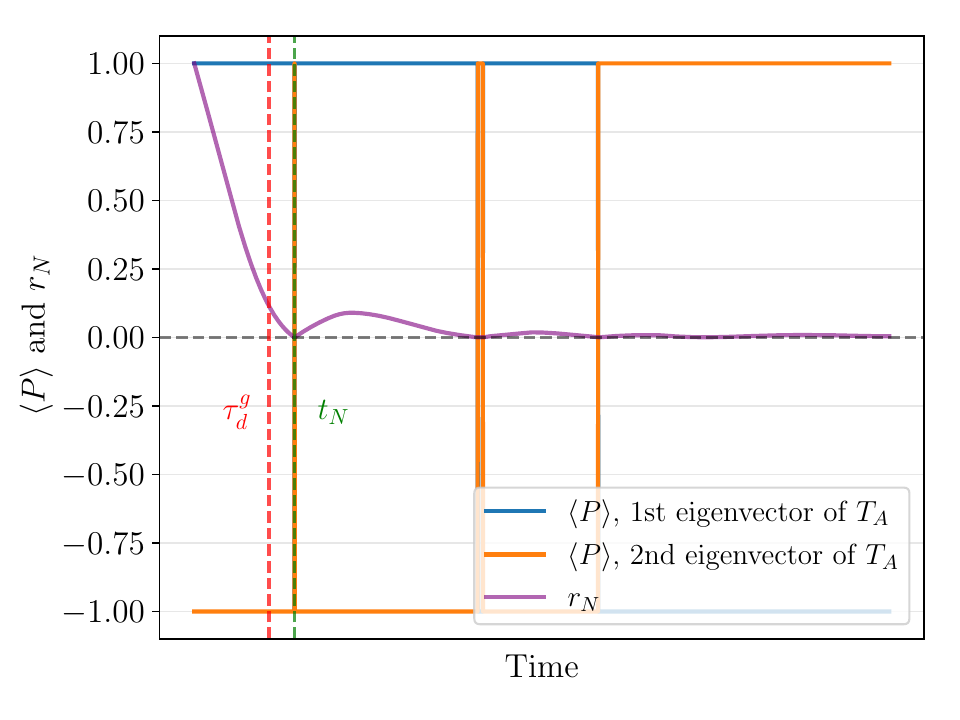}
	\end{subfigure}
	\caption{Schematic of the bipartition, the dissipative environment, and the temporal diagnostics studied in this work. \textbf{Top}: an open Kitaev chain is divided into nonempty regions $A$ and $B$. Each circle is a site $j$ and the two dots inside it are its Majorana operators $a_j$ and $b_j$, which carry a site index rather than a region label. At the compact Majorana point, the dark links pair $b_j$ with $a_{j+1}$ on neighboring sites, the orange markers denote the two unpaired physical end modes, and exactly one Majorana bond crosses the cut. The green arrows depict local particle loss and gain, which realize weak parity symmetry; the strong-symmetry density-dephasing bath acts on the same geometry and is not drawn. \textbf{Bottom}: schematic weak-symmetry traces of the normalized entanglement gap $r_E$ (\textbf{left}) and the normalized twisted-negativity gap $r_N$ (\textbf{right}) defined in \cref{eq:normalized_spectral_gaps}, shown together with the parities of the two leading eigenvectors, $P_A$ for $\rho_A$ and $P$ for $T_A$ as in \cref{eq:parity_labels}. The first zeros of the two gaps are temporal criticalities of the entanglement and twisted-negativity Hamiltonians, and they define the clocks $t_E$ and $t_N$; $\tau_d^g$ marks the geometric decoherence time~\cite{Jha2026Jul}. The exchange of the two leading parities at the closing in the left panel is the spectral signature of a simple crossing rather than a tangency. The time axis is not to scale. The position of $\tau_d^g$ relative to the two closings depends on the regime (\cref{tab:clock_summary}), whereas $t_N\leq t_E$ holds for every Gaussian evolution. The quantitative behavior is given in \cref{sec:temporal_criticalities_weak,sec:temporal_criticalities_strong} and in the sweeps of \cref{fig:weak_symmetry_clocks,fig:strong_symmetry_clocks}. Under strong symmetry, the same diagnostics remain well defined, but density dephasing changes their geometry dependence and, in the Zeno regime, separates coherence loss, geometric decoherence, and population relaxation on parametrically distinct scales, as derived in \cref{sec:temporal_criticalities_strong}.}
	\label{fig:model_setup}
\end{figure*}

We cut the chain of $L$ sites into $A=\{1,\ldots,L_A\}$ and $B=\{L_A+1,\ldots,L\}$, with $L_B=L-L_A$, and call site $L_A$ the cut site (\cref{fig:model_setup}). Both gaps are measured logarithmically, $\Delta_E=\ln(\lambda_0/\lambda_1)$ for the two largest entanglement eigenvalues and $\Delta_N$ likewise for the negativity spectrum (\cref{eq:spectral_gaps}). Unless stated otherwise, the chain starts in the Fock state with its first $N\leq L_A$ sites filled (\cref{eq:left_filled_initial_state}), and the Hamiltonian sits at the compact point, where without a bath both gaps close at $t_0=\pi/(4J)$. Loss and gain at rates $\gamma_\ell$ and $\gamma_g$ enter through their sum $\kappa=\gamma_\ell+\gamma_g$ and their imbalance $\xi=\gamma_\ell-\gamma_g$. \Cref{tab:clock_summary} lists, for a long chain cut far from its ends, whether each gap closes, in which order the closings come, and whether the negativity turns over first.

\begin{table*}[t]
	\caption{Existence and order of the two temporal criticalities for a long chain cut far from its ends, with the first $N$ sites initially filled. Here $t_0=\pi/(4J)$ is the closing time without a bath, $n_c$ is the initial occupation of the cut site, active sites are those that the bath drives through half filling at the scaled time $s_\star$ of \cref{eq:weak_local_half_occupation} (filled ones for dominant loss, empty ones for dominant gain), $t_q$ is the first zero of the parity of $A$, and ``before both'' means $\tau_d^g<\min(t_E,t_N)$. For weak dephasing this last entry is established for $L\leq10$ (\cref{tab:app_strong_geo_coefficients}).}
	\label{tab:clock_summary}
	\centering
	\small
	\def\tabcell#1#2{\parbox[c]{#1}{\raggedright\strut#2\strut}}
	\renewcommand{\arraystretch}{1.4}
	\setlength{\tabcolsep}{3pt}
	\begin{tabular}{|l!{\vrule width 1.4pt}l|l|l|l|}
		\hline
		Bath&Regime&Do the gaps close?&Order of $t_0$, $t_E$, $t_N$&$\tau_d^g$\\
		\Xhline{1.4pt}
		\tabcell{2.9cm}{None (closed system)}&Unitary evolution&Both&$t_E=t_N=t_0$&Undefined\\
		\hline
		\multirow{3}{*}{\tabcell{2.9cm}{Loss and gain: weak symmetry, Gaussian state}}&\tabcell{3.0cm}{Balanced rates}&Both, at every rate&\tabcell{4.9cm}{$t_E=t_N=t_0$; for any quadratic $H$ both keep their closed-system values}&Before both\\
		\cline{2-5}
		&\tabcell{3.0cm}{Imbalanced, weak rate}&Both&\tabcell{4.9cm}{$t_E=t_0+\xi(1-2n_c)/(8J^2)$, $t_N=\min[t_E,t_0+\xi/(8J^2)]$}&Before both\\
		\cline{2-5}
		&\tabcell{3.0cm}{Imbalanced, strong rate}&\tabcell{3.7cm}{Near $s_\star/\kappa$: $t_E$ if $A$ has an odd active block, $t_N$ if $A$ or $B$ has one}&\tabcell{4.9cm}{$s_\star/\kappa\ll t_0$; $t_N\leq t_E$, with equality under loss on this scale}&No fixed order\\
		\hline
		\multirow{2}{*}{\tabcell{2.9cm}{Dephasing: strong symmetry, non-Gaussian state}}&\tabcell{3.0cm}{Weak rate}&Both&$t_0<t_N<t_E<t_q$&Before both\\
		\cline{2-5}
		&\tabcell{3.0cm}{Strong rate, $\gamma\gg J$}&Neither, at leading order&\tabcell{4.9cm}{Gaps open at every finite scaled time, closing only asymptotically}&Before both\\
		\hline
	\end{tabular}
\end{table*}

Five results hold beyond the Kitaev chain.
\begin{enumerate}
	\item \emph{Pure states.} The negativity spectrum consists of the products $\sqrt{\lambda_i\lambda_j}$ of entanglement eigenvalues \cite{Shapourian2019Feb}, so $\Delta_N=\Delta_E/2$ and $t_N=t_E$ (\cref{eq:pure_state_clock_relation}).
	\item \emph{Entanglement clock of a Gaussian state.} The entanglement gap closes exactly when the subsystem parity $q_A=\operatorname{Tr}(\rho_AP_A)$ vanishes, with $P_A$ the fermion parity of $A$ (\cref{eq:weak_gaussian_equivalences}). The even and odd sectors of $A$ then carry equal weight, and at a simple zero the dominant state of $A$ changes parity. Measuring the single observable $P_A$ therefore locates $t_E$.
	\item \emph{Negativity clock of a Gaussian state.} The negativity spectrum consists of the moduli $|\eta_k|$ of the eigenvalues of the Hermitian operator $T_A=\rho^{R_A}P_A$, where the partial time reversal $\rho^{R_A}$ multiplies every product $\mathcal M$ of Majorana operators in $\rho$ by $i^{|\mathcal M|_A}$, with $|\mathcal M|_A$ the number of its factors in $A$ (\cref{eq:partial_time_reversal,eq:twisted_operator}). Let $p_k=\pm1$ be the global fermion parity of the eigenvector with eigenvalue $\eta_k$. For every physical state, which commutes with the global parity,
	\begin{equation}
		\begin{aligned}
			\sum_k\eta_k&=(-1)^{L_A}q_A,
			&\sum_kp_k\eta_k&=q_B,
		\end{aligned}
		\label{eq:summary_parity_sums}
	\end{equation}
	where $q_B=\operatorname{Tr}(\rho P_B)$ is the parity of $B$. Both identities hold because each trace picks out a single product of Majorana operators, $P_A$ or $P_B$, which partial time reversal multiplies by $i^{2L_A}$ or leaves unchanged. For a Gaussian state $T_A$ factorizes over fermionic modes $m$, with eigenvalues $\eta_m^\pm$ for the two occupations of each mode. Then $\sum_k\eta_k=\prod_m(\eta_m^++\eta_m^-)$ and $\sum_kp_k\eta_k=\pm\prod_m(\eta_m^+-\eta_m^-)$, and the two largest $|\eta_k|$ meet exactly when $|\eta_m^+|=|\eta_m^-|$ for some mode, that is, when one factor of one product vanishes. The negativity gap therefore closes exactly when $q_Aq_B=0$. The negativity clock is the earlier of the entanglement clocks of $A$ and of $B$, $t_N=\min(t_E^A,t_E^B)$ with $t_E^A\equiv t_E$, so $t_N\leq t_E$ for every Gaussian evolution (\cref{subsec:tnh_weak}).
	\item \emph{Balanced loss and gain.} When every site has loss and gain at the same rate $\kappa/2$, the bath damps every product of an even number $k$ of distinct Majorana operators at the rate $k\kappa/2$, and a quadratic Hamiltonian never changes $k$. The parities therefore only acquire positive factors, $q_A=e^{-L_A\kappa t}q_A^{\mathrm{cl}}$ and $q_B=e^{-L_B\kappa t}q_B^{\mathrm{cl}}$, where cl denotes the closed evolution. For any quadratic Hamiltonian and any Gaussian initial state, pure or mixed, both clocks keep their closed-system values at every rate (\cref{eq:weak_balanced_clock_locking}).
	\item \emph{Strong symmetry.} If the Hamiltonian and every jump commute with the global parity $P=P_AP_B$, a state of definite parity $p$ keeps it, so $q_B=pq_A$ at all times (\cref{eq:subsystem_parities}). Since a Gaussian state of definite parity is pure, every mixed state reached in this way is non-Gaussian. The closing criterion $q_Aq_B=0$ of item 3 is then no longer guaranteed, and that of item 2 only while $\rho_A$ stays Gaussian. For a single-site $B$, however, $t_N=t_E$ holds exactly for every state of definite global parity (\cref{subsec:tnh_strong}).
\end{enumerate}
Everything that follows is specific to the Kitaev chain.

\emph{Loss and gain.} Balanced rates leave both closings at $t_0$, yet the negativity already turns over at $\tau_d^g=t_0-\kappa/(2J^2)$ to first order (\cref{eq:weak_balanced_geometric_clock}). An imbalance moves the closings, because the bath now pushes both parities toward those of the vacuum (loss) or of the filled chain (gain). At weak rates only the site on either side of the cut matters, since the bath acting elsewhere shifts the two competing levels equally. For $L_A\geq2$ and with $n_c$ the initial occupation of the cut site, the parity of $A$ balances at $t_E=t_0+\xi(1-2n_c)/(8J^2)$ to first order (\cref{eq:weak_rate_entanglement_clock}). Loss thus delays the closing when the cut site starts empty and advances it when the site starts filled, and gain does the opposite. The parity of $B$ obeys the same rule with its own first site, which starts empty, so $t_N=\min[t_E,t_0+\xi/(8J^2)]$ (\cref{eq:weak_bulk_negativity_clock}). The two clocks agree unless gain meets a filled cut site, in which case $B$ balances first and $t_N<t_0<t_E$. In all these cases the negativity turns over before either gap closes, at $\tau_d^g\leq t_N-\kappa/(4J^2)$ to first order (\cref{eq:app_weak_geo_ordering}).

When the bath dominates, $\kappa\gg J$, each site relaxes on its own. When loss dominates ($\xi>0$), the sites that start filled pass through half filling, and when gain dominates ($\xi<0$), those that start empty do. We call these sites active. All of them reach half filling together, at the scaled time $s_\star=\kappa t=\ln[(1+b)/b]$ with $b=|\xi|/\kappa$, which is $t=\ln2/\gamma_\ell$ for pure loss (\cref{eq:weak_local_half_occupation}). Whether a parity actually balances there is decided by pairing. On a block of $M$ neighboring active sites, pairing forms standing waves with wave numbers $m\pi/(M+1)$, and at $s_\star$ each wave holds the gap open by an amount proportional to $|\cos[m\pi/(M+1)]|$ (\cref{eq:weak_entanglement_active_modes}). Only an odd block has the wave with $m=(M+1)/2$, for which this amount vanishes. Near $s_\star/\kappa$ a gap therefore closes only through an odd block, while an even block leaves a minimum of order $J/\kappa$ (\cref{eq:weak_strong_rate_entanglement_clock}). The entanglement gap feels the active block in $A$, and the negativity gap, through the parity of $B$, feels the block in $B$ as well (\cref{eq:weak_strong_rate_negativity_clock}). For $L_A=4$, $L=7$, and $N=2$ under gain, for example, the active sites form a block of two in $A$ and a block of three in $B$, so the entanglement gap stays open while the negativity gap closes. Under loss, by contrast, $B$ has no active site and the two clocks agree, and for even $N$ neither gap closes near $s_\star/\kappa$. Away from the compact point the odd-even rule holds at strong rates for any chemical potential $\mu$, and the weak-rate shift becomes an exact linear-response coefficient that depends on more than the cut site (\cref{eq:weak_finite_mu_simple_root,eq:weak_small_mu_clock,eq:weak_finite_mu_even_mass}).

\emph{Dephasing.} Dephasing at rate $\gamma$ never changes an occupation and only randomizes the phases between occupation patterns. At the compact point every history of kicks is a product of bond rotations by random angles (\cref{eq:strong_random_unitary}). Tracing out $B$ leaves exactly two blocks of opposite parity, the $\cos^2\phi$- and $\sin^2\phi$-weighted averages of the same evolution inside $A$, with $\phi$ the random angle of the bond crossing the cut (\cref{eq:strong_parity_block_decomposition}). The closing continued from $t_0$ occurs when the largest eigenvalues of the two blocks are equal, and the parity balance when their total weights are equal. Both spectra are independent of the initial occupation pattern, and the entanglement spectrum is also independent of the size of $B$ (\cref{eq:strong_pattern_universality,eq:strong_bath_length_universality}). Weak dephasing slows the rotation of the cut bond. For a cut far from the chain ends it delays the two closings and the parity balance in the ratio $2:3:4$,
\begin{equation}
	\begin{aligned}
		t_N-t_0&=\frac{2\gamma}{32J^2},
		&t_E-t_0&=\frac{3\gamma}{32J^2},
		&t_q-t_0&=\frac{4\gamma}{32J^2},
	\end{aligned}
	\label{eq:summary_weak_dephasing}
\end{equation}
to first order in $\gamma$ (\cref{eq:strong_weak_negativity_clock,eq:strong_weak_entanglement_clock,eq:strong_weak_parity_clock}). Here $t_q$ is the first zero of $q_A$, known exactly because $q_A$ obeys a damped-oscillator equation (\cref{eq:strong_parity_oscillator}). The global parity $p$ is conserved, so $q_B=pq_A$, and the parity rules of Gaussian states would place both closings at $t_q$. Instead, $t_E$ comes $\gamma/(32J^2)$ early, which certifies that $\rho_A$ is not Gaussian, and $t_N$ comes $\gamma/(16J^2)$ early, so the negativity gap closes first. Exact numerics for short chains with at least two sites on each side show that stronger dephasing moves $t_q$ ahead of both closings and, in most geometries, makes the negativity gap close last (\cref{fig:strong_symmetry_clocks}(a)), an order that no Gaussian evolution allows. All three times diverge together as $\gamma\to4J^-$, where the cut bond becomes critically damped (\cref{subsec:tnh_strong,fig:strong_symmetry_clocks}).

Strong dephasing, $\gamma\gg J$, nearly freezes the occupations, as in the quantum Zeno effect, and separates three time scales,
\begin{equation}
	\frac{1}{\gamma}\ll\tau_d^g\simeq\frac{2}{\gamma}\ln\frac{\gamma}{2J}\ll\frac{\gamma}{2J^2}.
	\label{eq:summary_zeno_scales}
\end{equation}
Coherences between neighboring occupation patterns die on the first scale, the negativity peaks at $\tau_d^g$ with the small value $2J/\gamma$, and the occupations move on the last scale through pair flips that conserve parity (\cref{eq:strong_time_scale_hierarchy}). On that slow scale each bond flips its two sites independently of the other bonds, and at the scaled time $\vartheta=2J^2t/\gamma$ its net effect is a flip with probability $x=(1-e^{-2\vartheta})/2$ (\cref{eq:strong_pair_flip_kinetics,eq:strong_zeno_bond_flips}). At leading order in $J/\gamma$ the spectra follow at once. The reduced state $\rho_A$ has the spectrum of $L_A$ independent coins that each come up heads with probability $x$, and $|T_A|$ has that of $L-1$ such coins. Both gaps equal $\ln\coth\vartheta$, which is positive at every finite $\vartheta$ and vanishes only as $\vartheta\to\infty$ (\cref{eq:strong_zeno_entanglement_clock,eq:strong_zeno_negativity_clock}).

\emph{Short chains.} A chain end at the cut doubles the loss-and-gain shift on that side, and taking the earlier of the two parity balances reproduces the complete first-order classification (\cref{tab:app_weak_negativity_branches}). Under dephasing a single-site $B$ gives $t_E=t_N$ and a single-site $A$ gives $t_E=t_q$, both exactly and at every rate (\cref{subsec:tnh_strong,eq:strong_one_site_entanglement_clock}).

Gaussianity turns the clocks into parity balances, which gives the order $t_N\leq t_E$ and, with the uniform damping of balanced loss and gain, the fixed closing times at every rate. The action of the bath on correlations gives the rest. A push toward the vacuum or the filled chain shifts the closings by amounts set by the sites at the cut, damping of the phases between sites without damping of the occupations delays them, and freezing of the occupations removes them. The symmetry label ties the parities of $A$ and $B$ together and fixes the long-time state, where a strong symmetry keeps the initial parity and leaves classical correlations between $A$ and $B$ (\cref{eq:strong_stationary_state}). Three properties of the dephasing solution are not signatures of strong symmetry: independence of the initial pattern, independence of $\rho_A$ from the size of $B$, and the exact kick picture. All three hold for balanced loss and gain as well, because this bath also treats particles and holes alike, acts on single sites, and amounts to random kicks by single Majorana operators~\footnote{With the Majorana operators of \cref{eq:majorana_operators} and the dissipator of \cref{eq:lindblad_dissipator}, balanced loss and gain at rate $\gamma_{\mathrm w}$ obey $\gamma_{\mathrm w}(\mathcal D[c_j]+\mathcal D[c_j^\dagger])=(\gamma_{\mathrm w}/2)(\mathcal D[a_j]+\mathcal D[b_j])$, with $\mathcal D[a_j]\rho=a_j\rho a_j-\rho$ because $a_j^2=1$. Flipping the occupation of a site maps these kicks onto themselves and commutes with the compact-point Hamiltonian, and at the compact point the covariance restricted to $A$ does not depend on $L_B$.}.

\section{Model and Setup}
\label{sec:model}

\subsection{Kitaev chain and the two baths}

We compare two dissipative environments acting on the same coherent system, so that every difference between the resulting clocks comes from the bath and not from the Hamiltonian. The coherent system is a Kitaev chain of $L$ spinless-fermion sites \cite{kitaev2001} with open boundaries, so it has two physical ends, at site $1$ and at site $L$. Its Hamiltonian is
\begin{equation}
	\begin{aligned}
		H={}&{-\mu}\sum_{j=1}^{L}n_j
		-J\sum_{j=1}^{L-1}
		\left(c_j^\dagger c_{j+1}+c_{j+1}^\dagger c_j\right)
		\\
		&+\Delta\sum_{j=1}^{L-1}
		\left(c_jc_{j+1}+c_{j+1}^\dagger c_j^\dagger\right),
	\end{aligned}
	\label{eq:kitaev_hamiltonian}
\end{equation}
where $c_j$ annihilates a fermion on site $j$, $n_j=c_j^\dagger c_j$ is the local occupation, $J$ is the hopping amplitude, $\Delta$ is the pairing amplitude, and $\mu$ is the chemical potential. The operators obey $\{c_i,c_j^\dagger\}=\delta_{ij}$ and $\{c_i,c_j\}=0$. We take $J$, $\Delta$, and $\mu$ to be real and set $\hbar=1$. Pairing changes the particle number by two, so the total number $\widehat N=\sum_j n_j$ need not be conserved when $\Delta\neq0$. Its parity is conserved:
\begin{equation}
	\begin{aligned}
		P&=(-1)^{\widehat N},
		&P^2&=1,
		&[H,P]&=0.
	\end{aligned}
	\label{eq:global_parity}
\end{equation}
Because $P^2=1$, the identity and parity form the two-element group $\mathbb{Z}_2=\{1,P\}$, and the Hilbert space splits into an even and an odd sector, labeled by the eigenvalue $p=+1$ or $p=-1$ of $P$. Equation~\eqref{eq:global_parity} therefore states that the Kitaev Hamiltonian has a global fermion-parity $\mathbb{Z}_2$ symmetry: since the pairing term creates and annihilates fermions two at a time, $H$ never connects the two sectors.

The density operator $\rho(t)$ evolves according to a Markovian Lindblad equation \cite{Gorini1976May,Lindblad1976Jun,Breuer2002}. For an operator $X$, we use the dissipator
\begin{equation}
	\begin{aligned}
		\mathcal D[X]\rho
		={}&X\rho X^\dagger
		-\frac{1}{2}\left\{X^\dagger X,\rho\right\},
	\end{aligned}
	\label{eq:lindblad_dissipator}
\end{equation}
where $\{X,Y\}=XY+YX$ is the anticommutator. We study two choices of local jump operators.

\emph{Weak symmetry.} Uniform particle loss and gain give
\begin{equation}
	\begin{aligned}
		\dot\rho=\mathcal L_{\mathrm w}(\rho)
		={}&{-i}[H,\rho]
		\\
		&+\gamma_\ell\sum_{j=1}^{L}\mathcal D[c_j]\rho+\gamma_g\sum_{j=1}^{L}\mathcal D[c_j^\dagger]\rho,
	\end{aligned}
	\label{eq:weak_lindblad}
\end{equation}
where the local Lindblad jump operators are $\sqrt{\gamma_\ell}\,c_j$ for loss and $\sqrt{\gamma_g}\,c_j^\dagger$ for gain, and $\gamma_\ell\geq0$ and $\gamma_g\geq0$ are the corresponding rates. It is useful to introduce the total rate $\kappa$ and the rate imbalance $\xi$,
\begin{equation}
	\begin{aligned}
		\kappa&\coloneq\gamma_\ell+\gamma_g,
		&\xi&\coloneq\gamma_\ell-\gamma_g.
	\end{aligned}
	\label{eq:loss_gain_parameters}
\end{equation}
Balanced loss and gain correspond to $\xi=0$. An individual loss or gain event changes the particle number by one and therefore flips parity. Explicitly,
\begin{equation}
	\begin{aligned}
		Pc_jP&=-c_j,
		&Pc_j^\dagger P&=-c_j^\dagger,
		\\
		\mathcal L_{\mathrm w}(P\rho P)
		&=P\mathcal L_{\mathrm w}(\rho)P.
	\end{aligned}
	\label{eq:weak_parity_covariance}
\end{equation}
Thus the complete evolution is unchanged under the $\mathbb{Z}_2$ action $\rho\mapsto P\rho P$, but the individual jump operators exchange the two parity sectors. This is a weak $\mathbb{Z}_2$ symmetry: the symmetry is respected by the full Lindblad evolution, but not by each jump separately \cite{Buca2012Jul,Albert2014Feb}. For example, one loss event maps an even state to an odd state, although the ensemble-averaged density operator remains parity symmetric, $P\rho P=\rho$. Because the Hamiltonian is quadratic and the jumps are linear in the fermion operators, an initial product-Fock state remains Gaussian \cite{Prosen2008Apr,Prosen2010Jul,Barthel2022Nov}. In this setting, Gaussian means that the state is fully determined by its two-point correlation matrix. This allows an exact covariance description and, through the parity rules of \cref{sec:summary}, ties both spectral clocks to the parities of $A$ and $B$ defined in \cref{eq:subsystem_parities} below.

\emph{Strong symmetry.} Uniform density dephasing instead gives
\begin{equation}
	\begin{aligned}
		\dot\rho=\mathcal L_{\mathrm s}(\rho)
		={}&{-i}[H,\rho]
		+\gamma\sum_{j=1}^{L}\mathcal D[n_j]\rho,
	\end{aligned}
	\label{eq:strong_lindblad}
\end{equation}
where the local Lindblad jump operators are $\sqrt{\gamma}\,n_j$, and $\gamma\geq0$ is the dephasing rate. Since every $n_j$ commutes with parity,
\begin{equation}
	\begin{aligned}
		0&=[H,P],
		&0&=[n_j,P]
		\quad\text{for every }j,
	\end{aligned}
	\label{eq:strong_parity_conservation}
\end{equation}
the Hamiltonian and every jump preserve each parity sector. This is a strong $\mathbb{Z}_2$ symmetry: the Hamiltonian and every jump operator commute with the generator $P$ \cite{Buca2012Jul,Albert2014Feb}. Physically, density dephasing suppresses coherence between occupation patterns that differ locally, but it never changes an occupation. It therefore preserves parity before any averaging over jump histories. Density dephasing also requires a different method. It generally produces a non-Gaussian mixed state, so two-point correlations alone do not determine its entanglement or negativity spectrum, and the parity rules no longer apply.

Most of our direct comparison is performed at the compact Majorana point. Define Majorana operators
\begin{equation}
	\begin{aligned}
		a_j&\coloneq c_j+c_j^\dagger,
		&b_j&\coloneq-i\left(c_j-c_j^\dagger\right).
	\end{aligned}
	\label{eq:majorana_operators}
\end{equation}
At
\begin{equation}
	\begin{aligned}
		\mu&=0,
		&\Delta&=J>0,
		&H&=iJ\sum_{j=1}^{L-1}b_j a_{j+1},
	\end{aligned}
	\label{eq:compact_majorana_point}
\end{equation}
every Majorana operator except $a_1$ and $b_L$ is paired with one on a neighboring site, which is the sense in which this point is compact. In particular, only one Majorana bond crosses a given spatial cut, and the bond terms $ib_ja_{j+1}$ commute with one another, so under $H$ each bond evolves on its own. Because $\Delta\neq0$, the pairing term reduces the $U(1)$ symmetry of particle-number conservation to the $\mathbb{Z}_2$ of \cref{eq:global_parity}. The compact point sits inside the topological phase of the closed system, which occupies $|\mu|<2J$ at $\Delta\neq0$ and carries the unpaired Majorana end modes drawn in \cref{fig:model_setup}. Except where a large chemical potential is taken explicitly, all Kitaev-chain results below refer to this phase. The parity rules stated in \cref{sec:summary} hold for any Gaussian state, and the balanced-rate result holds for any quadratic Hamiltonian and Gaussian initial state. Unless stated otherwise, all other exact results refer to \cref{eq:compact_majorana_point}. Whenever a small-$\mu$ expansion is used, we state explicitly that $|\mu|/J\ll1$ and give the retained order. Results that treat $\mu$ exactly are identified separately, and we claim no strong-symmetry result away from the compact point unless we state it explicitly.

We allow a general product-Fock initial state
\begin{equation}
	\begin{aligned}
		|\boldsymbol n\rangle
		&=\prod_{j=1}^{L}\left(c_j^\dagger\right)^{n_j}|0\rangle,
		&n_j&\in\{0,1\},
	\end{aligned}
	\label{eq:general_fock_state}
\end{equation}
where the product is ordered from left to right. For the direct comparison between weak and strong symmetry, our reference state contains $N$ consecutive particles at the left edge,
\begin{equation}
	\begin{aligned}
		|\psi_N(0)\rangle
		&=c_1^\dagger c_2^\dagger\cdots c_N^\dagger|0\rangle,
		&0&\leq N\leq L_A.
	\end{aligned}
	\label{eq:left_filled_initial_state}
\end{equation}
The empty product at $N=0$ is the vacuum, and $L_A$ is the size of the subsystem $A$ defined in \cref{eq:spatial_bipartition}. Some results extend to every occupation pattern $\boldsymbol n$; we state this stronger scope when those results are introduced.

We divide the chain into two regions, each containing at least one site,
\begin{equation}
	\begin{aligned}
		A&=\{1,\ldots,L_A\},
		&B&=\{L_A+1,\ldots,L\},
		\\
		L&=L_A+L_B,
		&\rho_A(t)&=\operatorname{Tr}_B\rho(t).
	\end{aligned}
	\label{eq:spatial_bipartition}
\end{equation}
The number and parity operators in $A$ are $\widehat N_A=\sum_{j\in A}n_j$ and $P_A=(-1)^{\widehat N_A}$, and $P_B$ is defined in the same way, so that $P=P_AP_B$. We call the expectation values
\begin{equation}
	\begin{aligned}
		q_A&\coloneq\operatorname{Tr}(\rho P_A),
		&q_B&\coloneq\operatorname{Tr}(\rho P_B)
	\end{aligned}
	\label{eq:subsystem_parities}
\end{equation}
the parities of $A$ and $B$. Under strong symmetry the global parity $p$ of the initial state is conserved, so $q_B=pq_A$ at all times, while loss and gain let the two parities evolve separately. The entanglement Hamiltonian probes the reduced state $\rho_A$, whereas fermionic negativity probes correlations across the full bipartition $A|B$. The bipartition and the three clocks developed below are summarized schematically in \cref{fig:model_setup}.

\subsection{Entanglement and negativity spectra}

To define the first clock, diagonalize the reduced state as
\begin{equation}
	\begin{aligned}
		\rho_A(t)&=\sum_k\lambda_k(t)
		|k_E(t)\rangle\langle k_E(t)|,
		\\
		\lambda_0(t)&\geq\lambda_1(t)\geq\lambda_2(t)\geq\cdots.
	\end{aligned}
	\label{eq:rho_A_spectrum}
\end{equation}
On the support of $\rho_A$, the entanglement Hamiltonian and its levels are \cite{Li2008Jul,Peschel2009Dec,Dalmonte2022Nov}
\begin{equation}
	\begin{aligned}
		K_E(t)&\coloneq-\ln\rho_A(t),
		&\epsilon_k^E(t)&\coloneq-\ln\lambda_k(t).
	\end{aligned}
	\label{eq:entanglement_hamiltonian}
\end{equation}
Because $\operatorname{Tr}\rho_A^n=\operatorname{Tr}e^{-nK_E}$, the R\'enyi index $n>0$ acts as a fictitious inverse temperature \cite{Calabrese2004Jun,Calabrese2009Dec,Dalmonte2022Nov,Kehrein2024Jun}. The name Hamiltonian reflects the spectral analogy; $K_E$ is not the generator of the physical time evolution. For a pure global state, the $\lambda_k$ are the squared Schmidt coefficients and directly describe bipartite entanglement. During open evolution, $\rho$ is mixed, so the same spectrum also reflects mixing produced by the environment. Its level crossings remain well defined, but they should not automatically be interpreted as extrema of a quantum coherence measure.

The second clock is built from fermionic partial time reversal. Collect the Majorana operators into $w_{2j-1}=a_j$ and $w_{2j}=b_j$, and let $\mathcal M$ be an ordered product of distinct $w_k$, a Majorana monomial, with $|\mathcal M|_A$ factors in $A$ and $|\mathcal M|_B$ factors in $B$. Partial time reversal on $A$ is the linear map \cite{Shapourian2017Apr,Shapourian2019Feb}
\begin{equation}
	\begin{aligned}
		\mathcal M^{R_A}=i^{|\mathcal M|_A}\mathcal M.
	\end{aligned}
	\label{eq:partial_time_reversal}
\end{equation}
Applying this rule term by term to the Majorana expansion of $\rho$ defines $\rho^{R_A}$~\footnote{One may equivalently apply partial time reversal to $B$. For every evolution considered here, $[\rho(t),P_AP_B]=0$. The initial product-Fock state has definite global parity. The closed Hamiltonian preserves this parity, as does every jump under strong symmetry. Under weak symmetry, individual loss and gain jumps flip parity, but Liouvillian covariance preserves $P\rho(t)P=\rho(t)$ for the ensemble. Let $R\equiv R_A\circ R_B$ denote full fermionic transposition. The phase rules then give $\rho^{R_B}=P_A(\rho^{R_A})^R P_A$. Full transposition and unitary parity conjugation preserve eigenvalues and singular values. Thus $\rho^{R_A}$ and $\rho^{R_B}$ are generally different operators but have identical eigenvalue and singular-value spectra. Their Hermitian twists, $T_A=\rho^{R_A}P_A$ and $T_B=\rho^{R_B}P_B$, may differ in their signed eigenvalues, but $\operatorname{spec}|T_A|=\operatorname{spec}|T_B|$. Hence $E_N$, $\Delta_N$, and $t_N$ are independent of which side is transformed. The choice of $A$ is a matter of convention.}. Despite its name, $R_A$ is a transformation of operators and does not reverse the physical evolution time. The chosen convention gives \cite{Shapourian2017Apr,Shapourian2019Feb}
\begin{equation}
	\begin{aligned}
		\left(\rho^{R_A}\right)^\dagger
		&=P_A\rho^{R_A}P_A,
		\\
		T_A&=\rho^{R_A}P_A=T_A^\dagger.
	\end{aligned}
	\label{eq:twisted_operator}
\end{equation}
The parity twist therefore converts the generally non-Hermitian operator $\rho^{R_A}$ into the Hermitian operator $T_A$ without changing its singular values. It also fixes which parity labels each spectrum. Parity covariance leaves $\rho$ with monomials of even $|\mathcal M|_A+|\mathcal M|_B$ only, \cref{eq:partial_time_reversal} rescales each monomial without changing $|\mathcal M|_A$ or $|\mathcal M|_B$, and the trace over $B$ discards every monomial with $|\mathcal M|_B\geq1$, so
\begin{equation}
	\begin{aligned}
		[T_A,P]&=0,
		&[\rho_A,P_A]&=0,
		&[T_A,P_A]&\neq0.
	\end{aligned}
	\label{eq:parity_labels}
\end{equation}
The eigenvectors of $T_A$ can therefore be chosen with definite global parity and those of $\rho_A$ with definite subsystem parity. The last relation holds whenever $\rho$ contains terms with $|\mathcal M|_A$ and $|\mathcal M|_B$ both odd, as it generically does once the bond crossing the cut has acted. Writing \cite{Shapourian2017Apr,Shapourian2019Feb}
\begin{equation}
	\begin{aligned}
		T_A&=\sum_k\eta_k|k_N\rangle\langle k_N|,
		&s_k&\coloneq|\eta_k|,
		\\
		s_0&\geq s_1\geq s_2\geq\cdots,
	\end{aligned}
	\label{eq:twisted_spectrum}
\end{equation}
the signed eigenvalues $\eta_k$ retain the information about positive and negative branches, while their magnitudes $s_k$ form the singular-value spectrum relevant to negativity. In particular,
\begin{equation}
	\begin{aligned}
		E_N(t)
		&=\ln\left\|\rho^{R_A}(t)\right\|_1
		=\ln\operatorname{Tr}|T_A(t)|
		\\
		&=\ln\sum_k s_k(t),
	\end{aligned}
	\label{eq:logarithmic_negativity}
\end{equation}
where $\|X\|_1=\operatorname{Tr}\sqrt{X^\dagger X}$ is the trace norm \cite{Vidal2002Feb,Plenio2005Aug,Shapourian2019Feb}. On the support of $T_A$, the twisted logarithmic operator separates into \cite{Murciano2022Apr,Rottoli2023Jan}
\begin{equation}
	\begin{aligned}
		\mathcal N_A^{\mathrm{tw}}
		&=-\ln T_A
		=K_N-i\pi\Pi_-,
		\\
		K_N&=-\ln|T_A|,
		\\
		\Pi_-&=\sum_{\eta_k<0}|k_N\rangle\langle k_N|.
	\end{aligned}
	\label{eq:twisted_negativity_hamiltonian}
\end{equation}
Here the principal logarithm is understood. Because $\operatorname{Tr}|T_A|^n=\operatorname{Tr}e^{-nK_N}=\sum_k s_k^n$, the R\'enyi index $n>0$ acts as a fictitious inverse temperature for the Hermitian modulus Hamiltonian $K_N$ \cite{Murciano2022Apr,Rottoli2023Jan}. The projector $\Pi_-$ records the negative-sign sector, while $K_N$ has real levels $\epsilon_k^N\coloneq-\ln s_k$. Since the temporal crossing is controlled by the ordering of the magnitudes $s_k$, the relevant real spectral gap belongs to $K_N$, and a closing of this gap is what we call a criticality of the twisted negativity Hamiltonian.

We can now define the two spectral clocks on equal footing. Their lowest gaps are
\begin{equation}
	\begin{aligned}
		\Delta_E(t)&=\ln\frac{\lambda_0(t)}{\lambda_1(t)},
		\\
		\Delta_N(t)&=\ln\frac{s_0(t)}{s_1(t)},
	\end{aligned}
	\label{eq:spectral_gaps}
\end{equation}
and their first closing times are
\begin{equation}
	\begin{aligned}
		t_E&=\inf\left\{t>0:\lambda_0(t)=\lambda_1(t)\right\},
		\\
		t_N&=\inf\left\{t>0:s_0(t)=s_1(t)\right\}.
	\end{aligned}
	\label{eq:spectral_clocks}
\end{equation}
If the set in either infimum is empty, so that the corresponding gap never closes, that clock is $+\infty$. We call a finite-time closing of the lowest spectral gap a temporal criticality of the corresponding entanglement or negativity Hamiltonian, in the same spirit as Refs.~\cite{Gong2018Dec,Poyhonen2021Nov,Sayyad2021Jul,Gadge2026Mar,wang2026criticaltouchingtemporalentanglement}. At a simple crossing the lowest level is nonanalytic, but because the crossing only relabels two levels, sums over the whole spectrum, such as the von Neumann and R\'enyi entropies or $E_N$, inherit no nonanalyticity from it. Two closings can also meet as a rate or the chemical potential is varied: they approach, merge into a tangency where the gap touches zero without reordering the levels, and annihilate, leaving a positive minimum behind. We call that event a fold, and \cref{app:weak_entanglement_asymptotics,app:weak_negativity_asymptotics} derive its two signatures for both spectra, a square-root separation of the pair of closings on one side and a linearly opening gap on the other. A closing can also recede to infinite time without meeting another one, as the entanglement closing of a one-site subsystem does when the dephasing rate approaches $4J$ from below (\cref{eq:strong_parity_clock,eq:strong_one_site_entanglement_clock}). We focus on the first closing because it provides a common reference point for systematic comparisons. This choice is also motivated by closed-system studies, including an analytically solvable model, in which later nonanalyticities can smear out in the thermodynamic limit \cite{Kehrein2024Jun,Jha2025Jun,Li2025Jul}.

An equivalent diagnostic, convenient in numerics, is
\begin{equation}
	\begin{aligned}
		r_E(t)
		&=\frac{\lambda_0(t)-\lambda_1(t)}
		{\lambda_0(t)+\lambda_1(t)}
		=\tanh\frac{\Delta_E(t)}{2},
		\\
		r_N(t)
		&=\frac{s_0(t)-s_1(t)}{s_0(t)+s_1(t)}
		=\tanh\frac{\Delta_N(t)}{2},
	\end{aligned}
	\label{eq:normalized_spectral_gaps}
\end{equation}
which vanish at exactly the same times as $\Delta_E$ and $\Delta_N$ but stay between $0$ and $1$, even where $\lambda_1$ or $s_1$ vanishes and the logarithmic gaps diverge.

\subsection{Unitary limit and the geometric clock}

In the unitary limit a pure initial state stays pure, and its twisted singular values are generated by pairs of reduced-state eigenvalues \cite{Vidal2002Feb,Shapourian2017Apr,Shapourian2019Feb}:
\begin{equation}
	\begin{aligned}
		s_{ij}&=\sqrt{\lambda_i\lambda_j},
		\\
		\Delta_N&=\frac{1}{2}\Delta_E,
		&t_N&=t_E.
	\end{aligned}
	\label{eq:pure_state_clock_relation}
\end{equation}
At the compact Majorana point of \cref{eq:compact_majorana_point}, only the bond crossing the cut entangles $A$ with $B$, so $\rho_A$ has at most two nonzero eigenvalues $\lambda_\pm$, and the closed evolution gives
\begin{equation}
	\begin{aligned}
		\lambda_\pm(t)&=\frac{1\pm|\cos(2Jt)|}{2},
		\\
		\Delta_E(t)&=2\operatorname{artanh}|\cos(2Jt)|,
		\\
		\Delta_N(t)&=\operatorname{artanh}|\cos(2Jt)|,
		\\
		E_N(t)&=\ln\left[1+|\sin(2Jt)|\right],
		\\
		t_E=t_N&=\frac{\pi}{4J}.
	\end{aligned}
	\label{eq:closed_compact_results}
\end{equation}
The derivation of \cref{eq:closed_compact_results} is given in Appendix~\ref{app:weak_gaussian_reduction}. At the first crossing $\lambda_+=\lambda_-=1/2$, so the two parity sectors of $A$ carry equal weight and the leading entanglement levels and the leading negativity-magnitude levels are both degenerate. Dissipation makes the global state mixed and can separate the two clocks. The entanglement clock then records a reordering within the reduced state of $A$, while the negativity clock records a reordering in an operator that probes both sides of the cut. For Gaussian states the order of the two clocks is fixed: the entanglement gap closes exactly when $q_A$ vanishes and the negativity gap as soon as either $q_A$ or $q_B$ does, so $t_N\leq t_E$ (\cref{sec:summary}).

Neither spectral closing should be confused with the loss of quantum coherence itself. The logarithmic negativity may begin to decrease before either gap closes. We use the geometric decoherence time $\tau_d^g$ \cite{Jha2026Jul} only as a secondary diagnostic of this distinction: it marks the first decrease of $E_N$ while the R\'enyi-$1/2$ entropy $S_{1/2}=2\ln\operatorname{Tr}\sqrt{\rho_A}$ is still nondecreasing, which a pure state, with $E_N=S_{1/2}$, cannot show. Its precise definition and derivations are given in Appendix~\ref{app:geometric_decoherence}. In the following sections we determine how each bath moves, preserves, splits, or removes these temporal crossings.

\section{Temporal Criticalities under Weak Symmetry: Gaussian Loss and Gain}
\label{sec:temporal_criticalities_weak}

\begin{figure*}[t]
	\centering
	\includegraphics[width=0.49\textwidth]{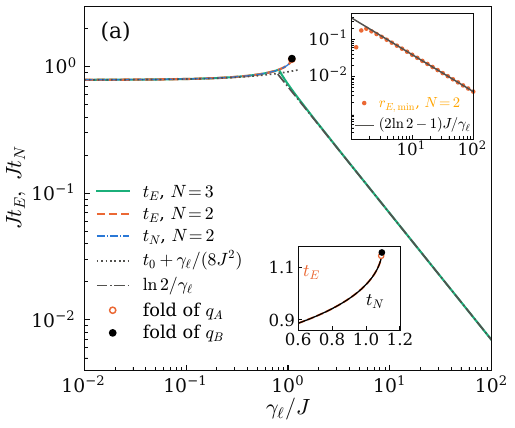}\hfill
	\includegraphics[width=0.49\textwidth]{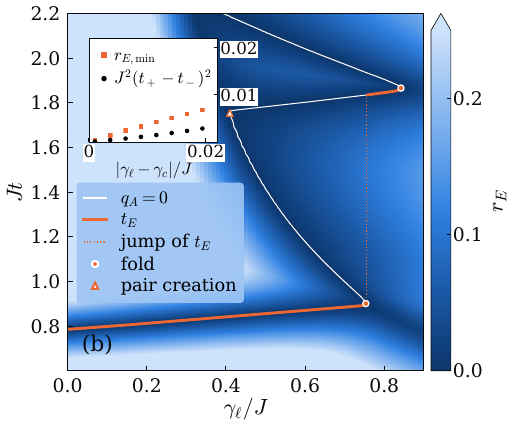}
	\caption{Entanglement clock from weak to strong loss. Exact covariance evolution, \cref{eq:weak_covariance_solution}, under pure loss, $\gamma_g=0$ (so $\kappa=\xi=\gamma_\ell$), at the compact point $\mu=0$, $\Delta=J$ (\cref{eq:compact_majorana_point}), with the first $N$ sites filled (\cref{eq:left_filled_initial_state}); $\rho_A$ does not depend on $L_B$. Pure gain gives the same clocks from the complemented initial pattern (\cref{eq:weak_gain_loss_duality}). The gap $r_E$ is the smallest $|\zeta_m|$ of $\Gamma_A$ (\cref{eq:weak_gaussian_entanglement_gap}), $t_E$ is the first zero of $q_A=(-1)^{L_A}\operatorname{Pf}\Gamma_A$ (\cref{eq:weak_gaussian_equivalences}), and $t_N$ is the first zero of $q_Aq_B$ (\cref{eq:weak_negativity_parity_rule}). \textbf{(a)} $t_E$ for $L_A=50$ with $N=3$ (odd) and $N=2$ (even), and $t_N$ for $N=2$ and $L=100$ (for $N=3$ it coincides with $t_E$). Weak loss delays all three from $t_0=\pi/(4J)$ by $\gamma_\ell/(8J^2)$ (dotted; \cref{eq:weak_rate_entanglement_clock,eq:weak_bulk_negativity_clock}). At strong loss the odd filling closes at $\ln2/\gamma_\ell$ (dash-dotted; \cref{eq:weak_local_half_occupation,eq:weak_strong_rate_entanglement_clock}), whereas the even branch ends at a fold of $q_A$ (open circle), beyond which no closing occurs up to $Jt=10$ and the minimum of $r_E$ over $t$ approaches $(2\ln2-1)J/\gamma_\ell$ (upper inset; \cref{eq:weak_strong_rate_entanglement_clock}). The zero of $q_B$ outlives this fold: for $N=2$, $t_N<t_E$ near the fold, and up to the fold of $q_B$ (filled circle) only the negativity gap closes (lower inset; both insets of (a) are plotted against $\gamma_\ell/J$). \textbf{(b)} Color: $r_E$ for $L_A=50$, $L=100$ and $N=48$ in the plane of loss rate and time, saturated above $0.25$. The closings lie on the zero line of $q_A$ (white), across which $q_A$ changes sign. Closings are created in pairs at left turning points of this line (triangle) and annihilated in pairs at right turning points, the folds (circles), with a square-root separation below a fold and a linearly opening gap above it (inset; $\gamma_c\approx0.738J$ is the rate at the first fold, $t_\pm$ are the two merging closings, and $r_{E,\min}$ is the minimum of $r_E$ near the fold time). The first fold removes the earliest pair while a later closing survives, so $t_E$ (thick line) jumps (dotted) before the branch ends at the second fold.}
	\label{fig:weak_symmetry_clocks}
\end{figure*}

Loss and gain keep the state Gaussian. In a Gaussian state every gap closing is a parity balance. The entanglement gap closes exactly when the parity $q_A$ of \cref{eq:subsystem_parities} vanishes (\cref{eq:weak_gaussian_equivalences}), and we prove that the negativity gap closes exactly when $q_A$ or $q_B$ vanishes, so the negativity clock is the earlier of the entanglement clocks of $A$ and $B$ and never comes after $t_E$ (\cref{eq:weak_negativity_parity_rule}). Both rules hold at every instant, for pure and mixed states alike, and it remains to find how the bath moves the zeros of $q_A$ and $q_B$. Uniform balanced loss and gain multiply both parities by positive factors, so for any quadratic Hamiltonian and Gaussian initial state both clocks keep their closed-system values at every rate (\cref{eq:weak_balanced_clock_locking}). In the Kitaev chain at the compact point, weak imbalanced rates shift the zero of each parity by an amount set by the initial occupation of the site next to the cut on that side, so $B$ can balance first (\cref{eq:weak_rate_entanglement_clock,eq:weak_rate_negativity_clock}; \cref{fig:weak_negativity_clocks}(a)). At strong rates, $\kappa\gg J$ with $\xi\neq0$, each site relaxes on its own, and the sites that the bath drives through half filling, which we call active, cross it together at $t=s_\star/\kappa$. A parity vanishes near that time only if its side of the cut holds a block of an odd number of active sites (\cref{eq:weak_strong_rate_entanglement_clock,eq:weak_strong_rate_negativity_clock}). An odd block in $B$ can then close the negativity gap while the entanglement gap stays open (\cref{fig:weak_negativity_clocks}(b)). At intermediate rates closings are created and annihilated in pairs (\cref{fig:weak_symmetry_clocks}(b)). At weak rates the logarithmic negativity starts to fall before either gap closes, while at strong rates the order depends on the imbalance and the filling (\cref{tab:weak_strong_geometric_branches}). \Cref{subsec:eh_weak} treats the entanglement clock, including finite chemical potential, and \cref{subsec:tnh_weak} the negativity clock and the comparison of the three clocks.

The product-Fock state in \cref{eq:left_filled_initial_state} is Gaussian, and the quadratic Hamiltonian with linear loss and gain in \cref{eq:weak_lindblad} preserves Gaussianity under the ensemble evolution. Every many-body correlation function then follows from two-point correlations, so one covariance matrix determines both the reduced density operator and the partially time-reversed operator. This common description lets us identify which features of the two clocks are shared and which arise from the different regions they probe.

Arrange the Majorana operators $w_k$ of \cref{sec:model} in the column vector $w=(w_1,\ldots,w_{2L})^T$, so that $w_{2j-1}=a_j$ and $w_{2j}=b_j$, and write $H=(i/4)w^T\mathsf A w+\text{constant}$, where $\mathsf A$ is a real antisymmetric matrix~\cite{kitaev2001}. The Majorana covariance matrix is
\begin{equation}
	\begin{aligned}
		\Gamma_{mn}(t)
		&\coloneq\frac{i}{2}\operatorname{Tr}
		\!\left(\rho(t)[w_m,w_n]\right).
	\end{aligned}
	\label{eq:weak_majorana_covariance}
\end{equation}
Substitution of \cref{eq:weak_lindblad} gives a closed linear equation and its variation-of-constants solution,
\begin{equation}
	\begin{aligned}
		\dot\Gamma
		&=\mathsf A\Gamma+\Gamma\mathsf A^T
		-\kappa\Gamma+\xi\Gamma_{\mathrm{vac}},
		\\
		\Gamma(t)
		&=e^{-\kappa t}R(t)\Gamma(0)R^T(t)
		\\
		&\quad+\xi\int_0^t\!du\,e^{-\kappa u}
		R(u)\Gamma_{\mathrm{vac}}R^T(u),
		\\
		R(t)&\coloneq e^{\mathsf A t}.
	\end{aligned}
	\label{eq:weak_covariance_solution}
\end{equation}
Here $\Gamma_{\mathrm{vac}}=\bigoplus_{j=1}^{L}\left(\begin{smallmatrix}0&-1\\1&0\end{smallmatrix}\right)$ is the vacuum covariance. Equation~\eqref{eq:weak_covariance_solution} is exact for every finite chain \cite{Prosen2008Apr,Prosen2010Jul,Barthel2022Nov}. The term controlled by the total rate $\kappa$ contracts the covariance inherited from closed evolution, while the source controlled by the imbalance $\xi$ drives it toward the covariance favored by loss or gain. The derivation and the equivalent particle-hole representation are given in Appendix~\ref{app:weak_gaussian_reduction}.

The two spectra use different parts of this same matrix. Let $\Gamma_A$ be the principal submatrix containing the $2L_A$ Majoranas in $A$. An orthogonal transformation $O_A$ brings it to canonical form,
\begin{equation}
	\begin{aligned}
		O_A\Gamma_AO_A^T
		&=\bigoplus_{m=1}^{L_A}
		\begin{pmatrix}
			0&\zeta_m\\
			-\zeta_m&0
		\end{pmatrix},
		&-1&\leq\zeta_m\leq1.
	\end{aligned}
	\label{eq:weak_canonical_covariance}
\end{equation}
The real numbers $\zeta_m$ are the covariance modes of subsystem $A$; we carry their signs and not only their magnitudes, assigning them by taking $\det O_A=+1$ and following each mode continuously in time. The transformation $O_A$ defines canonical fermionic modes in which the reduced state factorizes as \cite{Peschel2003Apr,Cheong2004Feb,Peschel2009Dec,Bravyi2005May}
\begin{equation}
	\begin{aligned}
		\rho_A
		=\bigotimes_{m=1}^{L_A}
		\left[
		\frac{1-\zeta_m}{2}|0_m\rangle\langle0_m|
		+\frac{1+\zeta_m}{2}|1_m\rangle\langle1_m|
		\right].
	\end{aligned}
	\label{eq:weak_gaussian_factorization}
\end{equation}
Here $f_m$ annihilates a fermion in canonical mode $m$, $|0_m\rangle$ and $|1_m\rangle$ are its empty and occupied states, and the convention of \cref{eq:weak_majorana_covariance} gives $\langle1-2f_m^\dagger f_m\rangle=-\zeta_m$, so that the vacuum of $A$ has every $\zeta_m=-1$. A mode therefore passes through zero when it reaches half occupation, and the more probable of its two local states switches there, exchanging the parity of the dominant eigenvector of $\rho_A$. Fermionic partial time reversal is equally simple at the covariance level~\cite{Shapourian2017Apr,Shapourian2019Feb}. Each Majorana index in $A$ acquires the phase $i$ prescribed by \cref{eq:partial_time_reversal}, so
\begin{equation}
	\begin{aligned}
		\mathsf D_A
		&\coloneq\operatorname{diag}
		(\underbrace{i,\ldots,i}_{2L_A},
		\underbrace{1,\ldots,1}_{2L_B}),
		\\
		\Gamma^{R_A}&=\mathsf D_A\Gamma\mathsf D_A^T.
	\end{aligned}
	\label{eq:weak_partial_time_reversal_covariance}
\end{equation}
The matrix $\Gamma^{R_A}$ is generally complex and specifies a Gaussian operator, which can be represented as the exponential of a quadratic Majorana form. Applying the parity twist in \cref{eq:twisted_operator} and bringing the result to canonical form gives the signed eigenvalues $\eta_k$ of $T_A$; taking $s_k=|\eta_k|$ then gives the singular values in \cref{eq:twisted_spectrum}. The reconstruction is detailed in Appendix~\ref{app:weak_gaussian_reduction}. Thus $K_E$ is fixed at each time by $\Gamma_A$ alone and $K_N$ by the full $\Gamma$, although the evolution of $\Gamma_A$ couples to $B$ through $\mathsf A$. \Cref{subsec:tnh_weak} uses the canonical form of $T_A$ to show that a negativity-gap closing is also a parity balance.

This reduction requires Gaussianity, which weak symmetry alone does not guarantee. A weakly symmetric jump need not be linear, as the correlated loss $c_jn_{j+1}$ shows, and closure of the two-point equations is not enough: density dephasing closes them but produces a non-Gaussian state, whose spectra require the many-body treatment of \cref{sec:temporal_criticalities_strong}.

\emph{Balanced loss and gain.} A result common to both spectra follows directly from \cref{eq:weak_covariance_solution}. Set $\gamma_\ell=\gamma_g\equiv\gamma_{\mathrm w}$, so $\xi=0$ and $\kappa=2\gamma_{\mathrm w}$. The covariance is then a uniform contraction of the covariance $\Gamma^{\mathrm{cl}}(t)$ generated by the closed evolution from the same initial state,
\begin{equation}
	\begin{aligned}
		f(t)&\coloneq e^{-2\gamma_{\mathrm w}t},
		&\Gamma(t)&=f(t)\Gamma^{\mathrm{cl}}(t).
	\end{aligned}
	\label{eq:weak_balanced_covariance}
\end{equation}
For a pure Gaussian initial state, the closed evolution keeps the state pure, and local canonical transformations decompose the bipartition into cut modes, each one half of an entangled pair straddling the cut, and confined modes, unentangled with the other side \cite{Botero2003May,Botero2004Oct,Peschel2009Dec}. Let $\zeta_m^{\mathrm{cl}}$ be the signed covariance mode of the $m$th cut pair. Diagonalizing its $4\times4$ open Gaussian block and using the gap definitions in \cref{eq:spectral_gaps,eq:normalized_spectral_gaps} gives
\begin{equation}
	\begin{aligned}
		\Delta_{E,m}
		&=2\operatorname{artanh}
		\!\left(f|\zeta_m^{\mathrm{cl}}|\right),
		\\
		\Delta_{N,m}
		&=\operatorname{artanh}
		\!\left(
		\frac{2f|\zeta_m^{\mathrm{cl}}|}{1+f^2}
		\right).
	\end{aligned}
	\label{eq:weak_balanced_pair_gaps}
\end{equation}
The many-body gap is the minimum of the mode-resolved gaps. A confined mode has $|\zeta_m^{\mathrm{cl}}|=1$ and retains a positive finite-time gap, while a cut mode closes when $\zeta_m^{\mathrm{cl}}=0$. Since $f(t)>0$ at every finite time, the contraction changes both gap functions but cannot move any of their zeros.

A mixed Gaussian initial state need not have such a pair decomposition, but the parity rules, stated in \cref{sec:summary} and derived below, give the same conclusion. Because $\Gamma_A$ is a $2L_A\times2L_A$ matrix, $\operatorname{Pf}\Gamma_A=f^{L_A}\operatorname{Pf}\Gamma_A^{\mathrm{cl}}$, so \cref{eq:weak_gaussian_equivalences} gives $q_A=f^{L_A}q_A^{\mathrm{cl}}$, and likewise $q_B=f^{L_B}q_B^{\mathrm{cl}}$. The positive factors leave the zeros of both parities in place, so by \cref{eq:weak_gaussian_equivalences,eq:weak_negativity_parity_rule} neither clock moves. Consequently, for uniform balanced rates, any quadratic Hamiltonian, any pure or mixed Gaussian initial state, and any spatial bipartition,
\begin{equation}
	\begin{aligned}
		t_E&=t_E^{\mathrm{cl}},
		&t_N&=t_N^{\mathrm{cl}}
		\qquad
		(\gamma_\ell=\gamma_g).
	\end{aligned}
	\label{eq:weak_balanced_clock_locking}
\end{equation}
The rates must be equal on every site, because balanced rates that vary from site to site contract different parts of the covariance by different factors and can move the closings. For a pure initial state, $t_N^{\mathrm{cl}}=t_E^{\mathrm{cl}}$ by \cref{eq:pure_state_clock_relation}, and at the compact point \cref{eq:closed_compact_results} gives $t_N=t_E=t_0$, where $t_0=\pi/(4J)$. The spectral zeros remain locked even though balanced noise reduces the logarithmic negativity and increases the mixedness of $\rho_A$.

The geometric decoherence time responds to this change before either spectral gap closes. At exactly $\gamma_{\mathrm w}=0$, the equality $E_N=S_{1/2}$ makes the geometric decoherence time undefined. For every $\gamma_{\mathrm w}>0$, $E_N$ turns over while $S_{1/2}$ is still increasing, and in the weak-noise limit
\begin{equation}
	\begin{aligned}
		\tau_d^g
		&=t_0-\frac{\gamma_{\mathrm w}}{J^2}
		+\mathcal O\!\left(
		\frac{\gamma_{\mathrm w}^2}{J^3}
		\right)
		<t_N=t_E.
	\end{aligned}
	\label{eq:weak_balanced_geometric_clock}
\end{equation}
Appendix~\ref{app:weak_gaussian_reduction} derives \cref{eq:weak_balanced_geometric_clock} from the exact balanced-rate negativity and entropy, and Appendix~\ref{app:geometric_decoherence} gives the definition of $\tau_d^g$ used there.

\subsection{Entanglement Hamiltonian}
\label{subsec:eh_weak}

The factorization in \cref{eq:weak_gaussian_factorization} converts the many-body entanglement spectrum into the motion of the signed modes $\zeta_m$. Define $\nu_A\coloneq\min_m|\zeta_m|$. The two largest eigenvalues, the normalized gap, and the logarithmic gap are
\begin{equation}
	\begin{aligned}
		\lambda_0
		&=\prod_{m=1}^{L_A}\frac{1+|\zeta_m|}{2},
		\\
		\lambda_1
		&=\lambda_0\frac{1-\nu_A}{1+\nu_A},
		\\
		r_E&=\nu_A,
		&\Delta_E&=2\operatorname{artanh}\nu_A.
	\end{aligned}
	\label{eq:weak_gaussian_entanglement_gap}
\end{equation}
This reduction also gives a direct physical interpretation of a closing. Let $\operatorname{Pf}\Gamma_A$ be the Pfaffian of the antisymmetric matrix $\Gamma_A$, defined by $(\operatorname{Pf}\Gamma_A)^2=\det\Gamma_A$ with its sign fixed by the Majorana ordering in \cref{eq:weak_majorana_covariance}. By \cref{eq:majorana_operators}, $1-2n_j=-ia_jb_j$, so the parity operator of $A$ is the ordered Majorana monomial $P_A=(-i)^{L_A}w_1\cdots w_{2L_A}$, and \cref{eq:app_weak_wick_pfaffian} evaluates its expectation. With the parity $q_A=\operatorname{Tr}(\rho_AP_A)$ of \cref{eq:subsystem_parities} and the parity weights $W_\pm\coloneq\operatorname{Tr}[(1\pm P_A)\rho_A/2]$, Gaussianity gives
\begin{equation}
	\begin{aligned}
		\lambda_0=\lambda_1
		&\Longleftrightarrow\nu_A=0
		\Longleftrightarrow\operatorname{Pf}\Gamma_A=0,
		\\
		q_A&=(-1)^{L_A}\operatorname{Pf}\Gamma_A,
		&W_\pm&=\frac{1\pm q_A}{2}.
	\end{aligned}
	\label{eq:weak_gaussian_equivalences}
\end{equation}
An entanglement-gap closing is therefore an exact balance between the even and odd fermion-parity weights of the subsystem $A$. At a simple zero, one signed mode changes sign and the dominant eigenvector of $\rho_A$ switches parity. At a tangency, the gap reaches zero without this exchange. Raising $\kappa$ at fixed $\xi/\kappa$ can also make two closings meet and annihilate at a fold, and at other rates it creates a new pair. \Cref{fig:weak_symmetry_clocks}(b) shows both events in the plane of rate and time, where the closings lie on the zero line of $q_A$, and \cref{app:weak_entanglement_asymptotics} derives the fold laws stated in \cref{sec:model}. We follow the first continuously signed mode that reaches zero, in accordance with \cref{eq:spectral_clocks}. When a fold removes the earliest pair while a later closing survives, $t_E$ jumps to that later closing; when none survives, the gap stays open. Later finite-size roots can merge or proliferate, and their nonanalyticities can smear in the thermodynamic free-fermion limit \cite{Kehrein2024Jun,Jha2025Jun,Li2025Jul}.

\emph{Weak dissipation at the compact point.} At the compact Majorana point of \cref{eq:compact_majorana_point}, the single Majorana bond crossing the cut gives
\begin{equation}
	\begin{aligned}
		q_A^{\mathrm{cl}}(t)
		&=(-1)^N\cos(2Jt),
		&t_0&=\frac{\pi}{4J}.
	\end{aligned}
	\label{eq:weak_compact_closed_parity}
\end{equation}
For $\kappa/J\ll1$ at fixed $\xi/\kappa$, write $t_E=t_0+\delta t$ and insert the exact solution \eqref{eq:weak_covariance_solution} into the parity condition in \cref{eq:weak_gaussian_equivalences}. Since the closed zero is simple,
\begin{equation}
	\begin{aligned}
		0={}&\dot q_A^{\mathrm{cl}}(t_0)\delta t
		+q_A^{(1)}(t_0)
		+\mathcal O\!\left(\frac{\kappa^2}{J^2}\right),
		\\
		q_A^{(1)}(t_0)
		={}&
		\begin{cases}
			\dfrac{\xi}{2J},&L_A=1,\\[5pt]
			\dfrac{(-1)^N(1-2n_{L_A})\xi}{4J},&L_A\geq2.
		\end{cases}
	\end{aligned}
	\label{eq:weak_compact_root_expansion}
\end{equation}
Here $q_A^{(1)}(t_0)$ is the first-order open-system correction evaluated at the closed root, and $n_j$ is the initial occupation of site $j$. The $\kappa$ term in \cref{eq:weak_covariance_solution} rescales the closed covariance and hence leaves its simple zero fixed at first order. At $t_0$, $\Gamma_A^{\mathrm{cl}}(t_0)$ has a two-dimensional kernel associated with the cut mode, while all other covariance modes remain nonzero. Because $q_A=(-1)^{L_A}\operatorname{Pf}\Gamma_A$, only the matrix element of the $\xi$-dependent source inside this kernel can change $q_A$ at first order. Evaluating that matrix element gives the displayed $q_A^{(1)}(t_0)$. The linearized root condition then gives $\delta t=-q_A^{(1)}(t_0)/\dot q_A^{\mathrm{cl}}(t_0)$ and yields
\begin{equation}
	\begin{aligned}
		t_E&=t_0+\frac{\xi}{J^2}C_E
		+\mathcal O\!\left(\frac{\kappa^2}{J^3}\right),
		\\
		C_E&=
		\begin{cases}
			\dfrac{1-2n_1}{4},&L_A=1,\\[5pt]
			\dfrac{1-2n_{L_A}}{8},&L_A\geq2.
		\end{cases}
	\end{aligned}
	\label{eq:weak_rate_entanglement_clock}
\end{equation}
For $L_A\geq2$, only the initial occupation $n_{L_A}$ of the cut site enters at this order; it is the $n_c$ of \cref{tab:clock_summary}. Loss delays the crossing when that site is empty and advances it when the site is occupied; gain reverses these shifts. For $L_A=1$, the subsystem consists entirely of the cut site and the coefficient doubles. \Cref{fig:weak_symmetry_clocks}(a) shows the delay for pure loss, where $\xi=\gamma_\ell$ and the cut site starts empty. Appendix~\ref{app:weak_entanglement_asymptotics} derives the projection for the left-filled state; with $(-1)^N$ replaced by the initial parity of $A$, the same derivation holds for every product-Fock pattern, as we also confirmed numerically.

Loss and gain are related by the exact staggered particle-hole transformation $c_j\mapsto(-1)^jc_j^\dagger$. It exchanges creation and annihilation, sends $\mu\mapsto-\mu$, and complements the occupation pattern. If $\overline{\boldsymbol n}=\boldsymbol 1-\boldsymbol n$, then
\begin{equation}
	\begin{aligned}
		\operatorname{spec}\rho_A
		(t;\gamma_\ell,\gamma_g,\mu,\boldsymbol n)
		={}&\operatorname{spec}\rho_A
		(t;\gamma_g,\gamma_\ell,-\mu,
		\overline{\boldsymbol n}).
	\end{aligned}
	\label{eq:weak_gain_loss_duality}
\end{equation}
Here $\operatorname{spec}$ denotes the eigenvalue multiset. The relation maps every loss-dominant branch to its gain-dominant partner, so pure loss suffices in \cref{fig:weak_symmetry_clocks} and Appendix~\ref{app:parameter_sweeps}, while \cref{fig:weak_negativity_clocks} uses gain, under which $B$ can balance first. To obtain the fixed-pattern symmetry, let $\mathcal K$ denote complex conjugation in the occupation basis and define the staggered gauge transformation $\mathcal S\coloneq\exp(i\pi\sum_{j=1}^{L}j\,n_j)$, so that $\mathcal S c_j\mathcal S^\dagger=(-1)^jc_j$. Because the matrices of $c_j$ and $c_j^\dagger$ are real in this basis, $\mathcal K$ maps $-i[H(\mu),\rho]$ to $-i[-H(\mu),\rho^*]$ and leaves the dissipators unchanged. The gauge transformation obeys $\mathcal S H(\mu)\mathcal S^\dagger=-H(-\mu)$ and changes each jump operator only by a sign. Their combination therefore gives $\rho(t;-\mu,\boldsymbol n)=\mathcal S\rho(t;\mu,\boldsymbol n)^*\mathcal S^\dagger$ with the same rates and initial product-Fock projector. Since $\mathcal S=\mathcal S_A\otimes\mathcal S_B$, the reduced states satisfy $\rho_A(t;-\mu,\boldsymbol n)=\mathcal S_A\rho_A(t;\mu,\boldsymbol n)^*\mathcal S_A^\dagger$ and have identical spectra. The reduced spectrum is therefore even in $\mu$. Appendices~\ref{app:weak_entanglement_asymptotics} and \ref{app:weak_finite_mu} derive the two relations.

\emph{Finite chemical potential at weak gain/loss rates.} Let $t_E^{\mathrm{cl}}(\mu)$ be a finite simple zero of $\operatorname{Pf}\Gamma_A^{\mathrm{cl}}(t,\mu)$. Choose an oriented orthonormal basis $u_E,v_E$ of the two-dimensional kernel at that zero and write $\delta t\coloneq t_E-t_E^{\mathrm{cl}}(\mu)$. If $\mathsf A_\mu$ is the closed Majorana generator at chemical potential $\mu$, then
\begin{equation}
	\begin{aligned}
		R_\mu(u)
		&\coloneq e^{\mathsf A_\mu u},
		\\
		Y(t,\mu)
		&\coloneq\int_0^t\!du\,
		R_\mu(u)\Gamma_{\mathrm{vac}}R_\mu^T(u),
		\\
		0
		={}&u_E^T\dot\Gamma_A^{\mathrm{cl}}
		(t_E^{\mathrm{cl}},\mu)v_E\,\delta t
		+\xi u_E^TY_A(t_E^{\mathrm{cl}},\mu)v_E,
		\\
		\chi_E(\mu)
		&\coloneq-
		\frac{u_E^T Y_A(t_E^{\mathrm{cl}},\mu)v_E}
		{u_E^T\dot\Gamma_A^{\mathrm{cl}}
			(t_E^{\mathrm{cl}},\mu)v_E}.
	\end{aligned}
	\label{eq:weak_finite_mu_susceptibility}
\end{equation}
The integral $Y(t,\mu)$ is the first-order covariance response per unit imbalance: the closed propagator evolves the fixed source $\Gamma_{\mathrm{vac}}$ for every duration $u$, and the integral sums these contributions. Its principal submatrix $Y_A$ is the part seen by subsystem $A$. Taking the matrix element between $u_E$ and $v_E$ isolates the zero block that controls the crossing. The $\kappa$ term is proportional to $\Gamma_A^{\mathrm{cl}}$ and has zero matrix element in this kernel at the closed root, so only the source proportional to $\xi$ appears in the linearized condition. The denominator is nonzero by the simple-root assumption. At a fold, two roots merge into a tangency, so both the selected kernel element and its time derivative vanish; the linear formula then fails and the expansion must retain the quadratic term in time. Away from such a fold, \cref{eq:weak_covariance_solution} gives
\begin{equation}
	\begin{aligned}
		t_E&=t_E^{\mathrm{cl}}(\mu)
		+\xi\chi_E(\mu)
		+\mathcal O\!\left(\frac{\kappa^2}{J^3}\right),
		\\
		\chi_E(\mu)&=\chi_E(-\mu).
	\end{aligned}
	\label{eq:weak_finite_mu_simple_root}
\end{equation}
For $|\mu|/J\ll1$, $\kappa/J\ll1$ at fixed $\xi/\kappa$, and $\Delta=J$, this exact simple-root formula expands as
\begin{equation}
	\begin{aligned}
		t_E={}&t_0
		+A_{L_A,L_B,N}\frac{\mu^2}{J^3}
		+C_E\frac{\xi}{J^2}
		\\
		&+X_{L_A,L_B,N}\frac{\xi\mu^2}{J^4}
		+\mathcal O\!\left(
		\frac{\mu^4}{J^5},
		\frac{|\xi|\mu^4}{J^6},
		\frac{\kappa^2}{J^3}
		\right).
	\end{aligned}
	\label{eq:weak_small_mu_clock}
\end{equation}
The coefficient $A_{L_A,L_B,N}$ is the closed quadratic displacement, while $X_{L_A,L_B,N}$ measures how chemical-potential rotation changes the linear response to loss-gain imbalance. With $r\coloneq L_A-N$, the tables in Appendix~\ref{app:weak_finite_mu} cover $L_A=1,2,\geq3$, $L_B=1,\geq2$, and $r=0,1,\geq2$, including every boundary and filling branch. The absence of odd powers of $\mu$ follows from the fixed-pattern $\mu\mapsto-\mu$ symmetry stated above. By \cref{eq:weak_negativity_parity_rule}, the same simple-root formulas applied to $B$ give the shifted zero of $q_B$, and the earlier of the two shifted zeros is $t_N$.

The opposite finite-$\mu$ hierarchy is $J/|\mu|\ll1$ with $\kappa/J\ll1$. In the rotating frame of the total number, pairing acquires the phase $e^{2i\mu t}$ and averages away on times $t=\mathcal O(J^{-1})$ \cite{Bukov2015}. If the corresponding number-conserving problem with $\Delta=0$ has an isolated simple first root $t_E^{\mathrm{nc}}$, where nc stands for number conserving, then
\begin{equation}
	\begin{aligned}
		t_E
		=t_E^{\mathrm{nc}}
		+\xi\chi_E^{\mathrm{nc}}
		+\mathcal O\!\left(
		\frac{\kappa^2}{J^3}
		+\frac{J}{\mu^2}
		\right).
	\end{aligned}
	\label{eq:weak_rapid_rotation_clock}
\end{equation}
The susceptibility $\chi_E^{\mathrm{nc}}$ is obtained from \cref{eq:weak_finite_mu_susceptibility} with $\Delta=0$. A pairing-supported root can disappear in this limit, while a simple hopping-supported root survives. Multiple roots and finite-rate folds require a separate expansion around the merged root and are treated in Appendix~\ref{app:weak_finite_mu}.

\emph{Strong dissipation.} We now return to the compact point $\mu=0$ and $\Delta=J>0$, and take $\kappa/J\gg1$ with nonzero imbalance. Introduce
\begin{equation}
	\begin{aligned}
		b&\coloneq\frac{|\xi|}{\kappa},
		&d&\coloneq\frac{\xi}{\kappa},
		&s&\coloneq\kappa t,
		&0&<b\leq1.
	\end{aligned}
	\label{eq:weak_strong_rate_variables}
\end{equation}
Here $b$ measures the magnitude of the loss-gain imbalance relative to the total rate, $d$ retains its sign, and $s$ measures time in units of the local relaxation time $\kappa^{-1}$. Thus $d=b$ under loss dominance and $d=-b$ under gain dominance. The balanced case $b=0$ is governed by \cref{eq:weak_balanced_clock_locking}. On the fast scale $s=\mathcal O(1)$, the coherent terms are subleading and each site relaxes independently. For a site that begins with occupation eigenvalue $n\in\{0,1\}$, define $\bar n_n(s)\coloneq\langle n_j\rangle_s$; the subscript $n$ labels the initial condition, so $\bar n_n(0)=n$. Loss empties an occupied site at rate $\gamma_\ell\bar n_n$, whereas gain fills an empty site at rate $\gamma_g(1-\bar n_n)$. Since $\gamma_g=(\kappa-\xi)/2=\kappa(1-d)/2$, one obtains
\begin{equation}
	\begin{aligned}
		\frac{d\bar n_n}{dt}
		&=-\gamma_\ell\bar n_n+\gamma_g(1-\bar n_n)
		=-\kappa\bar n_n+\gamma_g,
		\\
		\frac{d\bar n_n}{ds}
		&=-\bar n_n+\frac{1-d}{2},
		\\
		\bar n_n(s)
		&=\frac{1-d}{2}
		+\left(n-\frac{1-d}{2}\right)e^{-s},
		\\
		\bar p_n(s)
		&\coloneq\langle1-2n_j\rangle_s
		=1-2\bar n_n(s)
		=d+(1-2n-d)e^{-s}.
	\end{aligned}
	\label{eq:weak_local_kinetics}
\end{equation}
The quantity $\bar p_n(s)$ is the mean local fermion parity, and $\bar p_n(s)=0$ exactly when the site has mean occupation one half. The stationary occupation lies below one half for loss dominance, so only initially occupied sites cross half filling. It lies above one half for gain dominance, so only initially empty sites cross. We call these sites active. Inside $A$, the left-filled state therefore has one active block of length
\begin{equation}
	\begin{aligned}
		M_E&=
		\begin{cases}
			N,&\xi>0,\\
			L_A-N,&\xi<0.
		\end{cases}
	\end{aligned}
	\label{eq:weak_entanglement_active_length}
\end{equation}
For an active site, the loss- and gain-dominant solutions in \cref{eq:weak_local_kinetics} differ only by an overall sign. We therefore define their common signed distance from half filling by
\begin{equation}
	\begin{aligned}
		\alpha(s)
		&\coloneq\bar p_0(s)\big|_{d=-b}
		=-\bar p_1(s)\big|_{d=b}
		\\
		&=-b+(1+b)e^{-s},
		\\
		s_\star&\coloneq\ln\frac{1+b}{b},
		&\alpha(s_\star)&=0.
	\end{aligned}
	\label{eq:weak_local_half_occupation}
\end{equation}
Thus $|\alpha(s)|=|\bar p_n(s)|$ for either active type and measures the distance from half filling in units of the local parity. All active sites reach half filling simultaneously at $s=s_\star$, which provides the leading local reference time $s_\star/\kappa$. Restoring the Hamiltonian then determines whether an active covariance mode remains zero, giving a true many-body crossing, or acquires a finite minimum, giving an avoided crossing. The first variation of \cref{eq:weak_covariance_solution} through the pairing term carries the coefficient
\begin{equation}
	\begin{aligned}
		c(b)
		&\coloneq2\int_0^{s_\star}\!du\,
		e^{-(s_\star-u)}\alpha(u)
		\\
		&=2b\left(s_\star-\frac{1}{1+b}\right).
	\end{aligned}
	\label{eq:weak_local_pairing_coefficient}
\end{equation}
The exponential kernel damps covariance generated early in the relaxation, while $\alpha(u)$ records the instantaneous local parity available for pairing. Restricting this first-order covariance to the active block gives the usual sine standing waves of a finite nearest-neighbor chain, with momenta $m\pi/(M_E+1)$. Their positive covariance magnitudes are
\begin{equation}
	\begin{aligned}
		\nu_m(s)
		={}&\left[
		\alpha(s)^2
		+\frac{4c(b)^2J^2}{\kappa^2}
		\cos^2\!\left(\frac{m\pi}{M_E+1}\right)
		\right]^{1/2}
		\\
		&+\mathcal O_b\!\left(\frac{J^2}{\kappa^2}\right),
		\qquad1\leq m\leq M_E.
	\end{aligned}
	\label{eq:weak_entanglement_active_modes}
\end{equation}
The subscript in $\mathcal O_b$ means that the implicit constant may depend on the fixed imbalance $b$. An odd active block contains the standing wave at momentum $\pi/2$, for which the first-order pairing correction vanishes. Since $\alpha'(s_\star)=-b\neq0$, this mode produces a nearby true zero. An even block misses $\pi/2$, and pairing opens an avoided gap. Let $t_{E,\min}$ and $r_{E,\min}$ denote the time and value of this avoided minimum. The two active-block outcomes are
\begin{equation}
	\begin{aligned}
		t_E
		=&\frac{s_\star}{\kappa}
		+\mathcal O_b\!\left(\frac{J^2}{\kappa^3}\right),
		\qquad M_E\text{ odd},
		\\
		t_{E,\min}
		=&\frac{s_\star}{\kappa}
		+\mathcal O_b\!\left(\frac{J^2}{\kappa^3}\right),
		\qquad M_E\geq2\text{ even},
		\\
		r_{E,\min}
		=&\frac{2c(b)J}{\kappa}
		\sin\!\left[\frac{\pi}{2(M_E+1)}\right]
		\\
		&+\mathcal O_b\!\left(\frac{J^2}{\kappa^2}\right),
		\quad M_E\geq2\text{ even}.
	\end{aligned}
	\label{eq:weak_strong_rate_entanglement_clock}
\end{equation}
For even $M_E$, the positive $r_{E,\min}$ describes an avoided crossing, so no $t_E$ is assigned on the $\kappa^{-1}$ scale. When $M_E=0$, no covariance mode approaches zero on that scale, so the first candidate belongs to a later coherent or geometry-dependent scale. For pure loss or pure gain, $b=1$, $s_\star=\ln2$, and $c(1)=2\ln2-1$. \Cref{fig:weak_symmetry_clocks}(a) shows both outcomes under pure loss: the odd filling $N=3$ closes at $\ln2/\gamma_\ell$, and for the even filling $N=2$, where $M_E=2$, the minimum of $r_E$ approaches $c(1)J/\gamma_\ell$. Intermediate-rate root continuation, tangencies, and folds follow from the exact covariance condition in \cref{eq:weak_gaussian_equivalences} and are developed in Appendix~\ref{app:weak_entanglement_asymptotics}.

The chemical potential can be retained exactly in this strong-rate expansion. Assume $\kappa\gg|J|,|\Delta|$ with no restriction on $\mu/\kappa$, set $\omega\coloneq2\mu/\kappa$, and define
\begin{equation}
	\begin{aligned}
		c_b(\omega)
		&\coloneq
		\frac{2b\left[
			e^{i\omega s_\star}-1-i\omega/(1+b)
			\right]}
		{i\omega(1+i\omega)},
		&c_b(0)&=c(b).
	\end{aligned}
	\label{eq:weak_finite_mu_pairing_factor}
\end{equation}
The odd-block clock remains $s_\star/\kappa+\mathcal O_b[(J^2+\Delta^2)/\kappa^3]$, uniformly in $\mu$. For a nonempty even block,
\begin{equation}
	\begin{aligned}
		r_{E,\min}
		&=\frac{2|c_b(2\mu/\kappa)\Delta|}{\kappa}
		\sin\!\left[\frac{\pi}{2(M_E+1)}\right]
		\\
		&\quad+\mathcal O_b\!\left(
		\frac{J^2+\Delta^2}{\kappa^2}
		\right).
	\end{aligned}
	\label{eq:weak_finite_mu_even_mass}
\end{equation}
Rapid chemical-potential rotation suppresses this minimum by phase averaging without moving the local half-filling time. When $\Delta=0$, the leading minimum vanishes and higher-order hopping processes determine whether the even branch closes or remains avoided. Appendix~\ref{app:weak_finite_mu} gives the uniform expansion and its limiting regimes. Numerical evaluation of \cref{eq:weak_covariance_solution} confirms the odd-block clock and \cref{eq:weak_finite_mu_even_mass}. At $\mu=0.1J$, $L=20$, $L_A=10$, and $\gamma_\ell=20J$, the odd fillings $N=1,3,5$ close at $\ln2/\gamma_\ell$, and the minima of the even fillings $N=2,4,6$ agree with \cref{eq:weak_finite_mu_even_mass} to within $0.4\%$. At $\gamma_\ell=40J$, $L=8$, and $L_A=6$, the filling $N=3$ still closes at $\ln2/\gamma_\ell$ up to $\mu=100J$, and the minimum for $N=2$ agrees to within $0.2\%$ although it has dropped by $30\%$ at $\mu=100J$.

\subsection{Twisted negativity Hamiltonian}
\label{subsec:tnh_weak}

\emph{Parity rule for the negativity gap.} The twisted spectrum is built from the full transformed covariance in \cref{eq:weak_partial_time_reversal_covariance}, so both sides of the cut enter it. The two sum rules for the eigenvalues $\eta_k$ of $T_A$ in \cref{eq:summary_parity_sums}, with $p_k$ the global parity of the $k$th eigenvector, hold for every state that commutes with $P$, Gaussian or not, because each trace selects a single Majorana monomial of $\rho$. The sum $\sum_k\eta_k=\operatorname{Tr}(\rho^{R_A}P_A)$ selects $P_A=(-i)^{L_A}w_1\cdots w_{2L_A}$, which \cref{eq:partial_time_reversal} multiplies by $i^{2L_A}=(-1)^{L_A}$, and $\sum_kp_k\eta_k=\operatorname{Tr}(T_AP)=\operatorname{Tr}(\rho^{R_A}P_B)$ selects $P_B$, which has no factor in $A$ and is left unchanged. Gaussianity turns both sums into products.

\begin{figure*}[t]
	\centering
	\includegraphics[width=0.49\textwidth]{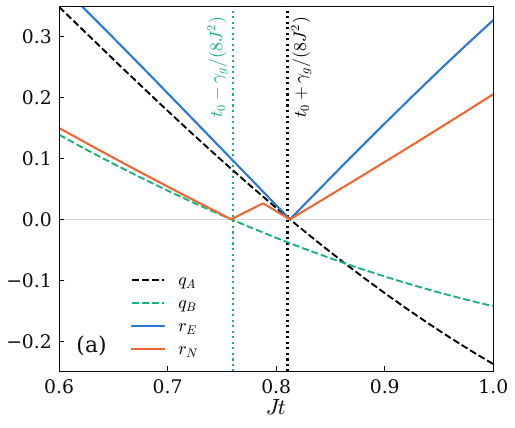}\hfill
	\includegraphics[width=0.49\textwidth]{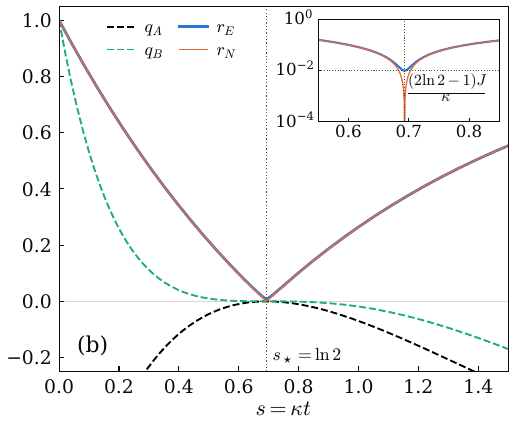}
	\caption{The negativity gap closes wherever $q_A$ or $q_B$ vanishes. Exact many-body evolution of \cref{eq:weak_lindblad}, which tests the rule without assuming Gaussianity, under pure gain, $\gamma_\ell=0$, at the compact point $\mu=0$, $\Delta=J$, with the first $N$ sites filled and $A$ the first $L_A$ sites. Solid lines show $r_E$ and $r_N$ of \cref{eq:normalized_spectral_gaps}, with $T_A$ of \cref{eq:partial_time_reversal,eq:twisted_operator}, and dashed lines $q_A$ and $q_B$ (\cref{eq:subsystem_parities}). \textbf{(a)} Weak gain, $\gamma_g=0.2J$, for $L=8$, $L_A=4$, and $N=4$. The parity $q_B$ vanishes first, before $t_0=\pi/(4J)$, and $r_N$ closes there; $q_A$ vanishes later, where $r_E$ closes and $r_N$ closes again. Dotted lines mark the first-order predictions $t_0-\gamma_g/(8J^2)$ for $q_B$ (\cref{eq:weak_bulk_negativity_clock}) and $t_0+\gamma_g/(8J^2)$ for $q_A$ (\cref{eq:weak_rate_entanglement_clock}). \textbf{(b)} Strong gain, $\gamma_g=40J$, for $L=8$, $L_A=5$, and $N=3$, against $s=\kappa t$ with $\kappa=\gamma_g$; $r_E$ and $r_N$ coincide on the linear scale. The three empty sites of $B$ form an odd active block, which drives $q_B$ through zero at $s_\star=\ln2$ (dotted; \cref{eq:weak_local_half_occupation}) and closes the negativity gap (\cref{eq:weak_strong_rate_negativity_clock}). The two empty sites of $A$ form an even block, so $q_A$ only approaches zero and the entanglement gap stays open with minimum $r_{E,\min}\approx(2\ln2-1)J/\kappa$ (dotted line in the logarithmic inset; \cref{eq:weak_strong_rate_entanglement_clock}).}
	\label{fig:weak_negativity_clocks}
\end{figure*}

The operator $T_A$ is Hermitian by \cref{eq:twisted_operator}, and it is Gaussian because $\rho^{R_A}$ and $P_A$ are Gaussian operators and products of Gaussian operators are Gaussian \cite{Bravyi2005May,Shapourian2017Apr,Shapourian2019Sep}. Since $T_A$ is Hermitian, its normalized covariance $(i/2)\operatorname{Tr}(T_A[w_m,w_n])/\operatorname{Tr}T_A$ is real and antisymmetric wherever $\operatorname{Tr}T_A\neq0$, and a real orthogonal matrix $O_N$ with $\det O_N=+1$ brings it to canonical form, as $O_A$ does for $\Gamma_A$ in \cref{eq:weak_canonical_covariance}. The rotated operators $\widehat w=O_Nw$ define $L$ canonical fermion modes with occupations $\widehat n_m$, where $1-2\widehat n_m=-i\widehat w_{2m-1}\widehat w_{2m}$, and Wick's theorem, which $T_A$ obeys as a Gaussian operator, makes $T_A$ a product of one-mode factors. Let $\eta_m^+$ and $\eta_m^-$ be the real values of the factor of mode $m$ at $\widehat n_m=0$ and $\widehat n_m=1$, fixed up to rescalings whose product is one. At the zeros of $\operatorname{Tr}T_A=(-1)^{L_A}q_A$ the normalized covariance diverges, but the product form survives as a limit, because $O_N$ ranges over a compact group and each factor can be normalized by $\max(|\eta_m^+|,|\eta_m^-|)$. At every time, therefore,
\begin{equation}
	\begin{aligned}
		T_A&=\prod_{m=1}^{L}
		\left[\eta_m^+(1-\widehat n_m)+\eta_m^-\widehat n_m\right],
		\\
		(-1)^{L_A}q_A&=\prod_{m=1}^{L}
		\left(\eta_m^++\eta_m^-\right),
		\\
		q_B&=\prod_{m=1}^{L}
		\left(\eta_m^+-\eta_m^-\right),
	\end{aligned}
	\label{eq:weak_twisted_factorization}
\end{equation}
where the last line uses $P=\prod_m(1-2\widehat n_m)$, which holds because $\det O_N=+1$. The largest singular value $s_0$ is the product of $\max(|\eta_m^+|,|\eta_m^-|)$ over all modes, and $s_1$ replaces one factor by the smaller modulus, in the mode where the two moduli are closest in ratio. Hence
\begin{equation}
	\begin{aligned}
		r_N&=\min_m
		\frac{\bigl||\eta_m^+|-|\eta_m^-|\bigr|}
		{|\eta_m^+|+|\eta_m^-|},
		\\
		s_0=s_1&\Longleftrightarrow q_Aq_B=0,
		\\
		t_N&=\min\left(t_E^A,t_E^B\right)\leq t_E.
	\end{aligned}
	\label{eq:weak_negativity_parity_rule}
\end{equation}
The first line is the negativity counterpart of $r_E=\nu_A$ in \cref{eq:weak_gaussian_entanglement_gap}. The second follows because $|\eta_m^+|=|\eta_m^-|$ holds exactly when $\eta_m^+=-\eta_m^-$, which makes one factor of $q_A$ vanish, or $\eta_m^+=\eta_m^-$, which makes one factor of $q_B$ vanish. In the third, $t_E^A\equiv t_E$, and $t_E^B$ is the first closing of the entanglement gap of $\rho_B$, which by \cref{eq:weak_gaussian_equivalences} applied to $B$ is the first zero of $q_B$. The negativity gap therefore closes as soon as either side of the cut balances its parity. \Cref{fig:weak_negativity_clocks}, computed from the full density matrix without the Gaussian formulas, shows both cases. At a zero of $q_A$ the two meeting levels of $T_A$ have opposite signs, and at a zero of $q_B$ they have the same sign; in both cases they have opposite global parity, because changing one $\widehat n_m$ flips $P$. For a state of definite global parity $p$, such as a pure state, $q_B=pq_A$, and the rule reduces to $t_N=t_E$ of \cref{eq:pure_state_clock_relation}. Each loss or gain jump flips the global parity, so the parities of $A$ and $B$ evolve separately and $B$ can balance first. The rule holds for every Gaussian state, so $t_N\leq t_E$ at every rate for every quadratic Hamiltonian, Gaussian initial state, and Gaussianity-preserving bath. Density dephasing makes the state non-Gaussian, and the rule then fails: for a cut away from the chain ends, weak dephasing closes both gaps before the parity balance (\cref{eq:summary_weak_dephasing}).

\emph{Weak dissipation at the compact point.} At $t_0$, the two equal reduced-state eigenvalues generate four equal leading twisted singular values through \cref{eq:pure_state_clock_relation}. For $\kappa/J\ll1$ at fixed $\xi/\kappa$, only first-order changes within this fourfold degeneracy need to be compared. Loss and gain on sites farther from the cut shift all four singular values equally and therefore cannot split them. The contribution controlled by $\kappa$ likewise shifts all four equally, so it cancels from their differences and leaves the crossing time unchanged at first order. Only the imbalance $\xi$, acting through the cut bond and, when present, an adjacent unpaired Majorana, can split the four values. It produces a bulk-cut offset of magnitude $|\xi|/8$ and a physical-edge offset of magnitude $|\xi|/4$, with signs fixed by the initial occupations of the two sites next to the cut. Let $\mathcal B_{L_A,L_B,N}(\xi)$ denote the resulting finite set of first-order offsets.

By the parity rule, these offsets are the first-order shifts of the zeros of $q_A$ and $q_B$. The zero of $q_A$ shifts as in \cref{eq:weak_rate_entanglement_clock}. Reflecting the chain, $a_j\mapsto b_{L+1-j}$ and $b_j\mapsto a_{L+1-j}$, together with complex conjugation in the occupation basis, which restores the sign of $H$ that the reflection alone reverses, maps the compact-point evolution onto itself with $\gamma_\ell$ and $\gamma_g$ exchanged, reverses and complements the occupation pattern, and turns $B$ into a left subsystem of $L_B$ sites. Since $\xi(1-2n)$ is unchanged, \cref{eq:weak_rate_entanglement_clock} for that subsystem gives the zero of $q_B$, whose shift is set by site $L_A+1$, and
\begin{equation}
	\begin{aligned}
		t_N
		&=t_0+\frac{1}{J^2}
		\min\mathcal B_{L_A,L_B,N}(\xi)
		+\mathcal O\!\left(\frac{\kappa^2}{J^3}\right),
		\\
		\mathcal B_{L_A,L_B,N}(\xi)
		&=\left\{\xi C_E,\,\xi C_E^B\right\},
		\\
		C_E^B&=
		\begin{cases}
			\dfrac{1-2n_L}{4},&L_B=1,\\[5pt]
			\dfrac{1-2n_{L_A+1}}{8},&L_B\geq2.
		\end{cases}
	\end{aligned}
	\label{eq:weak_rate_negativity_clock}
\end{equation}
Appendix~\ref{app:weak_negativity_asymptotics} derives \cref{tab:app_weak_negativity_branches} from the first-order splitting of the four leading singular values, without the parity rule, and every row agrees with this set. For the left-filled state, $n_{L_A+1}=n_L=0$, so $C_E^B=1/8$, or $1/4$ when $B$ is a single site. The minimum selects the first closing, as \cref{eq:spectral_clocks} prescribes, and the other element gives the later one. Later finite-size branches and their fold structure are recorded in Appendix~\ref{app:weak_negativity_asymptotics}; their associated nonanalyticities can smear in the thermodynamic free-fermion limit \cite{Kehrein2024Jun,Jha2025Jun,Li2025Jul}.

For a cut separated from both physical boundaries, the general branch set reduces to
\begin{equation}
	\begin{aligned}
		t_N
		&=
		\begin{cases}
			t_0+\dfrac{\xi}{8J^2}
			+\mathcal O(\kappa^2/J^3),&N<L_A,\\[6pt]
			t_0-\dfrac{|\xi|}{8J^2}
			+\mathcal O(\kappa^2/J^3),&N=L_A.
		\end{cases}
	\end{aligned}
	\label{eq:weak_bulk_negativity_clock}
\end{equation}
For $N<L_A$, sites $L_A$ and $L_A+1$ both start empty, so the two parities balance together and $t_N=t_E$ to first order. For $N=L_A$, site $L_A$ starts filled and site $L_A+1$ empty, so the two zeros move in opposite directions: loss makes $A$ balance first, $t_N=t_E<t_0$, while gain makes $B$ balance first, $t_N<t_0<t_E$ (\cref{fig:weak_negativity_clocks}(a)). When $A$ or $B$ is a single site, its shift doubles to $|\xi|/(4J^2)$, because the unpaired end Majorana then sits next to the cut.

\emph{Strong dissipation.} We retain the compact point and the strong-rate variables in \cref{eq:weak_strong_rate_variables}. By \cref{eq:weak_negativity_parity_rule} the negativity clock is the earlier of the entanglement clocks of $A$ and $B$, so the active blocks selected by \cref{eq:weak_local_kinetics} now matter on both sides of the cut. For the left-filled state, let $\mathcal L_N$ denote the list of their positive lengths,
\begin{equation}
	\begin{aligned}
		\mathcal L_N&=
		\begin{cases}
			(N),&\xi>0,\ N>0,\\
			\varnothing,&\xi>0,\ N=0,\\
			(L_A-N,L_B),&\xi<0.
		\end{cases}
	\end{aligned}
	\label{eq:weak_negativity_active_clusters}
\end{equation}
Here $L_B=L-L_A$, and each entry $\ell$ of $\mathcal L_N$ is the number of consecutive active sites in one region; $(N)$ denotes a one-entry list, and zero-length entries are omitted. Under loss dominance, the initially occupied sites $1,\ldots,N$ form the only active block. Under gain dominance, the empty sites $N+1,\ldots,L_A$ in $A$ and the $L_B$ empty sites in $B$ give block lengths $L_A-N$ and $L_B$. Although these two regions meet at the cut, the partially reversed cut-bond correction changes the relevant singular-value magnitudes only at second order, so they remain separate in the first-order problem. The parity rule makes this separation exact for the closings: each is a zero of $q_A$, fixed by the covariance of $A$, or of $q_B$, fixed by that of $B$. At $s=s_\star$, each block therefore has the standing-wave modes derived in \cref{eq:weak_entanglement_active_modes}. For a block of length $\ell$, let $m=1,\ldots,\ell$ label these modes and define
\begin{equation}
	\begin{aligned}
		\nu_{\ell,m}(s)
		={}&
		\left[
		\alpha(s)^2
		+\frac{4c(b)^2J^2}{\kappa^2}
		\cos^2\!\left(\frac{m\pi}{\ell+1}\right)
		\right]^{1/2}
		\\
		&+\mathcal O_b\!\left(\frac{J^2}{\kappa^2}\right),
		\qquad
		\ell\in\mathcal L_N,\quad1\leq m\leq\ell,
		\\
		r_N(s)
		={}&\min_{\ell\in\mathcal L_N}
		\min_{1\leq m\leq\ell}
		\nu_{\ell,m}(s).
	\end{aligned}
	\label{eq:weak_negativity_active_modes}
\end{equation}

In \cref{eq:weak_negativity_active_modes}, the outer minimum selects an active block and the inner minimum selects one of its standing-wave modes. If a block length $\ell$ is odd, the allowed value $m=(\ell+1)/2$ makes the cosine vanish, so that block supports a true zero near $s_\star$. If every entry of $\mathcal L_N$ is even, no allowed mode has a vanishing cosine. In that case, define $\ell_{\max}\coloneq\max\mathcal L_N$, let $t_{N,\min}$ denote the time at which the avoided minimum occurs, and let $r_{N,\min}$ denote its value. For $\mathcal L_N\neq\varnothing$, the alternatives are
\begin{equation}
	\begin{aligned}
		t_N
		=&\frac{s_\star}{\kappa}
		+\mathcal O_b\!\left(\frac{J^2}{\kappa^3}\right),
		\qquad\text{at least one entry odd},
		\\
		t_{N,\min}
		=&\frac{s_\star}{\kappa}
		+\mathcal O_b\!\left(\frac{J^2}{\kappa^3}\right),
		\qquad\text{all entries even},
		\\
		r_{N,\min}
		=&\frac{2c(b)J}{\kappa}
		\sin\!\left[\frac{\pi}{2(\ell_{\max}+1)}\right]
		\\
		&+\mathcal O_b\!\left(\frac{J^2}{\kappa^2}\right),
		\quad\text{all entries even}.
	\end{aligned}
	\label{eq:weak_strong_rate_negativity_clock}
\end{equation}
Here $t_{N,\min}$ is the location of an avoided minimum, not a negativity-gap closing. For the left-filled state in \cref{eq:weak_negativity_active_clusters}, the nonempty list $\mathcal L_N$ contains one or two positive block lengths. Under loss dominance, $\mathcal L_N=(N)$ for $N>0$, so odd $N$ gives a local closing and positive even $N$ an avoided crossing. Under gain dominance, both $L_A-N$ and $L_B$ matter. For example, $L_A=4$, $L=7$, and $N=2$ give $\mathcal L_N=(2,3)$, so the odd block in $B$ closes the negativity gap even though the active block in $A$ is even (\cref{fig:weak_negativity_clocks}(b)). By contrast, $L_A=5$, $L=9$, and $N=3$ give $\mathcal L_N=(2,4)$; both entries are even, and the longer block $\ell_{\max}=4$ sets the avoided minimum. If $\mathcal L_N=\varnothing$, no twisted singular-value branch approaches degeneracy on the $\kappa^{-1}$ scale. When both blocks are odd, they share the leading zero at $s_\star$ and generally split at order $J^2/\kappa^2$ in scaled time. The active-block expansion and its higher-order splittings are derived in Appendix~\ref{app:weak_negativity_asymptotics}.

\emph{Comparison of the temporal criticalities.} The parity rule gives $t_N\leq t_E$ at every rate, and the results above say when the inequality is strict. Balanced loss and gain lock $t_E=t_N=t_0$ at the compact point for every finite balanced rate (\cref{eq:weak_balanced_clock_locking}). With weak imbalance, both clocks stay within $\mathcal O(\kappa/J^2)$ of $t_0$. The zero of $q_A$, which fixes $t_E$, moves according to the initial occupation of site $L_A$ (\cref{eq:weak_rate_entanglement_clock}); the zero of $q_B$ moves according to that of site $L_A+1$, and $t_N$ is the earlier of the two (\cref{eq:weak_rate_negativity_clock}). For a bulk cut the two clocks agree to first order when $N<L_A$, whereas for $N=L_A$ gain lets $B$ balance first and separates them by $|\xi|/(4J^2)$.

Strong rates replace the coherent time $t_0$ by the local half-filling time $s_\star/\kappa$. Equations~\eqref{eq:weak_strong_rate_entanglement_clock} and \eqref{eq:weak_strong_rate_negativity_clock} show the same odd-even rule: an odd active block supports a true crossing near $s_\star/\kappa$, an even active block has an avoided minimum of order $J/\kappa$, and without an active block there is no candidate on this scale. The active lengths in \cref{eq:weak_entanglement_active_length,eq:weak_negativity_active_clusters} explain the difference between the spectra. Under loss dominance, $B$ starts empty and has no active site, so both spectra test the same occupied block of length $N$ and the two clocks agree on this scale. Under gain dominance, the entanglement Hamiltonian tests only the empty block of length $L_A-N$ in $A$, while the twisted negativity Hamiltonian also tests the empty block of length $L_B$ in $B$. An odd block in $B$ can therefore close the negativity gap even when the entanglement gap is avoided or has no local candidate, as in \cref{fig:weak_negativity_clocks}(b).

The weak- and strong-rate formulas are the two limits of one finite-chain crossover, shown in \cref{fig:weak_symmetry_clocks}(a). For $\kappa/J\ll1$, the displacement from $t_0=\mathcal O(J^{-1})$ is linear in the rates and has size $\mathcal O(\kappa/J^2)$. For $\kappa/J\gg1$ at fixed $b>0$, a local crossing instead follows the inverse-rate power law $t=\mathcal O(\kappa^{-1})$, while an avoided normalized gap scales as $\mathcal O(J/\kappa)$. This change of scaling is not a phase transition: the exact covariance in \cref{eq:weak_covariance_solution} remains smooth, while the first closing moves continuously, jumps to a later closing, or merges with another closing at a fold that leaves an avoided minimum. Along this crossover the two clocks separate even under pure loss, which leaves $B$ without active sites. For $N=2$ the zero of $q_B$ comes before that of $q_A$ near the fold and outlives it, so between the two folds only the negativity gap closes (\cref{fig:weak_symmetry_clocks}(a)). A sweep over the fillings of the left half of a $100$-site chain at $\mu=0.1J$ shows the same odd-even rule, folds, and jumps, and, for large even fillings, a range of rates in which only the negativity gap closes (Appendix~\ref{app:parameter_sweeps}).

\emph{Geometric decoherence time.} We use the geometric decoherence time only to determine whether the logarithmic negativity begins to decrease before the spectral criticalities. Appendix~\ref{app:geometric_decoherence} derives the complete filling and boundary classification. At sufficiently weak rates it proves $t_N-\tau_d^g\geq\kappa/(4J^2)+\mathcal O(\kappa^2/J^3)$ (\cref{eq:app_weak_geo_ordering}), so $\tau_d^g<t_N\leq t_E$ for every filling and boundary. At strong rates, if the first maximum of $E_N$ at which $S_{1/2}$ still rises lies at a scaled time $s_g^{\mathrm{loc}}=\mathcal O(1)$, then $\tau_d^g=s_g^{\mathrm{loc}}/\kappa+\mathcal O_b(J^2/\kappa^3)$; otherwise $\tau_d^g$ lies beyond the $\kappa^{-1}$ scale (\cref{eq:app_weak_geo_local_time,tab:weak_strong_geometric_branches}). For pure gain with $N<L_A$, for example, $\tau_d^g\simeq0.431/\gamma_g$ precedes any closing near $\ln2/\gamma_g$, whereas for pure loss $E_N$ has no such maximum on the $\kappa^{-1}$ scale.

Under loss and gain, Gaussianity makes both closings parity balances, with $t_E$ at the first zero of $q_A$ and $t_N\leq t_E$ at the first zero of $q_Aq_B$. The push of the bath toward the vacuum or the filled chain decides how these zeros move, while $\tau_d^g$ follows from the entropy-negativity trajectory. At weak rates the single bond across the cut keeps all three times close to $t_0$, and the negativity turns over first. At strong rates the active sites reach half filling together, and pairing decides whether each parity crosses zero there. The parity of $B$ can then vanish when that of $A$ cannot, and the position of $\tau_d^g$ relative to the closings depends on the branch.

\section{Temporal Criticalities under Strong Symmetry: Non-Gaussian Dephasing}
\label{sec:temporal_criticalities_strong}

\begin{figure*}[t]
	\centering
	\includegraphics[width=0.49\textwidth]{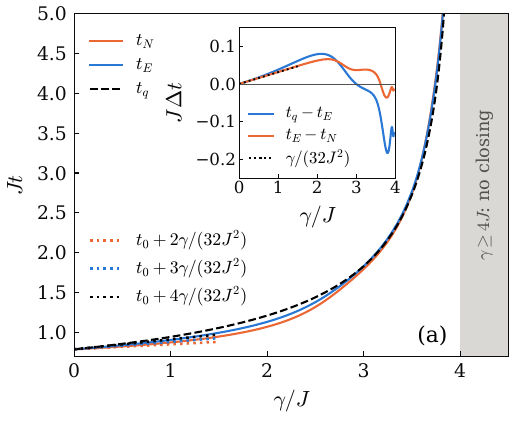}\hfill
	\includegraphics[width=0.49\textwidth]{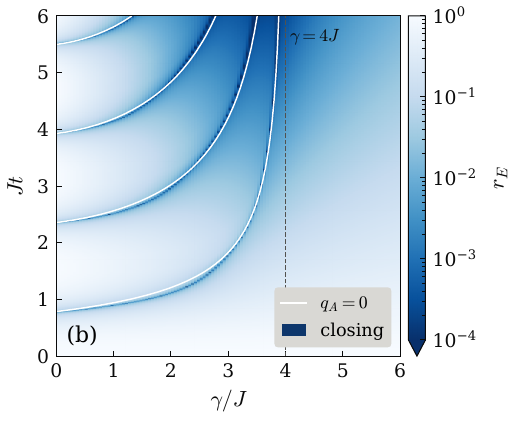}
	\caption{Closings and parity balance from weak dephasing to critical damping. Exact many-body evolution of \cref{eq:strong_lindblad}, from the eigenmodes of the Liouvillian, at the compact point $\mu=0$, $\Delta=J$, for $A$ the first $L_A=3$ sites and any product-Fock initial state (\cref{eq:strong_pattern_universality}); $\rho_A$ does not depend on $L_B$ (\cref{eq:strong_bath_length_universality}). \textbf{(a)} First closings $t_E$ and $t_N$ of \cref{eq:spectral_clocks} (solid), with $T_A$ of \cref{eq:partial_time_reversal,eq:twisted_operator} and $t_N$ for $L=8$, and the parity balance $t_q$ of \cref{eq:strong_parity_clock} (dashed). Dotted lines show the first-order clocks $t_0+2\gamma/(32J^2)$ for $t_N$ (\cref{eq:strong_weak_negativity_clock}), $t_0+3\gamma/(32J^2)$ for $t_E$ (\cref{eq:strong_weak_entanglement_clock}), and $t_0+4\gamma/(32J^2)$ for $t_q$ (\cref{eq:strong_weak_parity_clock}), with $t_0=\pi/(4J)$. All three clocks diverge together as $\gamma\to4J^-$, and at the sampled rates from $4.05J$ to $20J$ (shaded region $\gamma\geq4J$) neither gap closes up to $Jt=80$. Inset: $t_q-t_E$ and $t_E-t_N$ grow from zero with the common slope $1/(32J^2)$ (dotted; \cref{eq:strong_weak_parity_separation,eq:strong_weak_clock_splitting}) and change sign, so that close to $4J$ the order is $t_q<t_E<t_N$ for this geometry. \textbf{(b)} Color: $r_E$ of \cref{eq:normalized_spectral_gaps} in the plane of dephasing rate and time, on a logarithmic scale clipped below $10^{-4}$. The closings (dark valleys) run close to, but not on, the zero lines of $q_A$ (white; \cref{eq:strong_parity_clock}), where a Gaussian $\rho_A$ would place them (\cref{eq:weak_gaussian_equivalences}). None is created or annihilated as $\gamma$ grows, unlike under loss (\cref{fig:weak_symmetry_clocks}(b)), and each recedes to infinite time as $\gamma\to4J^-$ (vertical line).}
	\label{fig:strong_symmetry_clocks}
\end{figure*}

Density dephasing respects parity in every jump but makes the state non-Gaussian. The parity rules of \cref{sec:temporal_criticalities_weak} no longer hold, and each closing must be read off a many-body spectrum. At the compact point the bond terms commute, so every kick history is a product of bond rotations by random angles (\cref{eq:strong_random_unitary}). Both spectra are independent of the initial occupation pattern, and that of $\rho_A$ also of $L_B$ (\cref{eq:strong_pattern_universality,eq:strong_bath_length_universality}). In each history the bond across the cut splits the state into two branches of opposite subsystem parity. Traced over $B$, the branches form the two parity blocks of $\rho_A$ (\cref{eq:strong_parity_block_decomposition}); partially time reversed, their mixture forms a block $T_A^+$ of $T_A$ that holds its largest level in every case we computed, and their coherence a block $T_A^-$ that carries all of the negativity (\cref{eq:strong_twisted_blocks}). Every closing we find is therefore a crossing of the largest levels of two blocks of opposite subsystem parity, whereas the parity balance $t_q$, the first zero of $q_A$, compares only their total weights. At $t_E$ the dominant state of $A$ changes parity, and at $t_N$ the leading eigenvalue of $T_A$ changes sign at fixed global parity. The parity obeys a damped-oscillator equation, so $t_q$ is known exactly and is finite only for $\gamma<4J$ (\cref{eq:strong_parity_clock}). For a cut away from the chain ends, weak dephasing delays $t_N$, $t_E$, and $t_q$ from $t_0$ in the ratio $2:3:4$, so both closings precede the parity balance, at which the Gaussian parity rules would place them (\cref{eq:strong_weak_parity_separation,eq:strong_weak_clock_splitting}; \cref{fig:strong_symmetry_clocks}(a) and \cref{fig:strong_clock_traces}(a)). In exact numerics for short chains with at least two sites on each side, stronger dephasing moves the parity balance ahead of both closings, and all three times recede to infinity as the cut bond approaches critical damping, $\gamma\to4J^-$ (\cref{fig:strong_symmetry_clocks}). For $\gamma\gg J$, independent pair flips on the bonds make both spectra binomial at leading order in $J/\gamma$, and both gaps equal $\ln\coth(2J^2t/\gamma)$, which stays positive at all finite times (\cref{eq:strong_zeno_bond_flips,eq:strong_zeno_entanglement_clock,eq:strong_zeno_negativity_clock}). At weak dephasing the negativity turns over before either closing in every geometry evaluated (\cref{eq:strong_weak_geometric_clock}); at strong dephasing it turns over at $\tau_d^g\simeq(2/\gamma)\ln(\gamma/2J)$, long before the occupations move (\cref{eq:strong_time_scale_hierarchy}; \cref{fig:strong_clock_traces}(b)). Strong symmetry itself enters through the conserved global parity, which ties $q_B$ to $q_A$, splits $T_A$ into a separable block and a coherence block, and leaves one stationary state in each parity sector (\cref{eq:strong_stationary_state}). \Cref{subsec:eh_strong} treats the entanglement clock and the parity oscillator, and \cref{subsec:tnh_strong} the negativity clock, the comparison of the temporal criticalities, and the geometric decoherence time.

The dephasing jumps of \cref{eq:strong_lindblad} are quadratic in the fermions, so two-point functions still obey a closed equation but no longer fix the mixed state or its spectra (Appendix~\ref{app:strong_entanglement_asymptotics}). At the compact point of \cref{eq:compact_majorana_point}, a Jordan-Wigner transformation maps the bonds onto commuting spin couplings \cite{Jordan1928Sep,Lieb1961Dec}. With $Z_j=1-2n_j$, the parity of site $j$, and $X_j$ the Pauli operator that flips its occupation in the spin representation, the Hamiltonian and the dephasing terms become
\begin{equation}
	\begin{aligned}
		H&=-J\sum_{j=1}^{L-1}X_jX_{j+1},
		\\
		\dot\rho
		&=iJ\left[\sum_{j=1}^{L-1}X_jX_{j+1},\rho\right]
		+\frac{\gamma}{4}\sum_{j=1}^{L}
		\left(Z_j\rho Z_j-\rho\right).
	\end{aligned}
	\label{eq:strong_spin_lindblad}
\end{equation}
All bond operators $X_jX_{j+1}$ commute. Each dephasing term generates kicks $\rho\to Z_j\rho Z_j$ at the times of a Poisson process with rate $\gamma/4$. Let $N_j^{\mathrm d}(t')$ count the kicks at site $j$ up to time $t'$, and define the corresponding sign process and accumulated bond time by
\begin{equation}
	\begin{aligned}
		\sigma_j(t')&\coloneq(-1)^{N_j^{\mathrm d}(t')},
		\\
		\Theta_j(t)&\coloneq
		\int_0^t dt'\,
		\sigma_j(t')\sigma_{j+1}(t').
	\end{aligned}
	\label{eq:strong_telegraph_variables}
\end{equation}
Here $\sigma_j$ is a symmetric random telegraph process: it starts at $+1$ and switches sign at every kick on site $j$ \cite{kac1974telegraph,Apostolov2024Aug}. A kick on site $j$ reverses the sign of the bonds that touch $j$ and acts on a product-Fock state only by a phase. Moving all kicks to $t'=0$ therefore makes the sign of a bond at time $t'$ count the kicks after $t'$ rather than before it. A Poisson process has the same statistics read backward in time, so at each fixed $t$ the state is exactly the average of the pure states
\begin{equation}
	\begin{aligned}
		|\psi_{\boldsymbol\sigma}(t)\rangle
		&=
		\prod_{j=1}^{L-1}
		\exp\!\left[
		iJ\Theta_j(t)X_jX_{j+1}
		\right]
		|\boldsymbol n\rangle,
		\\
		\rho(t)
		&=
		\mathbb{E}_{\boldsymbol\sigma}
		\left[
		|\psi_{\boldsymbol\sigma}(t)\rangle
		\langle\psi_{\boldsymbol\sigma}(t)|
		\right],
	\end{aligned}
	\label{eq:strong_random_unitary}
\end{equation}
where $\mathbb{E}_{\boldsymbol\sigma}$ denotes the average over kick histories $\boldsymbol\sigma=(\sigma_1,\ldots,\sigma_L)$ and $|\boldsymbol n\rangle$ is any product-Fock state from \cref{eq:general_fock_state}. Equation~\eqref{eq:strong_random_unitary} is an exact random-unitary representation \cite{Dalibard1992Feb,PlenioKnight1998Jan,Daley2014Jul}, derived in Appendix~\ref{app:strong_exact_reduction}. Each history produces a pure Gaussian state, because the bond gates are Gaussian unitaries acting on a Fock state; their average, whose spectra define the clocks, is not Gaussian.

The same structure makes the spectra independent of the initial occupation pattern. Any two patterns are related by a product of local $X_j$ operators. At the compact point these operators commute with the Hamiltonian, and density dephasing does not distinguish particles from holes, $\mathcal D[1-n_j]=\mathcal D[n_j]$. Consequently,
\begin{equation}
	\begin{aligned}
		\operatorname{spec}\rho_A
		(t;\boldsymbol n)
		&=
		\operatorname{spec}\rho_A
		(t;\boldsymbol n'),
		\\
		\{s_k(t;\boldsymbol n)\}
		&=
		\{s_k(t;\boldsymbol n')\},
	\end{aligned}
	\label{eq:strong_pattern_universality}
\end{equation}
for any two product-Fock patterns $\boldsymbol n$ and $\boldsymbol n'$ at equal $J$, $\gamma$, $t$, and bipartition $(L_A,L_B)$. Thus $t_E$, $t_N$, $E_N$, and $\tau_d^g$ do not depend on the filling or on the occupation pattern. The entanglement spectrum is, in addition, independent of the size of $B$:
\begin{equation}
	\begin{aligned}
		\operatorname{spec}\rho_A(t;L_B)
		&=
		\operatorname{spec}\rho_A(t;L_B'),
		\qquad L_B,L_B'\geq1.
	\end{aligned}
	\label{eq:strong_bath_length_universality}
\end{equation}
Only the first site of $B$ enters the reduced state, because the gates inside $B$ drop out of the partial trace, whereas $T_A$ keeps both sides of the cut and retains $L_B$ as a parameter. Appendix~\ref{app:strong_exact_reduction} proves both statements, including the twisted line of \cref{eq:strong_pattern_universality}~\footnote{No Gaussian assumption is used here: $R_A$ is the full operator map defined in \cref{eq:partial_time_reversal}. For the unitary $X_{\mathcal I}$ that flips the occupations on a set of sites $\mathcal I$, the required identity is $(X_{\mathcal I}\rho X_{\mathcal I})^{R_A}=X_{\mathcal I}\rho^{R_A}X_{\mathcal I}$. Together with the parity twist, this preserves the complete twisted singular spectrum; see \cref{eq:app_strong_ptr_covariance_conjugation,eq:app_strong_twisted_conjugation}.}. Neither property, nor the kick solution itself, requires strong symmetry; all three hold for balanced loss and gain as well (\cref{sec:summary}).

Strong symmetry leaves one stationary state in each parity sector, and the initial state selects it. For a connected open chain with $J\neq0$ and $\gamma>0$, let $\Pi_p=(\mathbf{1}+pP)/2$ project onto the sector of the conserved global parity $p$ of the initial state. We write $\rho_{\infty,p}$ for the unique stationary state of the full chain in that sector, and $\rho_{A,\infty}\equiv\operatorname{Tr}_B\rho_{\infty,p}$ for its reduction to $A$ \cite{Evans1977Oct,Frigerio1978Oct,Buca2012Jul,Albert2014Feb}. Because $L_B\geq1$, the latter is independent of $p$:
\begin{equation}
	\begin{aligned}
		\rho_{\infty,p}
		&=\frac{\Pi_p}{2^{L-1}},
		\\
		\rho_{A,\infty}
		&=\frac{\mathbf{1}_A}{2^{L_A}}.
	\end{aligned}
	\label{eq:strong_stationary_state}
\end{equation}
The global stationary state is the uniform mixture over all Fock configurations of parity $p$, and each configuration of $A$ is completed to parity $p$ by exactly half of the $2^{L_B}$ configurations of $B$, which gives the uniform $\rho_{A,\infty}$ (Appendix~\ref{app:strong_exact_reduction}). Hence the entanglement gap approaches zero at long times, while the stationary state, which is diagonal, has vanishing logarithmic negativity. The memory of $p$ is the one feature of this state that requires strong symmetry, and it leaves classical correlations across the cut that none of the clocks detects.

\subsection{Entanglement Hamiltonian}
\label{subsec:eh_strong}

The random-unitary representation can be reduced exactly across the cut. Write the initial product-Fock state as $|\boldsymbol n\rangle=|\boldsymbol n_A\rangle\otimes|\boldsymbol n_B\rangle$, and define the internal evolution in $A$ and the random angle of the cut bond by
\begin{equation}
	\begin{aligned}
		U_A(t)
		&\coloneq
		\prod_{j=1}^{L_A-1}
		\exp\!\left[
		iJ\Theta_j(t)X_jX_{j+1}
		\right],
		\\
		\phi(t)&\coloneq J\Theta_{L_A}(t).
	\end{aligned}
	\label{eq:strong_cut_variables}
\end{equation}
For $L_A=1$, the product defining $U_A$ is the identity. The gate on the bond that crosses the cut splits a product configuration into two branches,
\begin{equation}
	\begin{aligned}
		e^{i\phi X_{L_A}X_{L_A+1}}
		|\boldsymbol n_A,\boldsymbol n_B\rangle &=
		\cos\phi\,
		|\boldsymbol n_A,\boldsymbol n_B\rangle
		\\
		&+i\sin\phi\,
		X_{L_A}|\boldsymbol n_A\rangle
		\otimes
		X_{L_A+1}|\boldsymbol n_B\rangle.
	\end{aligned}
	\label{eq:strong_cut_gate}
\end{equation}
The two states of $B$ appearing here are orthogonal, so their interference disappears under the partial trace, and all gates inside $B$ act as local unitaries and drop out. This leaves the mixture of the two branches,
\begin{equation}
	\begin{aligned}
		Q_A(t)
		={}&\cos^2\phi(t)\,
		|\boldsymbol n_A\rangle
		\langle\boldsymbol n_A|
		\\
		&+\sin^2\phi(t)\,
		X_{L_A}|\boldsymbol n_A\rangle
		\langle\boldsymbol n_A|X_{L_A},
	\end{aligned}
	\label{eq:strong_cut_mixture}
\end{equation}
and the reduced state is exactly
\begin{equation}
	\begin{aligned}
		\rho_A(t)
		&=\mathbb{E}_{\boldsymbol\sigma}
		\left[
		U_A(t)Q_A(t)U_A^\dagger(t)
		\right].
	\end{aligned}
	\label{eq:strong_reduced_trajectory}
\end{equation}
The two branches have opposite subsystem parity, because $X_{L_A}$ changes the occupation of one site while $U_A$ flips sites in pairs. Define the positive operators
\begin{equation}
	\begin{aligned}
		C_A(t)
		&\coloneq
		\mathbb{E}_{\boldsymbol\sigma}
		\left[
		\cos^2\phi\,
		U_A|\boldsymbol n_A\rangle
		\langle\boldsymbol n_A|U_A^\dagger
		\right],
		\\
		S_A(t)
		&\coloneq
		\mathbb{E}_{\boldsymbol\sigma}
		\left[
		\sin^2\phi\,
		U_A|\boldsymbol n_A\rangle
		\langle\boldsymbol n_A|U_A^\dagger
		\right].
	\end{aligned}
	\label{eq:strong_parity_blocks}
\end{equation}
In a basis of $A$ ordered by subsystem parity,
\begin{equation}
	\begin{aligned}
		\rho_A
		&=
		C_A\oplus X_{L_A}S_AX_{L_A}.
	\end{aligned}
	\label{eq:strong_parity_block_decomposition}
\end{equation}
Let $c_1\geq c_2\geq\cdots$ and $d_1\geq d_2\geq\cdots$ be the eigenvalues of $C_A$ and $S_A$, respectively. The two leading eigenvalues of $\rho_A$ are then
\begin{equation}
	\begin{aligned}
		\lambda_0
		&=\max(c_1,d_1),
		\\
		\lambda_1
		&=\max\!\left\{
		\min(c_1,d_1),c_2,d_2
		\right\}.
	\end{aligned}
	\label{eq:strong_entanglement_branch_map}
\end{equation}

By \cref{eq:strong_entanglement_branch_map}, the two leading levels can meet in two ways. Either the blocks cross, $c_1=d_1$, and at a simple crossing the dominant eigenvector of $\rho_A$ changes subsystem parity, as in a Gaussian state, or one block becomes degenerate, $c_1=c_2>d_1$ or $d_1=d_2>c_1$, and two levels of the same parity meet. Every closing we find is of the first kind, including the one that continues the closed-system crossing at $t_0$.

The parity $q_A$ of \cref{eq:subsystem_parities} weights the same two blocks with opposite signs. With $N_A(0)$ the initial number of fermions in $A$,
\begin{equation}
	\begin{aligned}
		q_A(t)
		&=
		(-1)^{N_A(0)}
		\left[
		\operatorname{Tr}C_A(t)
		-\operatorname{Tr}S_A(t)
		\right].
	\end{aligned}
	\label{eq:strong_subsystem_parity}
\end{equation}
The entanglement clock compares the largest eigenvalues of the two blocks, whereas $q_A$ compares their total weights. For a one-site subsystem each block is a single level, so the two comparisons agree, and for a Gaussian $\rho_A$ they agree for every subsystem (\cref{eq:weak_gaussian_equivalences}). Here the leading eigenvalues of $C_A$ and $S_A$ depend on the full non-Gaussian state, so the two comparisons can differ. A difference between $t_E$ and $t_q$ certifies that $\rho_A$ is not Gaussian.

The parity nevertheless obeys an exact closed equation at the compact point, because the cut bond is the only bond that anticommutes with $P_A$:
\begin{equation}
	\begin{aligned}
		\ddot q_A+\gamma\dot q_A+4J^2q_A&=0,
		\\
		q_A(0)&=(-1)^{N_A(0)},
		&\dot q_A(0)&=0.
	\end{aligned}
	\label{eq:strong_parity_oscillator}
\end{equation}
This is the Rabi oscillation of the cut bond, damped by the kicks on its two sites. For $0\leq\gamma<4J$, define
\begin{equation}
	\begin{aligned}
		\omega
		&\coloneq
		\sqrt{4J^2-\frac{\gamma^2}{4}}.
	\end{aligned}
	\label{eq:strong_parity_frequency}
\end{equation}
Throughout this section, $\omega$ denotes this frequency. The parity trajectory and its first zero $t_q$ are
\begin{equation}
	\begin{aligned}
		q_A(t)
		={}&
		(-1)^{N_A(0)}e^{-\gamma t/2}
		\left[
		\cos(\omega t)
		+\frac{\gamma}{2\omega}\sin(\omega t)
		\right],
		\\
		t_q
		={}&
		\frac{
			\pi/2+\arctan[\gamma/(2\omega)]
		}{\omega},
		\qquad0\leq\gamma<4J,
		\\
		t_q
		={}&+\infty,
		\qquad\gamma\geq4J.
	\end{aligned}
	\label{eq:strong_parity_clock}
\end{equation}
At $\gamma=4J$ the oscillator is critically damped, and for $\gamma\geq4J$ the parity of $A$ decays without changing sign. For $L_A=1$ the parity balance and the entanglement closing therefore coincide exactly:
\begin{equation}
	\begin{aligned}
		L_A=1:\qquad
		t_E=t_q.
	\end{aligned}
	\label{eq:strong_one_site_entanglement_clock}
\end{equation}
For $L_A\geq2$, $t_E$ must instead be read off the level ordering of \cref{eq:strong_entanglement_branch_map}. Appendix~\ref{app:strong_entanglement_asymptotics} derives \cref{eq:strong_parity_oscillator,eq:strong_parity_clock}.

\emph{Weak dephasing.} For $\gamma\ll J$ the crossing $c_1=d_1$ at the closed-system time $t_0=\pi/(4J)$ of \cref{eq:closed_compact_results} remains simple, so ordinary eigenvalue perturbation gives
\begin{equation}
	\begin{aligned}
		t_E
		&=
		t_0+
		\begin{cases}
			\dfrac{\gamma}{8J^2},
			&L_A=1,
			\\[6pt]
			\dfrac{3\gamma}{32J^2},
			&L_A\geq2,
		\end{cases}
		+\mathcal O\!\left(
		\frac{\gamma^2}{J^3}
		\right).
	\end{aligned}
	\label{eq:strong_weak_entanglement_clock}
\end{equation}
The correction is positive and, like the whole spectrum, independent of the initial pattern. If a kick occurs at either end site of the cut bond a time $u$ before the observation, it reverses the sign of the cut generator $X_{L_A}X_{L_A+1}$ for that remaining time. In the one-kick calculation at $t_0$, this changes the cut angle $\phi$ of \cref{eq:strong_cut_variables} from $Jt_0$ to $J(t_0-2u)$. A kick at the end site in $B$ has no further effect on $\rho_A$, because evolution inside $B$ disappears under the partial trace. When $L_A\geq2$, a kick at the end site in $A$ also reverses the adjacent internal bond $X_{L_A-1}X_{L_A}$, and only the part of the state of $A$ that returns to its kick-free form, with probability $\cos^2(2Ju)$, feeds the two leading levels. Integrating the two contributions gives the coefficients $1/8$ for $L_A=1$ and $3/32$ for $L_A\geq2$ (Appendix~\ref{app:strong_entanglement_asymptotics}).

Expanding the exact parity balance $t_q$ of \cref{eq:strong_parity_clock} gives
\begin{equation}
	\begin{aligned}
		t_q
		&=
		t_0+\frac{\gamma}{8J^2}
		+\mathcal O\!\left(
		\frac{\gamma^2}{J^3}
		\right).
	\end{aligned}
	\label{eq:strong_weak_parity_clock}
\end{equation}
Consequently, for a multi-site subsystem,
\begin{equation}
	\begin{aligned}
		L_A\geq2:\qquad
		t_q-t_E
		&=
		\frac{\gamma}{32J^2}
		+\mathcal O\!\left(
		\frac{\gamma^2}{J^3}
		\right).
	\end{aligned}
	\label{eq:strong_weak_parity_separation}
\end{equation}
The total weights of the two parity blocks balance at $t_q$, their largest levels earlier, at $t_E$ (\cref{fig:strong_clock_traces}(a)), so $\rho_A$ is not Gaussian. The inset of \cref{fig:strong_symmetry_clocks}(a) shows this offset growing and then changing sign at stronger dephasing.

\emph{Strong-dephasing Zeno regime.} For $\gamma\gg J$ we introduce
\begin{equation}
	\begin{aligned}
		\varepsilon
		&\coloneq\frac{J}{\gamma}\ll1,
		&
		t_{\mathrm f}
		&\coloneq\frac{1}{\gamma},
		\\
		t_{\mathrm s}
		&\coloneq\frac{\gamma}{2J^2},
		&
		\vartheta
		&\coloneq\frac{t}{t_{\mathrm s}}
		=\frac{2J^2}{\gamma}t.
	\end{aligned}
	\label{eq:strong_zeno_scales}
\end{equation}
On the fast scale $t_{\mathrm f}$ the kicks remove the coherences between occupation patterns. The occupations then move only on the much slower scale $t_{\mathrm s}$, through second-order virtual processes on single bonds. Let $\boldsymbol n=(n_1,\ldots,n_L)$ be a binary occupation configuration, let $\boldsymbol e_j$ denote the unit vector at site $j$, and define
\begin{equation}
	\begin{aligned}
		F_j\boldsymbol n
		&\coloneq
		\boldsymbol n
		\oplus\boldsymbol e_j
		\oplus\boldsymbol e_{j+1},
	\end{aligned}
	\label{eq:strong_pair_flip}
\end{equation}
where $\oplus$ between binary strings denotes addition modulo two. Thus $F_j$ flips the occupations of two adjacent sites and preserves global parity. Write $p(\boldsymbol n,t)\coloneq\rho_{\boldsymbol n,\boldsymbol n}(t)$ for the probability of configuration $\boldsymbol n$. Adiabatic elimination of the rapidly decaying bond coherences gives the leading population dynamics \cite{Facchi2008Dec,Reiter2012Mar,Kessler2012Jul,Znidaric2010Apr,Medvedyeva2016Sep}
\begin{equation}
	\begin{aligned}
		\partial_{\vartheta}
		p^{(0)}(\boldsymbol n,\vartheta)
		&=
		\sum_{j=1}^{L-1}
		\left[
		p^{(0)}(F_j\boldsymbol n,\vartheta)
		-p^{(0)}(\boldsymbol n,\vartheta)
		\right],
		\\
		p^{(0)}(\boldsymbol n,0)
		&=
		\delta_{\boldsymbol n,\boldsymbol n_0},
	\end{aligned}
	\label{eq:strong_pair_flip_kinetics}
\end{equation}
where the superscript $(0)$ marks the leading order in $J/\gamma$ at fixed $\vartheta$ and $\boldsymbol n_0$ is the initial configuration. Strong dephasing thus suppresses coherent hopping and pairing, and the Hamiltonian leaves behind a classical random walk among configurations connected by pair flips.

The complete leading distribution can be written in a binary Fourier basis. For binary vectors $\boldsymbol k$ and $\boldsymbol n$, define $\boldsymbol k\cdot\boldsymbol n=\sum_jk_jn_j$ modulo two and let
\begin{equation}
	\begin{aligned}
		d(\boldsymbol k)
		&\coloneq
		\sum_{j=1}^{L-1}
		\mathbf{1}_{\{k_j\neq k_{j+1}\}},
	\end{aligned}
	\label{eq:strong_binary_domain_walls}
\end{equation}
where $\mathbf{1}_{\{\cdots\}}$ equals one when its condition is satisfied and zero otherwise. Then
\begin{equation}
	\begin{aligned}
		p^{(0)}(\boldsymbol n,\vartheta)
		&=
		2^{-L}
		\sum_{\boldsymbol k\in\{0,1\}^{L}}
		(-1)^{
			\boldsymbol k\cdot
			(\boldsymbol n\oplus\boldsymbol n_0)}
		e^{-2d(\boldsymbol k)\vartheta}.
	\end{aligned}
	\label{eq:strong_zeno_full_probability}
\end{equation}
The integer $d(\boldsymbol k)$ counts domain walls in the binary Fourier mode $\boldsymbol k$. Modes with many domain walls decay rapidly, while the two uniform modes, which have none, fix the late-time uniform distribution within the conserved parity sector.

To obtain the entanglement spectrum, sum over all configurations in $B$. Let $\boldsymbol\eta$ and $\boldsymbol\eta_0$ be the final and initial binary configurations restricted to $A$, and define
\begin{equation}
	\begin{aligned}
		d_A(\boldsymbol k)
		&\coloneq
		\sum_{j=1}^{L_A-1}
		\mathbf{1}_{\{k_j\neq k_{j+1}\}}.
	\end{aligned}
	\label{eq:strong_A_domain_walls}
\end{equation}
The leading reduced probabilities are
\begin{equation}
	\begin{aligned}
		p_A^{(0)}(\boldsymbol\eta,\vartheta)
		={}
		2^{-L_A}
		\sum_{\boldsymbol k\in\{0,1\}^{L_A}}&
		(-1)^{
			\boldsymbol k\cdot
			(\boldsymbol\eta\oplus\boldsymbol\eta_0)}
		\\
		&\times
		e^{-2[
			d_A(\boldsymbol k)
			+\mathbf{1}_{\{k_{L_A}=1\}}
			]\vartheta}.
	\end{aligned}
	\label{eq:strong_zeno_marginal}
\end{equation}
The final indicator is the contribution of the cut: after the trace over $B$ the Fourier mode vanishes on $B$, so a domain wall remains on the cut bond whenever $k_{L_A}=1$.

For every nonzero displacement $\boldsymbol r$ in $A$, the leading-order probability margin is
\begin{equation}
	\begin{aligned}
		&p_A^{(0)}
		(\boldsymbol\eta_0,\vartheta)
		-p_A^{(0)}
		(\boldsymbol\eta_0\oplus\boldsymbol r,\vartheta)
		\\
		&\qquad=
		2^{1-L_A}
		\sum_{\substack{
				\boldsymbol k\in\{0,1\}^{L_A}\\
				\boldsymbol k\cdot\boldsymbol r=1}}
		e^{-2[
			d_A(\boldsymbol k)
			+\mathbf{1}_{\{k_{L_A}=1\}}
			]\vartheta}
		>0
	\end{aligned}
	\label{eq:strong_zeno_entanglement_margin}
\end{equation}
at every finite $\vartheta$, so the initial configuration of $A$ is the unique most probable one. The distribution itself factorizes over bonds, because pair flips on different bonds commute and each squares to the identity. At scaled time $\vartheta$, each bond has flipped its two sites an odd number of times with probability $x=(1-e^{-2\vartheta})/2$, independently of the other bonds. Every configuration in the parity sector of $\boldsymbol n_0$ is reached by exactly one record $\boldsymbol y\in\{0,1\}^{L-1}$ of flipped bonds, $\boldsymbol n=\boldsymbol n_0\oplus\bigoplus_{j}y_j(\boldsymbol e_j\oplus\boldsymbol e_{j+1})$. Every configuration of $A$ is likewise reached by exactly one record $\boldsymbol y_A=(y_1,\ldots,y_{L_A})$ of the bonds that touch $A$, the cut bond included, since the flips inside $B$ drop out of the trace. With $|\boldsymbol y|$ the number of flipped bonds,
\begin{equation}
	\begin{aligned}
		x&=\frac{1-e^{-2\vartheta}}{2},
		\\
		p^{(0)}(\boldsymbol n,\vartheta)
		&=x^{|\boldsymbol y|}(1-x)^{L-1-|\boldsymbol y|},
		\\
		p_A^{(0)}(\boldsymbol\eta,\vartheta)
		&=x^{|\boldsymbol y_A|}(1-x)^{L_A-|\boldsymbol y_A|}.
	\end{aligned}
	\label{eq:strong_zeno_bond_flips}
\end{equation}
The leading reduced state is diagonal, so its eigenvalues are these probabilities: $\rho_A$ has the spectrum of $L_A$ independent coins that each come up heads with probability $x$. The largest eigenvalue $\lambda_0^{(0)}=(1-x)^{L_A}$ belongs to the initial configuration, and the next, $\lambda_1^{(0)}=x(1-x)^{L_A-1}$, is $L_A$-fold degenerate and belongs to the configurations with a single flipped bond. Since $x<1/2$ at every finite $\vartheta$ and $(1-x)/x=\coth\vartheta$,
\begin{equation}
	\begin{aligned}
		\Delta_E^{(0)}(\vartheta)&=\ln\coth\vartheta>0,
		\qquad0<\vartheta<\infty,
		\\
		\lim_{\vartheta\rightarrow\infty}
		\Delta_E^{(0)}(\vartheta)&=0,
		\\
		\vartheta_E^{(0)}&=+\infty,
	\end{aligned}
	\label{eq:strong_zeno_entanglement_clock}
\end{equation}
where $\vartheta_E^{(0)}$ is the first scaled time at which the leading-order gap would close, and \cref{eq:normalized_spectral_gaps} gives $r_E^{(0)}=e^{-2\vartheta}$. Unlike under strong loss and gain, no site is driven through half filling, because each bond is more likely unflipped than flipped at every finite time.

The exact eigenvalues $\lambda_k$ of \cref{eq:rho_A_spectrum} approach the ordered leading-order probabilities $\lambda_k^{(0)}$ of \cref{eq:strong_zeno_bond_flips}. The density matrix itself contains coherences of order $J/\gamma$, but their first-order contribution to these eigenvalues vanishes, because each joins two configurations that differ by one pair flip and hence have different leading probabilities. Consequently,
\begin{equation}
	\begin{aligned}
		\lambda_k
		\left(
		t=\frac{\gamma\vartheta}{2J^2}
		\right)
		&=
		\lambda_k^{(0)}(\vartheta)
		+\mathcal O\!\left(
		\frac{J^2}{\gamma^2}
		\right).
	\end{aligned}
	\label{eq:strong_zeno_entanglement_control}
\end{equation}
For fixed $L$ the estimate is uniform on every interval $\vartheta_1\leq\vartheta\leq\vartheta_2$ with $0<\vartheta_1<\vartheta_2<\infty$, on which the leading-order gap is bounded away from zero, so at large $\gamma/J$ the exact gap stays open there. The closing in \cref{eq:strong_zeno_entanglement_clock} is therefore an asymptotic degeneracy, not a root at finite $\vartheta$. The estimate does not exclude an exact closing whose scaled time $\vartheta_E=2J^2t_E/\gamma$ grows without bound as $\gamma/J\to\infty$. Appendix~\ref{app:strong_zeno_reduction} proves the adiabatic elimination, the Fourier and product solutions, the positivity of the margin, and \cref{eq:strong_zeno_entanglement_control}.

Between the two limits, exact numerics for short chains show that the closing continued from $t_0$ moves later and, together with $t_q$, recedes to infinite time as $\gamma\to4J^-$ (\cref{fig:strong_symmetry_clocks}(a)); for $L_A=1$ this follows exactly from \cref{eq:strong_parity_clock,eq:strong_one_site_entanglement_clock}. Later closings behave in the same way (\cref{fig:strong_symmetry_clocks}(b)). Each runs close to a zero line of $q_A$ without lying on it, none is created or annihilated as $\gamma$ grows, and each recedes to infinite time at $\gamma=4J$, beyond which the gap stays open at every rate we sampled. Away from the compact point, where the kick solution no longer applies, the first entanglement closing still barely changes with the filling and with $L_B$ at small $\mu$ (Appendix~\ref{app:parameter_sweeps}).

\subsection{Twisted negativity Hamiltonian}
\label{subsec:tnh_strong}

\begin{figure*}[t]
	\centering
	\includegraphics[width=0.49\textwidth]{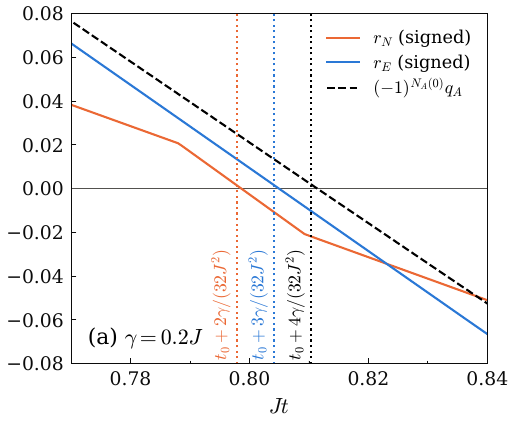}\hfill
	\includegraphics[width=0.49\textwidth]{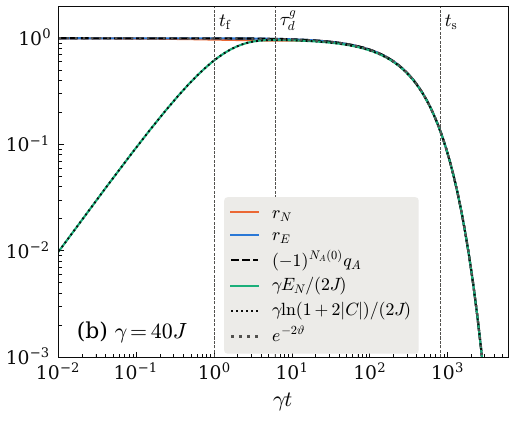}
	\caption{Closings as sign changes at weak dephasing, and their absence at strong dephasing. Exact many-body evolution of \cref{eq:strong_lindblad} at the compact point $\mu=0$, $\Delta=J$, for $L=8$, $A$ the first $L_A=4$ sites, and any product-Fock initial state, with $N_A(0)$ fermions initially in $A$. Solid: $r_E$ ($r_N$) of \cref{eq:normalized_spectral_gaps}, taken negative while the leading eigenvector of $\rho_A$ has subsystem parity (the leading eigenvalue of $T_A$ of \cref{eq:partial_time_reversal,eq:twisted_operator} has sign) opposite to its value at $t=0$; dashed: $(-1)^{N_A(0)}q_A$ (\cref{eq:strong_parity_clock,eq:strong_cut_aggregate_solution}). \textbf{(a)} $\gamma=0.2J$. The zeros $t_N<t_E<t_q$ lie close to $t_0+(2,3,4)\gamma/(32J^2)$, with $t_0=\pi/(4J)$ (dotted; Eqs.~(\ref{eq:strong_weak_negativity_clock}), (\ref{eq:strong_weak_entanglement_clock}), and (\ref{eq:strong_weak_parity_clock})), whereas the Gaussian parity rules of \cref{eq:weak_gaussian_equivalences,eq:weak_negativity_parity_rule} would give one common zero. The kinks of $r_N$ mark where the pair of $T_A^-$ passes the second level of $T_A^+$ (\cref{eq:strong_twisted_blocks}). \textbf{(b)} $\gamma=40J$, against $\gamma t$; no sign changes. The negativity $E_N$ of \cref{eq:logarithmic_negativity}, plotted as $\gamma E_N/(2J)$, follows, to leading order in $J/\gamma$, $\ln(1+2|C|)$ with $C(t)$ of \cref{eq:strong_cut_aggregate_solution} (dotted; \cref{eq:strong_twisted_blocks,eq:strong_boundary_negativity}). It peaks at $\tau_d^g$ while both gaps and $q_A$ have hardly moved. On the slow scale all four curves collapse onto $e^{-2\vartheta}$, with $\vartheta=2J^2t/\gamma$ (dotted; \cref{eq:strong_zeno_entanglement_clock,eq:strong_zeno_negativity_clock,eq:strong_zeno_negativity,eq:strong_cut_aggregate_solution}). Vertical lines: $t_{\mathrm f}=1/\gamma$; $\tau_d^g$, the first maximum of $E_N$, drawn at $t_{\mathrm{bl}}$ of \cref{eq:strong_boundary_layer_time}; and $t_{\mathrm s}=\gamma/(2J^2)$ (\cref{eq:strong_time_scale_hierarchy}).}
	\label{fig:strong_clock_traces}
\end{figure*}

\emph{Block structure of the twisted operator.} Strong symmetry gives $T_A$ a block structure that parallels \cref{eq:strong_parity_block_decomposition}. Split the state as $\rho=\rho_{\mathrm e}+\rho_{\mathrm o}$, where $\rho_{\mathrm e}=(\rho+P_A\rho P_A)/2$ collects the Majorana monomials with even $|\mathcal M|_A$ and $\rho_{\mathrm o}$ the rest, whose monomials are odd on both sides of the cut. Because $P\rho=p\rho$ and multiplication by $P$ maps a monomial with $|\mathcal M|_A$ factors in $A$, up to a constant factor, to one with $2L_A-|\mathcal M|_A$, \cref{eq:partial_time_reversal} gives $p\rho^{R_A}=(-1)^{L_A}P(\rho_{\mathrm e}^{R_A}-\rho_{\mathrm o}^{R_A})$, so the two parts lie in opposite sectors of $P$. At the compact point, each kick history in \cref{eq:strong_random_unitary} is the sum of the two branches of \cref{eq:strong_cut_gate}, each a product of a state of $A$ and a state of $B$ with definite parities. The part $\rho_{\mathrm e}$ is thus the average of the two branch projectors, a mixture of product states that we call the separable part of $\rho$, and $\rho_{\mathrm o}$ holds the coherences between the branches. On each product, $R_A$ acts on the factor in $A$ as the full fermionic transposition of $A$, which preserves its eigenvalues (see the footnote to \cref{eq:partial_time_reversal}), so $\rho_{\mathrm e}^{R_A}$ is Hermitian and positive. Altogether,
\begin{equation}
	\begin{aligned}
		T_A&=T_A^+\oplus T_A^-,
		&T_A^\pm&=\rho_{\mathrm{e/o}}^{R_A}P_A,
		\\
		PT_A^\pm&=\pm p(-1)^{L_A}T_A^\pm,
		&[\rho_{\mathrm e}^{R_A},P_A]&=0,
		\\
		P_AT_A^-P_A&=-T_A^-,
		&\rho_{\mathrm e}^{R_A}&\geq0,
		\\
		E_N&=\ln\left(1+\|T_A^-\|_1\right).
	\end{aligned}
	\label{eq:strong_twisted_blocks}
\end{equation}
Every relation except the last two holds for every state of definite global parity; positivity and the formula for $E_N$ use the kick solution and hold at the compact point. The eigenvalues of $T_A^+$ are those of $\rho_{\mathrm e}^{R_A}$, signed by their subsystem parity, and the total weights of $\rho_{\mathrm e}^{R_A}$ in the two parity sectors of $A$ are the pair $[1\pm(-1)^{L_A}q_A]/2$, the same pair as the traces of the two blocks of $\rho_A$. The spectrum of $T_A^-$ is symmetric about zero, and since $\operatorname{Tr}\rho_{\mathrm e}^{R_A}=1$, only the coherence between the branches carries negativity. A level of $T_A^-$ always comes with its partner of opposite sign, so if it were the largest, the negativity gap would vanish on a whole time interval. In all our computations the largest singular value lies strictly in $T_A^+$, and although the second often belongs to the pair of $T_A^-$, the two levels that meet at every closing are the largest eigenvalues of the two parity blocks of $\rho_{\mathrm e}^{R_A}$ (\cref{fig:strong_clock_traces}(a)). Each closing of the negativity gap is then a crossing between the blocks of $\rho_{\mathrm e}^{R_A}$, as each closing of the entanglement gap is one between the blocks of $\rho_A$. At $t_N$ the leading eigenvector of $T_A$ thus changes its subsystem parity, and with it the sign of its eigenvalue, while keeping its global parity; under loss and gain, by contrast, the two meeting levels differ in global parity (\cref{subsec:tnh_weak}).

For $L_B=1$ the two clocks coincide exactly, and the argument uses neither the compact point nor the kick solution. The occupation of $B$ is then fixed by the parity of $A$. Transforming $B$ instead of $A$, which leaves the twisted singular values unchanged (footnote to \cref{eq:partial_time_reversal}), makes their origin explicit: on a single site, $R_B$ multiplies $\rho_{\mathrm o}$ by $i$ and flips the occupation of $B$ in $\rho_{\mathrm e}$, so the twisted singular values are the eigenvalues of $\rho_A$ together with the singular values of the off-diagonal block of $\rho$ that joins the two parity sectors of $A$, each of the latter twice. With $c_1$ and $d_1$ the largest eigenvalues of the two parity blocks of $\rho_A$, positivity of $\rho$ bounds the singular values of the off-diagonal block by $\sqrt{c_1d_1}$, which is below $\lambda_0$ unless $c_1=d_1$. Hence $s_0=\lambda_0$, and $s_1=s_0$ exactly when $\lambda_1=\lambda_0$, so $t_N=t_E$ for every state of definite global parity, at every rate, for any $\mu$ and $\Delta$, and for any bath with strong parity symmetry.

\emph{Weak dephasing.} At $\gamma=0$, the compact crossing at $t_0$ is the pure-state crossing of \cref{eq:pure_state_clock_relation,eq:closed_compact_results}, where four nonzero twisted singular values meet, two in $T_A^+$ and a pair of opposite eigenvalues in $T_A^-$. With dephasing, the closing that continues this crossing is the crossing of the two levels of $T_A^+$, at which, in our numerics (\cref{fig:strong_clock_traces}(a)), the pair lies lower by an amount of order $\gamma$. A single kick changes the first-order shift only when it occurs at an end site of the cut bond. Define the boundary indicators
\begin{equation}
	\begin{aligned}
		\delta_A
		&\coloneq
		\begin{cases}
			1,&L_A=1,\\
			0,&L_A\geq2,
		\end{cases}
		&
		\delta_B
		&\coloneq
		\begin{cases}
			1,&L_B=1,\\
			0,&L_B\geq2.
		\end{cases}
	\end{aligned}
	\label{eq:strong_boundary_indicators}
\end{equation}
For every finite chain, every bipartition into nonempty $A$ and $B$, and every product-Fock initial state,
\begin{equation}
	\begin{aligned}
		t_N
		&=
		t_0+
		\frac{2+\delta_A+\delta_B}{32}
		\frac{\gamma}{J^2}
		+\mathcal O\!\left(
		\frac{\gamma^2}{J^3}
		\right).
	\end{aligned}
	\label{eq:strong_weak_negativity_clock}
\end{equation}
The constant term $2/32$ is the contribution of the two end sites of the cut bond when neither is a chain end. When a side is a single site, its end site has no further internal bond, and its contribution rises by $1/32$. Appendix~\ref{app:strong_negativity_asymptotics} gives the first-order perturbation theory within the four leading levels.

Combining \cref{eq:strong_weak_entanglement_clock,eq:strong_weak_parity_clock,eq:strong_weak_negativity_clock} gives
\begin{equation}
	\begin{aligned}
		t_E-t_N
		&=
		\frac{1-\delta_B}{32}
		\frac{\gamma}{J^2}
		+\mathcal O\!\left(
		\frac{\gamma^2}{J^3}
		\right),
		\\
		t_q-t_N
		&=
		\frac{2-\delta_A-\delta_B}{32}
		\frac{\gamma}{J^2}
		+\mathcal O\!\left(
		\frac{\gamma^2}{J^3}
		\right).
	\end{aligned}
	\label{eq:strong_weak_clock_splitting}
\end{equation}
The first line is independent of $L_A$. For $L_B\geq2$ the negativity closing precedes the entanglement closing by $\gamma/(32J^2)$ to first order, and for $L_B=1$ the two coincide at every rate, as shown above. The two coefficients differ because the negativity closing is a crossing inside the separable part, which keeps $B$: a kick at the end site of the cut bond in $B$ then also reverses the adjacent bond inside $B$, as a kick at the end site in $A$ does for both closings. The resulting symmetry of $t_N$ under exchanging $A$ and $B$ holds exactly at every rate: since $\mathcal D[1-n_j]=\mathcal D[n_j]$, reflecting the chain as in \cref{subsec:tnh_weak} maps the dephased compact chain onto itself with $A$ and $B$ exchanged and the initial pattern reversed and complemented, and the twisted spectrum depends neither on the pattern (\cref{eq:strong_pattern_universality}) nor on which side is transformed. The coefficient in \cref{eq:strong_weak_entanglement_clock}, by contrast, resolves only the geometry of $A$. The second line shows that the negativity closing also precedes the parity balance unless both $A$ and $B$ are single sites. If the parity rules of \cref{eq:weak_gaussian_equivalences,eq:weak_negativity_parity_rule} held with the exact parities, $q_B=pq_A$ would place both closings at $t_q$. The global state is not Gaussian here in any case: for a Gaussian state $|\operatorname{Tr}(\rho P)|$ is the product of the moduli of the canonical values of its covariance matrix, which equals one only for a pure state, whereas the dephased state has $|\operatorname{Tr}(\rho P)|=1$ and is mixed at every $t>0$. The offsets therefore measure how far the closings move from the parity balance that fixes them for Gaussian states. For a cut away from the chain ends their delays from $t_0$ stand in the ratio $2:3:4$ of \cref{eq:summary_weak_dephasing} (\cref{fig:strong_clock_traces}(a)).

\emph{Strong-dephasing Zeno regime.} The pair-flip kinetics of \cref{eq:strong_pair_flip_kinetics} also orders the twisted singular values. Fermionic partial time reversal acts simply on a density matrix that is diagonal in the occupation basis. For
\begin{equation}
	\begin{aligned}
		\rho_{\mathrm{diag}}
		&=
		\sum_{\boldsymbol n}
		p(\boldsymbol n)
		|\boldsymbol n\rangle
		\langle\boldsymbol n|,
	\end{aligned}
	\label{eq:strong_diagonal_state}
\end{equation}
partial time reversal complements the occupations in $A$, which merely permutes the diagonal probabilities. Hence
\begin{equation}
	\begin{aligned}
		\operatorname{sv}
		\left[
		\rho_{\mathrm{diag}}^{R_A}
		\right]
		&=
		\left\{
		p(\boldsymbol n)
		\right\}_{\boldsymbol n},
	\end{aligned}
	\label{eq:strong_diagonal_ptr_spectrum}
\end{equation}
where $\operatorname{sv}(\cdot)$ denotes the multiset of singular values, and the parity twist only attaches signs to the corresponding eigenvalues of $T_A$ (Appendix~\ref{app:strong_zeno_reduction}).

Applied to \cref{eq:strong_zeno_bond_flips}, this identity gives the leading twisted spectrum. The singular values of the leading diagonal state are the probabilities of the $L-1$ independent bonds, with $s_0^{(0)}=(1-x)^{L-1}$ for the initial configuration and the $(L-1)$-fold degenerate $s_1^{(0)}=x(1-x)^{L-2}$ for a single flipped bond. Therefore
\begin{equation}
	\begin{aligned}
		\Delta_N^{(0)}(\vartheta)&=\ln\coth\vartheta>0,
		\qquad0<\vartheta<\infty,
		\\
		\lim_{\vartheta\rightarrow\infty}
		\Delta_N^{(0)}(\vartheta)&=0,
		\\
		\vartheta_N^{(0)}&=+\infty.
	\end{aligned}
	\label{eq:strong_zeno_negativity_clock}
\end{equation}
Together with \cref{eq:strong_zeno_entanglement_clock}, this gives
\begin{equation}
	\begin{aligned}
		\vartheta_N^{(0)}
		&=
		\vartheta_E^{(0)}
		=+\infty.
	\end{aligned}
	\label{eq:strong_zeno_clock_locking}
\end{equation}
The two leading-order gaps are equal, $\Delta_N^{(0)}=\Delta_E^{(0)}$, but the spectra differ: the twisted singular values count all $L-1$ bonds, the entanglement levels only the $L_A$ bonds that touch $A$.

The full state contains bond coherences of order $J/\gamma$, and those across the cut form $T_A^-$, whose singular values are of that order. The ordered singular values therefore approach those of the leading diagonal state, $s_k^{(0)}(\vartheta)$, one order more slowly than the eigenvalues of $\rho_A$:
\begin{equation}
	\begin{aligned}
		s_k
		\left(
		t=\frac{\gamma\vartheta}{2J^2}
		\right)
		&=
		s_k^{(0)}(\vartheta)
		+\mathcal O\!\left(
		\frac{J}{\gamma}
		\right).
	\end{aligned}
	\label{eq:strong_zeno_negativity_control}
\end{equation}
As for $\rho_A$, this excludes a closing on any fixed interval $0<\vartheta_1\leq\vartheta\leq\vartheta_2<\infty$ once $\gamma/J$ is large enough, but not one at scaled times that grow without bound with $\gamma/J$, where the leading-order gap itself vanishes.

The coherence of order $J/\gamma$ barely affects the leading eigenvalues of $\rho_A$, but across the cut it forms $T_A^-$ and so carries all of the negativity. At fixed finite scaled time its contribution is universal:
\begin{equation}
	\begin{aligned}
		E_N
		\left(
		t=\frac{\gamma\vartheta}{2J^2}
		\right)
		&=
		\frac{2J}{\gamma}
		e^{-2\vartheta}
		+\mathcal O\!\left(
		\frac{J^2}{\gamma^2}
		\right).
	\end{aligned}
	\label{eq:strong_zeno_negativity}
\end{equation}
The coefficient is independent of $L_A$, $L_B$, and the initial product-Fock pattern. At leading order four quantities follow one curve: $r_E$, $r_N$, the relative parity $(-1)^{N_A(0)}q_A$ of the overdamped oscillator, and $\gamma E_N/(2J)$ all equal $e^{-2\vartheta}$ (\cref{fig:strong_clock_traces}(b)). The classical pair flips decide which levels lead and thus control both gaps, while a small coherence across the cut generates the negativity.

\emph{Comparison of the temporal criticalities.} Between the two controlled limits, exact numerics follow $t_N$, $t_E$, and $t_q$ through the whole underdamped range (\cref{fig:strong_symmetry_clocks}(a) and its inset). In every chain we computed with $L_A,L_B\geq2$, the offsets $t_q-t_E$ and $t_q-t_N$, positive at weak dephasing, change sign, so that close to $\gamma=4J$ the parity balance comes before both closings. The order of the two closings there depends on the geometry. In most geometries we computed, including a cut in the middle of the chain, $t_E-t_N$ changes sign as well, the weak-rate order $t_N<t_E<t_q$ reverses completely, and the negativity gap closes last, whereas for a long $A$ next to a two-site $B$ the two closings stay close and $t_N$ remains the earlier. A sweep at $\mu=0.1J$ also shows $t_E-t_N$ changing sign for a middle cut (Appendix~\ref{app:parameter_sweeps}). A one-site $A$ is exactly solvable here: by the exchange symmetry and the result for $L_B=1$, its $t_N$ equals the entanglement clock of a subsystem of $L_B$ sites, while its $t_E$ equals $t_q$, so $t_N>t_E$ as soon as that entanglement clock passes $t_q$. All three times diverge together as $\gamma\to4J^-$, at finite mutual offsets in our numerics, so both closings inherit the critical damping of the cut-bond oscillator of \cref{eq:strong_parity_oscillator}. For $\gamma\geq4J$ neither gap closes at any rate we sampled (shaded in \cref{fig:strong_symmetry_clocks}(a)), consistent with the Zeno results above.

\emph{Geometric decoherence time.} For every finite chain with nonempty $A$ and $B$, $J\neq0$, and $\gamma>0$, the logarithmic negativity first grows from zero, while the stationary state in \cref{eq:strong_stationary_state} has zero negativity and maximal R\'enyi-$1/2$ entropy of $A$. Continuity therefore guarantees
\begin{equation}
	\begin{aligned}
		0<\tau_d^g<\infty,
		\qquad\gamma>0.
	\end{aligned}
	\label{eq:strong_geometric_existence}
\end{equation}
Appendix~\ref{app:strong_geometric_decoherence} proves this, including the short-time and stationary-state bounds that enter the definition.

At weak dephasing the shift of $\tau_d^g$, the first local maximum of $E_N=\ln\sum_ks_k$ at which $S_{1/2}$ still increases, depends on the whole trace norm $\sum_ks_k$, including singular values that emerge from the kernel of the closed-system $T_A$, not only on the $s_0$ and $s_1$ that fix $t_N$:
\begin{equation}
	\begin{aligned}
		\tau_d^g
		&=
		t_0
		-\mathcal C_{L_A,L_B}^{\mathrm g}
		\frac{\gamma}{J^2}
		+o\!\left(
		\frac{\gamma}{J^2}
		\right),
	\end{aligned}
	\label{eq:strong_weak_geometric_clock}
\end{equation}
where $\mathcal C_{L_A,L_B}^{\mathrm g}$ is the exact first-order response of the trace norm, which depends on the whole finite geometry (Appendix~\ref{app:strong_geometric_decoherence}). The coefficient is positive for every geometry evaluated, $L\leq10$ (\cref{tab:app_strong_geo_coefficients}), so there $\tau_d^g<t_0<t_N\leq t_E$.

At strong dephasing, the negativity peaks in a boundary layer at $\vartheta\to0$ that \cref{eq:strong_zeno_negativity} does not resolve. Let $F\coloneq F_{L_A}$ be the pair flip across the cut and let $\mathcal R$ contain the configurations with the same subsystem parities as the initial state. Define the total population imbalance and the total cut coherence by
\begin{equation}
	\begin{aligned}
		D(t)
		&\coloneq
		\sum_{\boldsymbol n\in\mathcal R}
		\left[
		p(\boldsymbol n,t)
		-p(F\boldsymbol n,t)
		\right],
		\\
		C(t)
		&\coloneq
		\sum_{\boldsymbol n\in\mathcal R}
		\rho_{\boldsymbol n,F\boldsymbol n}(t).
	\end{aligned}
	\label{eq:strong_cut_aggregate}
\end{equation}
Every configuration of the parity sector lies in $\mathcal R$ or in $F\mathcal R$, so $D=(-1)^{N_A(0)}q_A$. Internal pair flips cancel from these sums, which leaves the exact closed equations
\begin{equation}
	\begin{aligned}
		\dot D&=4J\,\operatorname{Im}C,
		\\
		\dot C&=-\gamma C-iJD,
		\\
		D(0)&=1,
		&C(0)&=0.
	\end{aligned}
	\label{eq:strong_cut_aggregate_dynamics}
\end{equation}
Eliminating $C$ gives back the parity oscillator of \cref{eq:strong_parity_oscillator}. For $\gamma>4J$, define
\begin{equation}
	\begin{aligned}
		\bar\omega
		&\coloneq
		\sqrt{\frac{\gamma^2}{4}-4J^2},
	\end{aligned}
	\label{eq:strong_overdamped_frequency}
\end{equation}
the overdamped continuation of \cref{eq:strong_parity_frequency}, with $\omega=i\bar\omega$. The exact solution is
\begin{equation}
	\begin{aligned}
		D(t)
		={}&
		e^{-\gamma t/2}
		\left[
		\cosh(\bar\omega t)
		+\frac{\gamma}{2\bar\omega}\sinh(\bar\omega t)
		\right],
		\\
		C(t)
		={}&
		-\frac{iJ}{\bar\omega}
		e^{-\gamma t/2}
		\sinh(\bar\omega t).
	\end{aligned}
	\label{eq:strong_cut_aggregate_solution}
\end{equation}
In the boundary layer, the cut coherences contribute additively to the trace norm, giving
\begin{equation}
	\begin{aligned}
		E_N(t)
		&=
		\frac{2J}{\bar\omega}
		e^{-\gamma t/2}
		\sinh(\bar\omega t)
		+\mathcal O\!\left(
		\frac{J^2}{\gamma^2}
		\right).
	\end{aligned}
	\label{eq:strong_boundary_negativity}
\end{equation}
The cut coherence builds up on the fast scale $1/\gamma$ and then follows the slow decay of $D$ at the rate $\gamma/2-\bar\omega\simeq4J^2/\gamma$. Since $E_N=\ln(1+\|T_A^-\|_1)$ by \cref{eq:strong_twisted_blocks}, with $\|T_A^-\|_1=2|C|$ at leading order, the negativity peaks with $|C|$, once the fast transient has fallen to the relative size $4J^2/\gamma^2$, which takes the logarithmic time
\begin{equation}
	\begin{aligned}
		t_{\mathrm{bl}}
		&\coloneq
		\frac{1}{\bar\omega}
		\operatorname{artanh}
		\left(\frac{2\bar\omega}{\gamma}\right),
		\\
		t_{\mathrm{bl}}
		&=
		\frac{2}{\gamma}
		\ln\!\left(\frac{\gamma}{2J}\right)
		[1+o(1)],
	\end{aligned}
	\label{eq:strong_boundary_layer_time}
\end{equation}
the stationary point of \cref{eq:strong_boundary_negativity}. Writing $\sinh$ as a difference of exponentials,
\begin{equation}
	\begin{aligned}
		E_N(t)
		&=
		\frac{J}{\bar\omega}
		\left[
		e^{-(\gamma/2-\bar\omega)t}
		-e^{-(\gamma/2+\bar\omega)t}
		\right]
		+\mathcal O\!\left(
		\frac{J^2}{\gamma^2}
		\right).
	\end{aligned}
	\label{eq:strong_boundary_negativity_two_rates}
\end{equation}
At $t=t_{\mathrm{bl}}$, one has $\gamma/2-\bar\omega=4J^2/\gamma+\mathcal O(J^4/\gamma^3)$, so the first exponential equals $1+o(1)$, whereas the second is $o(1)$. Since $J/\bar\omega=(2J/\gamma)[1+o(1)]$, this gives $E_N(t_{\mathrm{bl}})=(2J/\gamma)[1+o(1)]$.

The same solution gives $q_{\mathrm{cut}}(t_{\mathrm{bl}})\coloneq[1-D(t_{\mathrm{bl}})]/2=\mathcal O[\varepsilon^2\ln(1/\varepsilon)]$ with $\dot q_{\mathrm{cut}}(t_{\mathrm{bl}})>0$. The other leading pair-flip probabilities are of the same order and still increasing, so $S_{1/2}$ is increasing at this first negativity maximum, which therefore gives the geometric decoherence time to leading order:
\begin{equation}
	\begin{aligned}
		\tau_d^g
		&=
		\frac{2}{\gamma}
		\ln\!\left(\frac{\gamma}{2J}\right)
		[1+o(1)],
		\\
		E_N(\tau_d^g)
		&=
		\frac{2J}{\gamma}
		[1+o(1)].
	\end{aligned}
	\label{eq:strong_geometric_zeno}
\end{equation}
For fixed finite $L$, the total weight of the reduced state outside the initial configuration is likewise $\mathcal O[\varepsilon^2\ln(1/\varepsilon)]$. Hence $\lambda_0=1-\mathcal O[\varepsilon^2\ln(1/\varepsilon)]$ and $\lambda_1=\mathcal O[\varepsilon^2\ln(1/\varepsilon)]$, which yields
\begin{equation}
	\begin{aligned}
		\lambda_0(\tau_d^g)
		-\lambda_1(\tau_d^g)
		&=
		1-\mathcal O\!\left[
		\varepsilon^2
		\ln\!\left(\frac{1}{\varepsilon}\right)
		\right]
		>0.
	\end{aligned}
	\label{eq:strong_geometric_entanglement_gap}
\end{equation}
Geometric decoherence thus occurs while both gaps are still wide open. In scaled time,
\begin{equation}
	\begin{aligned}
		\vartheta_g
		\coloneq
		\frac{2J^2}{\gamma}\tau_d^g
		\simeq
		\frac{4J^2}{\gamma^2}
		\ln\!\left(\frac{\gamma}{2J}\right)
		\longrightarrow0.
	\end{aligned}
	\label{eq:strong_geometric_zeno_time}
\end{equation}
Together with \cref{eq:strong_zeno_scales,eq:strong_zeno_clock_locking},
\begin{equation}
	\begin{aligned}
		t_{\mathrm f}
		&=\frac{1}{\gamma}
		\ll
		\tau_d^g
		\simeq
		\frac{2}{\gamma}
		\ln\!\left(\frac{\gamma}{2J}\right)
		\ll
		t_{\mathrm s}
		=\frac{\gamma}{2J^2},
		\\
		\vartheta_N^{(0)}
		&=\vartheta_E^{(0)}
		=+\infty.
	\end{aligned}
	\label{eq:strong_time_scale_hierarchy}
\end{equation}
The vertical lines in \cref{fig:strong_clock_traces}(b) mark the three scales. The kicks first remove local coherences on the scale $t_{\mathrm f}$. A residual coherence across the cut grows and turns over on the logarithmically longer scale $\tau_d^g$. Only on the scale $t_{\mathrm s}$ do the pair flips redistribute the occupations, and at leading order the initial configuration stays dominant at every finite scaled time, so both gaps remain open.

Under dephasing, the kick solution reduces both closings to crossings between the two branches of the cut bond, in $\rho_A$ for $t_E$ and in the partially time-reversed separable part for $t_N$; $t_q$ compares only their total weights, so both closings can precede it. At weak dephasing and away from the chain ends, the delays of $t_N$, $t_E$, and $t_q$ from $t_0$ stand in the ratio $2:3:4$, and in every geometry evaluated the negativity turns over first. Toward $\gamma=4J$, in every chain we computed with at least two sites on each side, the parity balance moves ahead of both closings, and all three times recede to infinite time. In the Zeno regime, independent pair flips keep both gaps open at every finite scaled time at leading order, and the negativity turns over at $\tau_d^g$, between $t_{\mathrm f}$ and $t_{\mathrm s}$. Strong symmetry enters only through the conserved global parity.

\section{Conclusion and Outlook}
\label{sec:conclusion}

Our results show that Gaussianity, rather than the parity symmetry of the bath, fixes the criterion for a temporal criticality. Such a criticality is an instant at which the two largest eigenvalues of an entanglement or negativity spectrum meet. At a simple crossing the dominant state changes and the lowest level of the entanglement or negativity Hamiltonian has a kink. Entropies and the logarithmic negativity $E_N$ stay smooth because they sum over the whole spectrum, so these instants reflect how correlations are organized rather than how much correlation is present. For a subsystem $A$ with reduced state $\rho_A$ and complement $B$, we followed three clocks. Two are spectral: the first closing $t_E$ of the entanglement gap of $\rho_A$ and the first closing $t_N$ of the gap of the negativity spectrum, obtained by fermionic partial time reversal \cite{Shapourian2017Apr,Murciano2022Apr}. The third is the geometric decoherence time $\tau_d^g$, at which $E_N$ first falls while the R\'enyi-$1/2$ entropy of $A$, equal to $E_N$ in a pure state, still rises \cite{Jha2026Jul}. A pure state also has $t_N=t_E$, so only a bath can separate the two spectral clocks of an initially pure state. We compared two baths acting on the same Hamiltonian. Loss and gain respect fermion parity only on average (weak symmetry) and keep the state Gaussian, that is, fixed by its two-point functions. Density dephasing respects parity in every quantum jump (strong symmetry) and makes the state non-Gaussian.

In a Gaussian state, pure or mixed, the entanglement gap closes exactly at a parity balance, where the even and odd parity sectors of $A$ carry equal weight and the expectation value $q_A$ of the fermion parity of $A$ vanishes (\cref{eq:weak_gaussian_equivalences}). The rule also serves as a test, since an entanglement gap that closes while $q_A\neq0$ certifies that $\rho_A$ is not Gaussian. The negativity gap closes exactly when $q_A$ or its counterpart $q_B$ for $B$ vanishes (\cref{eq:weak_negativity_parity_rule}), a rule that is, to our knowledge, new. The negativity clock is thus the earlier of the entanglement clocks $t_E$ of $A$ and $t_E^B$ of $B$, $t_N=\min(t_E,t_E^B)\leq t_E$, and two parity measurements locate both clocks without reconstructing either spectrum. Balanced loss and gain at a common rate on every site only multiply the two parities by positive factors. For any quadratic Hamiltonian and any Gaussian initial state both clocks therefore keep their closed-system values at every rate, although the state relaxes toward the maximally mixed one (\cref{eq:weak_balanced_clock_locking}).

A strong symmetry instead conserves the global parity. For an initial state of definite global parity $p=\pm1$ the two parities are then tied, $q_B=pq_A$, and the state never leaves its initial sector. Since a Gaussian state of definite parity is pure, every mixed state reached in this way is non-Gaussian, so the negativity rule no longer applies and the entanglement rule applies only while $\rho_A$ stays Gaussian. For a single-site $B$, however, $t_N=t_E$ holds exactly for every state of definite global parity (\cref{subsec:tnh_strong}).

The remaining results concern the Kitaev chain, with $A$ at its left end and, initially, the first few sites of $A$ filled and $B$ empty. We worked mostly at the compact point, where the chemical potential vanishes and the pairing equals the hopping amplitude $J$. There every Majorana mode except the two at the chain ends pairs with one on the neighboring site, a single bond crosses the cut, and without a bath both gaps close at $t_0=\pi/(4J)$. Weak imbalanced loss and gain shift the parity balance of each subsystem by an amount set by the initial occupation of its site next to the cut, twice as large when the subsystem is a single site (\cref{eq:weak_rate_entanglement_clock,eq:weak_rate_negativity_clock}). Under gain, for example, a filled last site of $A$ delays the balance of $A$, while the empty first site of $B$ advances that of $B$, so that $t_N<t_0<t_E$. When imbalanced rates far exceed $J$, each site relaxes on its own, and the sites that the bath drives through half filling, which we call active, cross it together. Near that moment a parity can vanish only through an odd block of neighboring active sites on its side of the cut, because pairing keeps an even block from balancing (\cref{eq:weak_strong_rate_entanglement_clock,eq:weak_strong_rate_negativity_clock}). Under gain, an odd block of active sites in $B$ can therefore close the negativity gap while the entanglement gap stays open. This odd-even rule persists at any fixed chemical potential as long as the pairing is nonzero (\cref{eq:weak_finite_mu_pairing_factor,eq:weak_finite_mu_even_mass}).

Dephasing never changes an occupation; it only randomizes the phases between occupation patterns. At the compact point we solved this dynamics exactly by writing dephasing at rate $\gamma$ as random kicks, each multiplying the state vector by the parity of one site. From a product occupation state each kick history then yields a pure Gaussian state, while their average, the physical state, is not Gaussian. We know of no earlier exact solution of non-Gaussian Lindblad dynamics in a fermionic chain with pairing that yields entanglement and negativity spectra. In each history the bond across the cut splits the state into two branches, one in which both sites of that bond have flipped and one in which neither has, and the branches have opposite parity of $A$. Averaged over histories, the branches give the two parity blocks of $\rho_A$ and, after partial time reversal, the two parity blocks of their incoherent mixture, the separable part of the state. The coherence between the branches carries all of the negativity. Each closing we found is a crossing of the largest levels of two such blocks, whereas the parity balance, at the first zero $t_q$ of $q_A$, only equates their traces. Weak dephasing slows the oscillation of the cut bond and, away from the chain ends, delays $t_N$, $t_E$, and $t_q$ from $t_0$ in the ratio $2:3:4$ to first order (\cref{eq:strong_weak_negativity_clock,eq:strong_weak_entanglement_clock,eq:strong_weak_parity_clock}). Both closings thus precede the parity balance, and the early entanglement closing shows that $\rho_A$ is not Gaussian. Close to $\gamma=4J$ the cut bond becomes critically damped, so $t_q$ diverges exactly (\cref{eq:strong_parity_clock}), and in our numerics $t_E$ and $t_N$ diverge with it. There, in every chain we computed with at least two sites on each side, the parity balance comes before both closings, and in most geometries the negativity gap closes last, which no Gaussian evolution allows (\cref{subsec:tnh_strong,fig:strong_symmetry_clocks}). Strong dephasing, $\gamma\gg J$, nearly freezes the occupations, as in the quantum Zeno effect. The remaining dynamics consists of slow, incoherent flips of the two sites of a bond, independent from bond to bond, and each bond stays more likely unflipped than flipped. At leading order in $J/\gamma$ both gaps therefore stay open at every finite time and close only asymptotically (\cref{eq:strong_zeno_entanglement_clock,eq:strong_zeno_negativity_clock}).

At weak rates $\tau_d^g$ precedes both closings, under loss and gain for every filling and cut and under dephasing in every chain of up to ten sites (\cref{eq:app_weak_geo_ordering,tab:app_strong_geo_coefficients}). Under strong dephasing $\tau_d^g\simeq(2/\gamma)\ln[\gamma/(2J)]$ lies long before the occupations move on the scale $\gamma/(2J^2)$ (\cref{eq:strong_time_scale_hierarchy}), whereas under strong imbalanced loss and gain its place relative to the closings depends on the imbalance and the filling (\cref{tab:weak_strong_geometric_branches}). Gaussianity ties the closings to parity balances and orders them, and balanced loss and gain leave them in place for any quadratic Hamiltonian. In the Kitaev chain a weak imbalance shifts them by amounts set by the occupations at the cut, dephasing delays them and detaches them from the parity balance, and the incoherent bond flips left by strong dephasing remove them at leading order. The symmetry label enters the clocks only through the global parity, and beyond the clocks it fixes the long-time state, which under dephasing encodes the initial parity in classical correlations that no clock detects (\cref{eq:strong_stationary_state}).

At the compact point a Jordan-Wigner transformation maps the chain onto commuting Ising couplings between neighboring qubits and dephasing onto local phase noise. Since $A$ sits at the left end, the Jordan-Wigner string of every site in $A$ stays in $A$, so the transformation leaves $\rho_A$ unchanged and turns the parity of $A$ into a product of Pauli $Z$ operators. The entanglement clock and the parity balance under dephasing can thus be followed on qubit platforms, on which entanglement spectra have been measured \cite{Choo2018Aug,Kokail2021Aug}. With cold atoms, for which a protocol for measuring the entanglement spectrum has been proposed \cite{Pichler2016Nov}, site-resolved snapshots of the occupations would give the two parities that locate both clocks for Gaussian dynamics. The negativity clock itself needs fermionic partial time reversal, which differs from the partial transpose of qubits. Full tomography reaches it at modest sizes, and randomized measurements of partial-transpose moments \cite{Elben2020Nov}, once adapted to the fermionic case, would avoid tomography. Our two baths change symmetry and Gaussianity together. A weakly symmetric but non-Gaussian bath would separate the two, for example loss with jump operator $c_jn_{j+1}$, which removes a fermion from site $j$ only when site $j+1$ is occupied, or dephasing combined with loss and gain. Such a bath would reveal whether the odd-even rule needs Gaussianity and whether the absence of closings under strong dephasing needs strong symmetry. In interacting chains, whose closed evolution is already non-Gaussian, the offset between a closing and the parity balance would serve as a witness of non-Gaussianity. In periodically driven chains, which show temporal entanglement transitions without a bath \cite{Gadge2026Mar,wang2026criticaltouchingtemporalentanglement}, the drive would compete with the bath. Whether the closings sharpen into transitions with a scaling form in the thermodynamic limit remains open.
	
\section*{Acknowledgments}
The authors thank Abhinav Prem for fruitful discussions and acknowledge the Center for Advanced Research Computing (CARC) at the University of Southern California (https://carc.usc.edu) for providing the computational resources used in this work. R.J. acknowledges partial support, and S.H. acknowledges support, from the U.S. Department of Energy, Office of Science, Office of Advanced Scientific Computing Research, through the Exploratory Research for Extreme Scale Science (EXPRESS) program under Award No. DE-SC0026337. S.C. acknowledges support from the same program under Award No. DE-SC0026216. ChatGPT 5.6-Sol was used to create the schematic of the Kitaev chain in Fig.~\ref{fig:model_setup} and aided in writing the code for this paper.

\textit{Disclaimer.---} This report was prepared as an account of work sponsored by an agency of the United States Government. Neither the United States Government nor any agency thereof, nor any of their employees, makes any warranty, express or implied, or assumes any legal liability or 	responsibility for the accuracy, completeness, or usefulness of any information, apparatus, product, or process disclosed, or represents that its use would not infringe privately owned rights. Reference herein to any specific commercial product, process, or service by trade name, trademark, manufacturer, or otherwise does not necessarily constitute or imply its endorsement, recommendation, or favoring by the United States Government or any agency thereof. The views and opinions of authors expressed herein do not necessarily state or reflect those of the United States Government or any agency thereof.

\section*{Data Availability}
The data that support the findings of this article are available upon reasonable request from the authors.
	
	\bibliography{refs_tet_kitaev}

	\appendix
	
	\include{kitaev_appendices_v2}

\end{document}

%% file: kitaev_appendices_v2.tex
\section{Gaussian Reduction for Weak Symmetry}
\label{app:weak_gaussian_reduction}

This appendix gives the algebra behind the covariance description used in \cref{sec:temporal_criticalities_weak}. We show why a quadratic Hamiltonian with linear loss and gain closes on two-point functions, how the reduced many-body spectrum is reconstructed from those functions, and why balanced loss and gain preserve every finite spectral zero \cite{Prosen2008Apr,Prosen2010Jul,Barthel2022Nov}.

\subsection{Closed covariance equation}

Order the Majoranas as $w=(a_1,b_1,\ldots,a_L,b_L)^T$ and write $H=(i/4)w^T\mathsf A w+\mathrm{constant}$, with $\mathsf A^T=-\mathsf A$. The Heisenberg equation for the closed system is $\dot w=\mathsf A w$. For the covariance in \cref{eq:weak_majorana_covariance}, the Hamiltonian contribution is therefore
\begin{equation}
	\begin{aligned}
		\left.\dot\Gamma\right|_H
		&=\mathsf A\Gamma+\Gamma\mathsf A^T.
	\end{aligned}
	\label{eq:app_weak_hamiltonian_covariance}
\end{equation}
For the dissipative part it is useful to work out one site explicitly. The covariance equation needs the rate at which the bath damps a product of two Majorana operators, and this rate is obtained most directly by letting the bath act on the operator instead of on the state. Both routes give the same expectation value, since cyclicity of the trace turns $\operatorname{Tr}(O\mathcal D[X]\rho)$ into $\operatorname{Tr}(\rho\mathcal D^\dagger[X]O)$ with the adjoint of \cref{eq:lindblad_dissipator} given by $\mathcal D^\dagger[X]O=X^\dagger OX-\{X^\dagger X,O\}/2$. On site $j$,
\begin{equation}
	\begin{aligned}
		i a_jb_j&=2n_j-1.
	\end{aligned}
	\label{eq:app_weak_local_bilinear}
\end{equation}
Loss and gain alone give $\dot n_j=-\kappa n_j+\gamma_g$, with $\kappa=\gamma_\ell+\gamma_g$. Hence
\begin{equation}
	\begin{aligned}
		\frac{d}{dt}\langle i a_jb_j\rangle
		&=-\kappa\langle i a_jb_j\rangle-\xi,
		\qquad \xi=\gamma_\ell-\gamma_g.
	\end{aligned}
	\label{eq:app_weak_local_bilinear_damping}
\end{equation}
A bilinear joining two different sites has no local source. Each endpoint Majorana is damped at rate $\kappa/2$, so the bilinear is damped at the total rate $\kappa$. Both cases are summarized by
\begin{equation}
	\begin{aligned}
		&\sum_j\gamma_\ell\mathcal D^\dagger[c_j](iw_mw_n)
		+\sum_j\gamma_g\mathcal D^\dagger[c_j^\dagger](iw_mw_n)
		\\
		&\qquad=-\kappa\,iw_mw_n
		+\xi\,(\Gamma_{\mathrm{vac}})_{mn},
		\qquad m\neq n,
	\end{aligned}
	\label{eq:app_weak_dissipative_covariance}
\end{equation}
where the source is nonzero only when $w_m,w_n$ are the two Majoranas on the same site, and
\begin{equation}
	\begin{aligned}
		\Gamma_{\mathrm{vac}}
		&=\bigoplus_{j=1}^{L}
		\begin{pmatrix}
			0&-1\\
			1&0
		\end{pmatrix}.
	\end{aligned}
	\label{eq:app_weak_vacuum_covariance}
\end{equation}
This also fixes the sign of the source: the vacuum has $\langle i a_jb_j\rangle=-1$.

Combining \cref{eq:app_weak_hamiltonian_covariance,eq:app_weak_dissipative_covariance} gives the first line of \cref{eq:weak_covariance_solution}.

To solve it, define the closed propagator $R(t)=e^{\mathsf A t}$. Since $\mathsf A$ is antisymmetric, $R(t)R^T(t)=\mathbf 1$. Introduce the interaction-picture covariance
\begin{equation}
	\begin{aligned}
		\widetilde\Gamma(t)
		&\coloneq e^{\kappa t}R^T(t)\Gamma(t)R(t).
	\end{aligned}
	\label{eq:app_weak_interaction_covariance}
\end{equation}
Differentiating and using the covariance equation cancels the Hamiltonian and damping terms:
\begin{equation}
	\begin{aligned}
		\frac{d\widetilde\Gamma}{dt}
		&=\xi e^{\kappa t}R^T(t)\Gamma_{\mathrm{vac}}R(t).
	\end{aligned}
	\label{eq:app_weak_interaction_covariance_derivative}
\end{equation}
Integrating from $0$ to $t$ and transforming back yields
\begin{equation}
	\begin{aligned}
		\Gamma(t)
		={}&e^{-\kappa t}R(t)\Gamma(0)R^T(t)
		\\
		&+\xi\int_0^t du\,e^{-\kappa u}R(u)\Gamma_{\mathrm{vac}}R^T(u),
	\end{aligned}
	\label{eq:app_weak_covariance_solution}
\end{equation}
which is \cref{eq:weak_covariance_solution}. The first term carries the closed covariance forward and damps its amplitude. The second term is the particle-hole asymmetric source. This separation is why the rate sum and rate imbalance play different roles in the weak-rate expansions.

An equivalent derivation starts from the Nambu vector $\Psi=(c_1,\ldots,c_L,c_1^\dagger,\ldots,c_L^\dagger)^T$ and the covariance $C_{ab}=\operatorname{Tr}(\rho\Psi_a\Psi_b^\dagger)$. If $H=(1/2)\Psi^\dagger H_{\mathrm{BdG}}\Psi+\mathrm{constant}$, then
\begin{equation}
	\begin{aligned}
		\dot C
		&=-i[H_{\mathrm{BdG}},C]-\kappa C+D,
		\\
		D&=
		\begin{pmatrix}
			\gamma_\ell\mathbf 1_L&0\\
			0&\gamma_g\mathbf 1_L
		\end{pmatrix}.
	\end{aligned}
	\label{eq:app_weak_nambu_covariance}
\end{equation}
If $H_{\mathrm{BdG}}=V\operatorname{diag}(E_1,\ldots,E_{2L})V^\dagger$, every matrix element in that basis obeys a scalar linear equation,
\begin{equation}
	\begin{aligned}
		\widetilde C_{mn}(t)
		={}&e^{-[\kappa+i(E_m-E_n)]t}\widetilde C_{mn}(0)
		\\
		&+\widetilde D_{mn}
		\frac{1-e^{-[\kappa+i(E_m-E_n)]t}}
		{\kappa+i(E_m-E_n)}.
	\end{aligned}
	\label{eq:app_weak_nambu_solution}
\end{equation}
This form is convenient for direct finite-size evaluation and is exactly equivalent to \cref{eq:app_weak_covariance_solution}.

\subsection{Canonical modes and the entanglement spectrum}

The reduced covariance $\Gamma_A$ is real and antisymmetric. Therefore there is a real orthogonal matrix $O_A$ for which \cref{eq:weak_canonical_covariance} holds. Let $\widetilde w=O_Aw_A$ be the transformed Majoranas. For each $2\times2$ block choose the canonical fermion orientation
\begin{equation}
	\begin{aligned}
		f_m
		&\coloneq\frac{\widetilde w_{2m-1}+i\widetilde w_{2m}}{2},
		& i\widetilde w_{2m-1}\widetilde w_{2m}
		&=2f_m^\dagger f_m-1.
		\end{aligned}
		\label{eq:app_weak_canonical_fermions}
	\end{equation}
The covariance block then gives $\langle1-2f_m^\dagger f_m\rangle=-\zeta_m$. Since all intermode covariances vanish in this basis and the state is Gaussian, Wick's theorem implies that different canonical modes are statistically independent \cite{Peschel2003Apr,Cheong2004Feb,Peschel2009Dec,Bravyi2005May}. Normalization and the displayed expectation fix the two probabilities of mode $m$ to $(1\pm\zeta_m)/2$. Hence the reduced Gaussian density operator factorizes as
\begin{equation}
	\begin{aligned}
		\rho_A
		&=\bigotimes_{m=1}^{L_A}\rho_m,
		\\
		\rho_m
		&=\frac{1-\zeta_m}{2}|0_m\rangle\langle0_m|
		+\frac{1+\zeta_m}{2}|1_m\rangle\langle1_m|.
	\end{aligned}
	\label{eq:app_weak_factorization}
\end{equation}
Every many-body eigenvalue is therefore
\begin{equation}
		\lambda_{\boldsymbol n}
		={}\prod_{m=1}^{L_A}
		\left(\frac{1-\zeta_m}{2}\right)^{1-n_m}
		\left(\frac{1+\zeta_m}{2}\right)^{n_m},
		\quad n_m\in\{0,1\}.
	\label{eq:app_weak_manybody_eigenvalues}
\end{equation}
For each mode, the larger local factor is $(1+|\zeta_m|)/2$. Hence the largest many-body eigenvalue takes the larger factor for every $m$,
\begin{equation}
	\begin{aligned}
		\lambda_0
		&=\prod_m\frac{1+|\zeta_m|}{2}.
	\end{aligned}
	\label{eq:app_weak_lambda0}
\end{equation}
The least expensive way to obtain a different many-body configuration is to flip the mode with the smallest $|\zeta_m|$. With $\nu_A=\min_m|\zeta_m|$,
\begin{equation}
	\begin{aligned}
		\lambda_1
		&=\lambda_0\frac{1-\nu_A}{1+\nu_A},
		\\
		r_E
		&=\nu_A,
		\\
		\Delta_E
		&=2\operatorname{artanh}\nu_A.
	\end{aligned}
	\label{eq:app_weak_entanglement_gap}
\end{equation}
This proves \cref{eq:weak_gaussian_entanglement_gap}.

The subsystem parity factorizes in the same canonical basis. With the Majorana ordering fixed in \cref{eq:weak_majorana_covariance},
\begin{equation}
	\begin{aligned}
		q_A
		&=\operatorname{Tr}(\rho_AP_A)
		=(-1)^{L_A}\prod_{m=1}^{L_A}\zeta_m
		=(-1)^{L_A}\operatorname{Pf}\Gamma_A.
	\end{aligned}
	\label{eq:app_weak_pfaffian_parity}
\end{equation}
Since $(\operatorname{Pf}\Gamma_A)^2=\det\Gamma_A$, the following statements are equivalent for a Gaussian state:
\begin{equation}
	\begin{aligned}
		\lambda_0=\lambda_1
		&\Longleftrightarrow\nu_A=0
		\Longleftrightarrow\det\Gamma_A=0,
		\\
		&\Longleftrightarrow\operatorname{Pf}\Gamma_A=0
		\Longleftrightarrow q_A=0.
	\end{aligned}
	\label{eq:app_weak_zero_equivalences}
\end{equation}
The parity weights are $W_\pm=(1\pm q_A)/2$. A simple zero of one signed mode $\zeta_m$ changes its sign and exchanges the parity of the dominant many-body eigenvector. A tangential zero reaches the same spectral degeneracy without exchanging the sign, the case analyzed in \cref{app:weak_entanglement_asymptotics}.

\subsection{Partial time reversal at the covariance level}

Expand a Gaussian operator in ordered Majorana monomials. Partial time reversal multiplies every Majorana in $A$ by $i$, so on the one-body Majorana vector its action is represented by the diagonal matrix $\mathsf D_A$ in \cref{eq:weak_partial_time_reversal_covariance} \cite{Shapourian2017Apr,Shapourian2019Feb}. Therefore
\begin{equation}
	\begin{aligned}
		\Gamma^{R_A}
		&=\mathsf D_A\Gamma\mathsf D_A^T.
	\end{aligned}
	\label{eq:app_weak_ptr_covariance}
\end{equation}
The covariance is state complete because Gaussianity fixes every higher Majorana moment by Wick's theorem. For distinct indices $i_1,\ldots,i_{2r}$,
\begin{equation}
	\begin{aligned}
		\operatorname{Tr}\!\left(
		\rho\,
		w_{i_1}\cdots w_{i_{2r}}\right)
		&=(-i)^r\operatorname{Pf}\Gamma_{I},
	\end{aligned}
	\label{eq:app_weak_wick_pfaffian}
\end{equation}
where $I=(i_1,\ldots,i_{2r})$ and $\Gamma_I$ is the corresponding $2r\times2r$ covariance submatrix. Odd moments vanish for the parity-even density operators considered here. Ordered Majorana monomials form a basis of the full operator space, so the complete set of moments uniquely fixes the operator.

Applying the same factor of $i$ inside each Wick contraction replaces $\Gamma$ by $\mathsf D_A\Gamma\mathsf D_A^T$, exactly as in \cref{eq:app_weak_ptr_covariance}. The transformed moments still satisfy the Pfaffian rule, now with the generally complex covariance $\Gamma^{R_A}$. Hence $\rho^{R_A}$ is completely determined without introducing any higher correlators. For explicit finite-size reconstruction one may either evaluate the Majorana-basis coefficients from \cref{eq:app_weak_wick_pfaffian} or reduce $\Gamma^{R_A}$ to complex $2\times2$ canonical blocks and tensor the corresponding one-mode Gaussian factors. Multiplication by $P_A$ then gives the Hermitian $T_A$, whose real eigenvalues are $\eta_k$ and whose singular values are $s_k=|\eta_k|$. This proves the state-complete covariance statement used in the weak-symmetry section.

\subsection{Closed compact benchmark}

At $\mu=0$ and $\Delta=J>0$, the Hamiltonian is a sum of commuting Majorana dimers,
\begin{equation}
	\begin{aligned}
		H
		&=iJ\sum_{j=1}^{L-1}b_ja_{j+1}.
	\end{aligned}
	\label{eq:app_compact_dimer_hamiltonian}
\end{equation}
Only the dimer $Q=i b_{L_A}a_{L_A+1}$ crosses the cut. It obeys $Q^2=1$ and anticommutes with $P_A$. Hence
\begin{equation}
	\begin{aligned}
		e^{iJQt}P_Ae^{-iJQt}
		&=P_A\cos(2Jt)+iQP_A\sin(2Jt).
	\end{aligned}
	\label{eq:app_compact_parity_rotation}
\end{equation}
The product-Fock initial state is an eigenstate of $P_A$ with eigenvalue $(-1)^N$, while $\langle iQP_A\rangle_0=0$. Therefore
\begin{equation}
	\begin{aligned}
		q_A^{\mathrm{cl}}(t)
		&=(-1)^N\cos(2Jt).
	\end{aligned}
	\label{eq:app_compact_parity}
\end{equation}
The single cut covariance mode has magnitude $|\cos(2Jt)|$, so \cref{eq:app_weak_entanglement_gap} gives
\begin{equation}
	\begin{aligned}
		\lambda_\pm
		&=\frac{1\pm|\cos(2Jt)|}{2},
		\\
		\Delta_E
		&=2\operatorname{artanh}|\cos(2Jt)|.
	\end{aligned}
	\label{eq:app_compact_entanglement}
\end{equation}
For completeness, we now derive the pure-state relation used in \cref{eq:pure_state_clock_relation} \cite{Vidal2002Feb,Shapourian2017Apr,Shapourian2019Feb}. Choose a Schmidt basis adapted to fermion parity and write
\begin{equation}
	\begin{aligned}
		|\psi\rangle
		&=\sum_i\sqrt{\lambda_i}\,|i_A\rangle|i_B\rangle,
		&\sum_i\lambda_i&=1.
	\end{aligned}
	\label{eq:app_pure_schmidt_decomposition}
\end{equation}
The density matrix contains terms $\sqrt{\lambda_i\lambda_j}|i_A\rangle\langle j_A|\otimes|i_B\rangle\langle j_B|$. Fermionic partial time reversal changes the $A$ matrix unit by a phase of unit modulus and exchanges the two $A$ labels in the same way as a transpose. After the parity twist, the $i=j$ term is a one-dimensional block of magnitude $\lambda_i$. For every unordered pair $i<j$, the two cross terms form a Hermitian block of the form
\begin{equation}
	\begin{aligned}
		\begin{pmatrix}
			0&e^{i\theta_{ij}}\sqrt{\lambda_i\lambda_j}\\
			e^{-i\theta_{ij}}\sqrt{\lambda_i\lambda_j}&0
		\end{pmatrix},
	\end{aligned}
	\label{eq:app_pure_ptr_pair_block}
\end{equation}
where the phase $\theta_{ij}$ depends on the fermion-parity convention but drops out of the eigenvalue magnitudes. The block eigenvalues are $\pm\sqrt{\lambda_i\lambda_j}$. Consequently the complete nonzero singular-value multiset can be indexed by ordered pairs,
\begin{equation}
	\begin{aligned}
		s_{ij}
		&=\sqrt{\lambda_i\lambda_j}.
	\end{aligned}
	\label{eq:app_pure_ptr_singular_values}
\end{equation}
This argument also shows why the parity twist can alter signs but not the magnitudes entering $E_N$ and $\Delta_N$. For the two nonzero eigenvalues in \cref{eq:app_compact_entanglement}, the four nonzero twisted singular values are $\lambda_+$, $\lambda_-$, and two copies of $\sqrt{\lambda_+\lambda_-}$. It follows that
\begin{equation}
	\begin{aligned}
		\Delta_N
		&=\operatorname{artanh}|\cos(2Jt)|,
		\\
		E_N
		&=\ln\!\left[1+|\sin(2Jt)|\right],
		\\
		t_E=t_N
		&=\frac{\pi}{4J}.
	\end{aligned}
	\label{eq:app_compact_negativity}
\end{equation}
This proves \cref{eq:pure_state_clock_relation,eq:closed_compact_results}.

\subsection{Balanced loss and gain}

Set $\gamma_\ell=\gamma_g=\gamma_{\mathrm w}$. Then $\xi=0$ and \cref{eq:app_weak_covariance_solution} becomes
\begin{equation}
	\begin{aligned}
		\Gamma(t)
		&=f(t)\Gamma^{\mathrm{cl}}(t),
		&f(t)&=e^{-2\gamma_{\mathrm w}t}>0.
	\end{aligned}
	\label{eq:app_balanced_contraction}
\end{equation}
Every canonical covariance mode is therefore multiplied by the same positive factor. The reduced mode $m$ has eigenvalues $(1\pm f|\zeta_m^{\mathrm{cl}}|)/2$, so
\begin{equation}
	\begin{aligned}
		\Delta_{E,m}
		&=2\operatorname{artanh}\!\left(f|\zeta_m^{\mathrm{cl}}|\right).
	\end{aligned}
	\label{eq:app_balanced_entanglement_gap}
\end{equation}
For a pure Gaussian initial state and one closed entangled pair across the cut, the balanced contraction leaves a $4\times4$ Gaussian problem. Its four twisted singular values are
\begin{equation}
	\begin{aligned}
		s_{m,+}&=\frac{1+f^2+2f|\zeta_m^{\mathrm{cl}}|}{4},
		\\
		s_{m,-}&=\frac{1+f^2-2f|\zeta_m^{\mathrm{cl}}|}{4},
		\\
		s_{m,0}&=\frac{1}{4}
		\sqrt{(1+f^2)^2-4f^2(\zeta_m^{\mathrm{cl}})^2},
	\end{aligned}
	\label{eq:app_balanced_pair_singular_values}
\end{equation}
where $s_{m,0}$ occurs twice. Since $s_{m,+}\geq s_{m,0}\geq s_{m,-}$, the pair gap is $\ln(s_{m,+}/s_{m,0})$. Factoring the radicand gives
\begin{equation}
	\begin{aligned}
		\Delta_{N,m}
		&=\operatorname{artanh}\!\left(
		\frac{2f|\zeta_m^{\mathrm{cl}}|}{1+f^2}
		\right).
	\end{aligned}
	\label{eq:app_balanced_negativity_gap}
\end{equation}
Both functions vanish exactly when $\zeta_m^{\mathrm{cl}}=0$. A mode confined to one side has $|\zeta_m^{\mathrm{cl}}|=1$ and keeps a positive finite-time gap. Taking the minimum over all mode pairs therefore proves
\begin{equation}
	\begin{aligned}
		t_N=t_E=t_E^{\mathrm{cl}}
	\end{aligned}
	\label{eq:app_balanced_clock_locking}
\end{equation}
for every finite closed-system zero, which is the pure-state case of \cref{eq:weak_balanced_clock_locking}; for mixed initial states the parity rules of the main text give $t_E=t_E^{\mathrm{cl}}$ and $t_N=t_N^{\mathrm{cl}}$.

At the compact point the single cut mode has $\zeta^{\mathrm{cl}}=\cos(2Jt)$. The balanced logarithmic negativity can be written explicitly as
\begin{equation}
	\begin{aligned}
		E_N(t)
		=\ln\!\left[
		\frac{1+f^2+
			\sqrt{(1+f^2)^2-4f^2\cos^2(2Jt)}}{2}
		\right].
	\end{aligned}
	\label{eq:app_balanced_exact_negativity}
\end{equation}
The reduced R\'enyi-$1/2$ entropy is
\begin{equation}
	\begin{aligned}
		S_{1/2}(t)
		={}&(L_A-1)\ln\!\left(1+\sqrt{1-f^2}\right)
		\\
		&+\ln\!\left(1+\sqrt{1-f^2\cos^2(2Jt)}\right).
	\end{aligned}
	\label{eq:app_balanced_entropy}
\end{equation}
On $0<t<t_0$, both $f$ and $f\cos(2Jt)$ decrease in magnitude, so $\dot S_{1/2}>0$. It remains only to locate the first maximum of $E_N$. At $t_0=\pi/(4J)$ one has $\cos(2Jt_0)=0$. Differentiating \cref{eq:app_balanced_exact_negativity} at this point gives
\begin{equation}
	\begin{aligned}
		\dot E_N(t_0)
		&=-\frac{4\gamma_{\mathrm w}f_0^2}{1+f_0^2}
		=-2\gamma_{\mathrm w}
		+\mathcal O\!\left(\frac{\gamma_{\mathrm w}^2}{J}\right),
	\end{aligned}
	\label{eq:app_balanced_slope_at_t0}
\end{equation}
where $f_0=f(t_0)=e^{-2\gamma_{\mathrm w}t_0}$. At zero rate, \cref{eq:app_compact_negativity} gives
\begin{equation}
	\begin{aligned}
		\ddot E_N^{\mathrm{cl}}(t_0)&=-2J^2.
	\end{aligned}
	\label{eq:app_balanced_closed_curvature}
\end{equation}
Write the maximum as $\tau_d^g=t_0+\delta t$ with $\delta t=\mathcal O(\gamma_{\mathrm w}/J^2)$. Taylor expansion of $\dot E_N(\tau_d^g)=0$ gives
\begin{equation}
	\begin{aligned}
		0
		&=\dot E_N(t_0)
		+\ddot E_N^{\mathrm{cl}}(t_0)\delta t
		+\mathcal O\!\left(\frac{\gamma_{\mathrm w}^2}{J}\right),
	\end{aligned}
	\label{eq:app_balanced_maximum_condition}
\end{equation}
and therefore
\begin{equation}
	\begin{aligned}
		\tau_d^g
		&=t_0-\frac{\gamma_{\mathrm w}}{J^2}
		+\mathcal O\!\left(\frac{\gamma_{\mathrm w}^2}{J^3}\right).
	\end{aligned}
	\label{eq:app_balanced_geometric}
\end{equation}
Because $\dot S_{1/2}>0$ throughout $(0,t_0)$, this negativity maximum satisfies the geometric first-passage condition. This proves \cref{eq:weak_balanced_geometric_clock}.

\section{Entanglement-Clock Asymptotics under Weak Symmetry}
\label{app:weak_entanglement_asymptotics}

This appendix derives the weak-rate displacement at the compact point and the strong-rate active-block spectrum. The same signed covariance zero controls both limits, but the physical origin of the zero changes from coherent cut rotation at weak rates to local half occupation at strong rates.

\subsection{Weak-rate displacement at the compact point}

Set $\mu=0$, $\Delta=J>0$, and $t_0=\pi/(4J)$. From \cref{eq:app_compact_parity},
\begin{equation}
	\begin{aligned}
		q_A^{\mathrm{cl}}(t_0)&=0,
		&\dot q_A^{\mathrm{cl}}(t_0)&=(-1)^{N+1}2J.
	\end{aligned}
	\label{eq:app_weak_closed_root_data}
\end{equation}
At times of order $J^{-1}$, expand \cref{eq:weak_covariance_solution} to first order in the rates:
\begin{equation}
	\begin{aligned}
		\Gamma(t)
		={}&\Gamma^{\mathrm{cl}}(t)-\kappa t\Gamma^{\mathrm{cl}}(t)
		\\
		&+\xi\int_0^t du\,R(u)\Gamma_{\mathrm{vac}}R^T(u)
		+\mathcal O\!\left(\frac{\kappa^2}{J^2}\right).
	\end{aligned}
	\label{eq:app_weak_rate_covariance_expansion}
\end{equation}
The term proportional to $\kappa$ is a scalar rescaling of the closed covariance. Since the selected Pfaffian vanishes at $t_0$, this term cannot move a simple zero at first order.

The remaining source, the term carrying $\xi$, is built from the closed propagator, so we record the Majorana evolution at the compact point. The Heisenberg equation is $\dot w=\mathsf Aw$, solved by $w(u)=R(u)w$ with $R(u)=e^{\mathsf Au}$ as in \cref{eq:weak_covariance_solution}. At the compact point $H=iJ\sum_jb_ja_{j+1}$, and the only nonvanishing commutators are $i[H,b_j]=2Ja_{j+1}$ and $i[H,a_{j+1}]=-2Jb_j$, so each dimerized pair $(b_j,a_{j+1})$ rotates into itself at angular frequency $2J$, while the unpaired end Majoranas $a_1$ and $b_L$ commute with $H$ and stay fixed. Integrating the two commutators gives the independent Majorana rotations
\begin{equation}
	\begin{aligned}
		a_1(u)&=a_1,
		\\
		b_j(u)&=\cos(2Ju)b_j+\sin(2Ju)a_{j+1},
		\\
		a_{j+1}(u)&=\cos(2Ju)a_{j+1}-\sin(2Ju)b_j.
	\end{aligned}
	\label{eq:app_compact_majorana_rotation}
\end{equation}
For $L_A\geq2$, the two-dimensional kernel of $\Gamma_A^{\mathrm{cl}}(t_0)$ is spanned by $b_{L_A-1}$ and $b_{L_A}$. Projecting the source term in \cref{eq:app_weak_rate_covariance_expansion} onto this kernel gives
\begin{equation}
	\begin{aligned}
		\delta\Gamma_{b_{L_A-1},b_{L_A}}(t_0)
		&=-\xi\int_0^{t_0}du\,\sin(2Ju)\cos(2Ju)
		\\
		&=-\frac{\xi}{4J}.
	\end{aligned}
	\label{eq:app_weak_kernel_source}
\end{equation}
Let $n_{L_A}\in\{0,1\}$ be the initial occupation next to the cut and define $z_A\coloneq2n_{L_A}-1=\pm1$. To see how the projected matrix element enters the full Pfaffian, keep the same kernel orientation and approach $t_0$ along the closed evolution. The kernel element is
\begin{equation}
	\begin{aligned}
		\left[\Gamma_A^{\mathrm{cl}}(t)\right]_{b_{L_A-1},b_{L_A}}
		&=z_A\sin(2Jt)\cos(2Jt).
	\end{aligned}
	\label{eq:app_weak_closed_kernel_element}
\end{equation}
All other canonical blocks are nonsingular at $t_0$. To first order in the kernel element, the Pfaffian is its value times the Pfaffian of those nonsingular blocks. Comparing \cref{eq:app_weak_closed_kernel_element} with $q_A^{\mathrm{cl}}=(-1)^N\cos(2Jt)$ fixes that nonzero prefactor without evaluating every internal block separately:
\begin{equation}
	\begin{aligned}
		q_A
		&=(-1)^N z_A\,
		\left[\Gamma_A\right]_{b_{L_A-1},b_{L_A}}
		+\mathcal O\!\left(
		\left[\Gamma_A\right]_{b_{L_A-1},b_{L_A}}^2
		\right)
	\end{aligned}
	\label{eq:app_weak_pfaffian_kernel_factor}
\end{equation}
near the simple zero. Substituting \cref{eq:app_weak_kernel_source} and using $z_A=-(1-2n_{L_A})$ gives
\begin{equation}
	\begin{aligned}
		q_A^{(1)}(t_0)
		&=\frac{(-1)^N(1-2n_{L_A})\xi}{4J},
		\qquad L_A\geq2.
	\end{aligned}
	\label{eq:app_weak_parity_correction_largeA}
\end{equation}
For $L_A=1$, $q_A=-\Gamma_{a_1b_1}$ and
\begin{equation}
	\begin{aligned}
		q_A^{(1)}(t_0)
		&=\xi\int_0^{t_0}du\,\cos(2Ju)
		=\frac{\xi}{2J}.
	\end{aligned}
	\label{eq:app_weak_parity_correction_oneA}
\end{equation}
Write $t_E=t_0+\delta t$. The simple-root condition is
\begin{equation}
	\begin{aligned}
		0
		&=\dot q_A^{\mathrm{cl}}(t_0)\delta t+q_A^{(1)}(t_0)
		+\mathcal O\!\left(\frac{\kappa^2}{J^2}\right).
	\end{aligned}
	\label{eq:app_weak_root_condition}
\end{equation}
Using \cref{eq:app_weak_closed_root_data,eq:app_weak_parity_correction_largeA,eq:app_weak_parity_correction_oneA} gives
\begin{equation}
	\begin{aligned}
		t_E
		&=t_0+\frac{\xi}{J^2}C_E
		+\mathcal O\!\left(\frac{\kappa^2}{J^3}\right),
		\\
		C_E
		&=
		\begin{cases}
			\dfrac{1-2n_1}{4},&L_A=1,\\[5pt]
			\dfrac{1-2n_{L_A}}{8},&L_A\geq2.
		\end{cases}
	\end{aligned}
	\label{eq:app_weak_entanglement_clock}
\end{equation}
This is \cref{eq:weak_rate_entanglement_clock}. For the left-filled state, $n_{L_A}=0$ when $N<L_A$ and $n_{L_A}=1$ when $N=L_A$.

\subsection{Gain-loss duality}

Define the staggered particle-hole map by
\begin{equation}
	\begin{aligned}
		\mathcal C c_j\mathcal C^\dagger
		&=(-1)^jc_j^\dagger.
	\end{aligned}
	\label{eq:app_weak_particle_hole_map}
\end{equation}
It sends $n_j\mapsto1-n_j$, exchanges $c_j$ and $c_j^\dagger$ up to the staggered sign, and transforms the Hamiltonian as
\begin{equation}
	\begin{aligned}
		\mathcal C H(J,\Delta,\mu)\mathcal C^\dagger
		&=H(J,\Delta,-\mu)-\mu L.
	\end{aligned}
	\label{eq:app_weak_particle_hole_hamiltonian}
\end{equation}
The additive constant has no effect on dynamics. The loss and gain dissipators are exchanged. Because $\mathcal C=\mathcal C_A\otimes\mathcal C_B$ with respect to the occupation tensor product, reduced states are related by unitary conjugation. Therefore
\begin{equation}
	\begin{aligned}
		\operatorname{spec}\rho_A
		(t;\gamma_\ell,\gamma_g,\mu,\boldsymbol n)
		={}&\operatorname{spec}\rho_A
		(t;\gamma_g,\gamma_\ell,-\mu,\boldsymbol 1-\boldsymbol n),
	\end{aligned}
	\label{eq:app_weak_gain_loss_duality}
\end{equation}
which proves \cref{eq:weak_gain_loss_duality}.

\subsection{Strong-rate local kinetics}

Take $\kappa/J\gg1$ with fixed $b=|\xi|/\kappa>0$, $d=\xi/\kappa$, and $s=\kappa t$. Measuring time in units of the local relaxation time $\kappa^{-1}$ leaves the coherent generator in the covariance equation with the small prefactor $J/\kappa$, so at zeroth order each site relaxes on its own. Since the initial product-Fock state carries no covariance between different sites and the zeroth-order evolution generates none, $\Gamma$ stays block diagonal in the sites and one site can be followed at a time. Let $n\in\{0,1\}$ be the initial occupation eigenvalue of site $j$, and let $\bar n_n(s)\coloneq\langle n_j\rangle_s$ be its mean occupation, the subscript recording the initial condition through $\bar n_n(0)=n$. Loss empties an occupied site at rate $\gamma_\ell\bar n_n$ and gain fills an empty one at rate $\gamma_g(1-\bar n_n)$, so with $\gamma_\ell+\gamma_g=\kappa$ and $\gamma_g=\kappa(1-d)/2$,
\begin{equation}
	\begin{aligned}
		\frac{d\bar n_n}{dt}
		&=-\gamma_\ell\bar n_n+\gamma_g(1-\bar n_n)
		=-\kappa\bar n_n+\gamma_g,
		\\
		\frac{d\bar n_n}{ds}
		&=-\bar n_n+\frac{1-d}{2},
		\\
		\bar n_n(s)
		&=\frac{1-d}{2}
		+\left(n-\frac{1-d}{2}\right)e^{-s}.
	\end{aligned}
	\label{eq:app_weak_local_kinetics}
\end{equation}
The site forgets its initial occupation on the scale $s=1$ and settles at $(1-d)/2$, which lies below one half when loss dominates and above one half when gain dominates.

The covariance involves the local fermion parity rather than the occupation. Its operator is $(-1)^{n_j}=1-2n_j$, and its mean is
\begin{equation}
	\begin{aligned}
		p_n(s)
		&\coloneq\langle1-2n_j\rangle_s
		=1-2\bar n_n(s)
		\\
		&=d+(1-2n-d)e^{-s},
	\end{aligned}
	\label{eq:app_weak_local_parity}
\end{equation}
so $p_n=+1$ on an empty site, $p_n=-1$ on a full one, and $p_n=0$ exactly at half occupation. With $ia_jb_j=2n_j-1$ from \cref{eq:app_weak_local_bilinear}, the two Majoranas of the site carry the covariance block
\begin{equation}
	\begin{aligned}
		\begin{pmatrix}
			\Gamma_{a_ja_j}&\Gamma_{a_jb_j}\\
			\Gamma_{b_ja_j}&\Gamma_{b_jb_j}
		\end{pmatrix}
		&=
		\begin{pmatrix}
			0&-p_n\\
			p_n&0
		\end{pmatrix},
	\end{aligned}
	\label{eq:app_weak_site_block}
\end{equation}
which is already in the canonical form of \cref{eq:weak_canonical_covariance} with $\zeta=-p_n$. A site driven to half occupation therefore contributes a vanishing canonical mode, and by \cref{eq:app_weak_entanglement_gap} a vanishing mode is a closing of the entanglement gap. The zeroth-order problem thus reduces to locating the sites that pass through half occupation.

The sign of the imbalance decides which sites those are. Equation~\eqref{eq:app_weak_local_parity} moves monotonically from $p_n(0)=1-2n$ to $p_n(\infty)=d$, so it crosses zero only when these two values have opposite signs. Under loss dominance $d=b>0$, and a site that starts occupied, with $p_1(0)=-1$, is the only one that crosses. Under gain dominance $d=-b<0$, and a site that starts empty, with $p_0(0)=+1$, is the only one that crosses. We call the crossing sites active. The two cases are mirror images, since passing from one to the other flips the initial occupation and the sign of $p_n$ together, so one function covers both:
\begin{equation}
	\begin{aligned}
		\alpha(s)
		&\coloneq p_0(s)\big|_{d=-b}
		=-p_1(s)\big|_{d=b}
		\\
		&=-b+(1+b)e^{-s},
		\\
		s_\star
		&=\ln\frac{1+b}{b},
		&\alpha(s_\star)&=0.
	\end{aligned}
	\label{eq:app_weak_alpha}
\end{equation}
With this definition $|\alpha(s)|=|p_n(s)|$ on every active site, and $\alpha$ falls monotonically from $\alpha(0)=1$ to $\alpha(\infty)=-b$, passing through zero once. Two features of that crossing are used below. It is simultaneous, because every active site in the chain reaches half occupation at the same scaled time $s_\star$, so the local time $s_\star/\kappa$ replaces the coherent time $t_0$ as the reference clock. It is also transverse, because $\alpha'(s_\star)=-b\neq0$, so a small coherent correction can displace the zero, and can spoil it only by lifting the whole mode away from zero.

For the left-filled state of \cref{eq:left_filled_initial_state}, sites $1,\ldots,N$ start occupied and the rest start empty, so $A$ contains a single active block: sites $1,\ldots,N$ under loss dominance and sites $N+1,\ldots,L_A$ under gain dominance, of length $M_E=N$ and $M_E=L_A-N$ respectively, which is \cref{eq:weak_entanglement_active_length}.

\subsection{First coherent correction and standing-wave diagonalization}

At $s_\star$ every active site sits exactly at half occupation, so at zeroth order all $M$ canonical modes of an active block of $M$ sites vanish together. This subsection determines whether any of them stays at zero once the Hamiltonian is restored. Inactive sites are far from half occupation at fixed $b$, so their modes keep an order-one distance from zero and affect the small modes only at second order in $J/\kappa$. Working to first order inside one block is enough.

In the scaled time $s=\kappa t$, the covariance equation in the first line of \cref{eq:weak_covariance_solution} reads
\begin{equation}
	\begin{aligned}
		\partial_s\Gamma
		&=-\Gamma+d\,\Gamma_{\mathrm{vac}}
		+\frac{1}{\kappa}\left(\mathsf A\Gamma+\Gamma\mathsf A^T\right),
	\end{aligned}
	\label{eq:app_weak_scaled_covariance}
\end{equation}
which displays the coherent generator as the small term. Write $\Gamma=\Gamma^{(0)}+\Gamma^{(1)}+\cdots$ with $\Gamma^{(1)}=\mathcal O_b(J/\kappa)$. The zeroth order is the site-diagonal solution of the previous subsection, and the first order obeys $\partial_s\Gamma^{(1)}+\Gamma^{(1)}=\kappa^{-1}(\mathsf A\Gamma^{(0)}+\Gamma^{(0)}\mathsf A^T)$ with $\Gamma^{(1)}(0)=0$, so that
\begin{equation}
	\begin{aligned}
		\Gamma^{(1)}(s)
		&=\frac{1}{\kappa}\int_0^s\!du\,e^{-(s-u)}
		\left[\mathsf A\Gamma^{(0)}(u)+\Gamma^{(0)}(u)\mathsf A^T\right].
	\end{aligned}
	\label{eq:app_weak_memory_integral}
\end{equation}
In the integrand, at time $u$ the Hamiltonian converts on-site parity into covariance between neighboring sites, and the dissipative factor $e^{-(s-u)}$ damps what was created, so only the recent past survives in the integral.

The block is now put into matrix form. Label its sites $1,\ldots,M$, where for a block inside $A$ the length $M$ is the $M_E$ of \cref{eq:weak_entanglement_active_length}, and order its $2M$ Majoranas with all $a$ operators first,
\begin{equation}
	\begin{aligned}
		w_{\mathrm{act}}
		&=(a_1,\ldots,a_M,b_1,\ldots,b_M)^T.
	\end{aligned}
	\label{eq:app_weak_active_majorana_order}
\end{equation}
Let $I_M$ be the $M\times M$ identity and define the real antisymmetric nearest-neighbor matrix $T_M$ by
\begin{equation}
	\begin{aligned}
		(T_M)_{j,j+1}&=-1,
		&(T_M)_{j+1,j}&=1,
		\qquad1\leq j<M,
	\end{aligned}
	\label{eq:app_weak_TM}
\end{equation}
with every other entry zero, so that for $M=3$
\begin{equation}
	\begin{aligned}
		T_3
		&=
		\begin{pmatrix}
			0&-1&0\\
			1&0&-1\\
			0&1&0
		\end{pmatrix}.
	\end{aligned}
	\label{eq:app_weak_T3}
\end{equation}
On the full active Majorana space define
\begin{equation}
	\begin{aligned}
		\mathsf J_M&\coloneq
		\begin{pmatrix}0&I_M\\-I_M&0\end{pmatrix},
		&
		\mathsf T_M&\coloneq
		\begin{pmatrix}T_M&0\\0&-T_M\end{pmatrix}.
	\end{aligned}
	\label{eq:app_weak_active_matrices}
\end{equation}
In this ordering the single-site block of \cref{eq:app_weak_site_block}, repeated on the $M$ active sites, is
\begin{equation}
	\begin{aligned}
		\Gamma^{(0)}_{\mathrm{act}}(s)
		&=\alpha(s)\mathsf J_M
	\end{aligned}
	\label{eq:app_weak_active_zeroth}
\end{equation}
under loss dominance. Under gain dominance this matrix and everything generated from it change sign together, which leaves every magnitude obtained below unchanged.

Only one matrix combination is needed to evaluate \cref{eq:app_weak_memory_integral} on the block. Writing the Kitaev Hamiltonian of \cref{eq:kitaev_hamiltonian} at $\mu=0$ in the Majorana variables of \cref{eq:majorana_operators},
\begin{equation}
	\begin{aligned}
		H
		&=\frac{i}{2}\sum_{j=1}^{L-1}
		\left[(\Delta-J)a_jb_{j+1}+(\Delta+J)b_ja_{j+1}\right],
	\end{aligned}
	\label{eq:app_weak_majorana_bilinear}
\end{equation}
and reading off the generator from $\dot w=\mathsf Aw$, its restriction to the block is
\begin{equation}
	\begin{aligned}
		\mathsf A_{\mathrm{act}}
		&=
		\begin{pmatrix}
			0&-\Delta T_M-JS_M\\
			-\Delta T_M+JS_M&0
		\end{pmatrix},
	\end{aligned}
	\label{eq:app_weak_active_generator}
\end{equation}
where $S_M$ is the symmetric nearest-neighbor matrix, $(S_M)_{j,j+1}=(S_M)_{j+1,j}=1$. Since $T_M^T=-T_M$ and $S_M^T=S_M$, the hopping part cancels in the combination that \cref{eq:app_weak_memory_integral} calls for and the pairing part doubles:
\begin{equation}
	\begin{aligned}
		\mathsf A_{\mathrm{act}}\mathsf J_M
		+\mathsf J_M\mathsf A_{\mathrm{act}}^T
		&=2\Delta\,\mathsf T_M.
	\end{aligned}
	\label{eq:app_weak_JT_identity}
\end{equation}
\Cref{eq:app_weak_JT_identity} shows that the splitting inside an active block is a pairing effect.

At the compact point $\Delta=J$, inserting \cref{eq:app_weak_active_zeroth,eq:app_weak_JT_identity} into \cref{eq:app_weak_memory_integral} leaves a scalar integral multiplying the fixed matrix $\mathsf T_M$. The root sought below lies within $\mathcal O_b(J^2/\kappa^2)$ of $s_\star$, so this smooth coefficient may be evaluated once at $s_\star$. Twice the integral there is
\begin{equation}
	\begin{aligned}
		c(b)
		&=2\int_0^{s_\star}du\,e^{-(s_\star-u)}\alpha(u)
		\\
		&=2b\left(s_\star-\frac{1}{1+b}\right),
	\end{aligned}
	\label{eq:app_weak_pairing_coefficient}
\end{equation}
with $c(1)=2\ln2-1$ for pure loss or pure gain. Collecting both orders,
\begin{equation}
	\begin{aligned}
		\Gamma_{\mathrm{act}}(s)
		&=\alpha(s)\mathsf J_M
		+\frac{J}{\kappa}c(b)\mathsf T_M
		+\mathcal O_b\!\left(\frac{J^2}{\kappa^2}\right).
	\end{aligned}
	\label{eq:app_weak_active_covariance}
\end{equation}
The first term is the local half-occupation clock, shared by every active site. The second is the pairing correlation between neighbors accumulated during the fast relaxation.

The two matrices anticommute and square to multiples of the identity,
\begin{equation}
	\begin{aligned}
		\mathsf J_M^2&=-I_{2M},
		&\mathsf J_M\mathsf T_M+\mathsf T_M\mathsf J_M&=0,
	\end{aligned}
	\label{eq:app_weak_active_matrix_algebra}
\end{equation}
so squaring \cref{eq:app_weak_active_covariance} cancels the cross term and leaves, to the order kept there,
\begin{equation}
	\begin{aligned}
		\Gamma_{\mathrm{act}}^2
		&=-\alpha(s)^2I_{2M}
		+\frac{c(b)^2J^2}{\kappa^2}\mathsf T_M^2,
		\\
		\mathsf T_M^2
		&=
		\begin{pmatrix}T_M^2&0\\0&T_M^2\end{pmatrix}.
	\end{aligned}
	\label{eq:app_weak_active_square}
\end{equation}
This step makes the block elementary. A real antisymmetric matrix with canonical values $\zeta_m$ has $\Gamma^2$ eigenvalues $-\zeta_m^2$, each appearing twice, so the canonical magnitudes follow from the spectrum of $T_M^2$ alone, and that spectrum is the standing-wave problem of a box of $M$ sites with hard walls. The vectors
\begin{equation}
	\begin{aligned}
		\varphi_m(j)
		&=i^j\sin\!\left(\frac{jm\pi}{M+1}\right),
		\qquad1\leq j,m\leq M,
	\end{aligned}
	\label{eq:app_weak_standing_waves}
\end{equation}
satisfy
\begin{equation}
	\begin{aligned}
		T_M\varphi_m
		&=-2i\cos\!\left(\frac{m\pi}{M+1}\right)\varphi_m,
	\end{aligned}
	\label{eq:app_weak_TM_eigenvalues}
\end{equation}
so $T_M^2$ has the eigenvalues $-4\cos^2[m\pi/(M+1)]$. The amplitude of $\varphi_m$ vanishes at the fictitious sites $j=0$ and $j=M+1$. This is the hard-wall condition, and it quantizes the momenta as $m\pi/(M+1)$. The phase $i^j$ turns a real standing wave into an eigenvector of an antisymmetric matrix, whose eigenvalues are imaginary. The $M$ positive canonical magnitudes of the block, the $|\zeta_m|$ of \cref{eq:weak_canonical_covariance}, are therefore
\begin{equation}
	\begin{aligned}
		\nu_m(s)
		={}&\left[
		\alpha(s)^2
		+\frac{4c(b)^2J^2}{\kappa^2}
		\cos^2\!\left(\frac{m\pi}{M+1}\right)
		\right]^{1/2}
		\\
		&+\mathcal O_b\!\left(\frac{J^2}{\kappa^2}\right).
	\end{aligned}
	\label{eq:app_weak_active_modes}
\end{equation}
This is \cref{eq:weak_entanglement_active_modes}, and by \cref{eq:app_weak_entanglement_gap} the normalized gap is the smallest of these together with the order-one inactive modes.

If $M$ is odd then $m=(M+1)/2$ is an integer, the corresponding momentum is exactly $\pi/2$, and its cosine vanishes, so the pairing correction leaves that one mode untouched. Since $\alpha'(s_\star)=-b\neq0$, the implicit-function theorem gives a nearby true zero,
\begin{equation}
	\begin{aligned}
		t_E
		&=\frac{s_\star}{\kappa}
		+\mathcal O_b\!\left(\frac{J^2}{\kappa^3}\right).
	\end{aligned}
	\label{eq:app_weak_odd_clock}
\end{equation}
If $M\geq2$ is even, no allowed momentum equals $\pi/2$. The two nearest ones, $m=M/2$ and $m=M/2+1$, sit at the distance $\pi/[2(M+1)]$ from $\pi/2$ on either side, so
\begin{equation}
	\begin{aligned}
		\min_m\left|\cos\!\left(\frac{m\pi}{M+1}\right)\right|
		&=\sin\!\left(\frac{\pi}{2(M+1)}\right),
	\end{aligned}
	\label{eq:app_weak_even_nearest_mode}
\end{equation}
and the crossing is rounded into the avoided minimum
\begin{equation}
	\begin{aligned}
		r_{E,\min}
		&=\frac{2c(b)J}{\kappa}
		\sin\!\left(\frac{\pi}{2(M+1)}\right)
		+\mathcal O_b\!\left(\frac{J^2}{\kappa^2}\right),
	\end{aligned}
	\label{eq:app_weak_even_mass}
\end{equation}
which shrinks as the block lengthens, so among even blocks the longest comes closest to a closing. For $M=0$ there is no active site and no local mode approaches zero on the $\kappa^{-1}$ scale. Equations \eqref{eq:app_weak_odd_clock} and \eqref{eq:app_weak_even_mass} prove \cref{eq:weak_strong_rate_entanglement_clock}.

\subsection{Finite-rate continuation and the fold theorem}

Every closing met so far is a simple root: a time at which the selected covariance mode crosses zero with nonzero velocity. When a parameter is varied slowly, a root of this kind moves but is neither created nor destroyed until it collides with a second root. That collision is a fold, and it is the way a closing is lost at finite rate. A minimal model shows the mechanism. Let the closing condition be $m=0$ for a real function $m$ of time and of one control parameter $\Lambda$, and expand it near a collision at $(t_{\mathrm f},\Lambda_{\mathrm f})$ in the model form $m\simeq\delta t^2-\delta\Lambda$, with $\delta t=t-t_{\mathrm f}$ and $\delta\Lambda$ proportional to $\Lambda-\Lambda_{\mathrm f}$. For $\delta\Lambda>0$ there are two roots $\delta t=\pm\sqrt{\delta\Lambda}$, an early closing and a late one. As $\delta\Lambda$ decreases they approach each other like a square root. At $\delta\Lambda=0$ they merge into one tangency, where $m$ touches zero while keeping its sign. For $\delta\Lambda<0$ both are gone and $m$ retains the positive minimum $|\delta\Lambda|$, which opens linearly.

Physically, the two roots bound a time window inside which the subsystem parity $q_A$ of \cref{eq:weak_gaussian_equivalences} carries the sign opposite to its initial one. Raising the dissipation rate shrinks that window until it closes, and past the fold the parity keeps its initial sign while the entanglement gap keeps a positive minimum. The first clock then jumps to whatever later root survives, or the branch ends; \cref{fig:weak_symmetry_clocks}(b) shows both events. The rest of this subsection makes the picture exact in three steps: a single analytic function whose zeros are exactly the closings of a finite chain, its continuation in an arbitrary parameter, and the two conditions under which the square-root law is the generic outcome.

For a fixed finite chain, every entry of the exact covariance in \cref{eq:app_weak_covariance_solution} is real analytic in time and in every finite Hamiltonian or rate parameter. The root can therefore be continued without diagonalizing the reduced density matrix. Let $\Lambda$ be one such control parameter, a rate or the chemical potential for instance, and suppose that exactly one canonical covariance pair vanishes at $(t_{\mathrm r},\Lambda_{\mathrm r})$. Let $P$ project onto the two-dimensional kernel of $\Gamma_A(t_{\mathrm r},\Lambda_{\mathrm r})$ and set $Q=\mathbf 1-P$. Here and in \cref{app:weak_finite_mu}, $P$ and $Q$ denote these two spectral projectors, while the fermion parity appears as $q_A$ and $P_A$. In a sufficiently small neighborhood, $Q\Gamma_AQ$ remains invertible. The exact reduction onto the kernel, obtained by eliminating the $Q$ block, is
\begin{equation}
	\begin{aligned}
		\mathcal F_E(t,\Lambda)
		&=P\Gamma_AP
		-P\Gamma_AQ(Q\Gamma_AQ)^{-1}Q\Gamma_AP.
	\end{aligned}
	\label{eq:app_weak_finite_rate_schur}
\end{equation}
Its second term collects the excursions from the kernel into the nonsingular sector and back, and makes the reduction exact rather than a plain projection.

Equation~\eqref{eq:app_weak_finite_rate_schur} reduces the closing problem to a single number. The matrix $\mathcal F_E$ inherits antisymmetry from $\Gamma_A$, and a real antisymmetric matrix on a two-dimensional space has one independent entry, so in an orthonormal basis $u_E,v_E$ of $P$ whose order is fixed once and held as $t$ and $\Lambda$ vary,
\begin{equation}
	\begin{aligned}
		\mathcal F_E(t,\Lambda)
		&=m_E(t,\Lambda)
		\begin{pmatrix}
			0&1\\
			-1&0
		\end{pmatrix},
		\\
		m_E(t,\Lambda)
		&=u_E^T\mathcal F_E(t,\Lambda)v_E.
	\end{aligned}
	\label{eq:app_weak_finite_rate_root_function}
\end{equation}
This fixed order is what we mean by ``oriented''. It gives the sign of $m_E$ a meaning, and the sign in turn separates a transverse crossing from a tangency. Block elimination of the $Q$ sector gives $\det\Gamma_A=\det(Q\Gamma_AQ)m_E^2$ with $\det(Q\Gamma_AQ)\neq0$ nearby, so $m_E=0$ is exactly the condition $\det\Gamma_A=0$, which by \cref{eq:app_weak_zero_equivalences} is exactly $\lambda_0=\lambda_1$. The same elimination for Pfaffians gives $\operatorname{Pf}\Gamma_A=\pm\operatorname{Pf}(Q\Gamma_AQ)m_E$, so the subsystem parity and $m_E$ share their roots and their local structure, and any nonvanishing analytic factor may be dropped from a root function. Unlike the smallest canonical magnitude $\nu_A$, which carries an absolute value and has a corner at the root, $m_E$ is analytic and passes smoothly through it. At a simple root, where $\partial_tm_E\neq0$, the implicit-function theorem gives the continuation equation
\begin{equation}
	\begin{aligned}
		\frac{dt_E}{d\Lambda}
		&=-\frac{\partial_\Lambda m_E}{\partial_tm_E}.
	\end{aligned}
	\label{eq:app_weak_finite_rate_continuation}
\end{equation}
All derivatives in this equation follow by differentiating the explicit covariance solution and the finite inverse in \cref{eq:app_weak_finite_rate_schur}. The continuation fails only when the root ceases to be simple or leaves the selected time interval.

\emph{Generic fold theorem \cite{Kuznetsov2004}.} Suppose two simple roots merge at $(t_{\mathrm f},\Lambda_{\mathrm f})$ and define
\begin{equation}
	\begin{aligned}
		m_E&=0,
		&\partial_tm_E&=0,
		\\
		m_\Lambda&\coloneq\partial_\Lambda m_E\neq0,
		&m_{tt}&\coloneq\partial_t^2m_E\neq0,
	\end{aligned}
	\label{eq:app_weak_fold_condition}
\end{equation}
with every quantity evaluated at the fold. The first two equalities state that the root is tangential there, and the last two are the nondegeneracy conditions: the control parameter moves the root function, and the touch is quadratic rather than flatter. Set $m_{t\Lambda}\coloneq\partial_t\partial_\Lambda m_E$, $m_{ttt}\coloneq\partial_t^3m_E$, and $\delta\Lambda=\Lambda-\Lambda_{\mathrm f}$. Taylor's theorem gives
\begin{equation}
	\begin{aligned}
		m_E(t,\Lambda)
		={}&m_\Lambda\delta\Lambda
		+\frac{m_{tt}}{2}\delta t^2
		+m_{t\Lambda}\delta t\delta\Lambda
		+\frac{m_{ttt}}{6}\delta t^3
		\\
		&+\mathcal O(\delta\Lambda^2,\delta t^2\delta\Lambda,\delta t^4),
	\end{aligned}
	\label{eq:app_weak_fold_expansion}
\end{equation}
where $\delta t=t-t_{\mathrm f}$. On the side $-2m_\Lambda\delta\Lambda/m_{tt}>0$, the two roots are
\begin{equation}
	\begin{aligned}
		t_\pm-t_{\mathrm f}
		={}&\pm\sqrt{-\frac{2m_\Lambda}{m_{tt}}\delta\Lambda}
		\\
		&+\left(\frac{m_\Lambda m_{ttt}}{3m_{tt}^2}
		-\frac{m_{t\Lambda}}{m_{tt}}\right)\delta\Lambda
		+\mathcal O(|\delta\Lambda|^{3/2}).
	\end{aligned}
	\label{eq:app_weak_fold_law}
\end{equation}
At $\delta\Lambda=0$ there is one tangential closing. On the opposite side there is no local root, and the extremal signed mode obeys $m_{E,\min}=m_\Lambda\delta\Lambda+\mathcal O(\delta\Lambda^2)$. The square-root separation of the two roots and the linearly opening avoided gap reproduce the model behavior described at the start of this subsection, now with coefficients fixed by the chain. This conclusion uses only finite-dimensional analyticity and the two nondegeneracy conditions in \cref{eq:app_weak_fold_condition}.

\section{Finite Chemical Potential under Weak Symmetry}
\label{app:weak_finite_mu}

The compact point is special because the Majorana dimers are exactly local. This appendix shows how a finite chemical potential moves the weak-rate root and modifies the strong-rate pairing mass. Two limits are useful: a local expansion in $|\mu|/J$ and a rapid-rotation expansion in $J/|\mu|$. They describe different asymptotic regimes and should not be extrapolated into one another.

\subsection{Simple-root susceptibility at arbitrary chemical potential}

Let $t_E^{\mathrm{cl}}(\mu)$ be a finite simple root of the closed reduced covariance,
\begin{equation}
	\begin{aligned}
		\operatorname{Pf}\Gamma_A^{\mathrm{cl}}
		(t_E^{\mathrm{cl}},\mu)&=0,
		\\
		\partial_t\operatorname{Pf}\Gamma_A^{\mathrm{cl}}
		(t_E^{\mathrm{cl}},\mu)&\neq0.
	\end{aligned}
	\label{eq:app_weak_mu_simple_root}
\end{equation}
The kernel is two dimensional. Choose an oriented orthonormal basis $u_E,v_E$ of that kernel. Write the covariance near the root as
\begin{equation}
	\begin{aligned}
		\Gamma_A(t,\xi)
		&=(1-\kappa t)\Gamma_A^{\mathrm{cl}}(t)
		+\xi Y_A(t,\mu)
		+\mathcal O(\kappa^2/J^2).
	\end{aligned}
	\label{eq:app_weak_mu_covariance_expansion}
\end{equation}
Here the label $\mu$ on $\mathsf A_\mu$ fixes the chemical potential at which the generator $\mathsf A$ of \cref{eq:app_weak_hamiltonian_covariance} is built, $R_\mu(u)\coloneq e^{\mathsf A_\mu u}$ is the closed propagator it generates, and
\begin{equation}
	\begin{aligned}
		Y(t,\mu)
		&=\int_0^t\!du\,R_\mu(u)\Gamma_{\mathrm{vac}}R_\mu^T(u).
	\end{aligned}
	\label{eq:app_weak_mu_Y}
\end{equation}
The factor $(1-\kappa t)$ is the first-order expansion of the exact contraction $e^{-\kappa t}$. At the closed root, the kernel matrix element of $(1-\kappa t)\Gamma_A^{\mathrm{cl}}$ vanishes because $u_E$ and $v_E$ lie in $\ker\Gamma_A^{\mathrm{cl}}$, so this term cannot displace a simple zero at first order. Only the source proportional to $\xi$ can do so. After an orthogonal change of basis that places $u_E,v_E$ first, the Pfaffian factors into the Pfaffian of the nonsingular block times the Pfaffian of the effective $2\times2$ kernel block. The root condition to first order is therefore
\begin{equation}
		0={}u_E^T\dot\Gamma_A^{\mathrm{cl}}
		(t_E^{\mathrm{cl}},\mu)v_E\,\delta t +\xi u_E^TY_A(t_E^{\mathrm{cl}},\mu)v_E.
	\label{eq:app_weak_mu_kernel_root}
\end{equation}
Solving gives
\begin{equation}
	\begin{aligned}
		\chi_E(\mu)
		&=-\frac{u_E^TY_A(t_E^{\mathrm{cl}},\mu)v_E}
		{u_E^T\dot\Gamma_A^{\mathrm{cl}}
			(t_E^{\mathrm{cl}},\mu)v_E},
		\\
		t_E
		&=t_E^{\mathrm{cl}}(\mu)+\xi\chi_E(\mu)
		+\mathcal O\!\left(\frac{\kappa^2}{J^3}\right).
	\end{aligned}
	\label{eq:app_weak_mu_susceptibility}
\end{equation}
This proves \cref{eq:weak_finite_mu_susceptibility,eq:weak_finite_mu_simple_root}. The denominator is the velocity of the selected signed covariance mode. It vanishes at a fold, which is why the simple-root formula fails there.

For evaluation without time stepping, diagonalize the generator as $\mathsf A_\mu=\mathsf S\operatorname{diag}(i\varepsilon_1,\ldots,i\varepsilon_{2L})\mathsf S^\dagger$, with $\mathsf S$ unitary and the real single-particle frequencies $\varepsilon_m$ running over the $2L$ Majorana modes; the explicit $i$ is there because a real antisymmetric matrix has imaginary eigenvalues. Set $\widetilde\Gamma_{\mathrm{vac}}=\mathsf S^\dagger\Gamma_{\mathrm{vac}}\mathsf S$. Then
\begin{equation}
	\begin{aligned}
		Y(t,\mu)
		&=\mathsf S\left(\widetilde\Gamma_{\mathrm{vac}}\circ F(t)\right)\mathsf S^\dagger,
		\\
		F_{mn}(t)
		&=
		\begin{cases}
			\dfrac{e^{i(\varepsilon_m-\varepsilon_n)t}-1}
			{i(\varepsilon_m-\varepsilon_n)},
			&\varepsilon_m\neq\varepsilon_n,\\[8pt]
			t,&\varepsilon_m=\varepsilon_n,
		\end{cases}
	\end{aligned}
	\label{eq:app_weak_mu_spectral_sum}
\end{equation}
where $\circ$ is entrywise multiplication.

The fixed-pattern evenness in $\mu$ follows from a different transformation that does not exchange loss and gain. Let $\mathcal K$ denote complex conjugation in the occupation basis and let $\mathcal S$ be the staggered gauge transformation
\begin{equation}
	\begin{aligned}
		\mathcal S c_j\mathcal S^\dagger
		&=(-1)^j c_j.
	\end{aligned}
	\label{eq:app_weak_mu_staggered_gauge}
\end{equation}
The hopping and pairing terms both change sign under $\mathcal S$, while the density term does not. Hence
\begin{equation}
	\begin{aligned}
		\mathcal S H(\mu)\mathcal S^\dagger
		&=-H(-\mu).
	\end{aligned}
	\label{eq:app_weak_mu_hamiltonian_symmetry}
\end{equation}
The matrices of $c_j$ and $c_j^\dagger$ can be chosen real in the occupation basis, so complex conjugation changes $-i[H(\mu),\rho]$ into $-i[-H(\mu),\rho^*]$ and leaves the dissipators unchanged. Conjugating the result by $\mathcal S$ therefore maps the master equation at $\mu$ to the same master equation at $-\mu$, with the same rates. A product-Fock projector is real and is invariant under the diagonal phase transformation $\mathcal S$. Uniqueness of the finite-dimensional initial-value problem then gives
\begin{equation}
	\begin{aligned}
		\rho(t;-\mu,\boldsymbol n)
		&=\mathcal S\rho(t;\mu,\boldsymbol n)^*\mathcal S^\dagger.
	\end{aligned}
	\label{eq:app_weak_mu_density_symmetry}
\end{equation}
Since $\mathcal S=\mathcal S_A\otimes\mathcal S_B$ and complex conjugation does not change eigenvalues of a Hermitian matrix, the reduced spectra at $\pm\mu$ are identical for the same initial pattern. Consequently
\begin{equation}
	\begin{aligned}
		t_E^{\mathrm{cl}}(\mu)&=t_E^{\mathrm{cl}}(-\mu),
		&\chi_E(\mu)&=\chi_E(-\mu).
	\end{aligned}
	\label{eq:app_weak_mu_evenness}
\end{equation}
This is why the local expansion below starts with $\mu^2$.

\subsection{Expansion about zero chemical potential}

Set $\tau=Jt$, $x=\mu/J$, and $\epsilon=\xi/J$. The closed Majorana generator $\mathsf A$ of \cref{eq:app_weak_hamiltonian_covariance}, fixed by $H=(i/4)w^T\mathsf Aw$ and $\dot w=\mathsf Aw$, splits according to where its terms come from, $\mathsf A=J(\mathsf A_{\mathrm c}+x\mathsf V)$: the compact dimers of \cref{eq:app_weak_majorana_bilinear} at $\Delta=J$ supply $J\mathsf A_{\mathrm c}$, and the chemical potential supplies the on-site $a_j$-$b_j$ coupling $\mu\mathsf V$, with the dimensionless $\mathsf A_{\mathrm c}$ and $\mathsf V$ listed entry by entry in \cref{eq:app_weak_mu_generators_explicit}. Expanding in $x$ therefore counts insertions of $\mathsf V$ between free compact rotations $R_0(\tau)=e^{\mathsf A_{\mathrm c}\tau}$, and the first two terms are
\begin{equation}
	\begin{aligned}
		R_1(\tau)
		&=\int_0^\tau ds\,R_0(\tau-s)\mathsf V R_0(s),
		\\
		R_2(\tau)
		&=\int_{0<s_2<s_1<\tau}ds_1ds_2\,
		R_0(\tau-s_1)\mathsf V
		\\
		&\qquad\times R_0(s_1-s_2)\mathsf V R_0(s_2).
	\end{aligned}
	\label{eq:app_weak_mu_dyson}
\end{equation}
Each further insertion of $\mathsf V$ costs one power of $x$, so the closed covariance inherits the same expansion, with a bracketed superscript counting those powers,
\begin{equation}
	\begin{aligned}
		\Gamma_A^{\mathrm{cl}}
		&=\Gamma_A^{[0]}+x\Gamma_A^{[1]}
		+x^2\Gamma_A^{[2]}+\mathcal O(x^3),
	\end{aligned}
	\label{eq:app_weak_mu_covariance_dyson}
\end{equation}
where, before restriction to $A$,
\begin{equation}
	\begin{aligned}
		\Gamma^{[0]}&=R_0\Gamma(0)R_0^T,
		\\
		\Gamma^{[1]}&=R_1\Gamma(0)R_0^T
		+R_0\Gamma(0)R_1^T,
		\\
		\Gamma^{[2]}&=R_2\Gamma(0)R_0^T
		+R_0\Gamma(0)R_2^T
		+R_1\Gamma(0)R_1^T.
	\end{aligned}
	\label{eq:app_weak_mu_covariance_coefficients}
\end{equation}
Here every $R_k$ is evaluated at the same dimensionless time $\tau$, and $R_k$ is the term carrying $k$ insertions of $\mathsf V$. Because the physical source $Y$ in \cref{eq:app_weak_mu_Y} has dimensions of time, define the dimensionless source $\mathcal Y(\tau,x)\coloneq JY(t,\mu)$. Then
\begin{equation}
	\begin{aligned}
		\mathcal Y_A(\tau,x)
		&=\mathcal Y_A^{[0]}+x\mathcal Y_A^{[1]}
		+x^2\mathcal Y_A^{[2]}+\mathcal O(x^3),
	\end{aligned}
	\label{eq:app_weak_mu_Y_expansion}
\end{equation}
with
\begin{equation}
	\begin{aligned}
		\mathcal Y^{[0]}&=\int_0^\tau ds\,R_0\Gamma_{\mathrm{vac}}R_0^T,
		\\
		\mathcal Y^{[1]}&=\int_0^\tau ds\,
		\left(R_1\Gamma_{\mathrm{vac}}R_0^T
		+R_0\Gamma_{\mathrm{vac}}R_1^T\right),
		\\
		\mathcal Y^{[2]}&=\int_0^\tau ds\,
		\left(R_2\Gamma_{\mathrm{vac}}R_0^T
		+R_0\Gamma_{\mathrm{vac}}R_2^T
		+R_1\Gamma_{\mathrm{vac}}R_1^T\right),
	\end{aligned}
	\label{eq:app_weak_mu_Y_coefficients}
\end{equation}
where in each integrand $R_k=R_k(s)$. Since $|\epsilon|\leq\kappa/J$, the full first-order covariance is
\begin{equation}
	\begin{aligned}
		\Gamma_A
		&=(1-\kappa\tau/J)\Gamma_A^{\mathrm{cl}}
		+\epsilon\mathcal Y_A
		+\mathcal O(\kappa^2/J^2).
	\end{aligned}
	\label{eq:app_weak_mu_dimensionless_open_covariance}
\end{equation}
The term proportional to $\kappa/J$ is a scalar multiple of the closed covariance and therefore leaves every closed zero at the same time. Mixed products of this contraction with the source are second order in the rates. For the first-order root displacement we may retain only $\Gamma_A^{\mathrm{cl}}+\epsilon\mathcal Y_A$ inside the effective kernel problem. These formulas reduce the small-$\mu$ calculation to finite products and elementary time integrals of the compact-dimer rotations.

The coefficients can be reproduced without evaluating nested Dyson integrals. In the Majorana order $(a_1,b_1,\ldots,a_L,b_L)$, the only independent nonzero entries of the dimensionless matrices at $\Delta=J$ are
\begin{equation}
	\begin{aligned}
		(\mathsf A_{\mathrm c})_{2j,2j+1}&=2,
		&1\leq j<L,
		\\
		\mathsf V_{2j-1,2j}&=-1,
		&1\leq j\leq L,
	\end{aligned}
	\label{eq:app_weak_mu_generators_explicit}
\end{equation}
together with their antisymmetric partners. The initial and vacuum blocks are
\begin{equation}
	\begin{aligned}
		\Gamma(0)_{2j-1,2j}&=2n_j-1,
		&(\Gamma_{\mathrm{vac}})_{2j-1,2j}&=-1.
	\end{aligned}
	\label{eq:app_weak_mu_covariances_explicit}
\end{equation}
Write $\Gamma^{\mathrm{cl}}=\sum_{k=0}^2x^kG_k+\mathcal O(x^3)$ and $\mathcal Y=\sum_{k=0}^2x^kY_k+\mathcal O(x^3)$, so that $G_k=\Gamma^{[k]}$ and $Y_k=\mathcal Y^{[k]}$ in the shorter notation used from here on. Direct coefficient matching in the covariance equations gives the triangular initial-value problems
\begin{equation}
	\begin{aligned}
		\dot G_0
&=\mathsf A_{\mathrm c}G_0+G_0\mathsf A_{\mathrm c}^T,
\\
\dot G_k
&=\mathsf A_{\mathrm c}G_k+G_k\mathsf A_{\mathrm c}^T
+\mathsf VG_{k-1}+G_{k-1}\mathsf V^T,
\\
\dot Y_0
&=\mathsf A_{\mathrm c}Y_0+Y_0\mathsf A_{\mathrm c}^T
+\Gamma_{\mathrm{vac}},
\\
\dot Y_k
&=\mathsf A_{\mathrm c}Y_k+Y_k\mathsf A_{\mathrm c}^T
+\mathsf VY_{k-1}+Y_{k-1}\mathsf V^T,
	\end{aligned}
	\label{eq:app_weak_mu_coefficient_recursions}
\end{equation}
for $k=1,2$, with $G_0(0)=\Gamma(0)$ and $G_k(0)=Y_0(0)=Y_k(0)=0$, where the dot denotes $d/d\tau$. Each line is a constant-coefficient linear matrix equation, and each order is driven by the one below it through $\mathsf V$. Solved in order from $G_0$ and $Y_0$, they determine every matrix needed below by elementary matrix exponentiation and integration.

At $\tau_\star=\pi/4$, let $P$ project onto the two-dimensional kernel of $\Gamma_A^{\mathrm{cl}}(\tau_\star,0)$ and let $Q=\mathbf 1-P$. The kernel reduction of \cref{eq:app_weak_finite_rate_schur}, applied here to $\Gamma_A$, is
\begin{equation}
	\begin{aligned}
		\mathcal F(\Gamma_A)
		&=P\Gamma_AP
		-P\Gamma_AQ(Q\Gamma_AQ)^{-1}Q\Gamma_AP.
	\end{aligned}
	\label{eq:app_weak_mu_schur}
\end{equation}
In the oriented kernel basis, $\mathcal F=m\left(\begin{smallmatrix}0&1\\-1&0\end{smallmatrix}\right)$, so
\begin{equation}
	\begin{aligned}
		m(\tau,x,\epsilon)
		&=u_E^T\mathcal F(\Gamma_A)v_E
	\end{aligned}
	\label{eq:app_weak_mu_scalar_root}
\end{equation}
is the scalar root function. The second term in \cref{eq:app_weak_mu_schur} accounts for virtual excursions from the zero-mode subspace $P$ into the nonsingular subspace $Q$ and back. Equation~\eqref{eq:app_weak_mu_evenness} makes $m$ even in $x=\mu/J$, so its Taylor coefficients, labeled by two subscripts that count the powers of $x$ and of $\epsilon$ in that order, may be defined directly by
\begin{equation}
	\begin{aligned}
		m_{20}&=\frac{1}{2}\partial_x^2m\big|_{0},
		&m_{01}&=\partial_\epsilon m\big|_{0},
		\\
		m_{21}&=\frac{1}{2}\partial_x^2\partial_\epsilon m\big|_{0},
	\end{aligned}
	\label{eq:app_weak_mu_m_derivatives}
\end{equation}
where the vertical bar means $x=\epsilon=0$ at fixed $\tau$. We also write $m_{00}(\tau)\coloneq m(\tau,0,0)$, and a prime below denotes $d/d\tau$. Equations \eqref{eq:app_weak_mu_covariance_coefficients}, \eqref{eq:app_weak_mu_Y_coefficients}, and \eqref{eq:app_weak_mu_schur} determine these derivatives without any additional dynamical approximation.
At $\tau=\tau_\star$, restrict $G_k$ and $Y_k$ to their principal $A$ blocks and cut each one into corners with the two projectors, labeling a corner by the projectors that bracket it:
\begin{equation}
	\begin{aligned}
		G_k^{PQ}&\coloneq PG_kQ,
		&G_k^{QQ}&\coloneq QG_kQ,
		\\
		Y_k^{PQ}&\coloneq PY_kQ,
		&Y_k^{QQ}&\coloneq QY_kQ,
		\\
		K_0&\coloneq\left(G_0^{QQ}\right)^{-1},
		&K_1&\coloneq-K_0G_1^{QQ}K_0.
	\end{aligned}
	\label{eq:app_weak_mu_block_data}
\end{equation}
The last line is the expansion of the inverse that \cref{eq:app_weak_mu_schur} calls for, since $(Q\Gamma_AQ)^{-1}=K_0+xK_1+\mathcal O(x^2)$ at $\epsilon=0$. Because $P$ projects onto the kernel of $G_0(\tau_\star)$, the corner $G_0^{PQ}$ vanishes, which removes every term that would need the covariance at order zero on that corner. Expanding the inverse in \cref{eq:app_weak_mu_schur} now gives the three kernel matrices explicitly:
\begin{equation}
	\begin{aligned}
		\mathcal F_{20}
		={}&PG_2P
		+G_1^{PQ}K_0\left(G_1^{PQ}\right)^T,
		\\
		\mathcal F_{01}
		={}&PY_0P,
		\\
		\mathcal F_{21}
		={}&PY_2P
		+Y_0^{PQ}K_0\left(G_2^{PQ}\right)^T
		\\
		&+Y_1^{PQ}K_0\left(G_1^{PQ}\right)^T
		+Y_0^{PQ}K_1\left(G_1^{PQ}\right)^T
		\\
		&+G_2^{PQ}K_0\left(Y_0^{PQ}\right)^T
		+G_1^{PQ}K_0\left(Y_1^{PQ}\right)^T
		\\
		&+G_1^{PQ}K_1\left(Y_0^{PQ}\right)^T
		\\
		&-G_1^{PQ}K_0Y_0^{QQ}K_0\left(G_1^{PQ}\right)^T.
	\end{aligned}
	\label{eq:app_weak_mu_kernel_coefficients_explicit}
\end{equation}
When $Q=0$, as for $L_A=1$, every corner carrying a $Q$ is absent and only $PG_2P$, $PY_0P$, and $PY_2P$ remain.
The scalar coefficients are then
\begin{equation}
	\begin{aligned}
		m_{20}&=u_E^T\mathcal F_{20}v_E,
		&m_{01}&=u_E^T\mathcal F_{01}v_E,
		\\
		m_{21}&=u_E^T\mathcal F_{21}v_E.
	\end{aligned}
	\label{eq:app_weak_mu_scalar_coefficients_explicit}
\end{equation}
Equations \eqref{eq:app_weak_mu_generators_explicit}--\eqref{eq:app_weak_mu_scalar_coefficients_explicit} are a closed finite-matrix construction of every entry in the coefficient tables. No fitted quantity enters them. For the six geometry rows of \cref{tab:app_weak_mu_X}, the smallest representatives are $(L_A,L_B)=(1,1),(1,2),(2,1),(2,2),(3,1),(3,2)$. The columns are labeled by the number $r=L_A-N$ of initially empty sites in $A$, and the class $r\geq2$ is represented by $N=0$. Two insertions of $\mathsf V$ cannot reach farther than two sites from the cut, so enlarging either saturated side leaves these matrix elements unchanged.

Thus
\begin{equation}
	\begin{aligned}
		m
		={}&m_{00}(\tau)+x^2m_{20}(\tau)
		+\epsilon m_{01}(\tau)
		+x^2\epsilon m_{21}(\tau)
		\\
		&+\mathcal O(x^4,\epsilon x^4,\epsilon^2).
	\end{aligned}
	\label{eq:app_weak_mu_effective_root}
\end{equation}
The root of $m$ carries the same two small parameters. Write it as
\begin{equation}
	\begin{aligned}
		\tau_E
		&=\tau_\star+Ax^2
		+\epsilon\left(C+Xx^2\right)+\cdots,
	\end{aligned}
	\label{eq:app_weak_mu_root_ansatz}
\end{equation}
where the three scalars are the coefficients of the main-text expansion \cref{eq:weak_small_mu_clock} with their geometry and filling labels suppressed, $A\equiv A_{L_A,L_B,N}$ for the closed quadratic displacement, $C\equiv C_E$ for the linear response to the imbalance, and $X\equiv X_{L_A,L_B,N}$ for the mixed term. Dividing $\tau_E$ by $J$ returns physical time, which is why the same three numbers multiply $\mu^2/J^3$, $\xi/J^2$, and $\xi\mu^2/J^4$ in that expansion.
Set $\delta\tau=A x^2+\epsilon(C+Xx^2)$ and Taylor expand \cref{eq:app_weak_mu_effective_root} about $\tau_\star$. Keeping the independent orders $x^2$, $\epsilon$, and $\epsilon x^2$ gives
\begin{equation}
	\begin{aligned}
		0={}&x^2\left(m_{00}'A+m_{20}\right)
		+\epsilon\left(m_{00}'C+m_{01}\right)
		\\
		&+\epsilon x^2\big(
		m_{00}'X+m_{21}+Am_{01}'
		\\
		&\qquad\quad
		+Cm_{20}'+ACm_{00}''
		\big)+\cdots.
	\end{aligned}
	\label{eq:app_weak_mu_order_matching}
\end{equation}
The three displayed monomials are independent, so each coefficient must vanish separately. Solving the three conditions in turn gives
\begin{equation}
	\begin{aligned}
		A&=-\frac{m_{20}}{m_{00}'},
		\\
		C&=-\frac{m_{01}}{m_{00}'},
		\\
		X&=-\frac{m_{21}+Am_{01}'+Cm_{20}'+ACm_{00}''}
		{m_{00}'},
	\end{aligned}
	\label{eq:app_weak_mu_coefficients}
\end{equation}
with all quantities on the right evaluated at $\tau=\tau_\star=\pi/4$. The value of $C$ produced this way is the compact weak-rate coefficient $C_E$ of \cref{eq:weak_rate_entanglement_clock}, which was obtained independently in \cref{app:weak_entanglement_asymptotics} and therefore checks the construction. By the locality argument above, the coefficients depend only on the boundary classes listed below.

For the left-filled state, the closed quadratic coefficient is
\begin{equation}
	\begin{aligned}
		A_{L_A,L_B,N}
		&=
		\begin{cases}
			\dfrac{4-\pi}{8},&L_A=1,\ L_B=1,\ N=0,\\[5pt]
			0,&L_A=1,\ L_B=1,\ N=1,\\[5pt]
			\dfrac{8+\pi}{128},&L_A\geq2,\ L_B=1,\ N<L_A,\\[5pt]
			\dfrac{24-7\pi}{128},&L_A\geq2,\ L_B=1,\ N=L_A,\\[5pt]
			\dfrac{8+\pi}{128},&L_A=1,\ L_B\geq2,\ N=0,\\[5pt]
			\dfrac{24-7\pi}{128},&L_A=1,\ L_B\geq2,\ N=1,\\[5pt]
			\dfrac{3\pi}{128},&L_A\geq2,\ L_B\geq2,\ N<L_A,\\[5pt]
			\dfrac{\pi}{128},&L_A\geq2,\ L_B\geq2,\ N=L_A.
		\end{cases}
	\end{aligned}
	\label{eq:app_weak_mu_A_table}
\end{equation}
\Cref{tab:app_weak_mu_X} lists the mixed coefficient $X_{L_A,L_B,N}$ for every geometry and filling class.
\begin{table*}[t]
	\caption{Mixed coefficient $X_{L_A,L_B,N}$ in the weak-rate small-$\mu$ expansion. A dash marks a filling class that cannot occur for that subsystem size.}
	\label{tab:app_weak_mu_X}
	\centering
	\small
	\begin{tabular}{@{}lccc@{}}
		\toprule
		Geometry&$r\geq2$&$r=1$&$r=0$\\
		\midrule
		$L_A=1,\ L_B=1$&---&$(\pi-2)/8$&$(3-\pi)/8$\\
		$L_A=1,\ L_B\geq2$&---&$(4+3\pi)/128$&$(4-3\pi)/128$\\
		$L_A=2,\ L_B=1$&$(60-24\pi+5\pi^2)/512$&$(44-16\pi+\pi^2)/512$&$(-76+48\pi-9\pi^2)/512$\\
		$L_A=2,\ L_B\geq2$&$(92-40\pi+9\pi^2)/1024$&$(12+8\pi-3\pi^2)/1024$&$(-92+40\pi-7\pi^2)/1024$\\
		$L_A\geq3,\ L_B=1$&$(72-16\pi+3\pi^2)/1024$&$(72-32\pi+5\pi^2)/1024$&$(-72+48\pi-11\pi^2)/1024$\\
		$L_A\geq3,\ L_B\geq2$&$(32+\pi^2)/1024$&$(8+\pi^2)/1024$&$-(24+\pi^2)/1024$\\
		\bottomrule
	\end{tabular}
\end{table*}
Combining \cref{eq:app_weak_mu_coefficients,eq:app_weak_mu_A_table,tab:app_weak_mu_X} gives
\begin{equation}
	\begin{aligned}
		t_E
		={}&t_0+A_{L_A,L_B,N}\frac{\mu^2}{J^3}
		+C_E\frac{\xi}{J^2}
		+X_{L_A,L_B,N}\frac{\xi\mu^2}{J^4}
		\\
		&+\mathcal O\!\left(
		\frac{\mu^4}{J^5},
		\frac{|\xi|\mu^4}{J^6},
		\frac{\kappa^2}{J^3}
		\right),
	\end{aligned}
	\label{eq:app_weak_mu_small_expansion}
\end{equation}
which is \cref{eq:weak_small_mu_clock}. There is no term linear in $\kappa$, including no $\kappa\mu^2$ term, because the rate-sum correction is proportional to the closed covariance at first order and therefore cannot move a simple covariance zero.

\subsection{Exact two-site chemical-potential fold}

The smallest system supplies a closed-form test of the construction and resolves the divergence of the simple-root susceptibility. Take $L=2$, $L_A=L_B=1$, and define $\Omega=\sqrt{J^2+\mu^2}$. Write $q_{A,n}$ for the subsystem parity of \cref{eq:app_weak_pfaffian_parity} when the single site of $A$ starts with occupation $n\in\{0,1\}$, and $q_{A,n}^{\mathrm{cl}}$ for its closed-system value. For the initial vacuum, $n=0$, direct exponentiation of the even sector $\{|00\rangle,|11\rangle\}$ gives
\begin{equation}
	\begin{aligned}
		q_{A,0}^{\mathrm{cl}}(t,\mu)
		&=\frac{\mu^2+J^2\cos(2\Omega t)}{J^2+\mu^2}.
	\end{aligned}
	\label{eq:app_weak_mu_two_site_vacuum_parity}
\end{equation}
Consequently, simple roots exist if and only if $|\mu|<J$, and the first one, which is the $t_E^{\mathrm{cl}}(\mu)$ of \cref{eq:app_weak_mu_simple_root} for this cell, is
\begin{equation}
	\begin{aligned}
		t_0(\mu)
		&=\frac{1}{2\Omega}
		\arccos\!\left(-\frac{\mu^2}{J^2}\right).
	\end{aligned}
	\label{eq:app_weak_mu_two_site_closed_root}
\end{equation}
At $|\mu|=J$, the two roots bordering the first negative lobe of $q_{A,0}^{\mathrm{cl}}$ merge at $t_{\mathrm f}=\pi/(2\sqrt2J)$, the subscript marking the fold. For $|\mu|>J$, this lobe stays positive. Writing $\chi_n^{(2)}$ for the susceptibility $\chi_E$ of \cref{eq:app_weak_mu_susceptibility} evaluated in this two-site cell at initial occupation $n$, the exact first-order imbalance susceptibility on the simple-root side is
\begin{equation}
	\begin{aligned}
		\chi_0^{(2)}(\mu)
		&=\frac{1}{4\Omega^2}
		+\frac{\mu^2t_0(\mu)}{2J^2\Omega\sin[2\Omega t_0(\mu)]},
		\\
		t_E
		&=t_0(\mu)+\xi\chi_0^{(2)}(\mu)
		+\mathcal O(\kappa^2/J^3).
	\end{aligned}
	\label{eq:app_weak_mu_two_site_vacuum_susceptibility}
\end{equation}
It diverges with the fold exponent
\begin{equation}
	\begin{aligned}
		\chi_0^{(2)}(\mu)
		&\sim\frac{\pi}{8J^2\sqrt{1-\mu^4/J^4}},
		\qquad |\mu|\to J^-.
	\end{aligned}
	\label{eq:app_weak_mu_two_site_susceptibility_divergence}
\end{equation}
The divergence is not a physical singularity of the covariance. It means that a linear expansion about one simple root is nonuniform when the two roots are only a square-root distance apart.

The joint chemical-potential and rate unfolding follows directly from the covariance, rather than from the divergent susceptibility. Set
\begin{equation}
	\begin{aligned}
		\varrho&=1-\frac{\mu^2}{J^2},
		&\theta&=2\Omega t-\pi,
		&\delta&=|\varrho|+\frac{\kappa}{J}.
	\end{aligned}
	\label{eq:app_weak_mu_two_site_fold_variables}
\end{equation}
Here $\varrho$ is the distance to the closed fold at $|\mu|=J$, positive on the side where the two roots exist, $\theta$ is the scaled time measured from the center of the first negative lobe of $q_{A,0}^{\mathrm{cl}}$, where those roots sit, and $\delta$ collects the two small quantities. In the fold scaling $\varrho=\mathcal O(\kappa/J)$ and $\theta=\mathcal O(\delta^{1/2})$, expanding the exact covariance solution puts the finite-rate parity of \cref{eq:app_weak_mu_two_site_vacuum_parity} into the normal form
\begin{equation}
	\begin{aligned}
		q_{A,0}(t,\mu)
		&=-\frac{\varrho}{2}+\frac{\theta^2}{4}
		+\frac{\pi}{4\sqrt2}\frac{\xi}{J}
		+\mathcal O(\delta^{3/2}).
	\end{aligned}
	\label{eq:app_weak_mu_two_site_fold_normal_form}
\end{equation}
The coefficient of the third term is the first-order displacement of the parity by the imbalance, the product $-\chi_0^{(2)}\partial_tq_{A,0}^{\mathrm{cl}}$. Both factors are evaluated at $|\mu|<J$, where each is finite, and only afterwards is the limit $|\mu|\to J$ taken: the divergence of the susceptibility in \cref{eq:app_weak_mu_two_site_susceptibility_divergence} is multiplied there by the speed at which the closed parity crosses zero, which vanishes at the same rate, so the product stays finite and tends to $\pi/(4\sqrt2J)$. Setting the normal form to zero makes $\theta^2$ equal to the discriminant
\begin{equation}
	\begin{aligned}
		\mathcal D_\mu
		&=2\varrho-\frac{\pi}{\sqrt2}\frac{\xi}{J}.
	\end{aligned}
	\label{eq:app_weak_mu_two_site_fold_discriminant}
\end{equation}
If $\mathcal D_\mu>0$, two roots remain and
\begin{equation}
	\begin{aligned}
		\theta_\pm
		&=\pm\sqrt{\mathcal D_\mu}+\mathcal O(\delta).
	\end{aligned}
	\label{eq:app_weak_mu_two_site_fold_roots}
\end{equation}
If $\mathcal D_\mu=0$, they form one tangential root; if $\mathcal D_\mu<0$, the local root pair is absent. Writing $\mu_{\mathrm f}$ for the chemical potential at which $\mathcal D_\mu$ vanishes at a given rate, the fold surface is
\begin{equation}
	\begin{aligned}
		1-\frac{\mu_{\mathrm f}^2}{J^2}
		&=\frac{\pi}{2\sqrt2}\frac{\xi}{J}
		+\mathcal O[(\kappa/J)^{3/2}].
	\end{aligned}
	\label{eq:app_weak_mu_two_site_fold_surface}
\end{equation}
Loss dominance therefore moves the fold below $|\mu|=J$, while gain dominance moves it above $|\mu|=J$.

For the initially occupied state, the same two-site calculation instead gives
\begin{equation}
	\begin{aligned}
		q_{A,1}^{\mathrm{cl}}(t,\mu)
		&=-\cos(2Jt),
		&t_0&=\frac{\pi}{4J},
		\\
		\chi_1^{(2)}(\mu)
		&=-\frac{\pi\mu^2}{8J^2\Omega^2}
		-\frac{J}{4\Omega^3}
		\sin\!\left(\frac{\pi\Omega}{2J}\right).
	\end{aligned}
	\label{eq:app_weak_mu_two_site_occupied}
\end{equation}
Thus the closed root is rigid for every $\mu$, although its open-system susceptibility is not. Expanding \cref{eq:app_weak_mu_two_site_vacuum_susceptibility,eq:app_weak_mu_two_site_occupied} gives the first row of \cref{tab:app_weak_mu_X}, including $X_{1,1,0}=(\pi-2)/8$ and $X_{1,1,1}=(3-\pi)/8$.

\subsection{Rapid chemical-potential rotation at weak rates}

Take $\Delta=J>0$, $\kappa/J\ll1$, and $J/|\mu|\ll1$ on a fixed time interval $t=\mathcal O(J^{-1})$. A large chemical potential makes every change of particle number expensive, and pairing changes it by two, so pairing is driven off resonance. The frame generated by the total number $\widehat N$ \cite{Bukov2015} makes this explicit: it absorbs the term $-\mu\widehat N$, it leaves the hopping part $H_J=-J\sum_{j=1}^{L-1}(c_j^\dagger c_{j+1}+c_{j+1}^\dagger c_j)$ untouched because that part commutes with $\widehat N$, and it gives the pairing part a fast phase,
\begin{equation}
	\begin{aligned}
		\widetilde H(t)
		&=H_J+J\left(e^{2i\mu t}\widehat B
		+e^{-2i\mu t}\widehat B^\dagger\right),
		\\
		\widehat B&=\sum_{j=1}^{L-1}c_jc_{j+1}.
	\end{aligned}
	\label{eq:app_weak_mu_rotating_hamiltonian}
\end{equation}
The loss and gain dissipators are invariant under this global number rotation. In the interaction picture of the number-conserving Liouvillian, the two oscillatory perturbations carry number charges $+2$ and $-2$. An operator of definite charge changes the total particle number by that charge. Since the initial product-Fock density matrix and $P_A$ both have charge zero, a single insertion of either $\widehat B$ or $\widehat B^\dagger$ cannot contribute to $\operatorname{Tr}(P_A\rho)$: the first Dyson correction has nonzero total charge and vanishes after the trace.

At second order, the only charge-zero terms contain one $\widehat B$ and one $\widehat B^\dagger$. The oscillatory integrals are controlled by
\begin{equation}
	\begin{aligned}
		\int_0^t ds\,e^{2i\mu s}F(s)
		={}&\frac{e^{2i\mu t}F(t)-F(0)}{2i\mu}
		\\
		&-\frac{1}{2i\mu}\int_0^t ds\,e^{2i\mu s}\dot F(s),
	\end{aligned}
	\label{eq:app_weak_mu_integration_by_parts}
\end{equation}
which follows from ordinary integration by parts and remains valid for the finite-dimensional operator-valued functions $F(s)$ appearing here. Each application trades an oscillating integrand for a boundary term plus a new integral carrying one more power of $1/\mu$, which is how the fast phase suppresses pairing. Applying it once to the time-ordered second-order term leaves a single piece whose phases cancel, so that its effect accumulates with time instead of averaging away,
\begin{equation}
	\begin{aligned}
		\frac{J^2}{2\mu}\left[\widehat B,\widehat B^\dagger\right].
	\end{aligned}
	\label{eq:app_weak_mu_secular_term}
\end{equation}
This is a virtual pair process: pairing creates a pair and reabsorbs it within a time of order $1/\mu$, and the denominator $2\mu$ is the energy cost of that excursion. The commutator appears because the two orderings, creation followed by reabsorption and the reverse, enter with opposite signs, and what survives conserves the particle number, which is why the fast phase does not average it away. This term changes sign under $\mu\mapsto-\mu$. The exact fixed-pattern parity and reduced spectrum are even in $\mu$ by \cref{eq:app_weak_mu_evenness}, so this odd contribution cannot shift the selected root. Applying \cref{eq:app_weak_mu_integration_by_parts} once more to the remaining oscillatory terms supplies a second factor $1/\mu$. On a fixed finite chain and fixed interval $0\leq t\leq T=\mathcal O(J^{-1})$, the relevant operator norms and their time derivatives are bounded. The remainder is therefore uniformly $\mathcal O(J^2/\mu^2)$. Writing nc for number conserving, we obtain
\begin{equation}
	\begin{aligned}
		\sup_{0\leq t\leq T}
		\left|q_A(t;\mu)-q_A^{\mathrm{nc}}(t)\right|
		&=\mathcal O\!\left(\frac{J^2}{\mu^2}\right),
	\end{aligned}
	\label{eq:app_weak_mu_rapid_bound}
\end{equation}
where $q_A^{\mathrm{nc}}$ is the parity for the same problem with $\Delta=0$. If the number-conserving problem has an isolated simple first root $t_E^{\mathrm{nc}}$ with nonzero velocity, uniform convergence and the implicit-function theorem give
\begin{equation}
	\begin{aligned}
		t_E
		&=t_E^{\mathrm{nc}}+\xi\chi_E^{\mathrm{nc}}
		+\mathcal O\!\left(
		\frac{\kappa^2}{J^3}+\frac{J}{\mu^2}
		\right),
	\end{aligned}
	\label{eq:app_weak_mu_rapid_clock}
\end{equation}
which is \cref{eq:weak_rapid_rotation_clock}. If the number-conserving gap stays strictly positive on the interval, the bound instead proves the absence of a large-$|\mu|$ root there.

\subsection{Strong rates at arbitrary chemical-potential ratio}

Now assume $\kappa\gg|J|,|\Delta|$ with no restriction on $\mu/\kappa$, and recall the strong-rate variables of \cref{eq:weak_strong_rate_variables}: the imbalance ratio $b=|\xi|/\kappa$, the scaled time $s=\kappa t$, the active-site detuning $\alpha(s)$ of \cref{eq:app_weak_alpha}, and its zero at $s_\star$. The same rotating frame leaves the local kinetics of \cref{eq:app_weak_local_kinetics} and the local clock $s_\star$ unchanged, while pairing created at scaled time $u$ now carries the phase $e^{i\omega u}$ with $\omega=2\mu/\kappa$. Carrying that phase through the memory integral of \cref{eq:app_weak_memory_integral} replaces the coefficient $c(b)$ of \cref{eq:app_weak_pairing_coefficient} by
\begin{equation}
	\begin{aligned}
		c_b(\omega)
		&=2\int_0^{s_\star}du\,
		e^{-(s_\star-u)}\alpha(u)e^{i\omega u}
		\\
		&=\frac{2b\left[e^{i\omega s_\star}-1-i\omega/(1+b)\right]}
		{i\omega(1+i\omega)}.
	\end{aligned}
	\label{eq:app_weak_mu_cb}
\end{equation}
The removable value at $\omega=0$ is $c_b(0)=c(b)$. Repeating the active-block calculation gives
\begin{equation}
	\begin{aligned}
		\nu_m(s)
		={}&\left[
		\alpha(s)^2
		+\frac{4|c_b(2\mu/\kappa)|^2\Delta^2}{\kappa^2}
		\cos^2\!\left(\frac{m\pi}{M+1}\right)
		\right]^{1/2}
		\\
		&+\mathcal O_b\!\left(\frac{J^2+\Delta^2}{\kappa^2}\right).
	\end{aligned}
	\label{eq:app_weak_mu_strong_modes}
\end{equation}
For odd $M$, the $m=(M+1)/2$ mode still has zero cosine, so
\begin{equation}
	\begin{aligned}
		t_E
		&=\frac{s_\star}{\kappa}
		+\mathcal O_b\!\left(\frac{J^2+\Delta^2}{\kappa^3}\right)
	\end{aligned}
	\label{eq:app_weak_mu_odd_clock}
\end{equation}
uniformly in $\mu$. For positive even $M$,
\begin{equation}
	\begin{aligned}
		r_{E,\min}
		&=\frac{2|c_b(2\mu/\kappa)\Delta|}{\kappa}
		\sin\!\left(\frac{\pi}{2(M+1)}\right)
		\\
		&\quad+\mathcal O_b\!\left(\frac{J^2+\Delta^2}{\kappa^2}\right),
	\end{aligned}
	\label{eq:app_weak_mu_even_mass}
\end{equation}
which proves \cref{eq:weak_finite_mu_pairing_factor,eq:weak_finite_mu_even_mass}. For $|\mu|\gg\kappa$, $|c_b(2\mu/\kappa)|\sim b\kappa/[(1+b)|\mu|]$, so the mass of an even-length block is suppressed by phase averaging. When $\Delta=0$, this leading mass vanishes and the outcome is controlled by higher-order hopping terms.

\section{Twisted-Negativity Asymptotics under Weak Symmetry}
\label{app:weak_negativity_asymptotics}

The negativity clock uses the full twisted operator, so its first-order problem is larger than the reduced covariance problem even at the compact point. Locality simplifies it, since only the cut dimer and an unpaired end Majorana adjacent to that dimer can split the four closing magnitudes at first order.

\subsection{Weak rates and the fourfold closing space}

At $\gamma_\ell=\gamma_g=0$ the reduced state has the two nonzero eigenvalues of \cref{eq:app_compact_entanglement}, which before the first root keep a fixed order and may be written without the absolute value,
\begin{equation}
	\begin{aligned}
		\lambda_+(t)&=\frac{1+\cos(2Jt)}{2},
		&\lambda_-(t)&=\frac{1-\cos(2Jt)}{2}.
	\end{aligned}
	\label{eq:app_weak_neg_closed_lambdas}
\end{equation}
By \cref{eq:app_pure_ptr_singular_values}, the four nonzero twisted singular values are
\begin{equation}
	\begin{aligned}
		\lambda_+,
		\quad\lambda_-,
		\quad\sqrt{\lambda_+\lambda_-},
		\quad\sqrt{\lambda_+\lambda_-}.
	\end{aligned}
	\label{eq:app_weak_neg_four_singular_values}
\end{equation}
They all equal $1/2$ at $t_0=\pi/(4J)$, which is the closing. To follow that closing at finite rate we need one scalar whose zero is the closing, analogous to $m_E$ for the entanglement gap in \cref{eq:app_weak_finite_rate_root_function}. Let $g_N(t,\kappa)$ be the signed difference between the largest and the second largest of the four magnitudes in \cref{eq:app_weak_neg_four_singular_values}, at time $t$ and total rate $\kappa$, with the ordering of the two levels fixed once and carried through the crossing, so that $g_N$ changes sign there instead of bouncing off zero. In the closed system the largest is $\lambda_+$ and the second is $\sqrt{\lambda_+\lambda_-}$, so
\begin{equation}
	\begin{aligned}
		g_N(t,0)
		&=\lambda_+(t)-\sqrt{\lambda_+(t)\lambda_-(t)},
	\end{aligned}
	\label{eq:app_weak_neg_oriented_branch}
\end{equation}
and this branch crosses zero transversally,
\begin{equation}
	\begin{aligned}
		g_N(t_0,0)&=0,
		&\partial_tg_N(t_0,0)&=-J.
	\end{aligned}
	\label{eq:app_weak_neg_branch_velocity}
\end{equation}

Strip the rates from the two dissipative terms of \cref{eq:weak_lindblad} and call what is left $\mathcal L_\ell\rho=\sum_j\mathcal D[c_j]\rho$ and $\mathcal L_g\rho=\sum_j\mathcal D[c_j^\dagger]\rho$. Superoperators are written here without brackets and act on everything to their right. To first order in the rates,
\begin{equation}
	\begin{aligned}
		\rho(t)
		={}&\rho_0(t)
		+\gamma_\ell\int_0^t\!du\,\mathcal U_{t-u}\mathcal L_\ell\mathcal U_u\rho(0)
		\\
		&+\gamma_g\int_0^t\!du\,\mathcal U_{t-u}\mathcal L_g\mathcal U_u\rho(0)
		+\mathcal O(\kappa^2/J^2),
	\end{aligned}
	\label{eq:app_weak_neg_dyson_density}
\end{equation}
where $\mathcal U_t$ is the closed evolution, $\mathcal U_tX=e^{-iHt}Xe^{iHt}$, so each term reads from the right as evolve to $u$, jump, evolve the rest of the way. Partial time reversal in \cref{eq:partial_time_reversal} and the parity twist in \cref{eq:twisted_operator} are both linear maps, so the same three terms carry over to $T_A$ one by one.

Two of the four closing levels of the closed $T_A(t_0)$ are positive and two are negative. Let $\Pi_+$ and $\Pi_-$ project onto those two eigenspaces, $\Pi_-$ being the negative-branch projector already used in \cref{eq:twisted_negativity_hamiltonian}. A Hermitian perturbation $\delta T_A$ changes the four magnitudes through the two projected matrices
\begin{equation}
	\begin{aligned}
		\Pi_+\delta T_A\Pi_+,
		\qquad
		-\Pi_-\delta T_A\Pi_-.
	\end{aligned}
	\label{eq:app_weak_neg_secular_blocks}
\end{equation}
The minus sign in the second block appears because a negative eigenvalue $\eta$ contributes the singular value $|\eta|=-\eta$. The remaining spectrum is separated from the closing value by an order-one gap, so mixing with it changes these four levels only at second order.

At the compact point every internal dimer factorizes. A dissipative event farther than one bond from the cut changes all four closing levels by the same amount or acts entirely on one side, and therefore cancels from their differences. Only the two Majoranas of the cut dimer, and an unpaired end Majorana next to it when the cut sits at a chain end, are left. Their contribution is one jump at some time $u$, after which the closed rotation of \cref{eq:app_compact_majorana_rotation} carries the disturbance to the observation time: a jump on a bond whose two Majoranas both rotate enters through the product $\sin(2Ju)\cos(2Ju)$, while a jump involving an end Majorana, which stands still, enters through a single $\cos(2Ju)$. Integrating over the time of the jump up to $t_0$ gives the only two numbers the first-order problem needs,
\begin{equation}
	\begin{aligned}
		I_{\mathrm{bulk}}
		&=\int_0^{t_0}du\,\sin(2Ju)\cos(2Ju)
		=\frac{1}{4J},
		\\
		I_{\mathrm{edge}}
		&=\int_0^{t_0}du\,\cos(2Ju)
		=\frac{1}{2J}.
	\end{aligned}
	\label{eq:app_weak_neg_elementary_integrals}
\end{equation}
These are the same two integrals that set the entanglement-clock offsets in \cref{eq:app_weak_kernel_source,eq:app_weak_parity_correction_oneA}, and the factor of two between them is the origin of the $1/8$ bulk-cut and $1/4$ physical-edge offsets in the clock.

The projected matrices can now be displayed explicitly. In each sign sector choose the basis that diagonalizes the local cut occupation; in this basis the displacement in time and the imbalance perturbation commute, so both are diagonal and the four magnitudes shift independently. Collect those four shifts, the two from $\Pi_+$ first and the two from $\Pi_-$ last, into one diagonal matrix, and subtract from it a scalar multiple of the identity. That scalar carries the whole $\kappa$ contribution, which moves all four levels together and cannot change a difference. The remaining matrix is
\begin{equation}
	\begin{aligned}
		\mathcal M_{\mathrm{split}}
		&=J\delta t\,\operatorname{diag}(\boldsymbol\tau)
		+\frac{\xi}{J}\operatorname{diag}(\boldsymbol\beta)
		+\mathcal O(\kappa^2/J^2),
	\end{aligned}
	\label{eq:app_weak_neg_secular_matrix_explicit}
\end{equation}
where $\delta t=t-t_0$, and the four-component vectors $\boldsymbol\tau$ and $\boldsymbol\beta$ hold the coefficients with which the time displacement and the imbalance move each of the four levels. Direct insertion of the two integrals in \cref{eq:app_weak_neg_elementary_integrals} gives their values, listed in \cref{tab:app_weak_neg_secular_matrices}.
\begin{table*}[t]
	\caption{First-order shifts of the four closing magnitudes, one row per geometry and filling class. The four entries of $\boldsymbol\tau$ multiply the time displacement $J\delta t$ and the four entries of $\boldsymbol\beta$ multiply the imbalance $\xi/J$, in the order of \cref{eq:app_weak_neg_secular_matrix_explicit}: the first two components are the two levels of the positive sector $\Pi_+$, the last two those of the negative sector $\Pi_-$. A common shift of all four entries has been removed, since it leaves every difference unchanged.}
	\label{tab:app_weak_neg_secular_matrices}
	\centering
	\small
	\begin{tabular}{@{}lll@{}}
		\toprule
		Geometry and filling&$\boldsymbol\tau$&$\boldsymbol\beta$\\
		\midrule
		$L_A=L_B=1$, $N=0$&$(0,1,-1,0)$&$(0,-1/4,1/4,0)$\\
		$L_A=L_B=1$, $N=1$&$(-1,0,0,1)$&$(0,-1/4,1/4,0)$\\
		$L_A=1$, $L_B\geq2$, $N=0$&$(0,1,-1,0)$&$(-1/16,-3/16,3/16,1/16)$\\
		$L_A=1$, $L_B\geq2$, $N=1$&$(-1,0,0,1)$&$(-1/16,-3/16,3/16,1/16)$\\
		$L_A\geq2$, $L_B=1$, $N<L_A$&$(-1,0,0,1)$&$(3/16,-1/16,1/16,-3/16)$\\
		$L_A\geq2$, $L_B=1$, $N=L_A$&$(-1,0,0,1)$&$(1/16,-3/16,3/16,-1/16)$\\
		$L_A,L_B\geq2$, $N<L_A$&$(-1,0,0,1)$&$(1/8,0,0,-1/8)$\\
		$L_A,L_B\geq2$, $N=L_A$&$(-1,0,0,1)$&$(0,-1/8,1/8,0)$\\
		\bottomrule
	\end{tabular}
\end{table*}
For example, the bulk row with $N<L_A$ gives the four centered corrections $-J\delta t+\xi/(8J),0,0,J\delta t-\xi/(8J)$. Their leading level is degenerate at $J^2\delta t=\xi/8$. For $N=L_A$, the entries are $-J\delta t,-\xi/(8J),\xi/(8J),J\delta t$, which give both signs of the root. Thus the bulk-cut conditions are
\begin{equation}
	\begin{aligned}
		J^2\delta t-\frac{\xi}{8}&=0,
		&&N<L_A,
		\\
		\left(J^2\delta t-\frac{\xi}{8}\right)
		\left(J^2\delta t+\frac{\xi}{8}\right)&=0,
		&&N=L_A.
	\end{aligned}
	\label{eq:app_weak_neg_bulk_factors}
\end{equation}
The rows containing a one-site side also yield the $\xi/4$ root. Its offset is doubled because $I_{\mathrm{edge}}/I_{\mathrm{bulk}}=2$. Equations \eqref{eq:app_weak_neg_dyson_density}, \eqref{eq:app_weak_neg_elementary_integrals}, and \eqref{eq:app_weak_neg_secular_matrix_explicit} together give every matrix entry and its physical origin.

Collecting these local factors gives the complete branch set in \cref{tab:app_weak_negativity_branches}.
\begin{table}[H]
	\caption{Complete first-order offset set $\mathcal B_{L_A,L_B,N}(\xi)$ for the weak-rate negativity clock. Every nonempty bipartition and every filling $0\leq N\leq L_A$ belongs to exactly one row.}
	\label{tab:app_weak_negativity_branches}
	\centering
	\small
	\begin{tabular}{@{}ll@{}}
		\toprule
		Geometry and filling&$\mathcal B_{L_A,L_B,N}(\xi)$\\
		\midrule
		$L_A=L_B=1$, $N=0$&$\{\xi/4\}$\\
		$L_A=L_B=1$, $N=1$&$\{-\xi/4,\xi/4\}$\\
		$L_A=1$, $L_B\geq2$, $N=0$&$\{\xi/8,\xi/4\}$\\
		$L_A=1$, $L_B\geq2$, $N=1$&$\{\xi/8,-\xi/4\}$\\
		$L_A\geq2$, $L_B=1$, $N<L_A$&$\{\xi/8,\xi/4\}$\\
		$L_A\geq2$, $L_B=1$, $N=L_A$&$\{-\xi/8,\xi/4\}$\\
		$L_A,L_B\geq2$, $N<L_A$&$\{\xi/8\}$\\
		$L_A,L_B\geq2$, $N=L_A$&$\{-\xi/8,\xi/8\}$\\
		\bottomrule
	\end{tabular}
\end{table}

Each element $B\in\mathcal B_{L_A,L_B,N}$ is the coefficient in a local branch
\begin{equation}
	\begin{aligned}
		t=t_0+\frac{B}{J^2}+\mathcal O(\kappa^2/J^3).
	\end{aligned}
	\label{eq:app_weak_neg_local_branch}
\end{equation}
The first clock is the smallest positive-time branch, so
\begin{equation}
	\begin{aligned}
		t_N
		&=t_0+\frac{1}{J^2}\min\mathcal B_{L_A,L_B,N}(\xi)
		+\mathcal O\!\left(\frac{\kappa^2}{J^3}\right),
	\end{aligned}
	\label{eq:app_weak_neg_clock}
\end{equation}
which proves \cref{eq:weak_rate_negativity_clock}. The remaining elements of $\mathcal B$ mark later closings of the same spectrum. For a bulk cut $L_A,L_B\geq2$, the table immediately reduces to \cref{eq:weak_bulk_negativity_clock}: the unique offset is $\xi/8$ when $N<L_A$, while $N=L_A$ has the pair $\{-\xi/8,\xi/8\}$ and hence the first displacement $-|\xi|/8$.

\subsection{Exact all-rate branches and their fold in the minimal cell}

The smallest nontrivial cut can be solved at every rate, which turns the weak-rate branches of \cref{tab:app_weak_negativity_branches} into curves that can be followed until they disappear. Take $L=2$ with $L_A=L_B=1$, set $\mu=0$ and $\Delta=J$, keep $t_0=\pi/(4J)$, and write $a\coloneq|\xi|=|\gamma_\ell-\gamma_g|$ for the size of the loss-gain imbalance of \cref{eq:loss_gain_parameters}. The bare letter $a$ is not to be confused with the Majorana operators $a_j$, which always carry a site index.

The state stays Gaussian, and the closing conditions involve only the two mean occupations, $\bar n_1(t)$ on the site in $A$ and $\bar n_2(t)$ on the site in $B$. Two combinations of them do the work below, the total occupation measured from half filling and the imbalance between the two sites,
\begin{equation}
	\begin{aligned}
		x&\coloneq\bar n_1+\bar n_2-1,
		&y&\coloneq\bar n_1-\bar n_2,
	\end{aligned}
	\label{eq:app_weak_neg_minimal_xy}
\end{equation}
because each on-site covariance entry of \cref{eq:app_weak_site_block} is their sum or difference, $\Gamma_{a_1b_1}=2\bar n_1-1=x+y$ and $\Gamma_{a_2b_2}=2\bar n_2-1=x-y$. For the initial state $|N,0\rangle$ of \cref{eq:left_filled_initial_state}, the exact solution \cref{eq:app_weak_covariance_solution} gives
\begin{equation}
	\begin{aligned}
		x(t)
		&=(N-1)e^{-\kappa t}\cos(2Jt)-\xi I_\kappa(t),
		\\
		y(t)
		&=Ne^{-\kappa t}\cos(2Jt),
		\\
		I_\kappa(t)
		&=\int_0^tdu\,e^{-\kappa u}\cos(2Ju)
		\\
		&=\frac{\kappa+e^{-\kappa t}[2J\sin(2Jt)-\kappa\cos(2Jt)]}
		{\kappa^2+4J^2}.
	\end{aligned}
	\label{eq:app_weak_neg_minimal_exact_xy}
\end{equation}
The first term of each line is the closed oscillation damped at the total rate, and $I_\kappa$ collects the vacuum source that the imbalance feeds in. Since $A$ holds a single site, its covariance has the one canonical mode $x+y$, so \cref{eq:app_weak_entanglement_gap} makes the normalized entanglement gap $r_E=|x+y|$.

The twisted operator $T_A$ of \cref{eq:twisted_operator} splits into two Hermitian $2\times2$ blocks, one on the odd global-parity states $|10\rangle$ and $|01\rangle$ and one on the even states $|00\rangle$ and $|11\rangle$. Direct evaluation gives these blocks the traces $x$ and $y$ and one common determinant, so writing $R_o>0$ and $R_e>0$ for the spacing between the two eigenvalues inside each block, the signed eigenvalues of \cref{eq:twisted_spectrum} are
\begin{equation}
	\begin{aligned}
		\eta_{o,\pm}&=\frac{x\pm R_o}{2},
		&\eta_{e,\pm}&=\frac{y\pm R_e}{2},
		\\
		R_o^2-R_e^2&=x^2-y^2,
	\end{aligned}
	\label{eq:app_weak_neg_minimal_twisted_eigenvalues}
\end{equation}
the last line being the equality of the two determinants. That common determinant is negative, so each block holds one positive and one negative eigenvalue, which is to say $R_o\geq|x|$ and $R_e\geq|y|$. The largest magnitude of each block is then $(R_o+|x|)/2$ and $(R_e+|y|)/2$, and the identity in the second line turns their difference into $(|x|-|y|)(1+(|x|+|y|)/(R_o+R_e))/2$, which carries the sign of $|x|-|y|$. The same inequalities give $|R_o-R_e|\leq\big||x|-|y|\big|$, so the runner-up is the leading magnitude of the other block rather than the small magnitude of the leading one. With $s_0$ and $s_1$ the two largest magnitudes of \cref{eq:twisted_spectrum},
\begin{equation}
	\begin{aligned}
		s_0=s_1
		&\Longleftrightarrow |x|=|y|
		\Longleftrightarrow (x+y)(x-y)=0,
	\end{aligned}
	\label{eq:app_weak_neg_minimal_exact_condition}
\end{equation}
so at every rate the negativity closing is carried by the two signed scalars $x+y$ and $x-y$, one for each site.

Both scalars are explicit in \cref{eq:app_weak_neg_minimal_exact_xy}, and setting either of them to zero gives an equation of the same form,
\begin{equation}
	\begin{aligned}
		e^{-\kappa t}\cos(2Jt)&=\pm aI_\kappa(t),
	\end{aligned}
	\label{eq:app_weak_neg_minimal_root_equations}
\end{equation}
with the sign fixed by the filling $N$ and by the sign of $\xi$. The upper sign has exactly one root before $t_0$. The difference between its two sides starts at $1$, its derivative $-e^{-\kappa t}[(\kappa+a)\cos(2Jt)+2J\sin(2Jt)]$ stays negative on $(0,t_0)$ where both trigonometric factors are positive, and at $t_0$ the difference has reached $-aI_\kappa(t_0)<0$. Call that root $t_<$,
\begin{equation}
	\begin{aligned}
		e^{-\kappa t_<}\cos(2Jt_<)
		&=aI_\kappa(t_<).
	\end{aligned}
	\label{eq:app_weak_neg_minimal_early_root}
\end{equation}
The lower sign needs a negative left-hand side, so it can only be met after $t_0$. Its first root, when one exists, is
\begin{equation}
	\begin{aligned}
		e^{-\kappa t_>}\cos(2Jt_>)
		&=-aI_\kappa(t_>).
	\end{aligned}
	\label{eq:app_weak_neg_minimal_late_root}
\end{equation}
Reading off which of $x\pm y$ vanishes in each case, and recalling that only $x+y$ controls $r_E$, gives the complete clock map
\begin{equation}
	\begin{aligned}
		N=0,\ \xi>0:&\quad t_E=t_N=t_>,
		\\
		N=0,\ \xi<0:&\quad t_E=t_N=t_<,
		\\
		N=1,\ \xi>0:&\quad t_E=t_N=t_<,
		\\
		N=1,\ \xi<0:&\quad t_N=t_<,\quad t_E=t_>.
	\end{aligned}
	\label{eq:app_weak_neg_minimal_branch_map}
\end{equation}
If the late root is absent, the clock assigned to it is $+\infty$. Expanding \cref{eq:app_weak_neg_minimal_early_root,eq:app_weak_neg_minimal_late_root} at weak rates gives $t_\lessgtr=t_0\mp|\xi|/(4J^2)+\mathcal O(\kappa^2/J^3)$ and reproduces the first two rows of \cref{tab:app_weak_negativity_branches}.

Whether the late root exists at all can be settled exactly, and this is where a branch is lost. Measure time and rate in units of the coherent frequency,
\begin{equation}
	\begin{aligned}
		q&=\frac{\kappa}{2J},
		&b&=\frac{|\xi|}{\kappa},
		\\
		\theta&=2Jt,
		&\theta_\ast&=\pi-\arctan[q(1-b)],
	\end{aligned}
	\label{eq:app_weak_neg_minimal_fold_variables}
\end{equation}
where $q$ is a scaled dissipation strength, unrelated to the subsystem parity $q_A$, and $b$ is the imbalance ratio of \cref{eq:weak_strong_rate_variables}. Multiplying \cref{eq:app_weak_neg_minimal_late_root} by $(1+q^2)$ and rewriting it in these variables puts the late-root condition in the form $F_b(\theta)=0$ with
\begin{equation}
	\begin{aligned}
		F_b(\theta)
		={}&e^{-q\theta}
		\{[1+q^2(1-b)]\cos\theta+bq\sin\theta\}
		+bq^2.
	\end{aligned}
	\label{eq:app_weak_neg_minimal_fold_function}
\end{equation}
Its derivative factorizes,
\begin{equation}
	\begin{aligned}
		F_b'(\theta)
		&=-(1+q^2)e^{-q\theta}
		[\sin\theta+q(1-b)\cos\theta],
	\end{aligned}
	\label{eq:app_weak_neg_minimal_fold_derivative}
\end{equation}
so $F_b$ turns exactly where $\tan\theta=-q(1-b)$, and the $\theta_\ast$ of \cref{eq:app_weak_neg_minimal_fold_variables} is that turning point in the first stretch where $F_b$ can dip below zero, just short of $\theta=\pi$. Substituting it gives the exact criterion for a late root,
\begin{equation}
	\begin{aligned}
		e^{-q\theta_\ast}
		\sqrt{1+q^2(1-b)^2}
		&\geq bq^2.
	\end{aligned}
	\label{eq:app_weak_neg_minimal_fold_criterion}
\end{equation}
Strict inequality gives two late roots, equality gives one tangential double root, and the reverse inequality gives none. The criterion covers all later times as well, because $F_b$ is a decaying oscillation sitting on the constant $bq^2$: one more period multiplies the oscillating part by $e^{-2\pi q}$ and leaves the constant where it is, so every later dip is shallower than the first. A first dip that fails to reach zero therefore ends the branch for good.

For pure loss or pure gain, $b=1$, the turning point is $\theta_\ast=\pi$, and the criterion collapses to $q^2\leq e^{-\pi q}$. Equality marks the fold,
\begin{equation}
	\begin{aligned}
		q_{\mathrm f}^2&=e^{-\pi q_{\mathrm f}},
		&q_{\mathrm f}&\simeq0.4745,
		\\
		\kappa_{\mathrm f}&=2Jq_{\mathrm f}\simeq0.949J,
		&t_\ast&=\frac{\pi}{2J},
	\end{aligned}
	\label{eq:app_weak_neg_minimal_pure_fold}
\end{equation}
so the late closing survives below $\kappa_{\mathrm f}$ and is gone above it. Near that point the two roots follow the general law of \cref{eq:app_weak_fold_law}, with the role of the root function $m_E$ played by $F_1$, the role of time by $\theta$, and the role of the control parameter by $q$. The two derivatives the law needs, evaluated at $b=1$, $\theta=\pi$, and $q=q_{\mathrm f}$, are $\partial_qF_1=q_{\mathrm f}(2+\pi q_{\mathrm f})$ and $\partial_\theta^2F_1=q_{\mathrm f}^2(1+q_{\mathrm f}^2)$, so
\begin{equation}
	\begin{aligned}
		\theta_\pm-\pi
		&=\pm\left[
		\frac{2(2+\pi q_{\mathrm f})}{q_{\mathrm f}(1+q_{\mathrm f}^2)}
		(q_{\mathrm f}-q)
		\right]^{1/2}
		+\mathcal O(q_{\mathrm f}-q).
	\end{aligned}
	\label{eq:app_weak_neg_minimal_pure_fold_law}
\end{equation}
This single cell therefore exhibits in closed form everything the general argument of \cref{app:weak_entanglement_asymptotics} predicts: a branch continued from weak to finite rate, two roots approaching each other, the tangency where they merge, the square root that governs their separation, and the disappearance of the late closing beyond it. The same reasoning carries over to a longer chain. Take the two eigenvalue branches of $T_A$ that meet at a closing, fix their signs, and follow them inside their own parity sectors, where they cannot mix; their signed difference is analytic until a third branch reaches it, and putting that difference in place of $m_E$ in \cref{eq:app_weak_fold_expansion,eq:app_weak_fold_law} gives the same square-root law for every isolated negativity branch.

\subsection{Strong rates and active clusters}

Take the strong-rate variables $b$, $d$, $s$, and $s_\star$ defined in \cref{eq:weak_strong_rate_variables,eq:weak_local_half_occupation}. The local kinetics selects active sites on both sides of the cut, and we call each block of consecutive active sites a cluster. For the left-filled state the positive cluster lengths are
\begin{equation}
	\begin{aligned}
		\mathcal L_N
		&=
		\begin{cases}
			(N),&\xi>0,\ N>0,\\
			\varnothing,&\xi>0,\ N=0,\\
			(L_A-N,L_B)\setminus\{0\},&\xi<0.
		\end{cases}
	\end{aligned}
	\label{eq:app_weak_neg_active_clusters}
\end{equation}
Throughout this subsection, calling a cluster odd or even refers to the number of sites it contains. At $s=s_\star$ every active site sits at half occupation, so the detuning $\alpha$ of \cref{eq:app_weak_alpha} vanishes, each active covariance mode vanishes with it, and the twisted magnitudes of \cref{eq:twisted_spectrum} that the cluster controls become degenerate, just as the reduced-state levels did in \cref{app:weak_entanglement_asymptotics}. The first coherent correction inside a cluster is the same active Majorana matrix derived in \cref{eq:app_weak_active_covariance}. The cut bond requires one extra observation. Keep the sign sectors of \cref{eq:app_weak_neg_secular_blocks}, that is, the projectors $\Pi_+$ and $\Pi_-$ onto the positive and the negative eigenvalues of $T_A$. In the local twisted basis the first-order correction carried by the cut bond, $\delta T_{\mathrm{cut}}$, maps one sector into the other and has no part inside either,
\begin{equation}
	\begin{aligned}
		\Pi_+\delta T_{\mathrm{cut}}\Pi_+&=0,
		&\Pi_-\delta T_{\mathrm{cut}}\Pi_-&=0.
	\end{aligned}
	\label{eq:app_weak_neg_cut_offdiagonal}
\end{equation}
A perturbation that connects two levels separated by an order-one distance shifts them only at second order, so the cut coupling leaves the leading magnitudes alone at first order and distinct active clusters can be diagonalized independently.

Inside one cluster the calculation is the one already done for the entanglement gap. Writing $\ell$ for the number of sites in the cluster, in place of the block length $M$ of \cref{eq:app_weak_active_modes}, the $\ell$ small twisted magnitudes it generates are
\begin{equation}
	\begin{aligned}
		\nu_{\ell,m}(s)
		={}&\left[
		\alpha(s)^2
		+\frac{4c(b)^2J^2}{\kappa^2}
		\cos^2\!\left(\frac{m\pi}{\ell+1}\right)
		\right]^{1/2}
		\\
		&+\mathcal O_b\!\left(\frac{J^2}{\kappa^2}\right),
		\qquad1\leq m\leq\ell.
	\end{aligned}
	\label{eq:app_weak_neg_cluster_mode}
\end{equation}
The normalized negativity gap is the smallest of these modes,
\begin{equation}
	\begin{aligned}
		r_N(s)
		&=\min_{\substack{\ell\in\mathcal L_N\\1\leq m\leq\ell}}
		\nu_{\ell,m}(s),
	\end{aligned}
	\label{eq:app_weak_neg_cluster_minimum}
\end{equation}
which proves \cref{eq:weak_negativity_active_modes} to first order. The absolute value in \cref{eq:app_weak_neg_cluster_mode} alone would not prove that an odd-cluster zero survives higher orders. The required protection and the second-order splitting are established next.

\emph{Odd-cluster crossing theorem.} For every fixed finite chain, every active cluster of odd length produces a true twisted singular-value crossing at
\begin{equation}
	\begin{aligned}
		\kappa t_{N,k}
		&=s_\star+h_k\frac{J^2}{\kappa^2}
		+\mathcal O_b(J^3/\kappa^3),
		\qquad1\leq k\leq n_o,
	\end{aligned}
	\label{eq:app_weak_neg_odd_cluster_resolved_clocks}
\end{equation}
where $n_o$ is the number of active clusters of odd length. The real coefficients $h_k$ are the eigenvalue differences of the finite second-order eigenvalue problem defined below, divided by the nonzero local slope. Distinct $h_k$ split simultaneous leading roots by order $J^2/\kappa^2$ in scaled time. If symmetry makes some $h_k$ equal, those roots remain coincident to this order.

The protection is exact, and it rests on the global fermion parity $P$ of \cref{eq:global_parity}. The initial Fock projector and the Liouvillian are invariant under $P$, partial time reversal preserves the parity of every Majorana monomial, and $P_A$ commutes with $P$, so the twisted operator of \cref{eq:twisted_operator} obeys
\begin{equation}
	\begin{aligned}
		[T_A(t),P]&=0
	\end{aligned}
	\label{eq:app_weak_neg_global_parity_blocks}
\end{equation}
at every time and every rate, and its eigenvectors carry a definite global parity. Only an odd cluster has the standing wave of \cref{eq:app_weak_standing_waves} at momentum $\pi/2$, the one with $m=(\ell+1)/2$, and occupying that mode changes the global parity. The two branches that meet through it therefore sit in opposite parity blocks, which no perturbation connects, so higher orders can move the time at which they meet but cannot open a gap between them. The first-order shift vanishes because $\cos(\pi/2)=0$, while the signed difference of the two branches passes through zero with the nonzero slope set by $\alpha'(s_\star)=-b$. A zero crossed with nonzero slope survives a small smooth perturbation and merely moves, which is what the implicit-function theorem states, and the displaced zero is the root in \cref{eq:app_weak_neg_odd_cluster_resolved_clocks}.

We now give a finite construction of $h_k$. Keep the scaled time $s=\kappa t$ and the signed imbalance $d=\xi/\kappa$ of \cref{eq:weak_strong_rate_variables}, and set $\epsilon=J/\kappa$. Dividing the master equation \cref{eq:weak_lindblad} by $\kappa$ separates it into a site-local part of order one and a coherent part of order $\epsilon$,
\begin{equation}
	\begin{aligned}
		\partial_s\rho
		&=\left(\mathcal L_{\mathrm{loc}}+\epsilon\mathcal H\right)\rho,
		\\
		\mathcal L_{\mathrm{loc}}
		&=\frac{1+d}{2}\mathcal L_\ell
		+\frac{1-d}{2}\mathcal L_g,
		\\
		\mathcal H\rho
		&=-\frac{i}{J}[H,\rho],
	\end{aligned}
	\label{eq:app_weak_neg_scaled_liouvillian}
\end{equation}
with $\mathcal L_\ell$ and $\mathcal L_g$ the rate-free loss and gain superoperators introduced above, and $H$ the compact Hamiltonian of \cref{eq:compact_majorana_point}. The weights $(1\pm d)/2$ are the loss and gain rates divided by $\kappa$. The expansion $\rho=\rho_0+\epsilon\rho_1+\epsilon^2\rho_2+\mathcal O(\epsilon^3)$ is then determined recursively by
\begin{equation}
	\begin{aligned}
		\rho_0(s)
		&=e^{s\mathcal L_{\mathrm{loc}}}\rho(0),
		\\
		\rho_k(s)
		&=\int_0^sdu\,
		e^{(s-u)\mathcal L_{\mathrm{loc}}}\mathcal H\rho_{k-1}(u),
		\qquad k=1,2,
	\end{aligned}
	\label{eq:app_weak_neg_duhamel_recursion}
\end{equation}
where each order is the previous one pushed once through the Hamiltonian at time $u$ and then relaxed for the time that remains, the same structure as \cref{eq:app_weak_memory_integral}.

Carrying each order through partial time reversal and the parity twist, $T_k=(\rho_k)^{R_A}P_A$, splits the twisted operator itself into orders, $T_A=T_0+\epsilon T_1+\epsilon^2T_2+\mathcal O(\epsilon^3)$. At $s=s_\star$ the leading eigenvalues of $T_0$ are degenerate with one common magnitude $\eta_\star>0$, some of them at $+\eta_\star$ and some at $-\eta_\star$. Label a sign by $\sigma=\pm1$ and a global parity by $p=\pm1$, let $\Pi_{\sigma p}$ project onto the eigenvalue $\sigma\eta_\star$ of $T_0$ inside the parity sector $p$, and let $\Pi_{\sigma p}^\perp$ project onto the rest of that sector. Multiplying by the sign, $\widehat T_{\sigma,k}=\sigma T_k$, turns the shift of a magnitude into an ordinary eigenvalue shift for both signs. Second-order degenerate perturbation theory then gives the finite matrix \cite{Kato1995,StewartSun1990}
\begin{equation}
	\begin{aligned}
		W_{\sigma p}
		={}&\Pi_{\sigma p}\widehat T_{\sigma,2}\Pi_{\sigma p}
		\\
		&-\Pi_{\sigma p}\widehat T_{\sigma,1}\Pi_{\sigma p}^\perp
		\\
		&\quad\times\left[\Pi_{\sigma p}^\perp
		\left(\widehat T_{\sigma,0}-\eta_\star\right)
		\Pi_{\sigma p}^\perp\right]^{-1}
		\\
		&\quad\times
		\Pi_{\sigma p}^\perp\widehat T_{\sigma,1}\Pi_{\sigma p},
	\end{aligned}
	\label{eq:app_weak_neg_second_order_secular_matrix}
\end{equation}
whose two subscripts are labels for the sign sector and the parity sector, and whose eigenvalues are the second-order shifts of the degenerate magnitudes there. The first term is the direct second-order contribution and the second is the excursion out of the degenerate subspace and back, weighted by the inverse distance to the levels visited.

First diagonalize $\Pi_{\sigma p}\widehat T_{\sigma,1}\Pi_{\sigma p}$. Its zero-splitting subspace contains one vector for each odd cluster. Restrict \cref{eq:app_weak_neg_second_order_secular_matrix} to these vectors and diagonalize it separately in the two parity sectors. Pair the opposite-parity branches that coincide at first order. If their second-order difference is $\delta w_k$ and the derivative of their zeroth-order signed difference is $v_k\neq0$, then
\begin{equation}
	\begin{aligned}
		h_k&=-\frac{\delta w_k}{v_k}.
	\end{aligned}
	\label{eq:app_weak_neg_second_order_shift}
\end{equation}
Equations \eqref{eq:app_weak_neg_duhamel_recursion}--\eqref{eq:app_weak_neg_second_order_shift} contain only finite matrix exponentials, integrals, projections, and inverses on spectrally separated subspaces. They are therefore a constructive higher-order calculation for any fixed geometry. Global parity makes the off-diagonal matrix element between the paired branches exactly zero, which is the step that rules out a second-order avoided gap.

Three geometries show what these formulas produce, including one with a single shifted root and one with a split pair. For pure loss or pure gain the imbalance ratio is $b=1$, the local clock of \cref{eq:app_weak_alpha} sits at $s_\star=\ln2$, and every coefficient collapses to a combination of that one number. Evaluating \cref{eq:app_weak_neg_duhamel_recursion,eq:app_weak_neg_second_order_secular_matrix} for the geometries $(L_A,L_B,N)$ gives
\begin{equation}
	\begin{aligned}
		(2,1,0),\ \text{gain}:&\quad
		\{h_k\}=\left\{2\left[1-\ln2-(\ln2)^2\right]\right\},
		\\
		(3,1,2),\ \text{gain}:&\quad
		\{h_k\}=\big\{2\left[1-\ln2-(\ln2)^2\right],
		\\
		&\qquad\qquad
		3-2\ln2-4(\ln2)^2\big\},
		\\
		(2,2,1),\ \text{loss}:&\quad
		\{h_k\}=\left\{1-2(\ln2)^2\right\},
	\end{aligned}
	\label{eq:app_weak_neg_second_order_examples}
\end{equation}
that is, $\{h_k\}\simeq\{-0.347\}$, $\{-0.347,-0.308\}$, and $\{0.039\}$. The first geometry has one odd cluster and its root is merely displaced; the second has two, and the two roots separate at order $J^2/\kappa^2$; the third shows that the sign of the displacement depends on the geometry.

A cluster of even length has no mode at $\pi/2$ and therefore no root of this kind; its gap stops at the positive minimum of \cref{eq:app_weak_neg_even_cluster_mass} instead. At leading order, the odd-cluster clock quoted in the main text follows,
\begin{equation}
	\begin{aligned}
		t_N
		&=\frac{s_\star}{\kappa}
		+\mathcal O_b\!\left(\frac{J^2}{\kappa^3}\right).
	\end{aligned}
	\label{eq:app_weak_neg_odd_cluster_clock}
\end{equation}
If all clusters are even,
\begin{equation}
	\begin{aligned}
		r_{N,\min}
		&=\frac{2c(b)J}{\kappa}
		\min_{\ell\in\mathcal L_N}
		\sin\!\left(\frac{\pi}{2(\ell+1)}\right)
		+\mathcal O_b\!\left(\frac{J^2}{\kappa^2}\right).
	\end{aligned}
	\label{eq:app_weak_neg_even_cluster_mass}
\end{equation}
For positive even lengths, the sine decreases as $\ell$ increases, so the longest active cluster gives the smallest avoided gap. If $\mathcal L_N=\varnothing$, no local twisted branch approaches degeneracy on the $\kappa^{-1}$ scale. Equations \eqref{eq:app_weak_neg_odd_cluster_clock} and \eqref{eq:app_weak_neg_even_cluster_mass} prove \cref{eq:weak_strong_rate_negativity_clock}.

Under loss dominance, $M_E=N$ and $\mathcal L_N=(N)$, so the entanglement and twisted spectra have the same odd, even, or inactive classification. Under gain dominance, $K_E$ sees the active block $L_A-N$ in $A$, while $K_N$ also sees the active block $L_B$ in $B$. This is the origin of the branch where an odd cluster in $B$ closes the twisted gap while the entanglement gap in $A$ is avoided or absent on the local scale.

\section{Geometric Decoherence Time under Weak Symmetry}
\label{app:geometric_decoherence}

The temporal criticalities locate degeneracies in two spectra. The geometric decoherence time instead marks when the logarithmic negativity first decreases while the R\'enyi-$1/2$ entropy of $A$ does not decrease. We use the first-passage definition of Ref.~\cite{Jha2026Jul}
\begin{equation}
	\begin{aligned}
		\tau_d^g
		&\coloneq\inf\Big\{t>0:\exists\,0\leq t_1<t_2\leq t,\\
		&\hspace{5.0em}S_{1/2}(t_2)\geq S_{1/2}(t_1),
		\ E_N(t_2)<E_N(t_1)\Big\}.
	\end{aligned}
	\label{eq:app_geometric_definition}
\end{equation}
This definition does not assume differentiability. If $E_N$ has a smooth first local maximum and $S_{1/2}$ is increasing there, the geometric decoherence time is the location of that maximum.

\subsection{Weak rates at the compact point}

The balanced case was settled in \cref{eq:app_balanced_exact_negativity,eq:app_balanced_geometric}; we now allow an arbitrary weak imbalance between loss and gain. The starting point is the closed-system negativity of \cref{eq:app_compact_negativity} on the branch that contains $t_0=\pi/(4J)$, together with its curvature there,
\begin{equation}
	\begin{aligned}
		E_N^{\mathrm{cl}}(t)
		&=\ln[1+\sin(2Jt)],
		\\
		\ddot E_N^{\mathrm{cl}}(t_0)
		&=-2J^2.
	\end{aligned}
	\label{eq:app_weak_geo_closed_curvature}
\end{equation}
To first order in the rates, use the Dyson correction in \cref{eq:app_weak_neg_dyson_density}. Compact factorization leaves only the cut dimer and any unpaired Majorana at a chain end adjacent to that dimer. For the left-filled state in \cref{eq:left_filled_initial_state}, sites $1,\ldots,N$ are occupied and $N\leq L_A$. Hence the site $L_A$ next to the cut is occupied if and only if $N=L_A$. Three indicators record this local geometry, written with the symbol $\mathbf 1_{\{\cdot\}}$, which equals one when the condition inside the braces holds and zero otherwise:
\begin{equation}
	\begin{aligned}
		n_c&\coloneq n_{L_A}(0)=\mathbf 1_{\{N=L_A\}},
		\\
	\delta_A&\coloneq\mathbf 1_{\{L_A=1\}},
	&\delta_B&\coloneq\mathbf 1_{\{L_B=1\}}.
	\end{aligned}
	\label{eq:app_weak_geo_boundary_indicators}
\end{equation}
Thus $n_c$ is the initial occupation of the site of $A$ that touches the cut: it equals one only for a completely filled subsystem, so $L_A=3$ with $N=3$ gives $n_c=1$, while $L_A=3$ with $N=1$ gives $n_c=0$. The indicators $\delta_A$ and $\delta_B$ equal one exactly when the corresponding side of the cut holds a single site, which is the case in which the cut dimer touches an end of the chain and shares a site with an unpaired Majorana, $a_1$ when $L_A=1$ and $b_L$ when $L_B=1$. For a bulk cut both end effects are absent and only $n_c$ survives.

The slope of the negativity at $t_0$ has two parts. The part that depends only on the total rate is already known: balanced loss and gain contract the covariance uniformly, and \cref{eq:app_balanced_slope_at_t0} gives $\dot E_N(t_0)=-2\gamma_{\mathrm w}+\mathcal O(\gamma_{\mathrm w}^2/J)$, which is $-\kappa$ in terms of the total rate. The imbalance adds the one-jump corrections of \cref{eq:app_weak_neg_dyson_density}, applied now to the trace norm itself rather than to differences of levels. Compact factorization confines them to the cut dimer and to an adjacent end Majorana, so they are built from the same two elementary integrals as in \cref{eq:app_weak_neg_elementary_integrals}: the cut dimer gives $\xi(1-n_c)/4$, which is present only when the cut site of $A$ starts empty, the end Majorana on the $A$ side gives $\xi\delta_A(1-2n_c)/8$, whose sign follows the occupation of that same site, and the end Majorana on the $B$ side gives $\xi\delta_B/8$, where the neighboring site always starts empty. Collecting the three,
\begin{equation}
	\begin{aligned}
		\dot E_N(t_0)
		&=-\kappa+2G_{L_A,L_B,N}\xi
		+\mathcal O\!\left(\frac{\kappa^2}{J}\right),
		\\
		G_{L_A,L_B,N}
		&=\frac{2(1-n_c)+\delta_A(1-2n_c)+\delta_B}{16}.
	\end{aligned}
	\label{eq:app_weak_geo_slope}
\end{equation}
Let $\tau_d^g=t_0+\delta t$. Expanding the maximum condition $\dot E_N(\tau_d^g)=0$ gives
\begin{equation}
	\begin{aligned}
		0
		&=\dot E_N(t_0)+\ddot E_N^{\mathrm{cl}}(t_0)\delta t
		+\mathcal O(\kappa^2/J),
	\end{aligned}
	\label{eq:app_weak_geo_max_condition}
\end{equation}
and therefore
\begin{equation}
	\begin{aligned}
		\tau_d^g
		&=t_0-\frac{\kappa}{2J^2}
		+G_{L_A,L_B,N}\frac{\xi}{J^2}
		+\mathcal O\!\left(\frac{\kappa^2}{J^3}\right).
	\end{aligned}
	\label{eq:app_weak_geo_clock}
\end{equation}
Equation~\eqref{eq:app_weak_geo_clock} is the required weak-rate expansion of the geometric decoherence time.

It remains to check that $S_{1/2}$ is not falling at this maximum, which is what the first-passage definition in \cref{eq:app_geometric_definition} demands. For $L_A=1$ the reduced state has the two eigenvalues $(1\pm q_A)/2$ fixed by the subsystem parity of \cref{eq:app_weak_pfaffian_parity}, so
\begin{equation}
	\begin{aligned}
		S_{1/2}
		&=\ln\!\left(1+\sqrt{1-q_A^2}\right).
	\end{aligned}
	\label{eq:app_weak_geo_one_site_entropy}
\end{equation}
Differentiating gives $\dot S_{1/2}=-q_A\dot q_A/[\sqrt{1-q_A^2}(1+\sqrt{1-q_A^2})]$, which is positive whenever $|q_A|$ is still falling. At the maximum located by \cref{eq:app_weak_geo_clock} the parity has not yet reached the zero that defines $t_E$, so $|q_A|$ is indeed still falling and the derivative is positive, of order $\kappa$, for every $|\xi|\leq\kappa$.

For $L_A\geq2$ a stronger statement follows from the square-root behavior of the entropy near a pure mode. A mode that the closed system leaves pure has $|\zeta|=1$, and the uniform contraction $e^{-\kappa t}$ in \cref{eq:app_weak_covariance_solution} displaces it to $\nu=1-\delta$ with $\delta=\mathcal O(\kappa/J)>0$ at times of order $t_0$. Its contribution to the R\'enyi-$1/2$ entropy is
\begin{equation}
	\begin{aligned}
		&2\ln\!\left[
		\sqrt{\frac{1+\nu}{2}}
		+\sqrt{\frac{1-\nu}{2}}
		\right]
		\\
		&\qquad=\sqrt{2\delta}+\mathcal O(\delta),
		\qquad\nu=1-\delta.
	\end{aligned}
	\label{eq:app_weak_geo_entropy_edge}
\end{equation}
The entropy of a nearly pure mode therefore grows like the square root of its displacement, and differentiating that square root gives a rate $\dot\delta/\sqrt{2\delta}=\mathcal O(\sqrt{\kappa J})$, which as $\kappa/J\to0$ overwhelms the $\mathcal O(\kappa)$ motion of the cut mode whatever its sign. The entropy is thus increasing at the first negativity maximum for every weak-rate boundary and filling class, and that maximum is the geometric decoherence time.

Subtracting \cref{eq:app_weak_geo_clock} from the branch formula \cref{eq:app_weak_neg_clock} gives
\begin{equation}
	\begin{aligned}
		J^2(t_N-\tau_d^g)
		={}&\frac{\kappa}{2}
		+\min\mathcal B_{L_A,L_B,N}(\xi)
		\\
		&-G_{L_A,L_B,N}\xi
		+\mathcal O(\kappa^2/J).
	\end{aligned}
	\label{eq:app_weak_geo_difference}
\end{equation}
Checking the eight entries in \cref{tab:app_weak_negativity_branches} and using $|\xi|\leq\kappa$ yields
\begin{equation}
	\begin{aligned}
		t_N-\tau_d^g
		&\geq\frac{\kappa}{4J^2}
		+\mathcal O\!\left(\frac{\kappa^2}{J^3}\right)>0
	\end{aligned}
	\label{eq:app_weak_geo_ordering}
\end{equation}
for sufficiently weak rates. This proves the ordering quoted in the main text. The possible separation of $t_E$ and $t_N$ follows from the parity rule of \cref{eq:weak_negativity_parity_rule}. The entanglement criticality is the zero of $q_A$, shifted at first order by the occupation of site $L_A$, and the negativity criticality is the earlier of this zero and the zero of $q_B$, shifted by the occupation of site $L_A+1$. When $L_A=1$ or $L_B=1$, the unpaired Majorana at that end doubles the shift on that side. For the left-filled state and a cut away from the chain ends, the two clocks therefore separate at first order only under gain with $N=L_A$.

\subsection{Strong-rate cut problem}

Take $b=|\xi|/\kappa>0$ fixed and use the scaled time $s=\kappa t$. At $J=0$, every site follows the independent local kinetics in \cref{eq:weak_local_kinetics}. For the left-filled state in \cref{eq:left_filled_initial_state}, the occupied sites are $1,\ldots,N$ and $N\leq L_A$. The site in $A$ next to the cut is site $L_A$, so the cut occupation $n_c$ of \cref{eq:app_weak_geo_boundary_indicators} is one only when $N=L_A$, which, written out, is
\begin{equation}
	\begin{aligned}
		n_c\coloneq n_{L_A}(0)
		&=\mathbf 1_{\{N=L_A\}}
		=\begin{cases}
			0,&N<L_A,\\
			1,&N=L_A.
		\end{cases}
	\end{aligned}
	\label{eq:app_weak_geo_cut_occupation}
\end{equation}
The site in $B$ next to the cut is site $L_A+1>N$ and is therefore empty. More generally, let $n\in\{0,1\}$ be the initial occupation of the $A$-side cut site. The functions $\bar n_n(s)$ and $\bar n_0(s)$ are then the mean occupations of the $A$- and $B$-side cut sites. With the cut bond switched off, their joint probabilities factorize. The probabilities that the pair contains an even or odd number of particles are therefore
\begin{equation}
	\begin{aligned}
		P_e^{(n)}(s;d)
		&=[1-\bar n_n(s)][1-\bar n_0(s)]
		+\bar n_n(s)\bar n_0(s),
		\\
		P_o^{(n)}(s;d)
		&=\bar n_n(s)[1-\bar n_0(s)]
		+[1-\bar n_n(s)]\bar n_0(s).
	\end{aligned}
	\label{eq:app_weak_geo_cut_weights}
\end{equation}
The two terms in $P_e^{(n)}$ are the probabilities of $00$ and $11$, while those in $P_o^{(n)}$ are the probabilities of $10$ and $01$. Hence $P_e^{(n)}+P_o^{(n)}=1$.

Turning on the cut bond to first order in $J/\kappa$ creates one coherence within each occupation-parity sector. Let $U_n$ describe the dimensionless pairing coherence $|00\rangle\leftrightarrow|11\rangle$ in the even sector, and let $V_n$ describe the dimensionless hopping coherence $|10\rangle\leftrightarrow|01\rangle$ in the odd sector. Projecting the first-order covariance equation onto these two matrix elements gives (recall $s=\kappa t$)
\begin{equation}
	\begin{aligned}
		\frac{dU_n}{ds}+U_n
		&=-d+(d+n-1)e^{-s},
		&U_n(0)&=0,
		\\
		\frac{dV_n}{ds}+V_n
		&=ne^{-s},
		&V_n(0)&=0.
	\end{aligned}
	\label{eq:app_weak_geo_cut_amplitude_odes}
\end{equation}
The term proportional to each amplitude is its dissipative damping, and the right-hand side is fixed by the uncoupled occupations in \cref{eq:app_weak_local_kinetics}. Multiplying each equation by $e^s$ and integrating from $0$ to $s$ gives
\begin{equation}
	\begin{aligned}
		U_n(s;d)
		&=-d+e^{-s}[d+s(d+n-1)],
		\\
		V_n(s)&=nse^{-s}.
	\end{aligned}
	\label{eq:app_weak_geo_cut_amplitudes}
\end{equation}
The corresponding physical coherences are $u=(J/\kappa)U_n+\mathcal O_b(J^3/\kappa^3)$ and $v=(J/\kappa)V_n+\mathcal O_b(J^3/\kappa^3)$. Partial time reversal complements the $A$ occupation of the diagonal weights but leaves both coherences in place, so after the parity twist the two-site operator separates into a $2\times2$ block pairing $u$ with the odd weight and one pairing $v$ with the even weight, the same structure met in \cref{eq:app_weak_neg_minimal_twisted_eigenvalues}, and we write $R_o$ and $R_e$ for their trace norms, now labeled by the parity of the weight rather than of the states. At zeroth order each block is diagonal and carries one positive and one negative eigenvalue whose magnitudes add up to the weight of that block, so $R_o=P_o^{(n)}$, $R_e=P_e^{(n)}$, and the negativity vanishes. Because the coherences sit off the diagonal inside a block, they move no eigenvalue at first order; at second order they push the two eigenvalues of a block apart by twice the squared coupling divided by their separation, and since one eigenvalue is positive and the other negative, both motions add to the trace norm,
\begin{equation}
	\begin{aligned}
		R_o
		&=P_o^{(n)}+\frac{2J^2}{\kappa^2}\frac{U_n^2}{P_o^{(n)}}
		+\mathcal O_b(J^4/\kappa^4),
		\\
		R_e
		&=P_e^{(n)}+\frac{2J^2}{\kappa^2}\frac{V_n^2}{P_e^{(n)}}
		+\mathcal O_b(J^4/\kappa^4).
	\end{aligned}
	\label{eq:app_weak_geo_cut_blocks}
\end{equation}
Each denominator is the zeroth-order weight of its block, which is also the separation between the two eigenvalues that the coherence pushes apart. Since $P_e^{(n)}+P_o^{(n)}=1$, the full trace norm is one plus the two second-order terms, so $E_N=\ln(R_e+R_o)$ expands as
\begin{equation}
	\begin{aligned}
		E_N(s/\kappa)
		&=\frac{J^2}{\kappa^2}\Phi_n(s;d)
		+\mathcal O_b\!\left(\frac{J^4}{\kappa^4}\right),
		\\
		\Phi_n(s;d)
		&\coloneq\frac{2U_n(s;d)^2}{P_o^{(n)}(s;d)}
		+\frac{2V_n(s)^2}{P_e^{(n)}(s;d)}.
	\end{aligned}
	\label{eq:app_weak_geo_phi}
\end{equation}
Thus $\Phi_n$ is the dimensionless leading coefficient of the cut-bond logarithmic negativity. If a denominator vanishes at $s=0$ or as $s\to\infty$, the quotient is defined by the one-sided limit of the exact $2\times2$ block before the expansion is taken.

At the same order, the sites in $A$ remain independent, so their R\'enyi-$1/2$ entropies add. A site occupied with probability $x$ contributes
\begin{equation}
	\begin{aligned}
		h_{1/2}(x)
		&\coloneq2\ln(\sqrt{x}+\sqrt{1-x}).
	\end{aligned}
	\label{eq:app_weak_geo_h}
\end{equation}
This is the same function that already appeared in \cref{eq:app_weak_geo_entropy_edge}. Region $A$ holds $N$ sites that start occupied and $L_A-N$ that start empty, and their mean occupations are the two solutions $\bar n_1(s)$ and $\bar n_0(s)$ of \cref{eq:app_weak_local_kinetics} at initial occupation $n=1$ and $n=0$. Therefore
\begin{equation}
	\begin{aligned}
		S_{1/2}(s/\kappa)
		&=\Sigma_{L_A,N}(s;d)
		+\mathcal O_b\!\left(\frac{J^2}{\kappa^2}\right),
		\\
		\Sigma_{L_A,N}(s;d)
		&\coloneq Nh_{1/2}[\bar n_1(s)]
		+(L_A-N)h_{1/2}[\bar n_0(s)].
	\end{aligned}
	\label{eq:app_weak_geo_sigma}
\end{equation}
The strong-rate first-passage problem is now explicit. First find the earliest $s>0$ at which $\partial_s\Phi_{n_c}$ changes from positive to negative. Then apply the entropy condition in \cref{eq:app_geometric_definition}. Away from a simultaneous entropy turning point, this requires $\partial_s\Sigma_{L_A,N}>0$ at the maximum; equality cases are resolved directly in the classification below. If the first qualifying maximum is at a finite $s_g^{\mathrm{loc}}$, the geometric decoherence time is
\begin{equation}
	\begin{aligned}
		\tau_d^g
		&=\frac{s_g^{\mathrm{loc}}}{\kappa}
		+\mathcal O_b\!\left(\frac{J^2}{\kappa^3}\right).
	\end{aligned}
	\label{eq:app_weak_geo_local_time}
\end{equation}
If no maximum passes both tests at finite $s$, then $\kappa\tau_d^g\to\infty$ within this local expansion and higher-order coherent or finite-geometry corrections determine the geometric decoherence time.

\subsection{Classification of the strong-rate branches}

The remaining task is to decide when the two tests above have a solution. The thresholds introduced below distinguish two ways of losing a finite local geometric decoherence time: $b_\star$ marks the disappearance of a local maximum, while $b_\ell$ marks the point at which an existing maximum no longer satisfies the entropy condition.

Gain dominance means $\gamma_g>\gamma_\ell$, so $\xi<0$ and the signed imbalance of \cref{eq:weak_strong_rate_variables} is $d=\xi/\kappa=-b$. With $N<L_A$ the cut site of $A$ starts empty, so $n_c=0$ and the function to study is $\Phi_0(s;-b)$. Differentiating it and removing factors that are strictly positive for $s>0$ shows that its nonzero stationary points satisfy
\begin{equation}
	\begin{aligned}
		0={}&-(1+b)^2
		+[-b^2s+2b^2-bs+3b]e^s
		\\
		&+[b^2s-b^2-b-s+1]e^{2s}.
	\end{aligned}
	\label{eq:app_weak_geo_sminus_equation}
\end{equation}
Let $s_-(b)$ be the first positive root at which $\partial_s\Phi_0$ changes from positive to negative. Direct substitution at the two ends of the interval gives
\begin{equation}
	\begin{aligned}
		0<s_-(b)<s_\star.
	\end{aligned}
	\label{eq:app_weak_geo_sminus_bound}
\end{equation}
Every initially empty site is still approaching half occupation on this interval, so $\partial_s\Sigma_{L_A,N}>0$ and the maximum qualifies. Hence
\begin{equation}
	\begin{aligned}
		\tau_d^g
		&=\frac{s_-(b)}{\kappa}
		+\mathcal O_b\!\left(\frac{J^2}{\kappa^3}\right),
		&\xi&<0,
		&N&<L_A.
	\end{aligned}
	\label{eq:app_weak_geo_gain_empty}
\end{equation}
For pure gain the loss rate vanishes, so $\kappa=\gamma_g$, $\xi=-\gamma_g$, and $b=|\xi|/\kappa=1$. At $b=1$ the three coefficients of \cref{eq:app_weak_geo_sminus_equation} collapse: the constant $-(1+b)^2$ becomes $-4$, the coefficient of $e^s$ becomes $5-2s$, and the coefficient of $e^{2s}$ becomes $-1$, so that the equation reads $-4+(5-2s)e^s-e^{2s}=0$. It has the endpoint root $s=0$, which is the start of the evolution rather than a maximum, and one strictly positive root. That positive root is the first maximum; denote it by $s_g$. Then
\begin{equation}
	\begin{aligned}
		e^{2s_g}+(2s_g-5)e^{s_g}+4&=0,
		\\
		s_g&\simeq0.4308,
		\\
		\tau_d^g
		&=\frac{s_g}{\gamma_g}
		+\mathcal O\!\left(\frac{J^2}{\gamma_g^3}\right),
		&N&<L_A.
	\end{aligned}
	\label{eq:app_weak_geo_pure_gain_root}
\end{equation}

For $N=L_A$ the cut site of $A$ starts occupied, so $n_c=1$ and the function to study is $\Phi_1$, the $n=1$ member of \cref{eq:app_weak_geo_phi}. This one function serves both protocols, because reversing the sign of $d$ exchanges the two cut occupations and complements them, $\bar n_1(s;d)=1-\bar n_0(s;-d)$, which leaves the weights $P_e^{(1)}$ and $P_o^{(1)}$ untouched and only flips the sign of $U_1$, so that $\Phi_1(s;-b)=\Phi_1(s;b)$. Let $s_1(b)$ be the first maximum of $\Phi_1$, meaning the first point where $\partial_s\Phi_1$ changes from positive to negative, whenever such a point exists. As $b$ increases that maximum approaches the neighboring minimum, and the two merge and vanish together. This is the collision that \cref{app:weak_entanglement_asymptotics} calls a fold, here between stationary points of $\Phi_1$ rather than between roots of a gap, and it happens where the first and second derivatives vanish at the same point:
\begin{equation}
	\begin{aligned}
		\partial_s\Phi_1(s_{\mathrm{fold}};b_\star)&=0,
		&\partial_s^2\Phi_1(s_{\mathrm{fold}};b_\star)&=0.
	\end{aligned}
	\label{eq:app_weak_geo_bstar_system}
\end{equation}
Solving these two equations gives
\begin{equation}
	\begin{aligned}
		b_\star&\simeq0.5213,
		&s_{\mathrm{fold}}&\simeq1.0965.
	\end{aligned}
	\label{eq:app_weak_geo_bstar}
\end{equation}
Both numbers come from a function of $s$ and $b$ alone, so neither depends on the chain length or on where the chain is cut; the geometry enters only through the occupation $n_c$ of the cut site, which selects $\Phi_1$ rather than $\Phi_0$. The threshold $b_\star$ therefore separates $0<b<b_\star$, where the local maximum exists, from $b\geq b_\star$, where it has been folded away. Under gain dominance the entropy of an initially occupied site increases throughout this branch, so every maximum that exists also qualifies.

Under loss dominance with $N=L_A$, the same function $\Phi_1$ controls the cut negativity, but the entropy of the initially occupied sites increases only until those sites reach half occupation at $s_\star$. The limiting value $b_\ell$ is therefore determined by requiring the maximum to occur exactly at $s_\star=\ln[(1+b_\ell)/b_\ell]$:
\begin{equation}
	\begin{aligned}
		\partial_s\Phi_1
		\!\left(\ln\frac{1+b_\ell}{b_\ell};b_\ell\right)
		&=0,
		&b_\ell&\simeq0.5209.
	\end{aligned}
	\label{eq:app_weak_geo_bell_equation}
\end{equation}
The local maximum qualifies for $0<b<b_\ell$ and fails the entropy test for $b\geq b_\ell$. For $b_\ell\leq b<b_\star$, the maximum still exists but occurs after the entropy has begun to decrease.

Finally, for loss dominance with $N<L_A$, one has $d=b$ and $n_c=0$. Let $s_+(b)$ be the first positive root at which $\partial_s\Phi_0(s;b)$ changes from positive to negative. Removing its positive factors gives
\begin{equation}
	\begin{aligned}
		0={}&-(1-b)^2
		+[-b^2s+bs+2b^2-3b]e^s
		\\
		&+[b^2s-b^2+b-s+1]e^{2s}.
	\end{aligned}
	\label{eq:app_weak_geo_splus_equation}
\end{equation}
For every $0<b<1$, this root is finite but not uniformly bounded in $b$: $s_+(b)$ increases without bound as $b\to1^-$ (for example $s_+\simeq1.09,\,2.23,\,5.7,\,10.7$ at $b=0.3,\,0.7,\,0.9,\,0.95$), so this branch joins continuously onto the pure-loss ``later scale'' entry below, and the local result is controlled only at fixed $b$ bounded away from $1$, where $s_+(b)=\mathcal O(1)$. This root gives the geometric decoherence time only when the entropy is still nondecreasing there:
\begin{equation}
	\begin{aligned}
		\partial_s\Sigma_{L_A,N}(s_+(b);b)&\geq0.
	\end{aligned}
	\label{eq:app_weak_geo_entropy_condition}
\end{equation}
For $N=0$, all sites in $A$ are initially empty and every one-site entropy is increasing at $s_+(b)$, so the condition holds for every $0<b<1$. For $N>0$, the initially occupied sites can already be moving away from half occupation while the initially empty sites are still moving toward it; their two contributions must then be compared using \cref{eq:app_weak_geo_sigma}.

\Cref{tab:weak_strong_geometric_branches} collects the resulting classification.
\begin{table}[t]
	\caption{Strong-rate branches of the geometric decoherence time for $0<b\leq1$, with $b=|\xi|/\kappa$ the imbalance ratio of \cref{eq:weak_strong_rate_variables}. Each leading result is the scaled time at which $\Phi_{n_c}$ of \cref{eq:app_weak_geo_phi} has its first qualifying maximum, divided by $\kappa$ as in \cref{eq:app_weak_geo_local_time}: $s_-(b)$ solves \cref{eq:app_weak_geo_sminus_equation}, $s_+(b)$ solves \cref{eq:app_weak_geo_splus_equation}, and $s_1(b)$ is the first maximum of $\Phi_1$, which exists below the fold $b_\star$ of \cref{eq:app_weak_geo_bstar} and passes the entropy test of \cref{eq:app_geometric_definition} below $b_\ell$ of \cref{eq:app_weak_geo_bell_equation}. The entry ``later scale'' means that no finite maximum qualifies within this local expansion.}
	\label{tab:weak_strong_geometric_branches}
	\centering
	\small
	\begin{tabular}{@{}ll@{}}
		\toprule
		Regime and condition&Leading result\\
		\midrule
		Gain, $N<L_A$, $0<b\leq1$&$s_-(b)/\kappa$\\
		Gain, $N=L_A$, $0<b<b_\star$&$s_1(b)/\kappa$\\
		Gain, $N=L_A$, $b_\star\leq b\leq1$&later scale\\
		Loss, $N=L_A$, $0<b<b_\ell$&$s_1(b)/\kappa$\\
		Loss, $N=L_A$, $b_\ell\leq b\leq1$&later scale\\
		\shortstack[l]{Loss, $N<L_A$, $0<b<1$,\\$\partial_s\Sigma_{L_A,N}(s_+(b);b)\geq0$}&$s_+(b)/\kappa$\\
		Loss, $N<L_A$, remaining cases&later scale\\
		\bottomrule
	\end{tabular}
\end{table}

For pure gain with $N<L_A$ the maximum sits at $s_g\simeq0.4308$, earlier than the half-occupation time $s_\star=\ln2\simeq0.6931$ of \cref{eq:app_weak_alpha}. The negativity therefore turns over while the cut sites are still filling, ahead of the local clock that sets the spectral closings of \cref{app:weak_negativity_asymptotics}, so the three clocks stay ordered on this branch. Pure loss at any filling, and pure gain with $N=L_A$, instead have no qualifying maximum at finite $s$. With $N=L_A$ the maximum of $\Phi_1$ has already been folded away, since $b=1$ exceeds $b_\star\simeq0.5213$; under pure loss with $N<L_A$ both cut sites stay empty, $P_o^{(0)}=0$, and the odd block has the eigenvalues $\pm|u|$, so that $E_N=2(J/\kappa)(1-e^{-s})+\mathcal O(J^2/\kappa^2)$ rises monotonically. The table entry for these cases reads ``later scale'', which is a statement about this expansion rather than about the physics. Coherence is still lost, and the entry means that the leading cut-bond negativity has no maximum while the entropy is still rising on the $\kappa^{-1}$ scale, so the geometric decoherence time is set by the next order in $J/\kappa$ or by the parts of the geometry that this two-site treatment of the cut leaves out.

\section{Exact Compact Reduction under Strong Symmetry}
\label{app:strong_exact_reduction}

Density dephasing preserves parity trajectory by trajectory but generally destroys Gaussianity after averaging. The compact point remains exactly tractable because the Jordan-Wigner Hamiltonian is a sum of commuting bonds. This appendix derives the random-unitary representation, the reduced parity blocks, the pattern and bath-length universality statements, and the stationary state.

\subsection{Jordan-Wigner form and Poisson representation}

At the compact point, use the Jordan-Wigner variables \cite{Jordan1928Sep,Lieb1961Dec} for which
\begin{equation}
	\begin{aligned}
		n_j&=\frac{1-Z_j}{2},
		&H&=-J\sum_{j=1}^{L-1}X_jX_{j+1}.
	\end{aligned}
	\label{eq:app_strong_jw}
\end{equation}
Since $n_j^2=n_j$,
\begin{equation}
	\begin{aligned}
		\mathcal D[n_j]\rho
		&=n_j\rho n_j-\frac{1}{2}\{n_j,\rho\}
		=\frac{1}{4}(Z_j\rho Z_j-\rho).
	\end{aligned}
	\label{eq:app_strong_dephasing_spin}
\end{equation}
Thus
\begin{equation}
	\begin{aligned}
		\dot\rho
		&=iJ\left[\sum_{j=1}^{L-1}X_jX_{j+1},\rho\right]
		+\frac{\gamma}{4}\sum_{j=1}^{L}(Z_j\rho Z_j-\rho).
	\end{aligned}
	\label{eq:app_strong_spin_master}
\end{equation}
For any unitary $U$, the generator $r(U\rho U^\dagger-\rho)$ is the ensemble generator of Poisson-distributed $U$ jumps at rate $r$ \cite{Dalibard1992Feb,PlenioKnight1998Jan,Daley2014Jul}. Equation \eqref{eq:app_strong_spin_master} is therefore an exact average over independent $Z_j$ jumps at rate $\gamma/4$, the kicks of \cref{sec:temporal_criticalities_strong}.

Let $N_j^{\mathrm d}(s)$ count the jumps at site $j$ and set $\sigma_j(s)=(-1)^{N_j^{\mathrm d}(s)}$. Thus $\sigma_j$ is a symmetric random telegraph process initialized at $+1$ and switching at rate $\gamma/4$ \cite{kac1974telegraph,Apostolov2024Aug}. Conjugating a bond by a jump gives
\begin{equation}
	\begin{aligned}
		Z_j(X_{j-1}X_j)Z_j&=-X_{j-1}X_j,
		\\
		Z_j(X_jX_{j+1})Z_j&=-X_jX_{j+1}.
	\end{aligned}
	\label{eq:app_strong_jump_sign_flip}
\end{equation}
All bond operators commute, so the jump history changes only their accumulated signs. Moving every $Z$ operator to the initial time, where it acts on the product-Fock state only by a phase, makes the sign of a bond at time $s$ count the jumps after $s$ rather than before it. A Poisson process has the same statistics read backward in time, so at each fixed $t$ the average over trajectories equals the average of the pure states
\begin{equation}
	\begin{aligned}
		|\psi_{\boldsymbol\sigma}(t)\rangle
		&=\prod_{j=1}^{L-1}
		\exp[iJ\Theta_j(t)X_jX_{j+1}]|\boldsymbol n\rangle,
		\\
		\Theta_j(t)
		&=\int_0^t ds\,\sigma_j(s)\sigma_{j+1}(s),
	\end{aligned}
	\label{eq:app_strong_random_trajectory}
\end{equation}
which is \cref{eq:strong_random_unitary}. No Gaussian assumption enters this construction.

\subsection{Exact reduction across the cut}

Split the trajectory unitary of \cref{eq:app_strong_random_trajectory} into gates inside $A$, the gate on the bond that crosses the cut, and gates inside $B$. The first two are the objects of \cref{eq:strong_cut_variables},
\begin{equation}
	\begin{aligned}
		U_A
		&=\prod_{j=1}^{L_A-1}e^{iJ\Theta_jX_jX_{j+1}},
		&\phi&=J\Theta_{L_A}.
	\end{aligned}
	\label{eq:app_strong_UA_phi}
\end{equation}
Since $(X_{L_A}X_{L_A+1})^2=1$, the cut gate acts on a product configuration as a rotation between leaving it alone and flipping the two sites that straddle the cut,
\begin{equation}
	\begin{aligned}
		&e^{i\phi X_{L_A}X_{L_A+1}}|\boldsymbol n_A,\boldsymbol n_B\rangle
		\\
		&\quad=\cos\phi|\boldsymbol n_A,\boldsymbol n_B\rangle
		\\
		&\quad+i\sin\phi\,X_{L_A}|\boldsymbol n_A\rangle
		\otimes X_{L_A+1}|\boldsymbol n_B\rangle.
	\end{aligned}
	\label{eq:app_strong_cut_gate}
\end{equation}
The two $B$ states are orthogonal because they differ on site $L_A+1$. The cross terms therefore vanish under $\operatorname{Tr}_B$. Gates entirely inside $B$ cancel by unitary invariance of the partial trace. Hence
\begin{equation}
	\begin{aligned}
		\rho_A
		=\mathbb E_{\boldsymbol\sigma}\Big[
		U_A\big(&\cos^2\phi|\boldsymbol n_A\rangle\langle\boldsymbol n_A|
		\\
		&+\sin^2\phi X_{L_A}|\boldsymbol n_A\rangle\langle\boldsymbol n_A|X_{L_A}\big)U_A^\dagger
		\Big],
	\end{aligned}
	\label{eq:app_strong_reduced_state}
\end{equation}
which is \cref{eq:strong_reduced_trajectory}.

The two terms in \cref{eq:app_strong_reduced_state} differ by a single flip of the cut site, and $X_{L_A}$ both changes the parity of $A$ and commutes with every internal bond gate in $U_A$, so the two terms never mix. They are the operators of \cref{eq:strong_parity_blocks}: the branch in which the cut gate left the two sides untouched, carrying the weight $\cos^2\phi$, and the branch in which it flipped them, carrying $\sin^2\phi$, each then propagated by the gates inside $A$,
\begin{equation}
	\begin{aligned}
		C_A
		&=\mathbb E[\cos^2\phi\,U_A|\boldsymbol n_A\rangle\langle\boldsymbol n_A|U_A^\dagger],
		\\
		S_A
		&=\mathbb E[\sin^2\phi\,U_A|\boldsymbol n_A\rangle\langle\boldsymbol n_A|U_A^\dagger].
	\end{aligned}
	\label{eq:app_strong_CA_SA}
\end{equation}
Ordering the Hilbert space by parity gives
\begin{equation}
	\begin{aligned}
		\rho_A
		&=C_A\oplus X_{L_A}S_AX_{L_A}.
	\end{aligned}
	\label{eq:app_strong_parity_blocks}
\end{equation}
If $c_1\geq c_2\geq\cdots$ and $d_1\geq d_2\geq\cdots$ are the spectra of $C_A$ and $S_A$, the two largest eigenvalues of the union of the two spectra are
\begin{equation}
	\begin{aligned}
		\lambda_0&=\max(c_1,d_1),
		\\
		\lambda_1&=\max\{\min(c_1,d_1),c_2,d_2\},
	\end{aligned}
	\label{eq:app_strong_top_two_levels}
\end{equation}
which proves \cref{eq:strong_entanglement_branch_map}.

The parity of $A$ in the initial product state is $(-1)^{N_A(0)}$. Every internal bond gate flips two sites at once, so $U_A$ preserves that parity and both $C_A$ and $S_A$ of \cref{eq:app_strong_CA_SA} sit in that sector, while the conjugation by $X_{L_A}$ in \cref{eq:app_strong_parity_blocks} carries the second block to the opposite sector. The subsystem parity $q_A=\operatorname{Tr}(\rho_AP_A)$ of \cref{eq:subsystem_parities} therefore weights the two blocks with opposite signs,
\begin{equation}
	\begin{aligned}
		q_A
		&=\operatorname{Tr}(\rho_AP_A)
		=(-1)^{N_A(0)}[\operatorname{Tr}C_A-\operatorname{Tr}S_A],
	\end{aligned}
	\label{eq:app_strong_parity_trace}
\end{equation}
which is \cref{eq:strong_subsystem_parity}. This makes the distinction between parity balance and a true leading-level crossing exact: the former compares traces, while the latter compares operator norms and internal degeneracies.

\subsection{Pattern universality}

Let $|\boldsymbol n'\rangle$ differ from $|\boldsymbol n\rangle$ on a set of sites $\mathcal I$. In spin variables,
\begin{equation}
	\begin{aligned}
		|\boldsymbol n'\rangle
		&=X_{\mathcal I}|\boldsymbol n\rangle,
		&X_{\mathcal I}&=\prod_{j\in\mathcal I}X_j.
	\end{aligned}
	\label{eq:app_strong_pattern_map}
\end{equation}
Every $X_j$ commutes with every bond $X_kX_{k+1}$, so $[X_{\mathcal I},H]=0$. Also
\begin{equation}
	\begin{aligned}
		X_{\mathcal I}n_jX_{\mathcal I}
		&=\begin{cases}
			1-n_j,&j\in\mathcal I,\\
			n_j,&j\notin\mathcal I,
		\end{cases}
	\end{aligned}
	\label{eq:app_strong_density_flip}
\end{equation}
and for any projector $n$,
\begin{equation}
	\begin{aligned}
		\mathcal D[1-n]\rho
		&=(1-n)\rho(1-n)-\frac{1}{2}\{1-n,\rho\}
		\\
		&=n\rho n-\frac{1}{2}\{n,\rho\}
		=\mathcal D[n]\rho.
	\end{aligned}
	\label{eq:app_strong_projector_dephasing}
\end{equation}
Dephasing is therefore blind to which of the two occupations a site carries, and the Liouvillian is covariant:
\begin{equation}
	\begin{aligned}
		\mathcal L_s(X_{\mathcal I}\rho X_{\mathcal I})
		&=X_{\mathcal I}\mathcal L_s(\rho)X_{\mathcal I}.
	\end{aligned}
	\label{eq:app_strong_L_covariance}
\end{equation}
Writing $\rho_{\mathcal I}$ for the state evolved from the flipped pattern, uniqueness of the finite-dimensional initial-value problem gives
\begin{equation}
	\begin{aligned}
		\rho_{\mathcal I}(t)&=X_{\mathcal I}\rho(t)X_{\mathcal I}.
	\end{aligned}
	\label{eq:app_strong_state_conjugation}
\end{equation}
Since $X_{\mathcal I}=X_{\mathcal I\cap A}\otimes X_{\mathcal I\cap B}$ in the occupation tensor product, the reduced states are unitarily conjugate and have the same eigenvalue spectrum.

For the twisted spectrum, one must also commute the conjugation through partial time reversal. The Jordan-Wigner identity
\begin{equation}
	\begin{aligned}
		X_j
		&=\left(\prod_{k<j}Z_k\right)a_j
	\end{aligned}
	\label{eq:app_strong_X_majorana_monomial}
\end{equation}
shows that $X_{\mathcal I}$ is, up to a phase, a Majorana monomial. Conjugation by a Majorana monomial acts diagonally by signs on the ordered Majorana-monomial operator basis. Partial time reversal also acts diagonally on the same basis, multiplying each monomial $\mathcal M$ by the phase $i^{|\mathcal M|_A}$ of \cref{eq:partial_time_reversal}. The two actions commute, so
\begin{equation}
	\begin{aligned}
		(X_{\mathcal I}\rho X_{\mathcal I})^{R_A}
		&=X_{\mathcal I}\rho^{R_A}X_{\mathcal I}.
	\end{aligned}
	\label{eq:app_strong_ptr_covariance_conjugation}
\end{equation}
Furthermore $X_{\mathcal I}P_AX_{\mathcal I}=\chi_{\mathcal I}P_A$ with $\chi_{\mathcal I}=(-1)^{|\mathcal I\cap A|}$, since each flipped site inside $A$ anticommutes with its own factor in $P_A$. Write $T_{A,\mathcal I}\coloneq(\rho_{\mathcal I})^{R_A}P_A$ for the twisted operator of \cref{eq:twisted_operator} built from the flipped state, the first subscript naming the region that is partially reversed and the second the set of flipped sites. Then
\begin{equation}
	\begin{aligned}
		T_{A,\mathcal I}
		&=\chi_{\mathcal I}X_{\mathcal I}T_AX_{\mathcal I}.
	\end{aligned}
	\label{eq:app_strong_twisted_conjugation}
\end{equation}
Unitary conjugation preserves eigenvalues and the common sign $\chi_{\mathcal I}$ disappears once absolute values are taken, so the complete singular-value multiset is the same for both patterns. This proves \cref{eq:strong_pattern_universality}.

\subsection{Bath-length universality of the reduced spectrum}

Equation~\eqref{eq:app_strong_reduced_state} contains the jump histories on sites $1,\ldots,L_A+1$ and no dynamical operator supported deeper inside $B$. The process on those first $L_A+1$ sites is the same for every $L_B\geq1$, because the jumps of \cref{eq:app_strong_spin_master} are independent and occur at the same rate on every site. Therefore $\rho_A(t)$ has exactly the same spectrum for all nonzero bath lengths, which proves \cref{eq:strong_bath_length_universality}. The twisted operator $T_A$ of \cref{eq:twisted_operator} is built from the full state before $B$ is traced out, so it keeps the jump histories of all $L$ sites and retains $L_B$ as a parameter.

\subsection{Stationary state in a fixed parity sector}

Assume $J\neq0$, $\gamma>0$, and a connected open chain; the general irreducibility and stationary-state framework is given in Refs.~\cite{Evans1977Oct,Frigerio1978Oct,Buca2012Jul,Albert2014Feb}. For a Hermitian stationary state $\rho_\ast$, take the Hilbert-Schmidt inner product with the master equation. The Hamiltonian term drops out by cyclicity, and
\begin{equation}
	\begin{aligned}
		\operatorname{Tr}[\rho_\ast\mathcal L_s(\rho_\ast)]
		&=-\frac{\gamma}{2}\sum_j
		\operatorname{Tr}\!\left([n_j,\rho_\ast]^\dagger[n_j,\rho_\ast]\right)
		\leq0.
	\end{aligned}
	\label{eq:app_strong_stationary_dissipation}
\end{equation}
Stationarity makes the left side zero. Every nonnegative term must therefore vanish, so $[n_j,\rho_\ast]=0$ for every site and $\rho_\ast$ is diagonal in the Fock basis. The stationary equation then requires $[H,\rho_\ast]=0$.

At the compact point, $X_jX_{j+1}$ maps a configuration to the one obtained by flipping the adjacent pair. Denote this map by $F_j$. Commutation with $H$ equates the diagonal probabilities of configurations connected by each $F_j$. The binary displacement vectors $e_j\oplus e_{j+1}$ span the even-weight subspace, so adjacent pair flips connect all configurations with the same global parity. Therefore the weights are uniform inside a fixed parity sector. Write $\rho_{\infty,p}$ for the stationary state in the sector of global parity $p=\pm1$, the two subscripts being labels for the long-time limit and for the sector rather than matrix indices, and let $\Pi_p=(\mathbf 1+pP)/2$ project onto that sector. Then
\begin{equation}
	\begin{aligned}
		\rho_{\infty,p}
		&=\frac{\Pi_p}{2^{L-1}}.
	\end{aligned}
	\label{eq:app_strong_stationary_global}
\end{equation}
For any fixed configuration of the nonempty region $A$, exactly $2^{L_B-1}$ configurations of the nonempty region $B$ carry the parity that the global sector demands, so every configuration of $A$ receives the same marginal weight $2^{L_B-1}/2^{L-1}=2^{-L_A}$. Tracing over $B$ therefore leaves the maximally mixed state, written $\rho_{A,\infty}$ with $\mathbf 1_A$ the identity on $A$,
\begin{equation}
	\begin{aligned}
		\rho_{A,\infty}
		&=\frac{\mathbf 1_A}{2^{L_A}}.
	\end{aligned}
	\label{eq:app_strong_stationary_reduced}
\end{equation}
Equations \eqref{eq:app_strong_stationary_global} and \eqref{eq:app_strong_stationary_reduced} prove \cref{eq:strong_stationary_state}. The global stationary state is diagonal and separable, so $E_N(\infty)=0$.

\section{Entanglement Clock under Strong Symmetry}
\label{app:strong_entanglement_asymptotics}

The strong-symmetry reduced state is not determined by its covariance for $L_A\geq2$, so the entanglement clock must be obtained from the parity blocks themselves. Two calculations remain tractable: the exact subsystem-parity oscillator at every rate and the true leading-level crossing to first order at weak dephasing.

\subsection{Exact subsystem-parity oscillator}

Write $P_A$ for the subsystem parity in spin variables and pair it with the single operator that the dynamics couples it to,
\begin{equation}
	\begin{aligned}
		P_A&=\prod_{j=1}^{L_A}Z_j,
		&\widehat Y_A&=-iX_{L_A}X_{L_A+1}P_A.
	\end{aligned}
	\label{eq:app_strong_parity_operators}
\end{equation}
The hat marks an operator, as it does for $\widehat B$ and $\widehat N$. Only the bond crossing the cut, $X_{L_A}X_{L_A+1}$, anticommutes with $P_A$, because it flips the occupation of exactly one site of $A$; every internal bond flips two and commutes. Expectation values evolve with the adjoint generator, the same device used in \cref{app:weak_gaussian_reduction}, which here reads
\begin{equation}
	\begin{aligned}
		\mathcal L_s^\dagger(O)
		&=i[H,O]+\frac{\gamma}{4}\sum_j(Z_jOZ_j-O),
	\end{aligned}
	\label{eq:app_strong_adjoint_spin}
\end{equation}
and we write $\dot O$ for $\mathcal L_s^\dagger(O)$. The coherent part rotates the two operators into each other,
\begin{equation}
	\begin{aligned}
		i[H,P_A]&=2J\widehat Y_A,
		&i[H,\widehat Y_A]&=-2JP_A,
	\end{aligned}
	\label{eq:app_strong_parity_hamiltonian_commutators}
\end{equation}
while the dephasing term leaves $P_A$ untouched, since it commutes with every $Z_j$, and damps $\widehat Y_A$, since $\widehat Y_A$ anticommutes with $Z_{L_A}$ and $Z_{L_A+1}$ and each of those two sites contributes $-\gamma\widehat Y_A/2$. Hence
\begin{equation}
	\begin{aligned}
		\dot P_A&=2J\widehat Y_A,
		&\dot{\widehat Y}_A&=-2JP_A-\gamma\widehat Y_A.
	\end{aligned}
	\label{eq:app_strong_parity_operator_pair}
\end{equation}
Now take expectation values in the initial product state. The first of them is the subsystem parity itself, $q_A(t)=\operatorname{Tr}[\rho(t)P_A]$ of \cref{eq:subsystem_parities}, and the second is fixed by the first through $\dot q_A=2J\langle\widehat Y_A\rangle$. Eliminating $\langle\widehat Y_A\rangle$ leaves one damped oscillator for the parity,
\begin{equation}
	\begin{aligned}
		\ddot q_A+\gamma\dot q_A+4J^2q_A&=0,
		\\
		q_A(0)&=(-1)^{N_A(0)},
		&\dot q_A(0)&=0.
	\end{aligned}
	\label{eq:app_strong_parity_oscillator}
\end{equation}
This proves \cref{eq:strong_parity_oscillator}.

For $0\leq\gamma<4J$, use the frequency $\omega=\sqrt{4J^2-\gamma^2/4}$ of \cref{eq:strong_parity_frequency}. The solution is
\begin{equation}
	\begin{aligned}
		q_A(t)
		&=(-1)^{N_A(0)}e^{-\gamma t/2}
		\left[\cos(\omega t)+\frac{\gamma}{2\omega}\sin(\omega t)\right].
	\end{aligned}
	\label{eq:app_strong_parity_underdamped}
\end{equation}
The first positive root, with $\omega t_q$ in the second quadrant, is
\begin{equation}
	\begin{aligned}
		t_q
		&=\frac{\pi/2+\arctan[\gamma/(2\omega)]}{\omega}.
	\end{aligned}
	\label{eq:app_strong_parity_root}
\end{equation}
At $\gamma=4J$,
\begin{equation}
	\begin{aligned}
		(-1)^{N_A(0)}q_A(t)
		&=e^{-2Jt}(1+2Jt)>0
	\end{aligned}
	\label{eq:app_strong_parity_critical}
\end{equation}
for every finite $t$. Above $\gamma=4J$ the radicand of $\omega$ changes sign, so $\omega=i\bar\omega$ with the overdamped rate $\bar\omega=\sqrt{\gamma^2/4-4J^2}$ of \cref{eq:strong_overdamped_frequency}, and $\cos$ and $\sin$ turn into $\cosh$ and $\sinh$,
\begin{equation}
	\begin{aligned}
		q_A(t)
		&=(-1)^{N_A(0)}e^{-\gamma t/2}
		\left[\cosh(\bar\omega t)+\frac{\gamma}{2\bar\omega}\sinh(\bar\omega t)\right],
	\end{aligned}
	\label{eq:app_strong_parity_overdamped}
\end{equation}
which also has no finite zero. This proves \cref{eq:strong_parity_clock}.

For $L_A=1$, the two parity blocks are one dimensional. Their eigenvalues are $(1\pm q_A)/2$, so
\begin{equation}
	\begin{aligned}
		\lambda_0-\lambda_1
		&=|q_A|.
	\end{aligned}
	\label{eq:app_strong_one_site_gap}
\end{equation}
Hence $t_E=t_q$ exactly for $L_A=1$, proving \cref{eq:strong_one_site_entanglement_clock}. For $L_A\geq2$, $q_A=0$ only equates the traces of the two parity blocks. It need not equate their largest eigenvalues.

\subsection{Covariance closure without state completeness}

The Majorana covariance still obeys a closed equation. A dephasing jump at site $j$ conjugates by $Z_j$, which reverses the sign of both Majoranas of that site and leaves every other Majorana alone, so on the $2L$ Majorana indices it acts as the diagonal matrix $\mathsf Z_j$ carrying $-1$ in the two slots of site $j$, $+1$ elsewhere, and satisfying $\mathsf Z_j^2=\mathbf 1$. With $\mathsf A$ the closed Majorana generator of \cref{eq:app_weak_hamiltonian_covariance},
\begin{equation}
	\begin{aligned}
		\dot\Gamma
		&=\mathsf A\Gamma+\Gamma\mathsf A^T
		+\frac{\gamma}{4}\sum_j(\mathsf Z_j\Gamma\mathsf Z_j-\Gamma).
	\end{aligned}
	\label{eq:app_strong_covariance_closure}
\end{equation}
Closure of this equation means only that two-point functions form a closed set. The jump $n_j$ is quadratic in fermions, and the ensemble average of the random Gaussian trajectories in \cref{eq:strong_random_unitary} is generally non-Gaussian. Wick factorization therefore fails. For $L_A\geq2$, neither $\operatorname{Pf}\Gamma_A=0$ nor $q_A=0$ is an exact criterion for $\lambda_0=\lambda_1$. This is the methodological reason the strong-symmetry calculation uses the full parity blocks.

\subsection{Weak-dephasing shift of the true leading levels}

Set $t_0=\pi/(4J)$ and define the leading block difference
\begin{equation}
	\begin{aligned}
		g(t,\gamma)
		&=c_1(t,\gamma)-d_1(t,\gamma),
	\end{aligned}
	\label{eq:app_strong_g_definition}
\end{equation}
where $c_1$ and $d_1$ are the largest eigenvalues of the two parity blocks $C_A$ and $S_A$ of \cref{eq:app_strong_CA_SA}. This $g$ plays for the entanglement gap the role that $g_N$ of \cref{eq:app_weak_neg_oriented_branch} plays for the negativity gap: its zero is the closing, and its sign says which parity leads. At $\gamma=0$ no jump occurs, the accumulated phase of \cref{eq:app_strong_UA_phi} is simply $\phi=Jt$, and \cref{eq:app_strong_CA_SA} gives
\begin{equation}
	\begin{aligned}
		c_1(t,0)&=\cos^2(Jt),
		&d_1(t,0)&=\sin^2(Jt),
		\\
		g(t,0)&=\cos(2Jt),
		&\partial_tg(t_0,0)&=-2J.
	\end{aligned}
	\label{eq:app_strong_g_closed}
\end{equation}
At $t_0$, each leading eigenvalue $1/2$ is simple within its parity block and separated from the remaining eigenvalues by $1/2$, so ordinary first-order eigenvalue perturbation is valid.

To first order in $\gamma$ only histories with no jump or with one jump survive, and three observations reduce them to two integrals. First, the probability that no jump occurs anywhere is $1-L\gamma t/4$, and it multiplies the closed answer; since $g(t_0,0)=0$, rescaling zero leaves the root exactly where it was. Second, a jump at a given site during a window $du$ carries probability $(\gamma/4)du$, so each one-jump history enters with that weight and must be integrated over the time at which it happens. Third, most jump locations cancel: a jump deep inside $A$ reverses bonds lying entirely in $A$, which acts identically on the two leading parity states and moves $c_1$ and $d_1$ together, while a jump deep inside $B$ acts as a unitary on $B$ alone and disappears under the partial trace. Only jumps on the two sites that the cut bond touches, $L_A$ and $L_A+1$, contribute.

Take the endpoint on the $B$ side first, and let $u$ be the time left between the jump and the observation time $t_0$. Conjugating the cut bond by $Z_{L_A+1}$ reverses its sign, as in \cref{eq:app_strong_jump_sign_flip}, so the last stretch of the evolution undoes what the first stretch built and the accumulated cut angle of \cref{eq:app_strong_UA_phi} is $J(t_0-u)-Ju=J(t_0-2u)$ in place of $Jt_0$. Feeding that angle into the closed branch difference $g(t,0)=\cos(2Jt)$ of \cref{eq:app_strong_g_closed}, and using $2Jt_0=\pi/2$, gives
\begin{equation}
	\begin{aligned}
		f_B(u)
		&=\cos[2J(t_0-2u)]
		=\sin(4Ju).
	\end{aligned}
	\label{eq:app_strong_fB}
\end{equation}
For $L_A\geq2$ a jump on the $A$ endpoint reverses the adjacent internal bond $X_{L_A-1}X_{L_A}$ as well, again for the remaining time $u$. Relative to the unjumped trajectory the state inside $A$ is then rotated by $e^{-2iJuX_{L_A-1}X_{L_A}}$, and only the part that returns to the unjumped leading vector feeds the same levels. That return probability is $\cos^2(2Ju)$, because a product-Fock state has zero expectation of $X_{L_A-1}X_{L_A}$, so the rotation splits into $\cos(2Ju)$ times the original state plus an orthogonal remainder. Thus
\begin{equation}
	\begin{aligned}
		f_A(u)
		&=\sin(4Ju)\cos^2(2Ju),
		\qquad L_A\geq2.
	\end{aligned}
	\label{eq:app_strong_fA}
\end{equation}
Both integrals are elementary, the second because the extra $\cos^2(2Ju)$ contributes a full period of $\sin(8Ju)$ that averages to zero,
\begin{equation}
	\begin{aligned}
		\int_0^{t_0}f_B(u)du&=\frac{1}{2J},
		&\int_0^{t_0}f_A(u)du&=\frac{1}{4J}.
	\end{aligned}
	\label{eq:app_strong_f_integrals}
\end{equation}
Adding the two endpoint contributions and multiplying by the jump rate $\gamma/4$ gives
\begin{equation}
	\begin{aligned}
		\partial_\gamma g(t_0,0)
		&=\frac{3}{16J},
		\qquad L_A\geq2.
	\end{aligned}
	\label{eq:app_strong_g_gamma_largeA}
\end{equation}
For $L_A=1$, there is no internal $A$ bond, so the two endpoints both contribute $f_B$ and
\begin{equation}
	\begin{aligned}
		\partial_\gamma g(t_0,0)
		&=\frac{1}{4J}.
	\end{aligned}
	\label{eq:app_strong_g_gamma_oneA}
\end{equation}
Differentiate the root condition $g[t_E(\gamma),\gamma]=0$ at $\gamma=0$:
\begin{equation}
	\begin{aligned}
		t_E'(0)
		&=-\frac{\partial_\gamma g(t_0,0)}{\partial_tg(t_0,0)}.
	\end{aligned}
	\label{eq:app_strong_implicit_clock}
\end{equation}
Using \cref{eq:app_strong_g_closed,eq:app_strong_g_gamma_largeA,eq:app_strong_g_gamma_oneA} gives
\begin{equation}
	\begin{aligned}
		t_E
		&=t_0+
		\begin{cases}
			\dfrac{\gamma}{8J^2},&L_A=1,\\[6pt]
			\dfrac{3\gamma}{32J^2},&L_A\geq2,
		\end{cases}
		+\mathcal O\!\left(\frac{\gamma^2}{J^3}\right),
	\end{aligned}
	\label{eq:app_strong_weak_entanglement_clock}
\end{equation}
which proves \cref{eq:strong_weak_entanglement_clock}.

Expanding the exact parity root \cref{eq:app_strong_parity_root} gives
\begin{equation}
	\begin{aligned}
		t_q
		&=t_0+\frac{\gamma}{8J^2}
		+\mathcal O\!\left(\frac{\gamma^2}{J^3}\right).
	\end{aligned}
	\label{eq:app_strong_weak_parity_clock}
\end{equation}
Subtracting \cref{eq:app_strong_weak_entanglement_clock} yields
\begin{equation}
	\begin{aligned}
		t_q-t_E
		&=\frac{\gamma}{32J^2}
		+\mathcal O\!\left(\frac{\gamma^2}{J^3}\right),
		\qquad L_A\geq2,
	\end{aligned}
	\label{eq:app_strong_weak_parity_separation}
\end{equation}
which proves \cref{eq:strong_weak_parity_clock,eq:strong_weak_parity_separation}.

\section{Weak-Dephasing Negativity Clock under Strong Symmetry}
\label{app:strong_negativity_asymptotics}

At zero dephasing the compact state is pure, so the four nonzero twisted singular values are the same four values used in \cref{eq:app_weak_neg_four_singular_values}. Density dephasing perturbs them through random $Z_j$ jumps rather than through linear loss or gain. This appendix derives the arbitrary-geometry coefficient in \cref{eq:strong_weak_negativity_clock}.

Two indicators record whether either side of the cut has shrunk to a single site, written with the symbol $\mathbf 1_{\{\cdot\}}$ of \cref{eq:app_weak_geo_boundary_indicators}, which equals one when the condition inside the braces holds and zero otherwise:
\begin{equation}
	\begin{aligned}
		\delta_A&=\mathbf 1_{\{L_A=1\}},
		&\delta_B&=\mathbf 1_{\{L_B=1\}}.
	\end{aligned}
	\label{eq:app_strong_neg_boundary_indicators}
\end{equation}
They are the same indicators that the weak-symmetry calculation used: $\delta_A=1$ says the cut bond touches the left end of the chain, $\delta_B=1$ that it touches the right end. A bulk cut such as $L_A=L_B=4$ has $\delta_A=\delta_B=0$, while the two-site chain $L_A=L_B=1$ has both equal to one. Let $e_+$ and $e_-$ be the largest eigenvalues of the two $P_A$ blocks of $\rho_{\mathrm e}^{R_A}$ in \cref{eq:strong_twisted_blocks}, labeled so that $e_\pm=\lambda_\pm$ at $\gamma=0$. In the numerics of \cref{subsec:tnh_strong} the closing continued from $t_0$ is their crossing, and the pair of opposite eigenvalues of $T_A^-$, which meets these two levels at $t_0$ when $\gamma=0$, lies lower by an amount of order $\gamma$. Use the oriented branch
\begin{equation}
	\begin{aligned}
		g_N(t,\gamma)
		&=\frac{e_+(t,\gamma)-e_-(t,\gamma)}{2},
		\\
		g_N(t,0)
		&=\frac{\lambda_+(t)-\lambda_-(t)}{2},
	\end{aligned}
	\label{eq:app_strong_neg_oriented_branch}
\end{equation}
with $\partial_tg_N(t_0,0)=-J$.

Expanding the jump average of \cref{eq:strong_random_unitary} in the number of jumps and keeping the first two terms gives the density matrix to first order in $\gamma$,
\begin{equation}
	\begin{aligned}
		\rho(t)
		={}&\left(1-\frac{L\gamma t}{4}\right)\rho_0(t)
		\\
		&+\frac{\gamma}{4}\sum_{j=1}^{L}\int_0^t\!du\,
		U(u)Z_j\rho_0(t-u)Z_jU^\dagger(u)
		+\mathcal O(\gamma^2),
	\end{aligned}
	\label{eq:app_strong_neg_one_jump_expansion}
\end{equation}
where $U(u)=e^{-iHu}$ and $\rho_0(t)=U(t)\rho(0)U^\dagger(t)$ is the closed evolution. The first term carries the probability that no jump occurred at all, $1-L\gamma t/4$, which multiplies the closed answer; since the closing branch vanishes there, $g_N(t_0,0)=0$, rescaling it cannot move the root. The second term places a single jump at site $j$ a time $u$ before the observation, with weight $\gamma/4$ per unit time.

A jump away from a cut endpoint reverses bond signs only inside $A$ or only inside $B$. Compact factorization then shifts all four closing magnitudes equally, so such a jump cancels from $g_N$. Consider one cut endpoint on side $X\in\{A,B\}$. Reversing the cut bond for the remaining duration $u$ changes the net angle from $Jt_0$ to $J(t_0-2u)$, exactly as in \cref{eq:app_strong_fB}. Everything in the branch depends on that angle through the single number $c$ appearing in the reduced eigenvalues $\lambda_\pm=(1\pm c)/2$ of \cref{eq:app_weak_neg_closed_lambdas}, which equals $\cos(2Jt)$ on the unjumped trajectory. In terms of $c$, \cref{eq:app_strong_neg_oriented_branch} reads $g_N=c/2$, which vanishes at the closing $c=0$ with slope $1/2$. The jumped history has $c=\cos[2J(t_0-2u)]=\sin(4Ju)$, so it shifts the branch by
\begin{equation}
	\begin{aligned}
		\frac{1}{2}\sin(4Ju).
	\end{aligned}
	\label{eq:app_strong_neg_bare_endpoint}
\end{equation}
If $L_X\geq2$, the same jump also reverses the internal bond attached to that endpoint. Relative to the unjumped trajectory, the local state is rotated by $e^{-2iJuXX}$, whose return probability in a product-Fock state is $\cos^2(2Ju)$. Thus
\begin{equation}
	\begin{aligned}
		h_X(u)
		&=\frac{1}{2}\sin(4Ju)
		\begin{cases}
			1,&L_X=1,\\
			\cos^2(2Ju),&L_X\geq2.
		\end{cases}
	\end{aligned}
	\label{eq:app_strong_neg_endpoint_function}
\end{equation}
The two possible integrals are
\begin{equation}
	\begin{aligned}
		\int_0^{t_0}h_X(u)du
		&=\begin{cases}
			\dfrac{1}{4J},&L_X=1,\\[5pt]
			\dfrac{1}{8J},&L_X\geq2.
		\end{cases}
	\end{aligned}
	\label{eq:app_strong_neg_endpoint_integrals}
\end{equation}
Multiplying by the jump rate $\gamma/4$ and adding the two endpoints gives a coefficient that the two indicators keep track of: a bulk cut contributes $1/(8J)$ from each endpoint, so one quarter of their sum is $1/(16J)$, and every side that shrinks to a single site replaces its $1/(8J)$ by $1/(4J)$ and so adds $1/(32J)$. Hence
\begin{equation}
	\begin{aligned}
		\partial_\gamma g_N(t_0,0)
		&=\frac{2+\delta_A+\delta_B}{32J}.
	\end{aligned}
	\label{eq:app_strong_neg_gamma_derivative}
\end{equation}
The fourfold nonzero magnitude level is separated from the zero singular values by a finite gap, so degenerate perturbation theory inside that four-dimensional support gives branches differentiable in $\gamma$ \cite{Kato1995,StewartSun1990}. Since $\partial_tg_N(t_0,0)=-J\neq0$, the closing is a simple root, which therefore moves smoothly with the rate instead of disappearing, exactly the continuation of \cref{eq:app_weak_finite_rate_continuation}. Differentiating $g_N[t_N(\gamma),\gamma]=0$ at $\gamma=0$ yields
\begin{equation}
	\begin{aligned}
		t_N'(0)
		&=-\frac{\partial_\gamma g_N}{\partial_tg_N}
		=\frac{2+\delta_A+\delta_B}{32J^2}.
	\end{aligned}
	\label{eq:app_strong_neg_clock_derivative}
\end{equation}
Hence
\begin{equation}
	\begin{aligned}
		t_N
		&=t_0+\frac{2+\delta_A+\delta_B}{32}
		\frac{\gamma}{J^2}
		+\mathcal O\!\left(\frac{\gamma^2}{J^3}\right),
	\end{aligned}
	\label{eq:app_strong_neg_weak_clock}
\end{equation}
which proves \cref{eq:strong_weak_negativity_clock}.

Combining this result with \cref{eq:app_strong_weak_entanglement_clock} gives
\begin{equation}
	\begin{aligned}
		t_E-t_N
		&=\frac{1-\delta_B}{32}\frac{\gamma}{J^2}
		+\mathcal O\!\left(\frac{\gamma^2}{J^3}\right),
	\end{aligned}
	\label{eq:app_strong_neg_clock_split}
\end{equation}
which is \cref{eq:strong_weak_clock_splitting}. The coefficient is independent of the product-Fock pattern because pattern universality is exact before the perturbative expansion is taken.

\section{Strong-Dephasing Zeno Reduction}
\label{app:strong_zeno_reduction}

The Zeno limit has two separated scales, the fast and slow times of \cref{eq:strong_zeno_scales}. Coherences decay on $t_{\mathrm f}=\gamma^{-1}$, while populations move only through second-order virtual processes on $t_{\mathrm s}=\gamma/(2J^2)$. The leading slow problem is classical, but order-$J/\gamma$ cut coherences remain essential for the negativity itself \cite{Facchi2008Dec,Reiter2012Mar,Kessler2012Jul,Znidaric2010Apr,Medvedyeva2016Sep}.

\subsection{Matrix-element equations and adiabatic elimination}

Let $|\boldsymbol n\rangle$ denote a Fock configuration, with $\boldsymbol n=(n_1,\ldots,n_L)\in\{0,1\}^L$ the binary occupation string of \cref{eq:strong_pair_flip}. Define the Hamming distance~\footnote{The Hamming distance between two binary strings is the number of places where they differ, and the Hamming weight $|\boldsymbol n|$ is the number of ones, that is, the distance from the all-zero string. The indicator $\mathbf 1_{\{\cdots\}}$ of \cref{eq:strong_binary_domain_walls} equals one at each disagreeing site and zero at the rest, so the sum counts exactly the disagreements. For $\boldsymbol n=(0,1,1,0)$ and $\boldsymbol m=(0,0,1,1)$ the two strings differ at sites $2$ and $4$, so $d_H(\boldsymbol n,\boldsymbol m)=2$, while $|\boldsymbol n|=|\boldsymbol m|=2$.}
\begin{equation}
	\begin{aligned}
		d_H(\boldsymbol n,\boldsymbol m)
		&=\sum_{j=1}^{L}\mathbf 1_{\{n_j\neq m_j\}}.
	\end{aligned}
	\label{eq:app_strong_hamming_distance}
\end{equation}
Since $Z_j|\boldsymbol n\rangle=(-1)^{n_j}|\boldsymbol n\rangle$, one has $(Z_j\rho Z_j)_{\boldsymbol n\boldsymbol m}=(-1)^{n_j+m_j}\rho_{\boldsymbol n\boldsymbol m}$, so each site of agreement contributes $1-1=0$ and each site of disagreement contributes $-1-1=-2$. Summing over sites, the dephasing part acts on a matrix element as
\begin{equation}
	\begin{aligned}
		\frac{\gamma}{4}\sum_j
		(Z_j\rho Z_j-\rho)_{\boldsymbol n\boldsymbol m}
		&=-\frac{\gamma}{2}d_H(\boldsymbol n,\boldsymbol m)
		\rho_{\boldsymbol n\boldsymbol m}.
	\end{aligned}
	\label{eq:app_strong_hamming_damping}
\end{equation}
Hence
\begin{equation}
	\begin{aligned}
		\dot\rho_{\boldsymbol n\boldsymbol m}
		&=-i(H\rho-\rho H)_{\boldsymbol n\boldsymbol m}
		-\frac{\gamma}{2}d_H(\boldsymbol n,\boldsymbol m)
		\rho_{\boldsymbol n\boldsymbol m}.
	\end{aligned}
	\label{eq:app_strong_matrix_element_equation}
\end{equation}
At the compact point, $H=-J\sum_jX_jX_{j+1}$. Each bond operator $X_jX_{j+1}$ flips the occupations of the two sites it touches, so it sends one configuration to exactly one other. That map is the adjacent pair flip of \cref{eq:strong_pair_flip},
\begin{equation}
	\begin{aligned}
		F_j\boldsymbol n
		&=\boldsymbol n\oplus\boldsymbol e_j\oplus\boldsymbol e_{j+1},
	\end{aligned}
	\label{eq:app_strong_pair_flip}
\end{equation}
where $\boldsymbol e_j$ is the string carrying a single one at site $j$ and $\oplus$ adds bit by bit modulo two. For $L=4$ and $\boldsymbol n=(0,1,1,0)$, the three flips give $F_1\boldsymbol n=(1,0,1,0)$, $F_2\boldsymbol n=(0,0,0,0)$, and $F_3\boldsymbol n=(0,1,0,1)$, and repeating any one of them returns $\boldsymbol n$, so $F_j^2=\mathbf 1$. The only nonzero off-diagonal elements of $H$ are therefore $H_{\boldsymbol n,F_j\boldsymbol n}=-J$, and such a pair differs at two sites, so $d_H(\boldsymbol n,F_j\boldsymbol n)=2$ and the corresponding coherence decays at rate $\gamma$.

Set
\begin{equation}
	\begin{aligned}
		p(\boldsymbol n,t)&=\rho_{\boldsymbol n\boldsymbol n}(t),
		&c_j(\boldsymbol n,t)&=\rho_{\boldsymbol n,F_j\boldsymbol n}(t).
	\end{aligned}
	\label{eq:app_strong_population_coherence}
\end{equation}
Both are functions of the configuration and of the time, the subscript $j$ on a coherence naming the bond that is flipped. The population equation is exact:
\begin{equation}
	\begin{aligned}
		\dot p(\boldsymbol n)
		&=2J\sum_{j=1}^{L-1}\operatorname{Im}c_j(\boldsymbol n).
	\end{aligned}
	\label{eq:app_strong_population_exact}
\end{equation}
The bond coherence obeys
\begin{equation}
	\begin{aligned}
		\dot c_j(\boldsymbol n)
		={}&-\gamma c_j(\boldsymbol n)
		+iJ[p(F_j\boldsymbol n)-p(\boldsymbol n)]
		\\
		&+iJ\sum_{k\neq j}
		\left(
		\rho_{F_k\boldsymbol n,F_j\boldsymbol n}
		-\rho_{\boldsymbol n,F_kF_j\boldsymbol n}
		\right).
	\end{aligned}
	\label{eq:app_strong_bond_coherence_exact}
\end{equation}
The matrix elements in the last line require two distinct bond flips and are order $(J/\gamma)^2$ after the fast transient. A useful parity of the expansion follows from the staggered unitary $\prod_{j\,\mathrm{even}}Z_j$, which maps $J\mapsto-J$ while leaving the dephasing term and the diagonal initial state invariant. Populations are therefore even in $J$, while one-bond coherences are odd. Thus no order-$(J/\gamma)^2$ term appears in $c_j$.

Solving \cref{eq:app_strong_bond_coherence_exact} on the fast scale with its slowly varying source gives
\begin{equation}
	\begin{aligned}
		c_j(\boldsymbol n,t)
		={}&\frac{iJ}{\gamma}
		[p(F_j\boldsymbol n,t)-p(\boldsymbol n,t)]
		\\
		&+\mathcal O\!\left(\frac{J^3}{\gamma^3}\right)
		+\mathcal O\!\left(\frac{J}{\gamma}e^{-\gamma t}\right).
	\end{aligned}
	\label{eq:app_strong_slaved_coherence}
\end{equation}
Substitution into \cref{eq:app_strong_population_exact} yields
\begin{equation}
	\begin{aligned}
		\dot p(\boldsymbol n)
		={}&\frac{2J^2}{\gamma}\sum_{j=1}^{L-1}
		[p(F_j\boldsymbol n)-p(\boldsymbol n)]
		\\
		&+\mathcal O\!\left(\frac{J^4}{\gamma^3}\right)
		+\mathcal O\!\left(\frac{J^2}{\gamma}e^{-\gamma t}\right).
	\end{aligned}
	\label{eq:app_strong_population_slow}
\end{equation}
Introduce $\varepsilon=J/\gamma$ and the Zeno time
\begin{equation}
	\begin{aligned}
		\vartheta
		&=\frac{2J^2}{\gamma}t.
	\end{aligned}
	\label{eq:app_strong_zeno_time}
\end{equation}
At fixed finite $\vartheta>0$ and $\varepsilon\to0$, the exponential transient is smaller than any power of $\varepsilon$. Write $p^{(0)}(\boldsymbol n,\vartheta)$ for what survives of $p(\boldsymbol n,t)$ in that limit. Throughout this appendix the superscript $(0)$ marks a leading Zeno object, the value reached as $\varepsilon\to0$ with $\vartheta$ held fixed, and the two arguments are the configuration and the Zeno time of \cref{eq:app_strong_zeno_time}. The system starts in the product Fock configuration $\boldsymbol n_0$, so at $\vartheta=0$ the Kronecker delta places the whole probability on that one string. Therefore
\begin{equation}
	\begin{aligned}
		\partial_\vartheta p^{(0)}(\boldsymbol n,\vartheta)
		&=\sum_{j=1}^{L-1}
		[p^{(0)}(F_j\boldsymbol n,\vartheta)-p^{(0)}(\boldsymbol n,\vartheta)],
		\\
		p^{(0)}(\boldsymbol n,0)
		&=\delta_{\boldsymbol n,\boldsymbol n_0}.
	\end{aligned}
	\label{eq:app_strong_zeno_markov}
\end{equation}
This proves \cref{eq:strong_pair_flip_kinetics}. Each term exchanges probability between a configuration and its pair-flip partner at unit rate in $\vartheta$, and the generator preserves global parity because every move flips exactly two bits.

\subsection{Binary Fourier solution}

For $\boldsymbol k,\boldsymbol n\in\{0,1\}^L$, define
\begin{equation}
	\begin{aligned}
		\chi_{\boldsymbol k}(\boldsymbol n)
		&=(-1)^{\boldsymbol k\cdot\boldsymbol n},
		&\boldsymbol k\cdot\boldsymbol n
		&=\sum_jk_jn_j\pmod 2.
	\end{aligned}
	\label{eq:app_strong_binary_characters}
\end{equation}
These $2^L$ sign functions are an orthogonal basis for functions on the binary configuration space. Under a pair flip,
\begin{equation}
	\begin{aligned}
		\chi_{\boldsymbol k}(F_j\boldsymbol n)
		&=(-1)^{k_j+k_{j+1}}\chi_{\boldsymbol k}(\boldsymbol n).
	\end{aligned}
	\label{eq:app_strong_character_flip}
\end{equation}
A bond carries a domain wall of $\boldsymbol k$ when its two endpoints differ, and \cref{eq:strong_binary_domain_walls} counts them,
\begin{equation}
	\begin{aligned}
		d(\boldsymbol k)
		&=\sum_{j=1}^{L-1}\mathbf 1_{\{k_j\neq k_{j+1}\}}.
	\end{aligned}
	\label{eq:app_strong_k_domain_walls}
\end{equation}
For $L=5$, the wavevector $\boldsymbol k=(0,0,1,1,0)$ has walls on bonds $2$ and $4$, so $d(\boldsymbol k)=2$, while the two uniform strings $(0,0,0,0,0)$ and $(1,1,1,1,1)$ have $d(\boldsymbol k)=0$. In \cref{eq:app_strong_character_flip} the factor $(-1)^{k_j+k_{j+1}}-1$ equals $-2$ on a wall and zero on every other bond, so the Markov generator $G=\sum_j(F_j-\mathbf 1)$ acts diagonally:
\begin{equation}
	\begin{aligned}
		G\chi_{\boldsymbol k}
		&=-2d(\boldsymbol k)\chi_{\boldsymbol k}.
	\end{aligned}
	\label{eq:app_strong_fourier_eigenvalue}
\end{equation}
Write $\boldsymbol n\oplus\boldsymbol n_0$ for the bitwise difference of the two configurations, the string carrying a one at every site where they disagree. Summing a character of \cref{eq:app_strong_binary_characters} over all $2^L$ wavevectors gives $2^L$ when that difference vanishes, since every term is then $+1$, and gives zero otherwise, since the terms cancel in pairs. This orthogonality expands the point probability by binary Fourier inversion,
\begin{equation}
	\begin{aligned}
		\delta_{\boldsymbol n,\boldsymbol n_0}
		&=2^{-L}\sum_{\boldsymbol k}
		(-1)^{\boldsymbol k\cdot(\boldsymbol n\oplus\boldsymbol n_0)}.
	\end{aligned}
	\label{eq:app_strong_binary_delta}
\end{equation}
Evolving each mode independently gives
\begin{equation}
	\begin{aligned}
		p^{(0)}(\boldsymbol n,\vartheta)
		&=2^{-L}\sum_{\boldsymbol k\in\{0,1\}^L}
		(-1)^{\boldsymbol k\cdot(\boldsymbol n\oplus\boldsymbol n_0)}
		e^{-2d(\boldsymbol k)\vartheta},
	\end{aligned}
	\label{eq:app_strong_full_fourier_distribution}
\end{equation}
which is \cref{eq:strong_zeno_full_probability}.

To trace $B$, split both strings into their two parts, $\boldsymbol n=(\boldsymbol\eta,\boldsymbol n_B)$ and $\boldsymbol k=(\boldsymbol k_A,\boldsymbol k_B)$, where $\boldsymbol\eta$ is the configuration restricted to $A$ as in \cref{eq:strong_zeno_marginal}. Tracing sums over every $\boldsymbol n_B$. One site at a time, $\sum_{n_j\in\{0,1\}}(-1)^{k_jn_j}$ equals $1+1=2$ for $k_j=0$ and $1-1=0$ for $k_j=1$, and the product of these $L_B$ factors gives
\begin{equation}
	\begin{aligned}
		\sum_{\boldsymbol n_B}(-1)^{\boldsymbol k_B\cdot\boldsymbol n_B}
		&=2^{L_B}\delta_{\boldsymbol k_B,0}.
	\end{aligned}
	\label{eq:app_strong_trace_B_fourier}
\end{equation}
Only $\boldsymbol k_B=0$ survives. Write $p_A^{(0)}(\boldsymbol\eta,\vartheta)=\sum_{\boldsymbol n_B}p^{(0)}((\boldsymbol\eta,\boldsymbol n_B),\vartheta)$ for the marginal probability of an $A$ pattern, the subscript naming the region that is kept and the arguments carrying the same meaning as in \cref{eq:app_strong_zeno_markov}. The first site of $B$ now carries wavevector zero, so the cut bond holds a domain wall whenever $k_{L_A}=1$, and the last indicator adds it to the count. Hence
\begin{equation}
	\begin{aligned}
		p_A^{(0)}(\boldsymbol\eta,\vartheta)
		={}&2^{-L_A}
		\sum_{\boldsymbol k\in\{0,1\}^{L_A}}
		(-1)^{\boldsymbol k\cdot(\boldsymbol\eta\oplus\boldsymbol\eta_0)}
		\\
		&\times e^{-2[d_A(\boldsymbol k)+\mathbf 1_{\{k_{L_A}=1\}}]\vartheta},
	\end{aligned}
	\label{eq:app_strong_marginal_fourier}
\end{equation}
with $d_A(\boldsymbol k)=\sum_{j=1}^{L_A-1}\mathbf 1_{\{k_j\neq k_{j+1}\}}$. This proves \cref{eq:strong_zeno_marginal}.

\subsection{Strict no-closing theorem for the full distribution}

The pair-flip displacement vectors $\boldsymbol e_j\oplus\boldsymbol e_{j+1}$ are linearly independent over $\mathbb Z_2$ and span the even-weight subspace. Thus a configuration is reachable from $\boldsymbol n_0$ exactly when the displacement $\boldsymbol r=\boldsymbol n\oplus\boldsymbol n_0$ has even Hamming weight. Fix such a nonzero $\boldsymbol r$ and subtract \cref{eq:app_strong_full_fourier_distribution} at $\boldsymbol n_0\oplus\boldsymbol r$ from its value at $\boldsymbol n_0$. Every term with $\boldsymbol k\cdot\boldsymbol r=0$ is common to the two sums and drops out, while a term with $\boldsymbol k\cdot\boldsymbol r=1$ enters with opposite signs and therefore doubles. At the Zeno time $\vartheta$ of \cref{eq:app_strong_zeno_time},
\begin{equation}
	\begin{aligned}
		&p^{(0)}(\boldsymbol n_0,\vartheta)
		-p^{(0)}(\boldsymbol n_0\oplus\boldsymbol r,\vartheta)
		\\
		&\qquad=2^{1-L}
		\sum_{\boldsymbol k:\boldsymbol k\cdot\boldsymbol r=1}
		e^{-2d(\boldsymbol k)\vartheta}>0.
	\end{aligned}
	\label{eq:app_strong_full_probability_margin}
\end{equation}
The sum is nonempty: pick a site $q$ where the bit $r_q$ of the displacement equals one and take the unit string $\boldsymbol k=\boldsymbol e_q$, for which $\boldsymbol k\cdot\boldsymbol r=r_q=1$. Every summand is strictly positive at finite $\vartheta$. Configurations of the opposite parity have zero probability. Hence the initial configuration is the unique most probable configuration at every finite Zeno time.

The only wavevectors with $d(\boldsymbol k)=0$ are the all-zero and all-one vectors. Taking $\vartheta\to\infty$ therefore gives the uniform distribution in the selected parity sector,
\begin{equation}
	\begin{aligned}
		p^{(0)}(\boldsymbol n,\infty)
		&=\begin{cases}
			2^{1-L},&P(\boldsymbol n)=P(\boldsymbol n_0),\\
			0,&P(\boldsymbol n)\neq P(\boldsymbol n_0),
		\end{cases}
	\end{aligned}
	\label{eq:app_strong_full_stationary_probability}
\end{equation}
where $P(\boldsymbol n)=(-1)^{|\boldsymbol n|}$ is the global parity of \cref{eq:global_parity} read off a configuration. The leading probability gap is therefore strictly positive at every finite $\vartheta$ and vanishes only asymptotically.

\subsection{Strict no-closing theorem for the reduced distribution}

For any nonzero binary displacement $\boldsymbol r_A$ in $A$, subtract the two marginal probabilities in \cref{eq:app_strong_marginal_fourier}:
\begin{equation}
	\begin{aligned}
		&p_A^{(0)}(\boldsymbol\eta_0,\vartheta)
		-p_A^{(0)}(\boldsymbol\eta_0\oplus\boldsymbol r_A,\vartheta)
		\\
		&\quad=2^{1-L_A}
		\sum_{\boldsymbol k:\boldsymbol k\cdot\boldsymbol r_A=1}
		e^{-2[d_A(\boldsymbol k)+\mathbf 1_{\{k_{L_A}=1\}}]\vartheta}>0.
	\end{aligned}
	\label{eq:app_strong_marginal_probability_margin}
\end{equation}
Here $\boldsymbol r_A$ is the displacement inside $A$ and $r_{A,j}$ is its bit at site $j$, the first subscript naming the region and the second the site. Again the sum is nonempty, since it contains the unit string on any site where $r_{A,j}=1$. Thus the initial $A$ pattern is uniquely most probable at every finite $\vartheta$. As $\vartheta\to\infty$, only $\boldsymbol k=0$ survives and every marginal probability approaches $2^{-L_A}$. Since the leading reduced state is diagonal, its eigenvalues are these probabilities, and the largest of them stays strictly above the next one. The lowest entanglement gap $\Delta_E$ of \cref{eq:spectral_gaps}, evaluated on that leading state and written $\Delta_E^{(0)}$, therefore stays open, and $\vartheta_E^{(0)}$ denotes the first Zeno time at which it would close. Hence
\begin{equation}
	\begin{aligned}
		\Delta_E^{(0)}(\vartheta)&>0,
		\qquad0<\vartheta<\infty,
		\\
		\lim_{\vartheta\to\infty}\Delta_E^{(0)}(\vartheta)&=0,
		\\
		\vartheta_E^{(0)}&=+\infty.
	\end{aligned}
	\label{eq:app_strong_zeno_entanglement_no_closing}
\end{equation}
This proves \cref{eq:strong_zeno_entanglement_margin} and the absence of a closing in \cref{eq:strong_zeno_entanglement_clock}; the value of the gap follows from the product form below.

\subsection{Diagonal partial time reversal}

Let
\begin{equation}
	\begin{aligned}
		\rho_{\mathrm{diag}}
		&=\sum_{\boldsymbol n}p(\boldsymbol n)
		|\boldsymbol n\rangle\langle\boldsymbol n|.
	\end{aligned}
	\label{eq:app_strong_diag_state}
\end{equation}
On one site, the Majorana operators $a_j$ and $b_j$ of \cref{eq:majorana_operators} give the occupation as
\begin{equation}
	\begin{aligned}
		n_j
		&=\frac{1+ia_jb_j}{2}.
	\end{aligned}
	\label{eq:app_strong_n_majorana}
\end{equation}
The partial time reversal $R_A$ of \cref{eq:partial_time_reversal} multiplies $a_j$ and $b_j$ by $i$ on every site of $A$, so there $ia_jb_j\mapsto-ia_jb_j$ and hence $n_j\mapsto1-n_j$, while a site in $B$ keeps its occupation. Therefore
\begin{equation}
	\begin{aligned}
		(|\boldsymbol n\rangle\langle\boldsymbol n|)^{R_A}
		&=|\boldsymbol n^{A}\rangle\langle\boldsymbol n^{A}|,
	\end{aligned}
	\label{eq:app_strong_diag_ptr_projector}
\end{equation}
where $\boldsymbol n^A$ carries the complemented bits on $A$ and the original bits on $B$. For $L=4$ with $L_A=2$, the configuration $\boldsymbol n=(1,0,1,1)$ gives $\boldsymbol n^A=(0,1,1,1)$. Complementing twice returns the original string, so $\boldsymbol n\mapsto\boldsymbol n^A$ is one to one and partial time reversal merely permutes the diagonal probabilities:
\begin{equation}
	\begin{aligned}
		\operatorname{sv}(\rho_{\mathrm{diag}}^{R_A})
		&=\{p(\boldsymbol n)\}_{\boldsymbol n}.
	\end{aligned}
	\label{eq:app_strong_diag_ptr_singular_values}
\end{equation}
Here $\operatorname{sv}(X)$ is the multiset of singular values of $X$, the eigenvalues of $|X|=\sqrt{X^\dagger X}$ counted with multiplicity, which for the twisted operator are the $s_k$ of \cref{eq:twisted_spectrum}. This proves \cref{eq:strong_diagonal_ptr_spectrum}.

Combining \cref{eq:app_strong_full_probability_margin,eq:app_strong_diag_ptr_singular_values} gives the same conclusion for the lowest twisted gap $\Delta_N$ of \cref{eq:spectral_gaps}, again evaluated on the leading Zeno state,
\begin{equation}
	\begin{aligned}
		\Delta_N^{(0)}(\vartheta)&>0,
		\qquad0<\vartheta<\infty,
		\\
		\lim_{\vartheta\to\infty}\Delta_N^{(0)}(\vartheta)&=0,
		\\
		\vartheta_N^{(0)}&=+\infty,
	\end{aligned}
	\label{eq:app_strong_zeno_negativity_no_closing}
\end{equation}
which proves \cref{eq:strong_zeno_clock_locking} and the absence of a closing in \cref{eq:strong_zeno_negativity_clock}.

\subsection{Control of the exact spectra at fixed Zeno time}

At fixed finite $L$ and $\vartheta$ in a compact interval $K\subset(0,\infty)$, eliminating the fast coherences as in \cref{eq:app_strong_slaved_coherence} controls the state itself,
\begin{equation}
	\begin{aligned}
		\rho\!\left(t=\frac{\gamma\vartheta}{2J^2}\right)
		&=\rho_{\mathrm{diag}}^{(0)}(\vartheta)
		+\mathcal O\!\left(\frac{J}{\gamma}\right)
	\end{aligned}
	\label{eq:app_strong_state_zeno_control}
\end{equation}
in operator norm, where $\rho_{\mathrm{diag}}^{(0)}$ is the diagonal state built from the leading probabilities of \cref{eq:app_strong_zeno_markov}. Eigenvalues need one step more. A perturbation moves a nondegenerate eigenvalue by its own diagonal matrix element, which vanishes here, whereas it moves a degenerate level by the eigenvalues of the perturbation projected onto that level's eigenspace. The leading spectrum is degenerate, so the projected block must be examined.

The map that carries the probability vector forward by $\vartheta$ is $e^{\vartheta G}$, so that $p^{(0)}(\vartheta)=e^{\vartheta G}p^{(0)}(0)$ and two successive steps compose into one, $e^{\vartheta G}e^{\vartheta'G}=e^{(\vartheta+\vartheta')G}$. This family is the Markov semigroup of the pair-flip kinetics. Commuting generators exponentiate one factor at a time, and inside a single factor $F_j^2=\mathbf 1$ collects the even powers of $\vartheta F_j$ into $\cosh\vartheta$ and the odd ones into $\sinh\vartheta$, so that $e^{\vartheta(F_j-\mathbf 1)}=e^{-\vartheta}(\cosh\vartheta\,\mathbf 1+\sinh\vartheta\,F_j)$. The semigroup therefore factorizes bond by bond:
\begin{equation}
	\begin{aligned}
		e^{\vartheta G}
		&=\prod_{j=1}^{L-1}
		\left(a_\vartheta\mathbf 1+b_\vartheta F_j\right),
		\\
		a_\vartheta&=\frac{1+e^{-2\vartheta}}{2},
		&b_\vartheta&=\frac{1-e^{-2\vartheta}}{2}.
	\end{aligned}
	\label{eq:app_strong_zeno_factorized_semigroup}
\end{equation}
Let $y_j\in\{0,1\}$ record whether bond $j$ is flipped in this product. After tracing $B$, the relevant variables are $\boldsymbol y_A=(y_1,\ldots,y_{L_A})$, including the cut bond $y_{L_A}$. Site $j$ of $A$ lies between bonds $j-1$ and $j$, so its occupation changes when exactly one of those two bonds is flipped. Setting $y_0=0$ for the absent bond to the left of site $1$, writing $\eta_{0,j}$ for the initial value at site $j$, and with $\oplus$ again addition modulo two,
\begin{equation}
	\begin{aligned}
		\eta_j\oplus\eta_{0,j}
		&=y_{j-1}\oplus y_j,
		\qquad1\leq j\leq L_A.
	\end{aligned}
	\label{eq:app_strong_zeno_edge_toggle_map}
\end{equation}
This map is one to one because it inverts to $y_j=\bigoplus_{i=1}^j(\eta_i\oplus\eta_{0,i})$, where the large symbol adds the first $j$ differences modulo two. For $L_A=3$ and $\boldsymbol\eta\oplus\boldsymbol\eta_0=(1,1,0)$, the toggles are $y_1=1$, $y_2=1\oplus1=0$, and $y_3=0$. Every bond variable strictly inside $B$ is summed over when $B$ is traced, and $a_\vartheta+b_\vartheta=1$, so those bonds contribute a factor of one. Therefore the reduced probability has the product form
\begin{equation}
	\begin{aligned}
		p_A^{(0)}(\boldsymbol\eta,\vartheta)
		&=a_\vartheta^{L_A-|\boldsymbol y_A|}
		b_\vartheta^{|\boldsymbol y_A|}.
	\end{aligned}
	\label{eq:app_strong_zeno_reduced_product_probability}
\end{equation}
With $b_\vartheta=x$ and $a_\vartheta=1-x$ this is $p_A^{(0)}$ of \cref{eq:strong_zeno_bond_flips}, and the same bookkeeping over all $L-1$ bonds, whose flip record $\boldsymbol y$ is fixed uniquely by $\boldsymbol n$, gives $p^{(0)}$. For every finite $\vartheta>0$, $0<b_\vartheta<a_\vartheta<1$. In both distributions the two largest probabilities therefore belong to records that differ by one flipped bond, so $\lambda_0^{(0)}/\lambda_1^{(0)}=s_0^{(0)}/s_1^{(0)}=a_\vartheta/b_\vartheta=\coth\vartheta$ and both gaps equal $\ln\coth\vartheta$, as stated in \cref{eq:strong_zeno_entanglement_clock,eq:strong_zeno_negativity_clock}. The probability in \cref{eq:app_strong_zeno_reduced_product_probability} depends on $\boldsymbol y_A$ only through its Hamming weight $|\boldsymbol y_A|$, the number of flipped bonds, so two leading eigenvalues are equal exactly when their toggle strings carry the same weight. Write $w$ for that shared weight and let $\Pi_w$ project onto the span of all eigenvectors with $|\boldsymbol y_A|=w$, in the projector notation of \cref{eq:app_weak_neg_secular_blocks}.

Write the reduced-state expansion on a compact $K\subset(0,\infty)$ as
\begin{equation}
	\begin{aligned}
		\rho_A
		&=\rho_A^{(0)}+\varepsilon\rho_A^{(1)}
		+\mathcal O(\varepsilon^2),
		\qquad \varepsilon=\frac{J}{\gamma}.
	\end{aligned}
	\label{eq:app_strong_zeno_reduced_expansion}
\end{equation}
Populations are even in $J$, so $\rho_A^{(1)}$ has no diagonal part. A one-bond coherence survives the trace over $B$ only when its bond lies strictly inside $A$. Such a coherence joins $\boldsymbol\eta$ to $F_j\boldsymbol\eta$ with $j<L_A$. Under the one-to-one map of \cref{eq:app_strong_zeno_edge_toggle_map}, $F_j$ toggles the single variable $y_j$. It changes $|\boldsymbol y_A|$ by exactly one, and hence joins two different eigenvalues in \cref{eq:app_strong_zeno_reduced_product_probability}. A cut-bond coherence flips a $B$ occupation and vanishes under the trace. It follows that
\begin{equation}
	\begin{aligned}
		\Pi_w\rho_A^{(1)}\Pi_w&=0,
		\qquad0\leq w\leq L_A.
	\end{aligned}
	\label{eq:app_strong_zeno_degenerate_block_lemma}
\end{equation}
This is the required statement $\Pi_\lambda\rho_A^{(1)}\Pi_\lambda=0$ for every degenerate eigenspace of $\rho_A^{(0)}$. Degenerate perturbation theory now gives a vanishing first derivative for every eigenvalue \cite{Kato1995,StewartSun1990}. On a compact $K$, distinct weight sectors in \cref{eq:app_strong_zeno_reduced_product_probability} have a uniform positive separation, so the remainder is uniform. Hence
\begin{equation}
	\begin{aligned}
		\lambda_r
		\!\left(\frac{\gamma\vartheta}{2J^2}\right)
		&=\lambda_r^{(0)}(\vartheta)
		+\mathcal O\!\left(\frac{J^2}{\gamma^2}\right),
	\end{aligned}
	\label{eq:app_strong_schmidt_zeno_control}
\end{equation}
which proves \cref{eq:strong_zeno_entanglement_control}.

Partial time reversal is a bounded linear map in fixed dimension. Let $s_r(X)$ denote the singular values of $X$ in decreasing order, as in \cref{eq:twisted_spectrum}. For any two matrices of the same size they obey the standard perturbation bound \cite{StewartSun1990}
\begin{equation}
	\begin{aligned}
		|s_r(X)-s_r(Y)|
		&\leq\|X-Y\|_2,
	\end{aligned}
	\label{eq:app_strong_singular_lipschitz}
\end{equation}
where $\|X\|_2$ is the operator norm, the largest singular value of $X$, and the dual of the trace norm in \cref{eq:logarithmic_negativity}. Applied to \cref{eq:app_strong_state_zeno_control}, with $s_r^{(0)}$ the singular values of the leading diagonal state supplied by \cref{eq:app_strong_diag_ptr_singular_values}, this gives
\begin{equation}
	\begin{aligned}
		s_r
		\!\left(\frac{\gamma\vartheta}{2J^2}\right)
		&=s_r^{(0)}(\vartheta)
		+\mathcal O\!\left(\frac{J}{\gamma}\right),
	\end{aligned}
	\label{eq:app_strong_twisted_zeno_control}
\end{equation}
which proves \cref{eq:strong_zeno_negativity_control}. Since the leading gaps are continuous and strictly positive on any compact $K$, they have a positive minimum there. For sufficiently small $J/\gamma$, the exact gaps therefore cannot close anywhere on $K$. This statement leaves open a root on a scaled time $\vartheta(J/\gamma)$ that diverges as $J/\gamma\to0$.

\subsection{Universal fixed-Zeno-time negativity}

The negativity is carried by the single bond that crosses the cut, so we separate the flip on that bond from the rest. Let $F=F_{L_A}$ be the pair flip across the cut, as in \cref{eq:strong_cut_aggregate}, and split the generator $G$ of \cref{eq:app_strong_fourier_eigenvalue} as
\begin{equation}
	\begin{aligned}
		G
		&=G_{\mathrm{int}}+(F-\mathbf 1),
	\end{aligned}
	\label{eq:app_strong_G_split}
\end{equation}
where $G_{\mathrm{int}}$ collects the flips whose two sites both lie in $A$ or both lie in $B$. All pair flips are translations on the binary configuration space and commute, so the two pieces exponentiate independently. Let
\begin{equation}
	\begin{aligned}
		r(\vartheta)
		&=e^{\vartheta G_{\mathrm{int}}}\delta_{\boldsymbol n_0}
	\end{aligned}
	\label{eq:app_strong_r_distribution}
\end{equation}
be the distribution produced by the internal flips alone, starting from the point distribution $\delta_{\boldsymbol n_0}$ that places unit weight on $\boldsymbol n_0$ and zero elsewhere, exactly the initial condition of \cref{eq:app_strong_zeno_markov}. An internal flip moves two occupations on the same side of the cut, so it preserves the parity of $A$ and the parity of $B$ separately. Let $\mathcal R$ be the set of configurations sharing both subsystem parities with $\boldsymbol n_0$, as in \cref{eq:strong_cut_aggregate}. Then $r$ lives on $\mathcal R$. The cut flip moves one occupation on each side, so it reverses both subsystem parities and carries $\mathcal R$ onto the disjoint set $F\mathcal R$, where $Fr$ lives. Since $F^2=\mathbf 1$, the same resummation as in \cref{eq:app_strong_zeno_factorized_semigroup} gives
\begin{equation}
	\begin{aligned}
		e^{\vartheta(F-\mathbf 1)}
		&=\frac{1+e^{-2\vartheta}}{2}\mathbf 1
		+\frac{1-e^{-2\vartheta}}{2}F.
	\end{aligned}
	\label{eq:app_strong_cut_flip_exponential}
\end{equation}
Therefore
\begin{equation}
	\begin{aligned}
		p^{(0)}(\vartheta)
		&=\frac{1+e^{-2\vartheta}}{2}r(\vartheta)
		+\frac{1-e^{-2\vartheta}}{2}Fr(\vartheta).
	\end{aligned}
	\label{eq:app_strong_probability_split}
\end{equation}
Take $\boldsymbol n\in\mathcal R$. Then $r(F\boldsymbol n,\vartheta)=0$ and $(Fr)(\boldsymbol n,\vartheta)=0$, since those arguments lie in the other set, so one term survives on each side of the difference and the two weights subtract,
\begin{equation}
	\begin{aligned}
		p^{(0)}(\boldsymbol n,\vartheta)
		-p^{(0)}(F\boldsymbol n,\vartheta)
		&=e^{-2\vartheta}r(\boldsymbol n,\vartheta).
	\end{aligned}
	\label{eq:app_strong_cut_pair_difference}
\end{equation}
The distribution $r(\vartheta)$ carries total weight one on $\mathcal R$, so summing one representative from every cut-flip pair gives
\begin{equation}
	\begin{aligned}
		\sum_{\boldsymbol n\in\mathcal R}
		[p^{(0)}(\boldsymbol n,\vartheta)-p^{(0)}(F\boldsymbol n,\vartheta)]
		&=e^{-2\vartheta}.
	\end{aligned}
	\label{eq:app_strong_cut_pair_sum}
\end{equation}

Evaluating \cref{eq:app_strong_slaved_coherence} on the cut bond and writing $c_{\boldsymbol n}=c_{L_A}(\boldsymbol n)$ for it, the cut coherence for $\boldsymbol n\in\mathcal R$ is
\begin{equation}
	\begin{aligned}
		c_{\boldsymbol n}
		&=\rho_{\boldsymbol n,F\boldsymbol n}
		=-\frac{iJ}{\gamma}e^{-2\vartheta}r(\boldsymbol n,\vartheta)
		+\mathcal O\!\left(\frac{J^3}{\gamma^3}\right).
	\end{aligned}
	\label{eq:app_strong_cut_coherence_fixed_theta}
\end{equation}
Let $T_0$ be the twisted operator $T_A=\rho^{R_A}P_A$ of \cref{eq:twisted_operator} evaluated on the leading diagonal state. It is Hermitian, and by \cref{eq:app_strong_diag_ptr_singular_values} its eigenvectors are the occupation states, split into those carrying a nonzero eigenvalue and those in the kernel. Let $\Pi$ project onto the first group, which is the support, and let $\Pi^\perp=\mathbf 1-\Pi$ project onto the kernel. On the support, $\operatorname{sgn}(T_0)$ is the Hermitian operator returning $+1$ on each eigenvector with a positive eigenvalue and $-1$ on each eigenvector with a negative one, the signs being those of the twisted eigenvalues in \cref{eq:twisted_spectrum}. For a Hermitian perturbation $\delta T=\mathcal O(J/\gamma)$, the trace norm of \cref{eq:logarithmic_negativity} expands as
\begin{equation}
	\begin{aligned}
		\|T_0+\delta T\|_1
		={}&\|T_0\|_1
		+\operatorname{Tr}[\operatorname{sgn}(T_0)\Pi\delta T\Pi]
		\\
		&+\|\Pi^\perp\delta T\Pi^\perp\|_1
		+\mathcal O(J^2/\gamma^2).
	\end{aligned}
	\label{eq:app_strong_trace_norm_expansion}
\end{equation}
Internal bond coherences lie inside the support block and are off diagonal in the occupation basis, so their trace term vanishes. A cut coherence joins $\boldsymbol n$ to $F\boldsymbol n$, and partial time reversal moves that pair into the kernel block. There each pair contributes the $2\times2$ Hermitian matrix with zeros on the diagonal and $c_{\boldsymbol n}$ off it, whose eigenvalues are $\pm|c_{\boldsymbol n}|$, so it adds $2|c_{\boldsymbol n}|$ to the trace norm. Therefore
\begin{equation}
	\begin{aligned}
		\|\rho^{R_A}\|_1
		&=1+2\sum_{\boldsymbol n\in\mathcal R}|c_{\boldsymbol n}|
		+\mathcal O(J^2/\gamma^2)
		\\
		&=1+\frac{2J}{\gamma}e^{-2\vartheta}
		+\mathcal O(J^2/\gamma^2),
	\end{aligned}
	\label{eq:app_strong_fixed_theta_trace_norm}
\end{equation}
where \cref{eq:app_strong_cut_pair_sum} was used in the second line. Finally, $\ln(1+z)=z+\mathcal O(z^2)$ gives
\begin{equation}
	\begin{aligned}
		E_N
		\!\left(t=\frac{\gamma\vartheta}{2J^2}\right)
		&=\frac{2J}{\gamma}e^{-2\vartheta}
		+\mathcal O\!\left(\frac{J^2}{\gamma^2}\right),
	\end{aligned}
	\label{eq:app_strong_fixed_theta_negativity}
\end{equation}
which proves \cref{eq:strong_zeno_negativity}.

\section{Geometric Decoherence Time under Strong Symmetry}
\label{app:strong_geometric_decoherence}

This appendix proves three statements used in the strong-symmetry section. The first is that the geometric decoherence time $\tau_d^g$ of \cref{eq:app_geometric_definition}, the first moment at which the negativity turns down while the R\'enyi-$1/2$ entropy $S_{1/2}=2\ln\operatorname{Tr}\sqrt{\rho_A}$ still rises, is finite at every nonzero dephasing rate. The second is that its weak-rate displacement is an exact trace-norm susceptibility, meaning the derivative of the trace norm $\|\rho^{R_A}\|_1$ of \cref{eq:logarithmic_negativity} with respect to the rate, evaluated on finite matrices. The third is that at strong dephasing $\tau_d^g$ falls inside the coherence boundary layer, the early window, lasting of order $\ln(\gamma/J)$ times the fast time $1/\gamma$, in which the cut coherence builds up, before the slow Zeno regime of \cref{app:strong_zeno_reduction} sets in.

\subsection{Existence at nonzero dephasing}

The definition in \cref{eq:app_geometric_definition} asks for one pair of times at which the negativity has fallen while the entropy has risen, so exhibiting such a pair proves existence. At $t=0$, the state is a product state, so $E_N(0)=S_{1/2}(0)=0$. The cut Hamiltonian connects the initial configuration $|\boldsymbol n_0\rangle$ to the pair-flipped configuration $F_{L_A}|\boldsymbol n_0\rangle$ with amplitude $iJt+\mathcal O(t^2)$. Density dephasing first modifies this amplitude at order $\gamma Jt^2$. After partial time reversal, the cut coherence produces a positive-negative twisted pair, so
\begin{equation}
	\begin{aligned}
		\|\rho^{R_A}(t)\|_1
		&=1+2Jt+\mathcal O(t^2),
		\\
		E_N(t)
		&=2Jt+\mathcal O(t^2)>0
	\end{aligned}
	\label{eq:app_strong_geo_short_time}
\end{equation}
for sufficiently small $t>0$.

By \cref{eq:app_strong_stationary_global,eq:app_strong_stationary_reduced}, the long-time state in the selected global parity sector is diagonal and separable, while the reduced state is maximally mixed:
\begin{equation}
	\begin{aligned}
		E_N(\infty)&=0,
		&S_{1/2}(\infty)&=L_A\ln2.
	\end{aligned}
	\label{eq:app_strong_geo_stationary_values}
\end{equation}
Choose a sufficiently small $t_1>0$ for which $E_N(t_1)>0$ and $S_{1/2}(t_1)<L_A\ln2$. Convergence to the stationary state gives a finite $t_2>t_1$ such that
\begin{equation}
	\begin{aligned}
		E_N(t_2)&<E_N(t_1),
		&S_{1/2}(t_2)&\geq S_{1/2}(t_1).
	\end{aligned}
	\label{eq:app_strong_geo_exist_pair}
\end{equation}
Thus the set in \cref{eq:app_geometric_definition} is nonempty and
\begin{equation}
	\begin{aligned}
		0<\tau_d^g<\infty
	\end{aligned}
	\label{eq:app_strong_geo_exists}
\end{equation}
for every finite chain, nontrivial cut, $J\neq0$, and $\gamma>0$. This proves \cref{eq:strong_geometric_existence}.

\subsection{Exact weak-rate susceptibility}

Weak dephasing displaces the maximum of the negativity, and this subsection computes that displacement exactly. Measure time and rate against the coherent scale, writing $x=Jt$ for the dimensionless time and taking the dimensionless rate $\gamma/J$ small. Since the twisted operator $T_A$ of \cref{eq:twisted_operator} is linear in the state, it inherits the expansion of the state in that rate,
\begin{equation}
	\begin{aligned}
		T_A(x,\gamma/J)
		&=T_0(x)+\frac{\gamma}{J}T_1(x)+\mathcal O\!\left(\frac{\gamma^2}{J^2}\right),
	\end{aligned}
	\label{eq:app_strong_geo_T_expansion}
\end{equation}
where $T_0$ is the closed-system twisted operator and $T_1$ collects the first response to the bath. The one-jump expansion of \cref{app:strong_entanglement_asymptotics}, in which a second jump costs another power of $\gamma$, gives
\begin{equation}
	\begin{aligned}
		T_1(x)
		&=\frac{1}{4}\sum_{j=1}^{L}\int_0^xdu\,
		[T_{j,u}(x)-T_0(x)],
	\end{aligned}
	\label{eq:app_strong_geo_T1}
\end{equation}
where $T_{j,u}$ is the twisted pure-trajectory operator with one $Z_j$ jump a dimensionless time $u$ before observation.

The negativity is a trace norm, so the susceptibility named above is the response of that norm to $T_1$. Fix $x$, let $\Pi$ project onto the support of $T_0(x)$, the span of its eigenvectors with nonzero eigenvalue, and let $\Pi^\perp=\mathbf 1-\Pi$ project onto its kernel, exactly the splitting used in \cref{eq:app_strong_trace_norm_expansion}. On the support, $\operatorname{sgn}(T_0)$ is again the operator returning $+1$ on each eigenvector with a positive eigenvalue and $-1$ on each eigenvector with a negative one. A nonzero eigenvalue of the Hermitian $T_0$ moves by an amount given by ordinary first-order perturbation theory, and since the trace norm adds absolute values, that eigenvalue changes the norm by its own sign times its shift. An eigenvalue leaving the kernel starts from zero, so its absolute value grows linearly and the kernel block contributes its own trace norm. Cross terms between support and kernel enter only at second order. Write $D\|T_0\|_1[T_1]$ for the resulting directional derivative, the initial rate of growth of $\|T_0+(\gamma/J)T_1\|_1$ as the rate increases from zero. Then \cite{Watson1992Jun,Lewis1995,StewartSun1990}
\begin{equation}
	\begin{aligned}
		D\|T_0\|_1[T_1]
		&=\operatorname{Tr}[\operatorname{sgn}(T_0)\Pi T_1\Pi]
		+\|\Pi^\perp T_1\Pi^\perp\|_1,
	\end{aligned}
	\label{eq:app_strong_geo_directional_derivative}
\end{equation}
which is the derivative form of the expansion in \cref{eq:app_strong_trace_norm_expansion}. Measuring this response relative to the norm itself defines
\begin{equation}
	\begin{aligned}
		e_1(x)
		&=\frac{D\|T_0(x)\|_1[T_1(x)]}{\|T_0(x)\|_1},
	\end{aligned}
	\label{eq:app_strong_geo_e1}
\end{equation}
a function of the dimensionless time alone, the subscript recording that it is the first-order coefficient. Since $\|T_A\|_1=\|T_0\|_1[1+(\gamma/J)e_1(x)+o(\gamma/J)]$ and $E_N=\ln\|T_A\|_1$, the logarithm separates the two orders,
\begin{equation}
	\begin{aligned}
		E_N(x,\gamma/J)
		&=E_N^{(0)}(x)+\frac{\gamma}{J}e_1(x)+o\!\left(\frac{\gamma}{J}\right),
	\end{aligned}
	\label{eq:app_strong_geo_EN_split}
\end{equation}
where the superscript $(0)$ marks the closed-system value, the term that survives at $\gamma=0$. On the first branch $0<x<\pi/2$ the sine in \cref{eq:closed_compact_results} is positive, so its absolute value may be dropped and
\begin{equation}
	\begin{aligned}
		E_N^{(0)}(x)
		&=\ln[1+\sin(2x)],
		&E_N^{(0)\prime\prime}(\pi/4)&=-2.
	\end{aligned}
	\label{eq:app_strong_geo_closed_dimensionless}
\end{equation}
The closed maximum sits at $x=\pi/4$, where the first derivative vanishes and the second equals $-2$. Write the displaced maximum as $x_g=\pi/4+\delta x$ with $\delta x=\mathcal O(\gamma/J)$, and differentiate \cref{eq:app_strong_geo_EN_split} at that point. The closed part contributes its second derivative times $\delta x$ and the correction contributes its own slope, so the stationarity condition reads
\begin{equation}
	\begin{aligned}
		0
		&=-2\delta x+\frac{\gamma}{J}e_1'(\pi/4)+o\!\left(\frac{\gamma}{J}\right),
	\end{aligned}
	\label{eq:app_strong_geo_max_expansion}
\end{equation}
where the prime is the derivative with respect to $x$. Hence
\begin{equation}
	\begin{aligned}
		\delta x
		&=\frac{\gamma}{2J}e_1'(\pi/4)+o\!\left(\frac{\gamma}{J}\right).
	\end{aligned}
	\label{eq:app_strong_geo_delta_x}
\end{equation}
Returning to physical time yields
\begin{equation}
	\begin{aligned}
		\tau_d^g
		&=t_0-\mathcal C_{L_A,L_B}^{\mathrm g}\frac{\gamma}{J^2}
		+o(\gamma/J^2),
		\\
		\mathcal C_{L_A,L_B}^{\mathrm g}
		&=-\frac{1}{2}e_1'(\pi/4).
	\end{aligned}
	\label{eq:app_strong_geo_weak_susceptibility}
\end{equation}
This is the exact finite-dimensional susceptibility quoted in \cref{eq:strong_weak_geometric_clock}. The coefficients in \cref{eq:strong_weak_entanglement_clock,eq:strong_weak_negativity_clock} are built from shifts of individual levels and depend only on the two cut endpoints. The kernel term in \cref{eq:app_strong_geo_directional_derivative} adds the levels that grow out of zero, which sample the whole finite geometry, and that is why $\mathcal C_{L_A,L_B}^{\mathrm g}$ carries more than endpoint data.

Evaluating the finite matrices in \cref{eq:app_strong_geo_weak_susceptibility} gives the coefficients collected in \cref{tab:app_strong_geo_coefficients}.
\begin{table}[t]
	\caption{Weak-rate geometric coefficient $\mathcal C_{L_A,L_B}^{\mathrm g}$ of \cref{eq:app_strong_geo_weak_susceptibility} under strong symmetry for every geometry with $L\leq10$, with the geometries that share a value grouped on one row.}
	\label{tab:app_strong_geo_coefficients}
	\centering
	\small
	\begin{tabular}{@{}lc@{}}
		\toprule
		Geometry&$\mathcal C_{L_A,L_B}^{\mathrm g}$\\
		\midrule
		$(1,1),(1,2),(2,1),(2,2)$&$0.1250$\\
		$(1,3),(3,1),(2,3),(3,2)$&$0.0759$\\
		$\min(L_A,L_B)\leq2$, $\max(L_A,L_B)\geq4$&$0.0742$\\
		$(3,3)$&$0.0268$\\
		$(3,4),(4,3)$&$0.0252$\\
		$\min(L_A,L_B)=3$, $\max(L_A,L_B)\geq5$&$0.0251$\\
		$(4,4)$&$0.0235$\\
		$(4,5),(5,4),(4,6),(6,4),(5,5)$&$0.0234$\\
		\bottomrule
	\end{tabular}
\end{table}
Every geometry listed there has a positive coefficient, so $\tau_d^g$ lies below $t_0$ at sufficiently weak positive $\gamma$, which is the advance seen in \cref{eq:strong_weak_geometric_clock}. The values saturate once both sides have at least four sites, although positivity at every size remains an open question.

\subsection{Exact aggregate variables in the strong-dephasing boundary layer}

The fixed-$\vartheta$ expansion of \cref{app:strong_zeno_reduction} describes times at which the fast transient $e^{-\gamma t}$ of \cref{eq:app_strong_slaved_coherence} has already died out, being smaller than any power of $J/\gamma$ at fixed $\vartheta>0$. The geometric decoherence time arrives before that, inside the coherence boundary layer, the window
\begin{equation}
	\begin{aligned}
		s&=\gamma t=\mathcal O\!\left(\ln\frac{\gamma}{J}\right),
		&\vartheta&=2\left(\frac{J}{\gamma}\right)^2s\to0.
	\end{aligned}
	\label{eq:app_strong_geo_boundary_window}
\end{equation}
In this window the cut coherence is still building on the fast scale $1/\gamma$, while the population difference that feeds it has only begun to move. Both are therefore kept, as one exact pair of coupled equations.

Let $F=F_{L_A}$ be the pair flip across the cut of \cref{eq:app_strong_pair_flip}, and let $\mathcal R$ collect the configurations whose $A$ parity and whose $B$ parity both match those of the initial state, as in \cref{eq:strong_cut_aggregate}. With the populations $p(\boldsymbol n,t)$ and the coherences $\rho_{\boldsymbol n,F\boldsymbol n}(t)$ of \cref{eq:app_strong_population_coherence}, define the two aggregates
\begin{equation}
	\begin{aligned}
		D(t)
		&=\sum_{\boldsymbol n\in\mathcal R}
		[p(\boldsymbol n,t)-p(F\boldsymbol n,t)],
		\\
		C(t)
		&=\sum_{\boldsymbol n\in\mathcal R}
		\rho_{\boldsymbol n,F\boldsymbol n}(t).
	\end{aligned}
	\label{eq:app_strong_geo_aggregate_definitions}
\end{equation}
Here $D$ is the population still held on the initial side of the cut flip minus the population carried across, equal to one at $t=0$ and falling to zero once the two sides are equally filled, while $C$ is the total coherence across the cut. Summing over $\mathcal R$ collapses the whole chain onto these two numbers. Consider first the exact population equations summed over $\mathcal R$. Every bond internal to $A$ or $B$ maps $\mathcal R$ onto itself, so its probability currents cancel pairwise. Only the cut bond transfers probability between $\mathcal R$ and $F\mathcal R$, giving
\begin{equation}
	\begin{aligned}
		\dot D&=4J\operatorname{Im}C.
	\end{aligned}
	\label{eq:app_strong_geo_D_equation}
\end{equation}
Now sum the exact equations for the cut coherences, the matrix elements joining a configuration to its cut-flipped partner. Terms from a bond $k$ inside $A$ or $B$ cancel after relabeling $\boldsymbol n\mapsto F_k\boldsymbol n$ with the pair flip of \cref{eq:app_strong_pair_flip}, since that relabeling permutes $\mathcal R$. Dephasing damps each cut coherence at rate $\gamma$, as in \cref{eq:app_strong_bond_coherence_exact}, and the population difference $D$ is its source, giving
\begin{equation}
	\begin{aligned}
		\dot C&=-\gamma C-iJD.
	\end{aligned}
	\label{eq:app_strong_geo_C_equation}
\end{equation}
For the product initial state, $D(0)=1$ and $C(0)=0$. Equations \eqref{eq:app_strong_geo_D_equation} and \eqref{eq:app_strong_geo_C_equation} are exactly \cref{eq:strong_cut_aggregate_dynamics}.

Eliminating $C$ gives
\begin{equation}
	\begin{aligned}
		\ddot D+\gamma\dot D+4J^2D&=0.
	\end{aligned}
	\label{eq:app_strong_geo_D_oscillator}
\end{equation}
For $\gamma>4J$, use the overdamped rate $\bar\omega=\sqrt{\gamma^2/4-4J^2}$ of \cref{eq:strong_overdamped_frequency}. The initial-value solution is
\begin{equation}
	\begin{aligned}
		D(t)
		&=e^{-\gamma t/2}
		\left[\cosh(\bar\omega t)+\frac{\gamma}{2\bar\omega}\sinh(\bar\omega t)\right],
		\\
		C(t)
		&=-\frac{iJ}{\bar\omega}e^{-\gamma t/2}\sinh(\bar\omega t),
	\end{aligned}
	\label{eq:app_strong_geo_aggregate_solution}
\end{equation}
which proves \cref{eq:strong_cut_aggregate_solution}.

Each cut coherence is purely imaginary with one common sign, by \cref{eq:app_strong_cut_coherence_fixed_theta}, so the magnitudes add without cancellation and $\sum_{\boldsymbol n\in\mathcal R}|\rho_{\boldsymbol n,F\boldsymbol n}|=|C|$. By \cref{eq:app_strong_trace_norm_expansion} each such coherence sits in the kernel block after partial time reversal and contributes twice its magnitude to the trace norm, so the total is $2|C|$,
\begin{equation}
	\begin{aligned}
		E_N(t)
		&=2|C(t)|+\mathcal O(J^2/\gamma^2)
		\\
		&=\frac{2J}{\bar\omega}e^{-\gamma t/2}\sinh(\bar\omega t)
		+\mathcal O\!\left(\frac{J^2}{\gamma^2}\right),
	\end{aligned}
	\label{eq:app_strong_geo_boundary_negativity}
\end{equation}
which is \cref{eq:strong_boundary_negativity}. Equivalently,
\begin{equation}
	\begin{aligned}
		2|C(t)|
		&=\frac{J}{\bar\omega}
		\left[e^{-(\gamma/2-\bar\omega)t}-e^{-(\gamma/2+\bar\omega)t}\right].
	\end{aligned}
	\label{eq:app_strong_geo_two_exponentials}
\end{equation}
The two exponentials carry the two rates of \cref{eq:strong_cut_aggregate_dynamics}: the second is the fast buildup on the scale $1/\gamma$, and the first decays slowly, at the rate
\begin{equation}
	\begin{aligned}
		\frac{\gamma}{2}-\bar\omega
		&=\frac{4J^2}{\gamma}
		+\mathcal O\!\left(\frac{J^4}{\gamma^3}\right),
	\end{aligned}
	\label{eq:app_strong_geo_slow_rate}
\end{equation}
so the first exponential is the slow Zeno decay and becomes $e^{-2\vartheta}$ at fixed $\vartheta$.

\subsection{Location of the boundary-layer maximum}

The negativity rises while the coherence builds and falls once dephasing wins, so its first maximum is where the leading $|C(t)|$ of \cref{eq:app_strong_geo_boundary_negativity} is stationary. Setting the derivative of $e^{-\gamma t/2}\sinh(\bar\omega t)$ to zero and dividing by that same product, the stationary time $t_C$ satisfies
\begin{equation}
	\begin{aligned}
		-\frac{\gamma}{2}+\bar\omega\coth(\bar\omega t_C)&=0.
	\end{aligned}
	\label{eq:app_strong_geo_stationary_condition}
\end{equation}
Thus
\begin{equation}
	\begin{aligned}
		t_C
		&=\frac{1}{\bar\omega}\operatorname{artanh}\!\left(\frac{2\bar\omega}{\gamma}\right).
	\end{aligned}
	\label{eq:app_strong_geo_exact_tC}
\end{equation}
For $\gamma/J\to\infty$,
\begin{equation}
	\begin{aligned}
		t_C
		&=\frac{2}{\gamma}\ln\!\left(\frac{\gamma}{2J}\right)[1+o(1)].
	\end{aligned}
	\label{eq:app_strong_geo_asymptotic_tC}
\end{equation}
At this time, the slow exponent satisfies $(\gamma/2-\bar\omega)t_C=\mathcal O[(J/\gamma)^2\ln(\gamma/J)]\to0$, while the fast exponential is negligible. Therefore
\begin{equation}
	\begin{aligned}
		2|C(t_C)|
		&=\frac{2J}{\gamma}[1+o(1)].
	\end{aligned}
	\label{eq:app_strong_geo_peak_height}
\end{equation}

It remains to check the entropy condition in \cref{eq:app_geometric_definition}, namely that $S_{1/2}$ is still rising where the negativity turns. Since $D$ is the difference of the two populations and they sum to one, the probability carried into the opposite subsystem-parity set $F\mathcal R$ is
\begin{equation}
	\begin{aligned}
		q_{\mathrm{cut}}(t)
		&=\frac{1-D(t)}{2}.
	\end{aligned}
	\label{eq:app_strong_geo_qcut}
\end{equation}
In the boundary window at $t=t_C$,
\begin{equation}
	\begin{aligned}
		q_{\mathrm{cut}}(t_C)
		&=\mathcal O\!\left[
		\left(\frac{J}{\gamma}\right)^2
		\ln\!\left(\frac{\gamma}{J}\right)
		\right],
		&\dot q_{\mathrm{cut}}(t_C)&>0.
	\end{aligned}
	\label{eq:app_strong_geo_qcut_scale}
\end{equation}
Internal pair flips in $A$ add further small nonnegative eigenvalues to $\rho_A$, while flips internal to $B$ disappear under the partial trace. One reduced eigenvalue is therefore $1-\mathcal O[(J/\gamma)^2\ln(\gamma/J)]$ and the small ones are still growing, so the reduced state is moving away from a pure state and $S_{1/2}$ is increasing in a neighborhood of $t_C$. The two leading reduced eigenvalues of \cref{eq:spectral_gaps} meanwhile stay far apart,
\begin{equation}
	\begin{aligned}
		\lambda_0(t_C)-\lambda_1(t_C)
		&=1-\mathcal O\!\left[
		\left(\frac{J}{\gamma}\right)^2
		\ln\!\left(\frac{\gamma}{J}\right)
		\right]>0.
	\end{aligned}
	\label{eq:app_strong_geo_schmidt_gap_at_peak}
\end{equation}
Both conditions of \cref{eq:app_geometric_definition} are therefore met at the first negativity maximum, which is the geometric decoherence time to leading order. Substituting \cref{eq:app_strong_geo_asymptotic_tC,eq:app_strong_geo_peak_height},
\begin{equation}
	\begin{aligned}
		\tau_d^g
		&=\frac{2}{\gamma}\ln\!\left(\frac{\gamma}{2J}\right)[1+o(1)],
		\\
		E_N(\tau_d^g)
		&=\frac{2J}{\gamma}[1+o(1)],
	\end{aligned}
	\label{eq:app_strong_geo_final_asymptotics}
\end{equation}
which proves \cref{eq:strong_geometric_zeno,eq:strong_geometric_entanglement_gap}. Measuring this time in Zeno units by multiplying by $2J^2/\gamma$, as in \cref{eq:strong_zeno_scales}, gives
\begin{equation}
	\begin{aligned}
		\vartheta_g
		&\sim\frac{4J^2}{\gamma^2}
		\ln\!\left(\frac{\gamma}{2J}\right)\to0,
	\end{aligned}
	\label{eq:app_strong_geo_theta_g}
\end{equation}
where $\vartheta_g$ denotes the Zeno time at the geometric decoherence time. This proves \cref{eq:strong_geometric_zeno_time}. The geometric clock therefore sits above the fast time $t_{\mathrm f}=1/\gamma$ by a logarithm and far below the slow time $t_{\mathrm s}=\gamma/(2J^2)$, which is the scale hierarchy in \cref{eq:strong_time_scale_hierarchy}.

\section{Parameter Sweeps at Finite Chemical Potential}
\label{app:parameter_sweeps}

\begin{figure*}[t]
	\centering
	\includegraphics[width=0.49\textwidth]{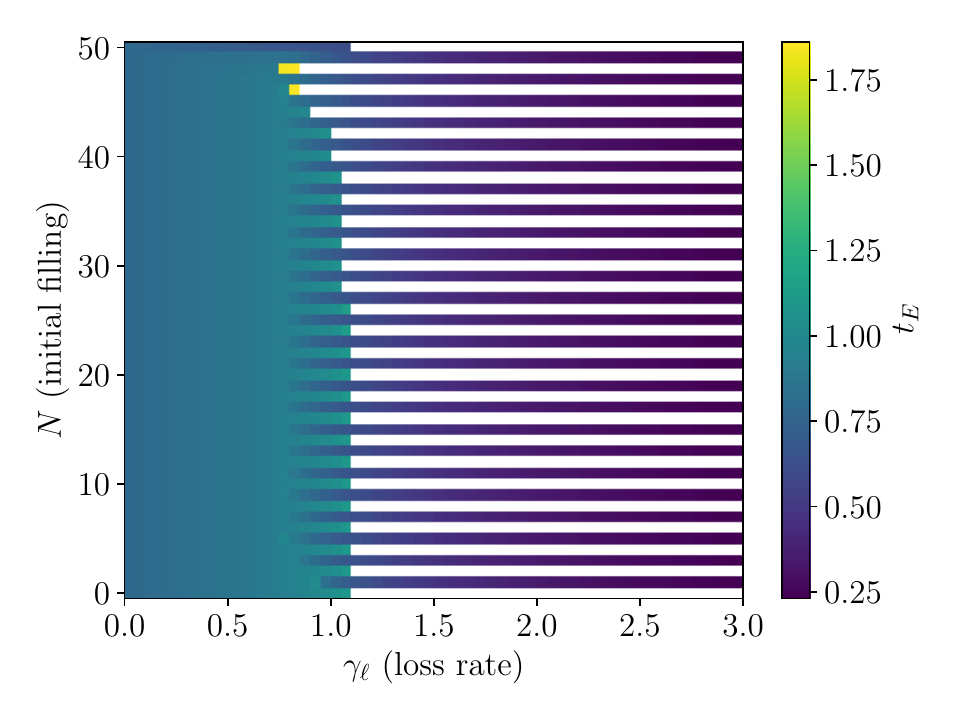}\hfill
	\includegraphics[width=0.49\textwidth]{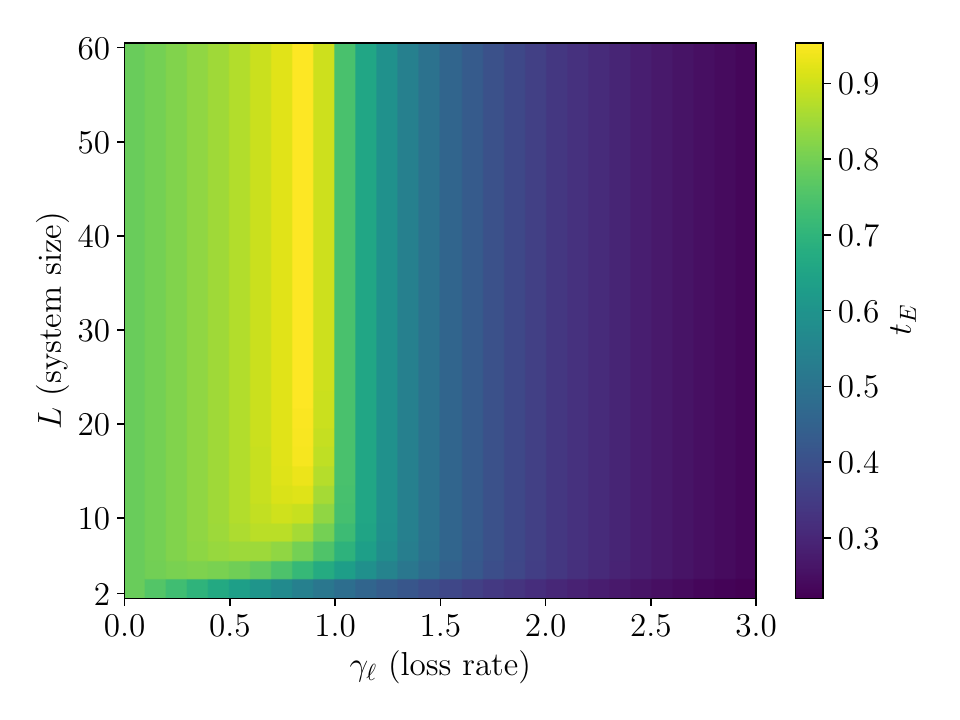}
	\caption{Entanglement clock under pure loss at $\mu=0.1J$ and $\Delta=J$, from exact covariance evolution, \cref{eq:weak_covariance_solution}, with $\gamma_g=0$. Color: first closing time $t_E$ in the search window, on an independent scale in each panel; white: no closing in the window. \textbf{Left}: $L=100$, $L_A=50$, fillings $0\leq N\leq50$, rate step $0.05J$, and window $0<Jt\leq10$. At weak rates every filling follows \cref{eq:weak_small_mu_clock}, with $t_E$ rising from $t_0=\pi/(4J)$ by $\gamma_\ell/(8J^2)$, except $N=L_A$, whose initially filled site $L_A$ reverses the sign of the shift (\cref{eq:weak_rate_entanglement_clock}). Above $\gamma_\ell\approx1.1J$ only odd fillings close. The bright points near $\gamma_\ell\approx0.8J$, at $N=46$ and $48$, mark jumps of $t_E$ to a later closing, at about $1.85/J$. \textbf{Right}: $2\leq L\leq60$, $L_A=\lfloor L/2\rfloor$, $N=1$, rate step $0.1J$, and window $0<Jt\leq2$. The single filled site is an odd active block for every $L$ (\cref{eq:weak_entanglement_active_length}), and a closing is found at every rate. The clock peaks near $\gamma_\ell\approx0.85J$ and approaches $\ln2/\gamma_\ell$ (\cref{eq:weak_strong_rate_entanglement_clock}) to within a few percent for $\gamma_\ell\gtrsim2J$, except at the smallest $L$; the rows with $L\gtrsim20$ coincide.}
	\label{fig:app_weak_sweeps}
\end{figure*}

\begin{figure*}[t]
	\centering
	\includegraphics[width=0.49\textwidth]{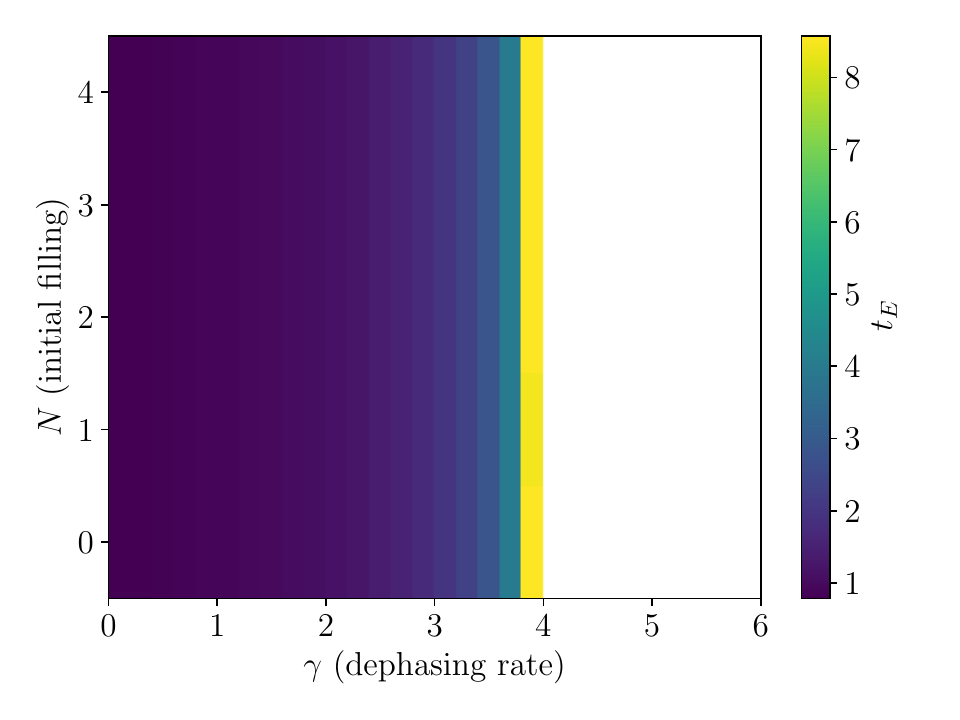}\hfill
	\includegraphics[width=0.49\textwidth]{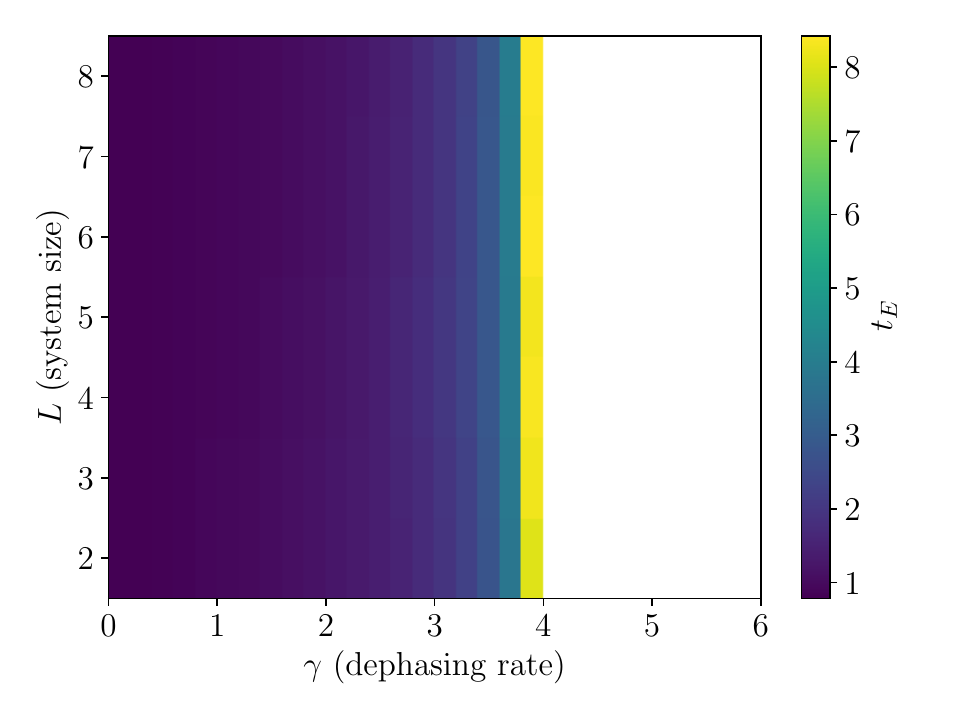}
	\caption{Entanglement clock under density dephasing at $\mu=0.1J$ and $\Delta=J$, from exact evolution of \cref{eq:strong_lindblad} with rate step $0.2J$ and window $0<Jt\leq 15$. Color: first closing time $t_E$ in the window; white: no closing in the window. \textbf{Left}: $L=8$, $L_A=4$, and fillings $0\leq N\leq4$. \textbf{Right}: $2\leq L\leq8$, $L_A=\lfloor L/2\rfloor$, and $N=1$. In both panels $t_E$ starts near $t_0=\pi/(4J)$, grows with $\gamma$, and diverges near $\gamma=4J$, as at $\mu=0$ (\cref{fig:strong_symmetry_clocks}(a)). To the resolution of the color scale, the rows of the left panel coincide, and so do the rows with equal $L_A$ in the right panel ($L=2,3$; $4,5$; $6,7$), except $N=1$ and $L=2$ at the largest rate below $4J$. Thus $t_E$ depends only weakly on $L_A$ and hardly on the filling or on $L_B$.}
	\label{fig:app_strong_sweeps}
\end{figure*}

\begin{figure*}[t]
	\centering
	\includegraphics[width=0.49\textwidth]{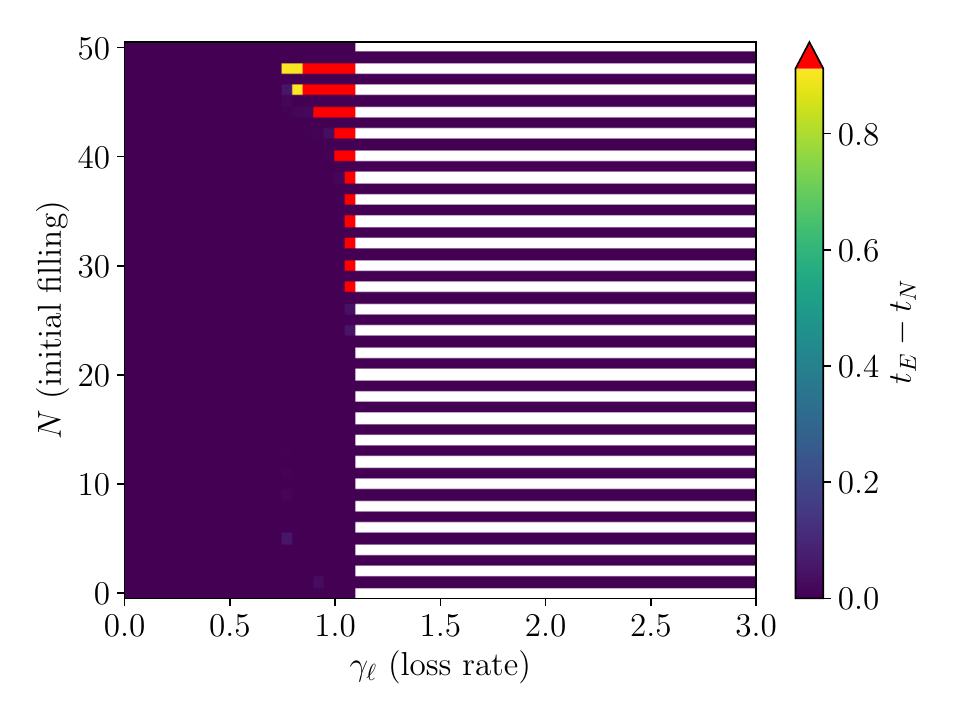}\hfill
	\includegraphics[width=0.49\textwidth]{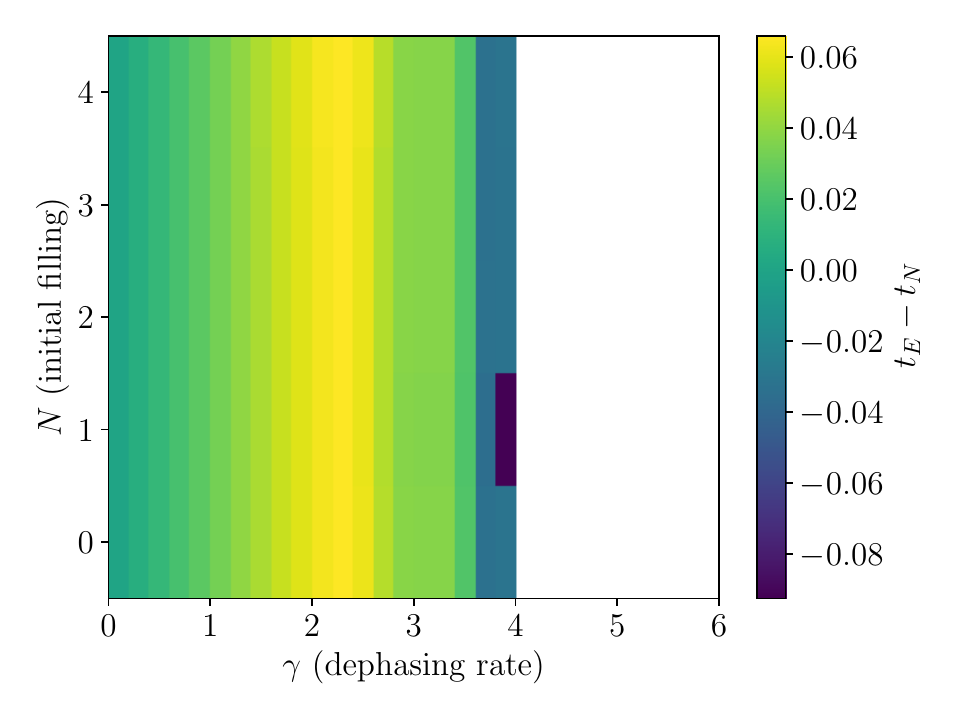}
	\caption{Difference $t_E-t_N$ of the two clocks of \cref{eq:spectral_clocks}, for pure loss with $L=100$ and $L_A=50$ (left) and for dephasing with $L=8$ and $L_A=4$ (right); the other parameters and the windows are those of the left panels of \cref{fig:app_weak_sweeps,fig:app_strong_sweeps}. White: neither clock closes in the window; red: only the negativity gap closes, recorded as $t_E-t_N=+\infty$. The two color scales are independent. \textbf{Left}: Under loss the filled sites of $A$ form the only active block of both spectra (\cref{eq:weak_negativity_active_clusters}), so $t_E=t_N$ over almost the whole sweep, and every nonzero difference is positive, as $t_N\leq t_E$ (\cref{eq:weak_negativity_parity_rule}) requires. The bright points are the jumps of $t_E$ in \cref{fig:app_weak_sweeps}, the red cells mark rates between the folds of $q_A$ and $q_B$ for even $28\leq N\leq48$, and a few faint cells, below $0.06/J$, lie between $\gamma_\ell\approx0.8J$ and $1.1J$. \textbf{Right}: The difference is the same for all fillings except $N=1$ at the largest rate below $4J$. It grows as $\gamma/(32J^2)$ (\cref{eq:strong_weak_clock_splitting}), peaks at about $0.065/J$ between $\gamma=2J$ and $2.5J$, and is negative, below the $0.00$ tick of the color bar, at the two largest rates below $4J$. No red appears.}
	\label{fig:app_clock_difference}
\end{figure*}

The figures of the main text are computed at the compact point, $\mu=0$ and $\Delta=J$. Here we test their main features at $\mu=0.1J$ and $\Delta=J$, where the chemical-potential term no longer commutes with the bond terms and the kick solution of \cref{sec:temporal_criticalities_strong} does not apply. All runs start from the left-filled states of \cref{eq:left_filled_initial_state}, with $A$ the first $L_A$ sites. Each clock of \cref{eq:spectral_clocks} is the first closing found in a fixed search window $0<t\leq t_{\max}$. The axes of \cref{fig:app_weak_sweeps,fig:app_strong_sweeps,fig:app_clock_difference} give rates in units of $J$ and times in units of $1/J$.

Under loss (\cref{fig:app_weak_sweeps}), the odd-even rule of \cref{eq:weak_strong_rate_entanglement_clock,eq:weak_finite_mu_even_mass} holds for all fillings: odd $N$ close at every sampled rate, whereas even $N$ stop closing between $\gamma_\ell\approx0.85J$ and $1.1J$. As at the folds of \cref{fig:weak_symmetry_clocks}(b), each even branch ends at a finite $t_E$, and for $N=46$ and $48$ the first closing jumps to a later one just before the end. The two clocks nearly coincide (\cref{fig:app_clock_difference}). They separate at the ends of the even branches with $28\leq N\leq48$, where the zero of $q_B$ (\cref{eq:subsystem_parities}) outlives that of $q_A$ and only the negativity gap closes, as for $N=2$ and $L_A=8$ in \cref{fig:weak_symmetry_clocks}(a).

Under dephasing (\cref{fig:app_strong_sweeps}), the first closing is independent of the filling and of $L_B$ to the resolution of the sweep, except for $N=1$ and $L=2$ next to $\gamma=4J$, so the universalities of \cref{eq:strong_pattern_universality,eq:strong_bath_length_universality}, proved at $\mu=0$, still hold approximately. The clock diverges near $\gamma=4J$, and $t_E-t_N$ changes sign just below $4J$ (\cref{fig:app_clock_difference}), as at $\mu=0$ in \cref{fig:strong_symmetry_clocks}(a).

